\documentclass[11pt,a4paper]{article}
\usepackage[top=2cm,left=1.6cm,footskip=1cm,marginparwidth=0cm]{geometry}

\usepackage[T1]{fontenc}

\usepackage[round]{natbib}

\raggedright
\usepackage{changepage}

\usepackage[aboveskip=1pt,labelfont=bf,labelsep=period,singlelinecheck=off]{caption}

\usepackage{graphicx}
\usepackage[graphicx]{realboxes}
\usepackage{capt-of}
\usepackage{subfig,caption}
\usepackage[flushleft]{threeparttable}

\usepackage{color,amsmath,amssymb,mathastext,xfrac}
\usepackage{lmodern,upgreek}
\usepackage{booktabs}
\usepackage{morefloats}
\usepackage{txfonts}
\usepackage{hyperref}
\usepackage{xcolor}
\usepackage{placeins}
\usepackage{url}
\usepackage{rotating}

\usepackage{lastpage,fancyhdr,graphicx}
\usepackage{csvsimple}
\usepackage{aas_macros}

\usepackage[none]{hyphenat}

\newcommand{\lya}{Ly$\upalpha$}

\begin{document}
\title{The Lyman Alpha Reference Sample XI:\\[0.1in]
	Efficient Turbulence Driven Ly$\upalpha$ Escape\\[0.1in]
	and the Analysis of IR, CO and [C~II]158$\upmu$m
	\thanks{Based on observations with {\it Herschel}, an ESA space observatory with science instruments provided by
             European-led Principal Investigator consortia and with important participation from NASA.
             Based on observations made with the NASA/DLR Stratospheric Observatory for Infrared Astronomy
             ({\it SOFIA}). SOFIA is jointly operated by the Universities Space Research Association, Inc. (USRA), under
             NASA contract NNA17BF53C, and the Deutsches SOFIA Institut (DSI) under DLR contract 50 OK 0901 to the
             University of Stuttgart.
             Based on observations carried out with the {\it IRAM 30m} Telescope. IRAM is supported by INSU/CNRS (France), MPG (Germany) and IGN (Spain).
             This publication is based on data acquired with the Atacama Pathfinder Experiment ({\it APEX}). APEX is a collaboration
             between the Max-Planck-Institut f\"ur Radioastronomie, the European Southern Observatory, and the Onsala Space Observatory.
             Based on observations made with the NASA/ESA {\it Hubble Space Telescope}, obtained at the Space Telescope Science Institute,
             which is operated by the Association of Universities for Research in Astronomy, Inc., under NASA contract NAS 5-26555.
             These observations are associated with program GO 12310.
	}
}
\date{Accepted for publication in A\&A, 22 April 2020}

\maketitle

\begin{flushleft}
J. Puschnig,$^{1,2,*}$
M. Hayes,$^{1}$
G. \"Ostlin,$^{1}$
J. Cannon,$^{3}$
I. Smirnova-Pinchukova,$^{4}$
B. Husemann,$^{4}$
D. Kunth,$^{5}$
J. Bridge,$^{6}$
E. C. Herenz,$^{7}$
M. Messa,$^{8}$
I. Oteo$^{9,10}$
\\
\bigskip
* e-mail: johannes.puschnig@uni-bonn.de\\
\bigskip
\small{
1 Department of Astronomy, University of Stockholm, AlbaNova, Stockholm\\
2 Argelander-Institut f\"ur Astronomie, Auf dem H\"ugel 71, 53121 Bonn, Germany\\
3 Department of Physics \& Astronomy, Macalester College, 1600 Grand Avenue, Saint Paul, MN 55105, USA\\
4 Max Planck Institute for Astronomy, Königstuhl 17, 69117 Heidelberg, Germany\\
5 Institut d’Astrophysique de Paris (IAP), 98bis Boulevard Arago, 75014 Paris, France\\
6 University of Louisville, Natural Science Bldg. 208, Louisville KY 40292, USA\\
7 European Southern Observatory, Av. Alonso de C\'ordova 3107, 763 0355 Vitacura, Santiago, Chile\\
8 Department of Astronomy, University of Massachusetts–Amherst, Amherst, MA 01003, USA\\
9 Institute for Astronomy, University of Edinburgh, Royal Observatory, Blackford Hill, Edinburgh EH9 3HJ, UK\\
10 European Southern Observatory, Karl-Schwarzschild-Str. 2, 85748 Garching, Germany\\
}
\bigskip

\end{flushleft}

\section*{Abstract}
Lyman-$\upalpha$ (\lya) is intrinsically the brightest emission line in star forming galaxies.
However, its interpretation in terms of physical properties is hampered by the resonant nature of \lya\ photons.
In order to remedy this complicated situation, 
the \textit{Lyman Alpha Reference Sample} (LARS) was defined,
allowing to study
\lya\ production and escape mechanisms in 14 local star-forming galaxies.   
With this paper, we complement
our efforts and study the global dust and (molecular) gas content as well as properties of gas associated with
photon-dominated regions (PDRs). We aim to characterize the interstellar medium of LARS galaxies,
allowing us to relate these newly derived properties to quantities relevant for \lya\ escape.
We observed LARS galaxies with Herschel/PACS, SOFIA/FIFI-LS,
the IRAM 30m telescope and APEX, targeting far-infrared (FIR) continuum
and emission lines of [C~II]158$\upmu$m, [O~I]63$\upmu$m, [O~III]88$\upmu$m
and low-J CO lines.
Using Bayesian methods we derive Draine \& Li (2007) model parameters
and estimate total gas masses for all LARS galaxies,
taking into account
a metallicity-dependent gas-to-dust ratio. Star formation rates were estimated
from FIR, [C~II]158$\upmu$m and [O~I]63$\upmu$m luminosities.
LARS covers a wide dynamic range in the derived properties, with FIR-based star formation rates
from $\sim$0.5--100 $M_{\odot}\ yr^{-1}$, gas fractions between $\sim$15--80\% and
gas depletion times ranging from a few hundred Myr up to more than 10~Gyr.
The distribution of LARS galaxies in the $\Sigma_{gas}$ vs. $\Sigma_{SFR}$ (Kennicutt-Schmidt plane)
is thus quite heterogeneous. However, we find that LARS galaxies with the longest gas depletion
times, i.e. relatively high gas surface densities ($\Sigma_{gas}$) and low star formation rate densities ($\Sigma_{SFR}$),
have by far the highest \lya\ escape fraction. A strong $\sim$linear relation is found between \lya\ escape fraction
and the total gas (HI+H$_2$) depletion time.
We argue that the \lya\ escape in those galaxies is driven by
turbulence in the star-forming gas that shifts the \lya\ photons out of resonance
close to the places where they originate.
We further report on an extreme [C~II]158$\upmu$m excess in LARS 5, corresponding to
$\sim$14$\pm$3\% of the
FIR luminosity, i.e. the most extreme [C~II]-to-FIR ratio observed in a non-AGN galaxy to date.



\section{Introduction}

\subsection{Challenges of Lyman Alpha Astrophysics}
Already more than fifty years ago, \cite{Partridge1967} suggested to
search for galaxies at high redshifts using a very narrow spectral window
around the (redshifted) Lyman Alpha (hereafter \lya) hydrogen emission line
at 1216\AA\ rest wavelength.
According to their prediction, bulk of the ionizing photons in young galaxies
are ultimately converted to a single line, \lya, making it the brightest spectral feature,
that is holding a substantial fraction of the bolometric flux.

In practice, however, it turned out to be very difficult to find \lya\ emitters and it
was not until 1981, when \cite{Meier1981} discovered weak \lya\ emission in a
local blue compact galaxy. It took another 17 years until \cite{Cowie1998} found a
substantial population of \lya\ emitters at z$\sim$3.4.
Although the number of high-z galaxies detected through the \lya\ line
has increased ever since
\citep{Ouchi2008,Ouchi2010,Adams2011,Matthee2015,Santos2016,Herenz2017,Sobral2018,Urrutia2019},
the decades long endeavour made evident that
some physical processes are at play that may suppress or enhance the \lya\ emission.

Yet, it is known that resonant scattering of \lya\ photons leads to a complex radiative
transport problem
\citep{Neufeld1990,Ahn2001,Ahn2002,Dijkstra2006a,Dijkstra2006b,Verhamme2006,Schaerer2008,Verhamme2008}.
The line strength and visibility of \lya\ depends on many factors such as
dust content \citep{Hayes2010}, dust geometry \citep{Scarlata2009},
neutral gas content \citep{Zheng2017}, kinematics \citep{Kunth1998,Cannon2004,Wofford2013,Pardy2014} and
gas geometry \citep{TenorioTagle1999,MasHesse2003,Duval2014,Jaskot2014}.
See \cite{Hayes2015} for a review.

While, for example, large scale galaxy winds may enhance \lya\ escape via a large velocity gradient
along which the photons are shifted to wavelengths less prone to further scattering,
dust works in the other direction. Since the resonant nature of the line
drastically increases the path length out of the galaxy, the chance for
\lya\ photons to be destroyed by dust is significantly enhanced. Remarkably, \lya\ photons do not appear
to diffuse very far in space, because their re-distribution scales are small \citep{Bridge2018}, i.e. \lya\ photons
get stuck locally over many scattering events.
As a result, in some galaxies,
the interstellar medium is optically thin even for UV photons
(e.g. at a hydrogen column density of $10^{16}\ cm^{-2}$),
but opaque for \lya\ \citep{Verhamme2015,Puschnig2017}.

Given the fact that the total escape fraction of \lya\ photons depends
on such a large variety of effects that all act together in a complex way,
predictions of \lya\ emission are difficult.
However, progress has been made recently \citep{Trainor2019,Sobral2019}.
In particular, \cite{Runnholm2020} reproduced the \lya\ output of local
\lya\ emitters with a dispersion of $\sim0.3$~dex only.

One path to remedy this complicated situation is to study the
\lya\ production and escape mechanisms in a sample of star-forming galaxies on a spatially resolved basis. To do so,
\textit{LARS} -- the Lyman Alpha Reference Sample -- was defined \citep{Hayes2013,Hayes2014,Oestlin2014}.
It is a sample of 14 galaxies (see Table \ref{tab:lars_general} for basic properties)
which is statistically meaningful enough to observe trends
(i.e. it covers a wide range in stellar mass, star formation rate and metallicity)
and, most importantly,
comparable in selection to galaxies observed at high-z \citep{Ouchi2008,Cowie2011}.
LARS galaxies are selected by two main parameters: UV luminosity and H$\upalpha$ equivalent width. The former
is comparable to Lyman Break galaxies at redshifts $\sim$3--4 and the latter is $>$100\AA,
limiting the sample to galaxies with currently ongoing (young) bursts of star formation.
Multi-band imaging with the Hubble Space Telescope (HST) forms the 
groundwork of our study which includes HST ultraviolet spectroscopy \citep{RiveraThorsen2015}, optical broadband imaging \citep{Micheva2018},
PMAS/LArr IFU data \citep{Herenz2016}, Green Bank Telescope 21 cm spectra \citep{Pardy2014} and JVLA data (Le Reste et al. in prep.).

\subsection{The Kennicutt-Schmidt Relation}
The Kennicutt-Schmidt relation or KS law \citep{Schmidt1959,Kennicutt1998}
is key for studies of star formation, because it relates
gas surface densities to star formation rate (SFR) surface densities.
In its classical form, i.e. \textit{total} gas (H~I+H$_2$) vs. SFR, the slope is typically found
to be non-linear with power-law indices of $\sim$ 1.4--1.5 \citep{Kennicutt1998}.
In contrast, the \textit{molecular} gas relation is often
found to be \textit{linear} \citep[e.g.][]{Wong2002,Bigiel2008,Bigiel2011,Leroy2008,Leroy2013,
Bolatto2017,Utomo2018,Colombo2018},
and it holds over eight orders of magnitude in mass, on a wide range
of spatial scales -- from tens of parsec \citep{Wu2005} to entire galaxies -- 
and even in low-density disk outskirts of galaxies \citep{Schruba2011} as well as in
low-metallicity dwarf galaxies \citep{Bolatto2011,Jameson2016}.
The observed linear slope has far reaching consequences:
1) the time scale to consume the gas at the current SFR, i.e. $\sfrac{M_{gas}}{SFR}$, or the gas depletion time, is constant
and 2) the driving physics behind the KS law is not
self-gravity alone \citep{Semenov2019}, but other processes (e.g. feedback) must play a fundamental role as well.

However, non-linear molecular KS relations are also observed, namely on even smaller scales
\citep[e.g.][]{Evans2009,Evans2014,Heiderman2010,Gutermuth2011},
in starburst galaxies \citep[e.g.][]{Genzel2010,Genzel2015}, in the dwarf magellanic irregular galaxy NGC4449 \citep{Calzetti2018}
and also in galactic centers \citep[e.g.][]{Leroy2013}.

It was further shown by \cite{Gao2004b,Gao2004a} and \cite{Lada2010} that star formation linearly scales with the \textit{dense gas} traced by HCN~(1--0).
However, it is still a matter of debate which of these scaling relations (molecular or dense gas) is the more fundamental one.
See e.g. \cite{Lada2012} for a discussion.
In recent years, however, large efforts have been undertaken to study the dense gas
in external galaxies \citep{Bigiel2016,JimenezDonaire2019}, showing that the environment and feedback
plays a fundamental role \citep{Kruijssen2019}.

Observations of galaxies at redshifts between 0 and 4 have shown that the level of star formation
is mainly dictated by stellar mass, and regulated by secular processes \citep{Popesso2019}.
This is seen as a tight relation between stellar mass and SFR, the so called
\textit{main sequence of star forming galaxies}
\citep{Brinchmann2004,Noeske2007,Daddi2007,Elbaz2007,Peng2010,Wuyts2011,Whitaker2012,Whitaker2014,Tomczak2016}.
While the slope of the relation does not vary with redshift, its intercept shifts towards higher SFRs,
see for example \cite{Wuyts2011,Rodighiero2011}. However, for main sequence galaxies it was also found
that even out to redshift $z\sim4$ galaxies have constant gas depletion times due to an increase of the gas fraction
\citep{Tacconi2013,Genzel2015,Scoville2017}.
It was also realized that the consumption time scales are too short compared to the age of the Universe.
And thus, galaxies must accrete gas, e.g. from the circumgalactic medium or
other galaxies by mergers, in order to sustain star formation.

\subsection{This Work -- KS law for LARS galaxies}
In this paper, we establish the KS relation, calculate gas fractions, gas depletion times and star
formation efficiencies in the Lyman Alpha Reference Sample. This allows us to study dependencies between
previously derived quantities related to the \lya\ emission and global properties of the star forming gas,
giving new insights into physical processes and environmental properties that
efficiently drive the escape of \lya\ photons.

We have performed low-J CO spectroscopy with the IRAM 30m telescope,
far-infrared imaging and spectroscopy with Herschel/PACS and SOFIA/FIFI-LS, targeting continuum and fine structure emission lines
of [C~II]158$\upmu$m and [O~I]63$\upmu$m.

We utilize the aforementioned radio and far infrared observations (see a description of the observations
in Section \ref{sec:data}) to derive properties of the ISM from dust models (see Section \ref{sec:DL07})
and to establish the KS law for LARS galaxies and
study the environment, i.e. molecular and total gas fractions, star formation rates and efficiencies, as well as
gas depletion times (see Section \ref{sec:KS}). We further use far-infrared photometry and spectroscopy to explore
photon-dominated regions (hereafter PDR) within LARS (see Section \ref{sec:PDR}), i.e. FUV radiation field and densities.
In Section \ref{sec:discussion} we finally dicuss which of the derived properties influence
\lya\ escape and bring our results in context with early phases of disk formation.

Throughout the paper, we adopt a cosmology with H$_0$=70, $\Omega_M$=0.3 and $\Omega_{vac}$=0.7.

\begin{table*}
\resizebox{\textwidth}{!}{%
\begin{threeparttable}
\caption{Compilation of properties of LARS galaxies relevant for this work}
\label{tab:lars_general}
\begin{tabular}{lrrrrrrrrrrr}
\hline \hline
ID & z     & $log_{10}M_*$   & Z                      & $log_{10}M_{HI}$  & $L_{LyA}$         & $f_{esc}^{LyA}$ & $D_{scatt}$ & $D_{25_{SDSS}}$  & $\sfrac{b}{a}_{SDSS}$ & $R_{cut_{M18}}$ & $\sfrac{b}{a}_{M18}$ \\
   &       & $[M_{\odot}]$   & $[12+log\sfrac{O}{H}]$ & $[M_{\odot}]$     & $[10^{42}cgs]$    &                 & $[kpc]$     & $["]$            &                       & $[kpc]$         &  \\
1    & 0.02782 & 9.785  & 8.24 & 9.3979  & 0.85   & 0.119     & 0.72      & 14.94     & 0.69     & 2.69      & 0.71    \\
2    & 0.02982 & 9.371  & 8.23 & 9.4472  & 0.81   & 0.521     & 0.36      & 25.67     & 0.43     & 2.00       & 0.39    \\
3    & 0.03143 & 10.303 & 8.41 & 9.8388  & 0.10   & 0.003     & 0.35      & 85.80      & 0.22     & 2.34      & 0.29    \\
4    & 0.03249 & 10.111 & 8.19 & 9.8865  & 0.00    & 0.000       & 0.70      & 19.07     & 0.89     & 4.21      & 0.64    \\
5    & 0.03375 & 9.630  & 8.12 & <9.4624  & 1.11   & 0.174     & 0.40      & 13.11     & 0.73     & 2.20       & 0.71    \\
6    & 0.03461 & 9.320  & 8.08 & *<10.3617 & 0.00   & 0.000       & 0.41      & 20.32     & 0.60      & 2.05      & 0.55    \\
7    & 0.03774 & 9.677  & 8.35 & 9.4914  & 1.01   & 0.100       & 0.37      & 19.03     & 0.48     & 2.33      & 0.55    \\
8    & 0.03825 & 10.970 & 8.51 & 10.3424 & 1.00   & 0.025     & 0.44      & 30.84     & 0.59     & 5.10       & 1.00     \\
9    & 0.04677 & 10.708 & 8.37 & 10.1139 & 0.33   & 0.007     & 0.56      & 7.37      & 0.86     & 6.70       & 0.50     \\
10   & 0.05765 & 10.332 & 8.51 & 9.6532  & 0.16   & 0.026     & 0.52      & 22.44     & 0.43     & 3.12      & 0.60     \\
11   & 0.08461 & 11.083 & 8.44 & 10.415  & 1.20   & 0.036     & 1.56      & 33.52     & 0.28     & 8.40       & 0.24    \\
12   & 0.10210 & 9.870  & 8.34 & --     & 0.93   & 0.009     & 1.03      & 8.71      & 0.81     & 4.44      & 0.88    \\
13   & 0.14670 & 10.772 & 8.50 & --     & 0.72   & 0.010      & 2.24      & 10.26     & 0.59     & 7.40       & 0.71    \\
14   & 0.18070 & 9.243   & 8.06 & --     & 4.46   & 0.163     & 1.68      & 5.46      & 0.82     & 5.03      & 1.00    
\end{tabular}
\begin{tablenotes}
        \item The redshifts were derived from radial velocities (given in the optical definition) in \cite{Pardy2014} and are thus related to the H~I
        systemic velocities for all LARS galaxies but 12-14 (which have no H~I detection). For those, \cite{Pardy2014}
        report optical SDSS velocities.
        Stellar masses ($log_{10}M_*$), \lya\ luminosities ($L_{LyA}$) and escape fractions
        ($f_{esc}^{LyA}$) were previously published by \cite{Hayes2014}. The metallicities ($Z$) derived through the
        empirical O3N2 relation are taken from \cite{Oestlin2014},
        and atomic gas masses ($log_{10}M_{HI}$) from \cite{Pardy2014}. *Note that the atomic mass reported for LARS 6 is contaminated
	by the field spiral UGC10028 and thus only an upper limit. Typical \lya\ scattering distances ($D_{scatt}$) for the galaxies
        were derived by \cite{Bridge2018} and the radii at which the star formation rate density has decreased to
        0.01M$_{\odot}\ yr^{-1}\ kpc^2$ ($R_{cut_{M18}}$), together with axis ratios ($\sfrac{b}{a}_{M18}$)
        were calculated by \cite{Micheva2018}.
        The blue-band diameters of the 25$mag\ arcsec^{2}$ isophote (D$_{25_{SDSS}}$) as well as $\sfrac{b}{a}_{SDSS}$ were
        derived from SDSS g-band observations using an SQL query on SDSS DR7. Since the SDSS g-band isophote
        typically gives sizes that are $\sim$1.3 times larger than those measured in the Johnson B band \citep{Hakobyan2012},
        we divided the g-band diameters by that factor and report estimated values for D$_{25_{SDSS}}$ here.
\end{tablenotes}
\end{threeparttable}}
\end{table*}


\section{Observations and Archival Data}\label{sec:data}
\subsection{Observations of Molecular Lines of CO, HCN and HCO$^+$}
Observations with the IRAM
30m telescope were carried out under programs 082-14, 064-15 and 178-15.
Data were acquired during the heterodyne pool weeks in September, October and November 2014 as well
as in January and February 2016. Observing runs in visitor mode were carried out
in June and September 2015, as well as during May 2016. In total, more than 110h
of observing time were scheduled at the telescope, from which roughly 1/3 was lost
due to poor weather conditions.
Using SFR and metallicity as indicators for the chance to detect CO,
we first select 10 galaxies out of our sample for molecular line observations,
i.e. we initially rejected LARS 4, 5, 6 and 14 as targets. Finally, we include
LARS~5, given the fact that the galaxy has strong winds \citep{Duval2016} that might trigger
molecular gas formation.

\begin{figure*}[!htbp]
   \subfloat{\includegraphics[width=3.1cm]{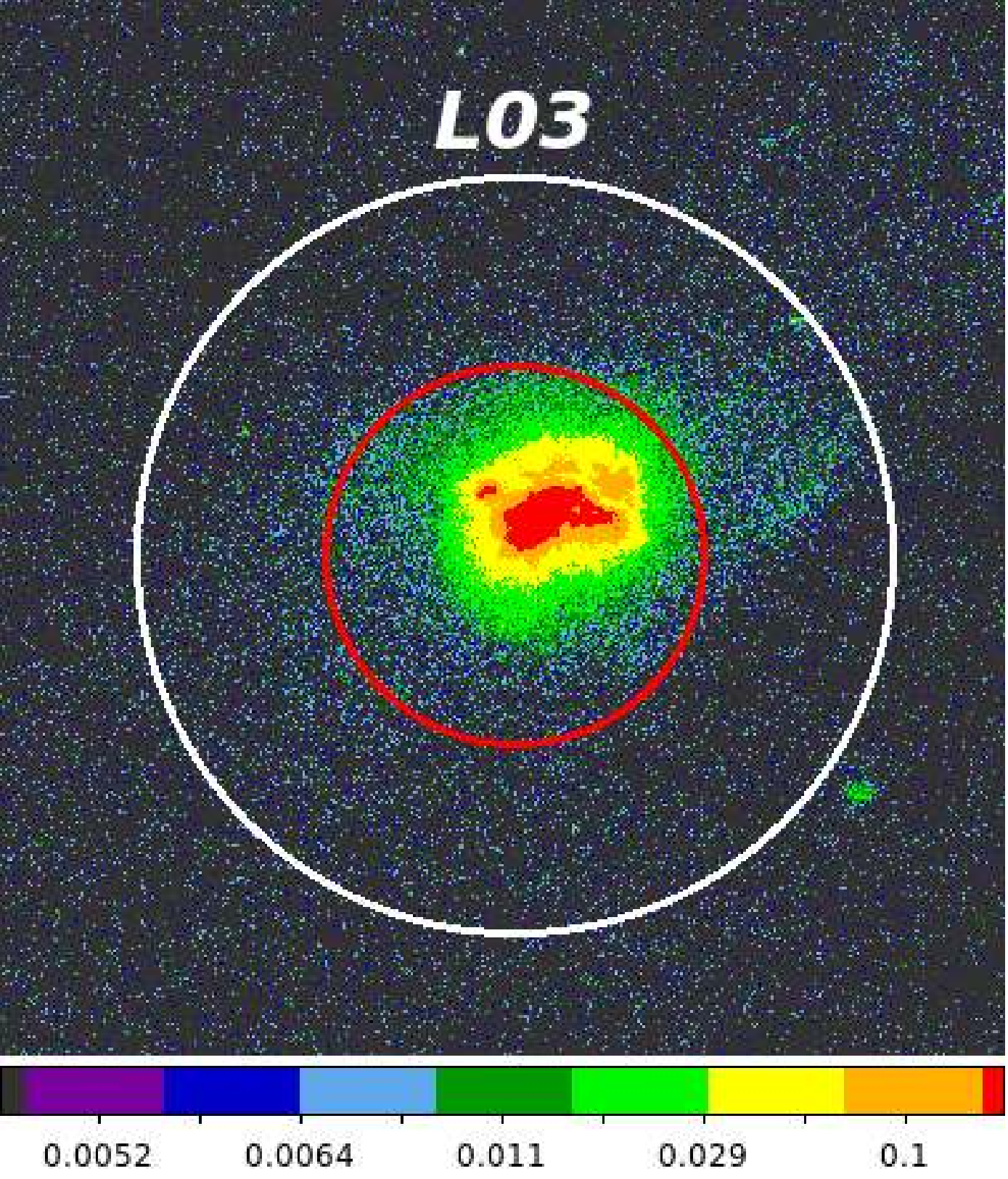}}\hspace*{\fill}
   \subfloat{\includegraphics[width=3.7cm]{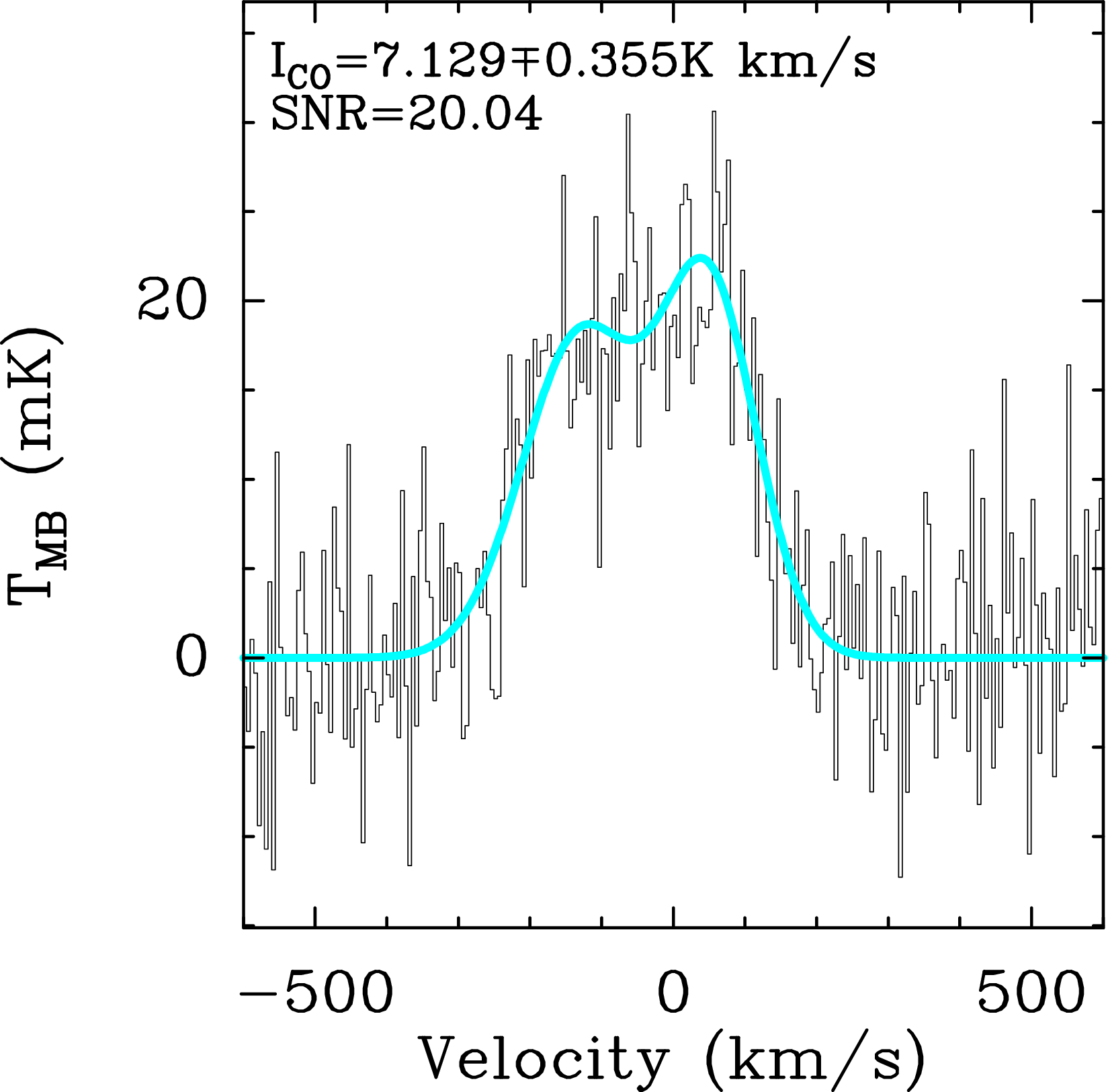}}\hspace*{\fill}
   \subfloat{\includegraphics[width=3.7cm]{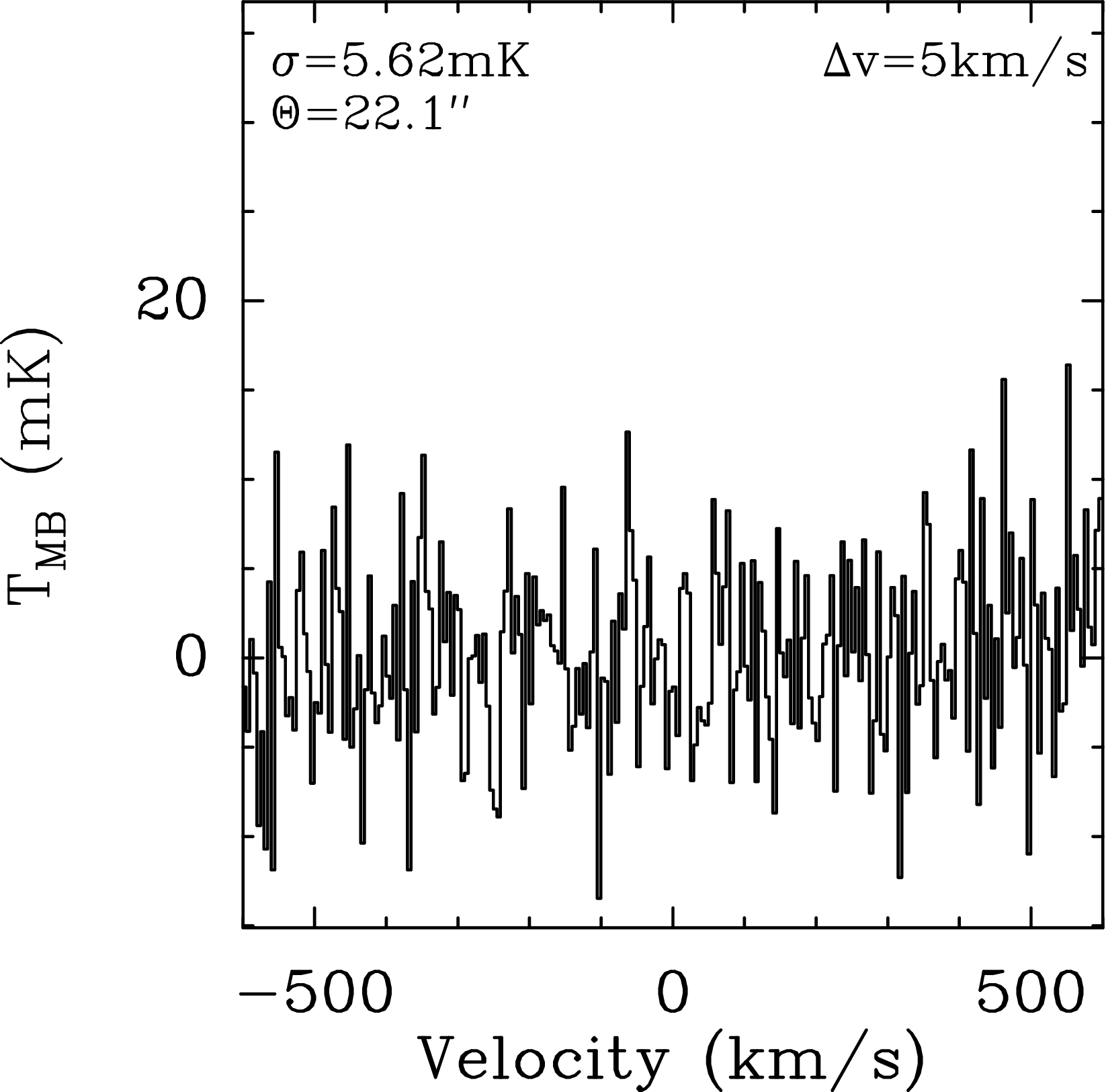}}\hspace*{\fill}
   \subfloat{\includegraphics[width=3.7cm]{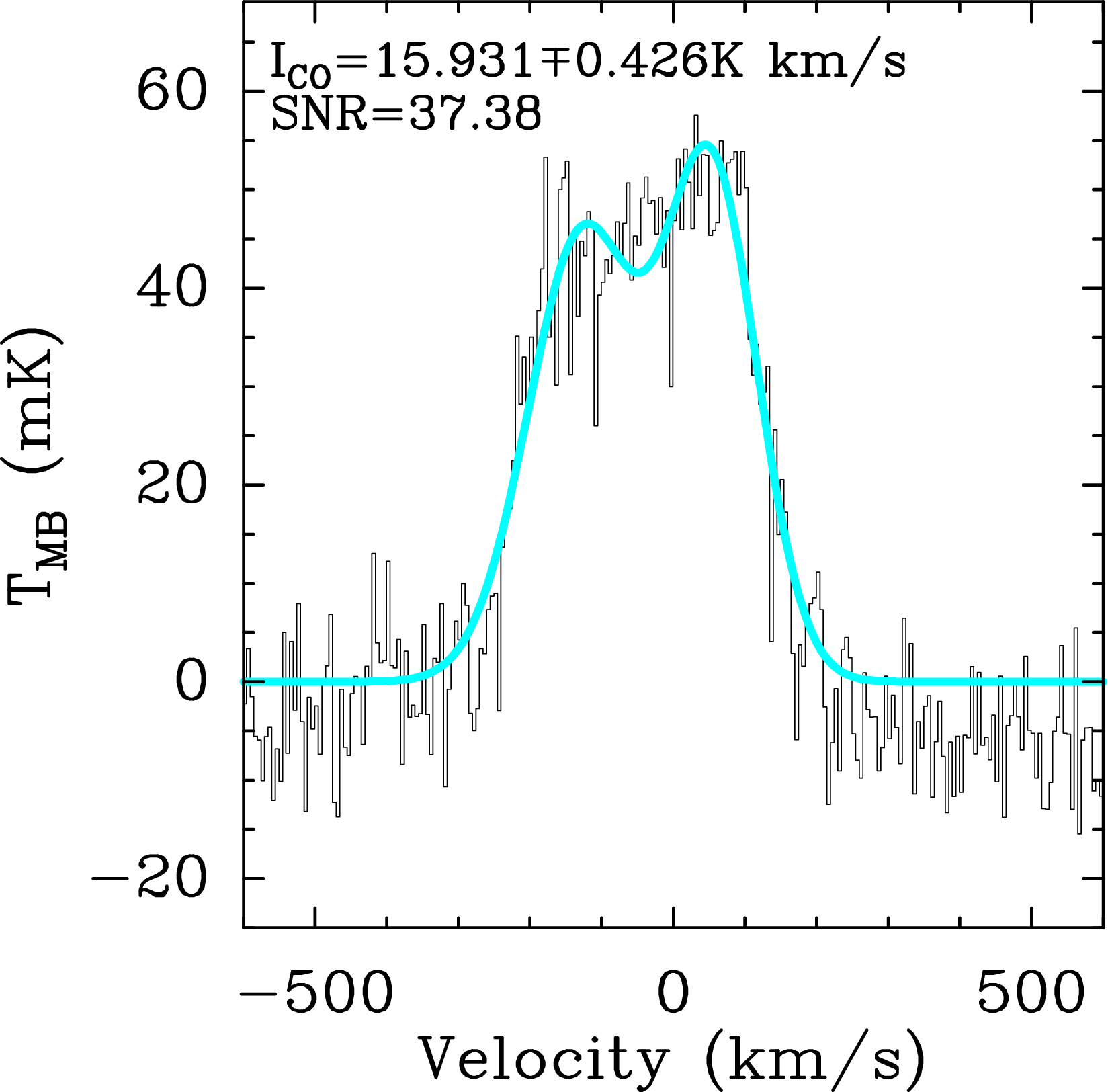}}\hspace*{\fill}
   \subfloat{\includegraphics[width=3.7cm]{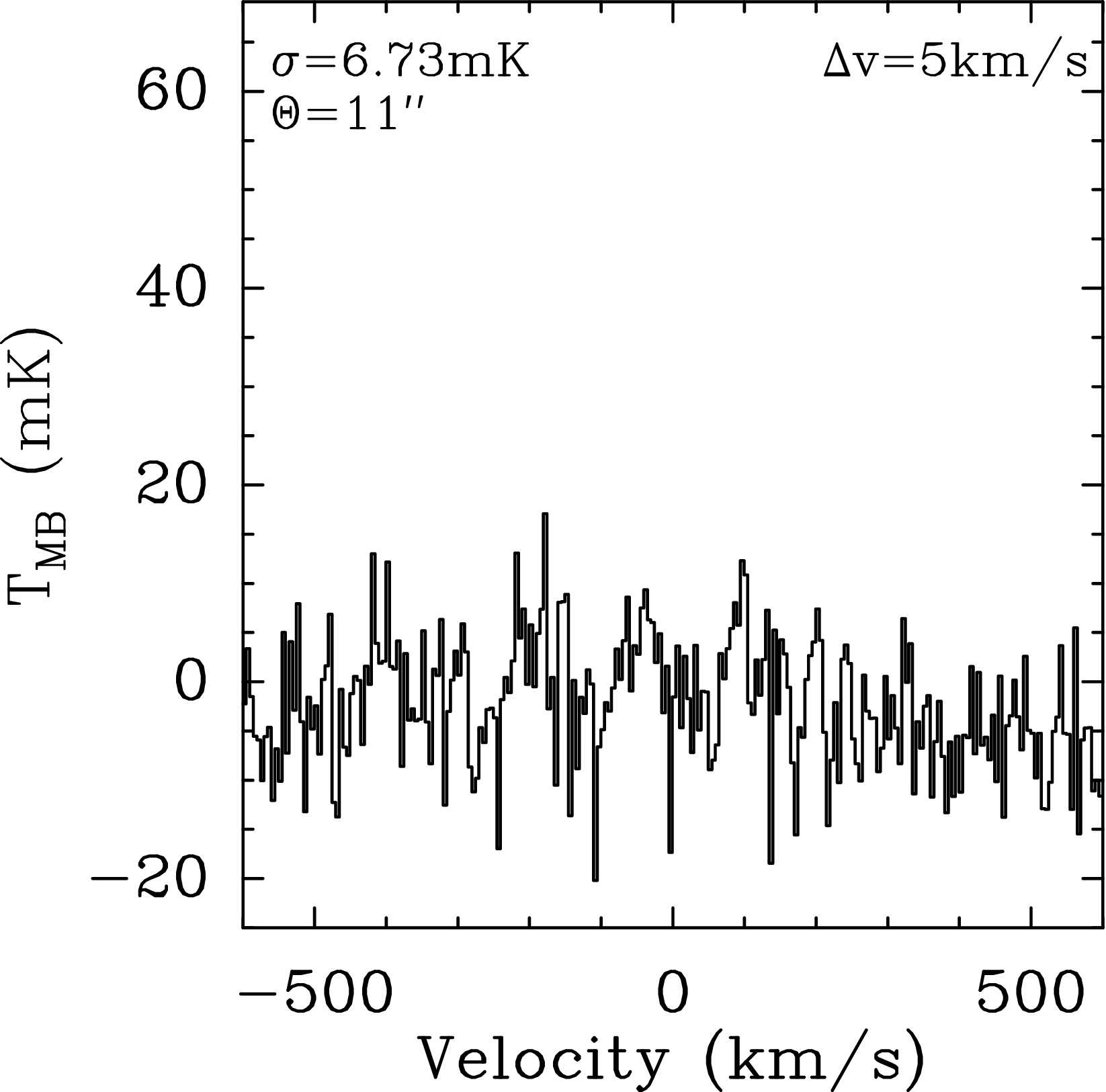}}\hspace*{\fill}

   \subfloat{\includegraphics[width=3.1cm]{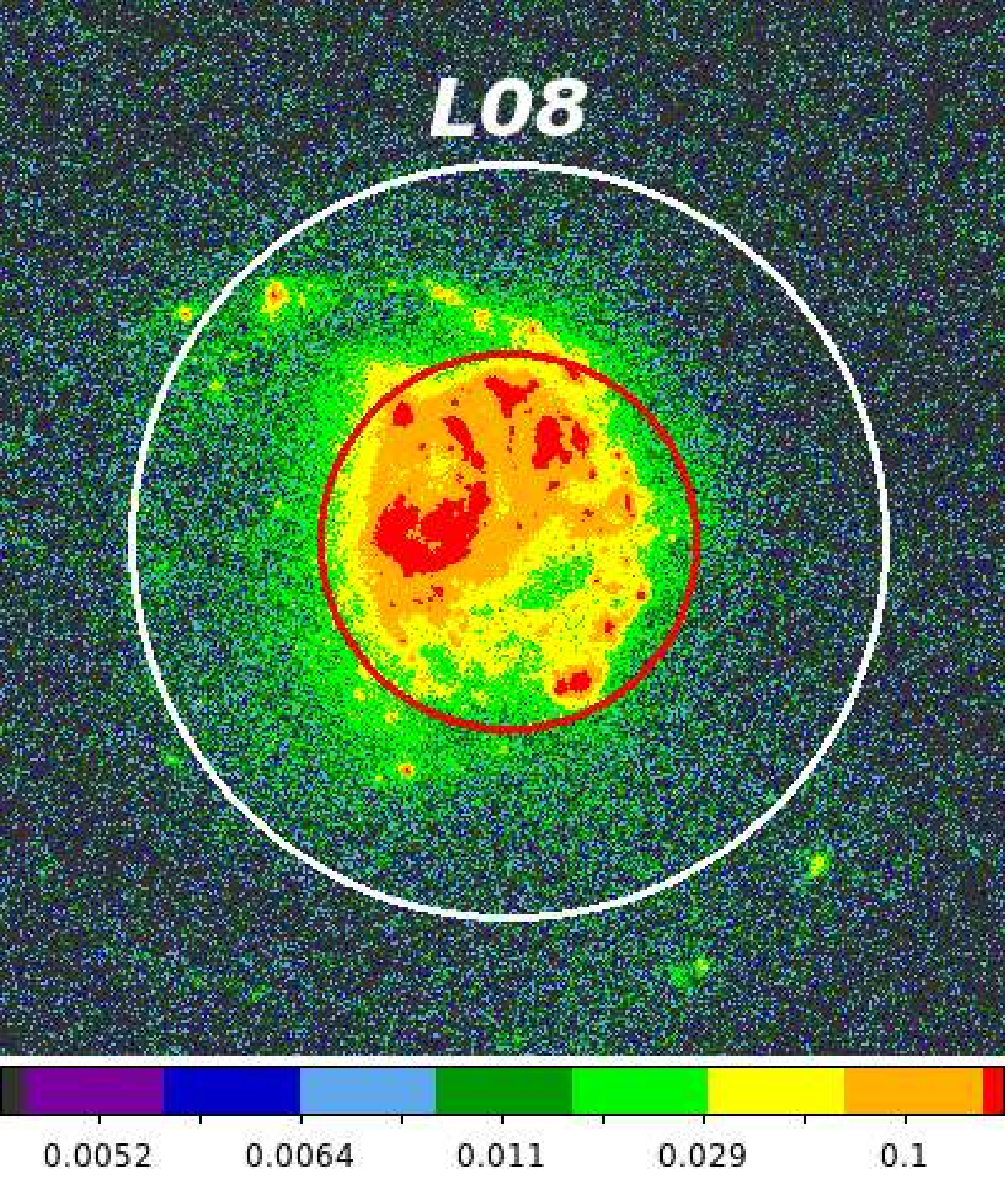}}\hspace*{\fill}
   \subfloat{\includegraphics[width=3.7cm]{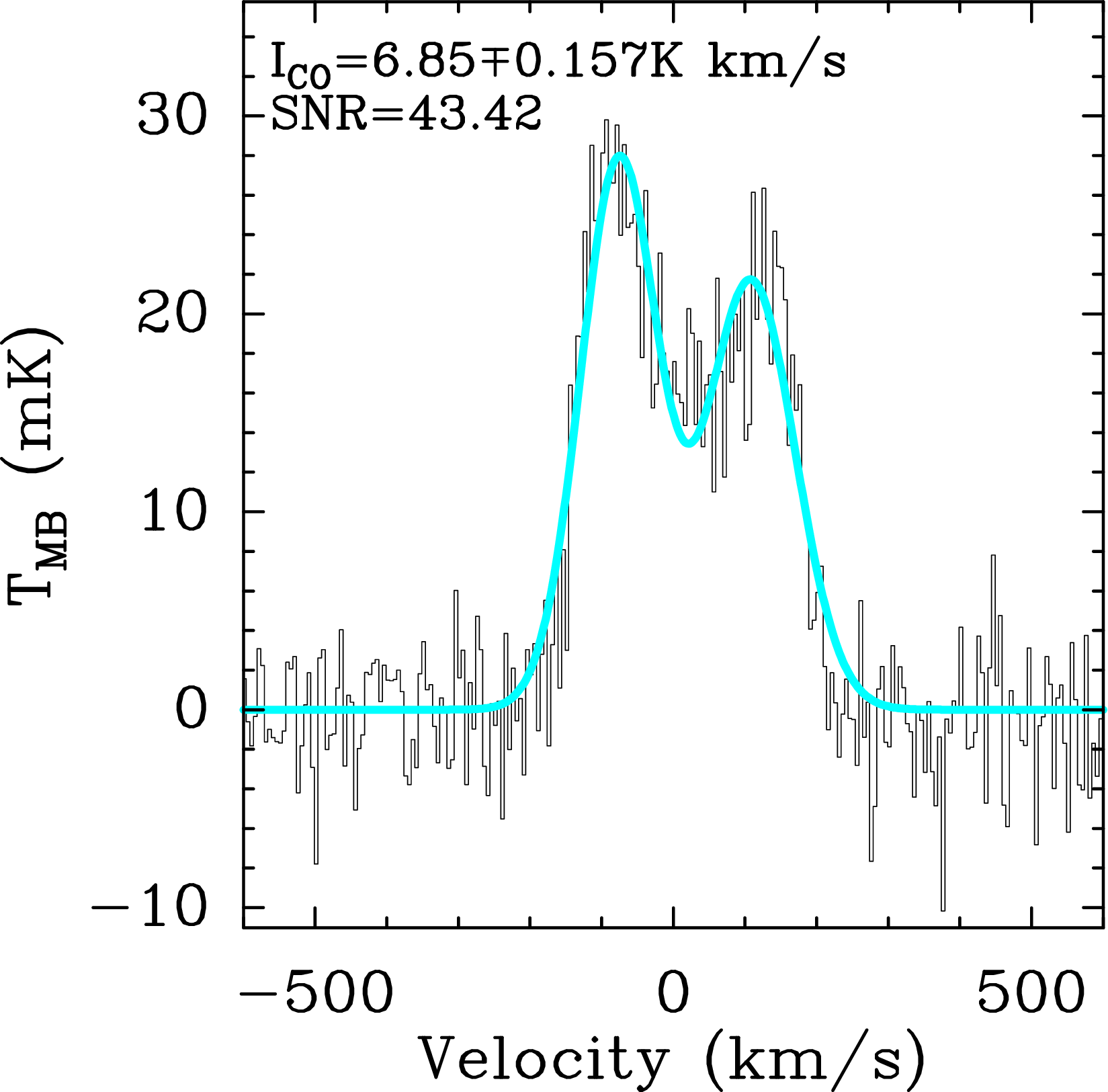}}\hspace*{\fill}
   \subfloat{\includegraphics[width=3.7cm]{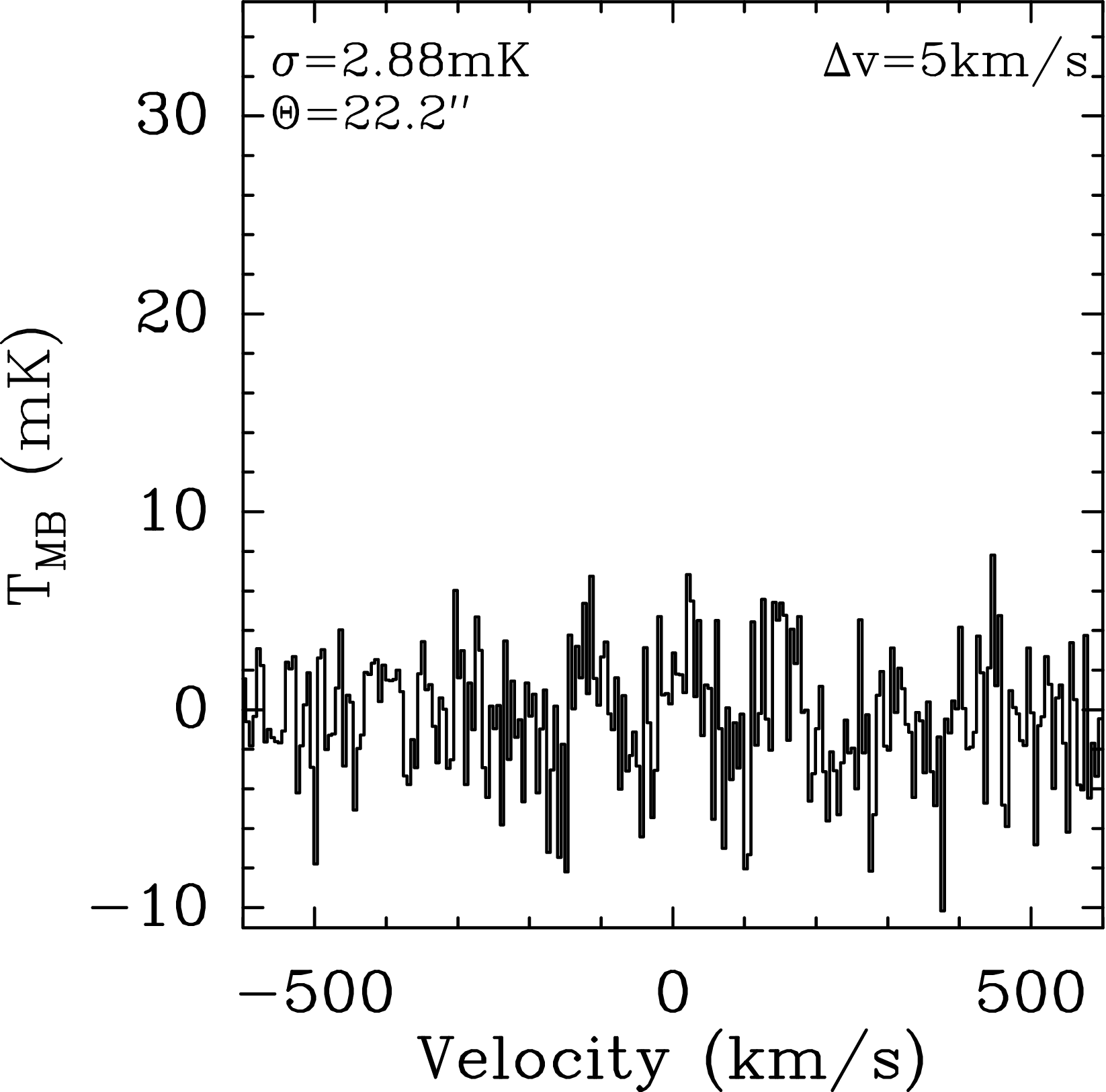}}\hspace*{\fill}
   \subfloat{\includegraphics[width=3.7cm]{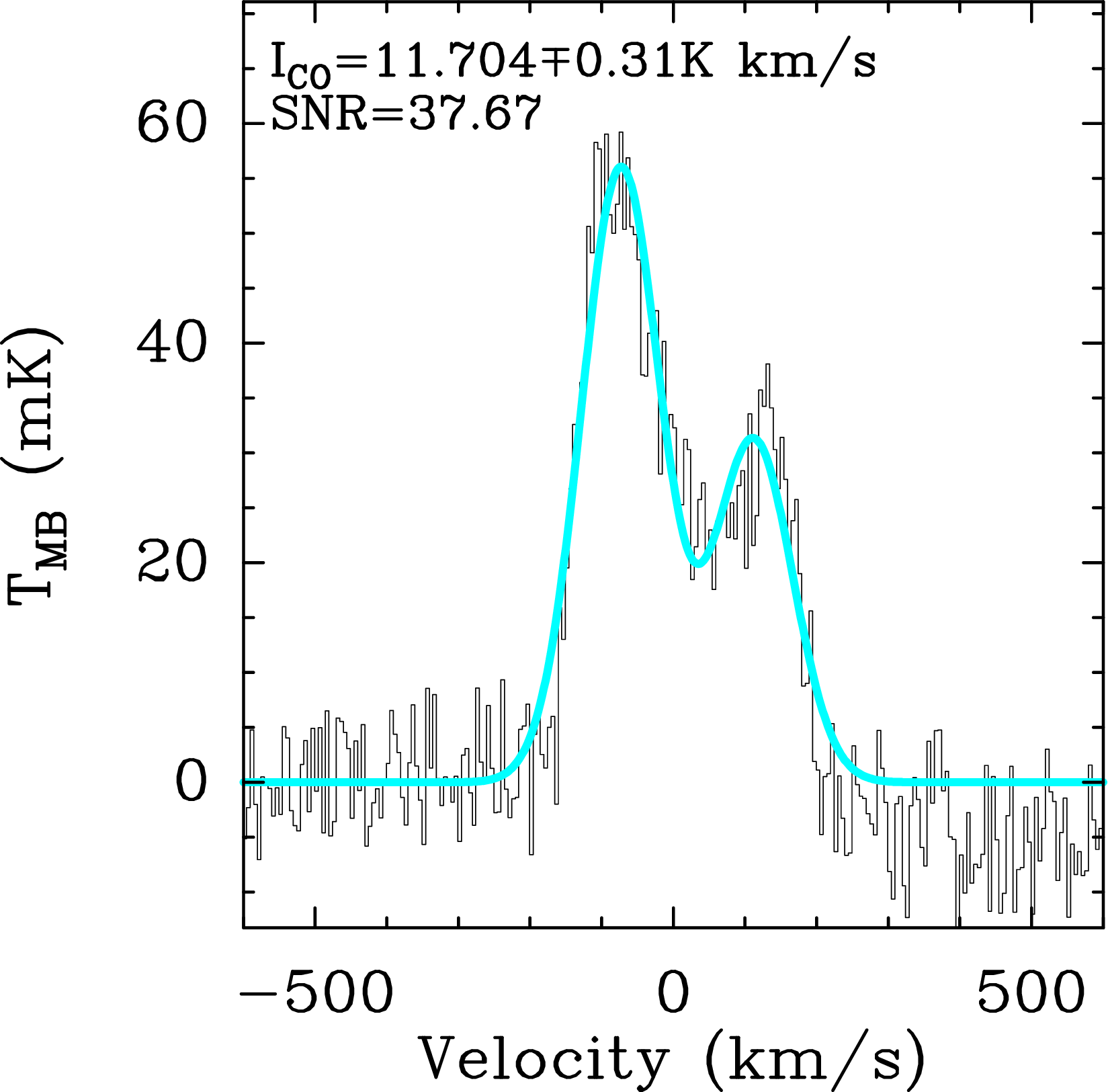}}\hspace*{\fill}
   \subfloat{\includegraphics[width=3.7cm]{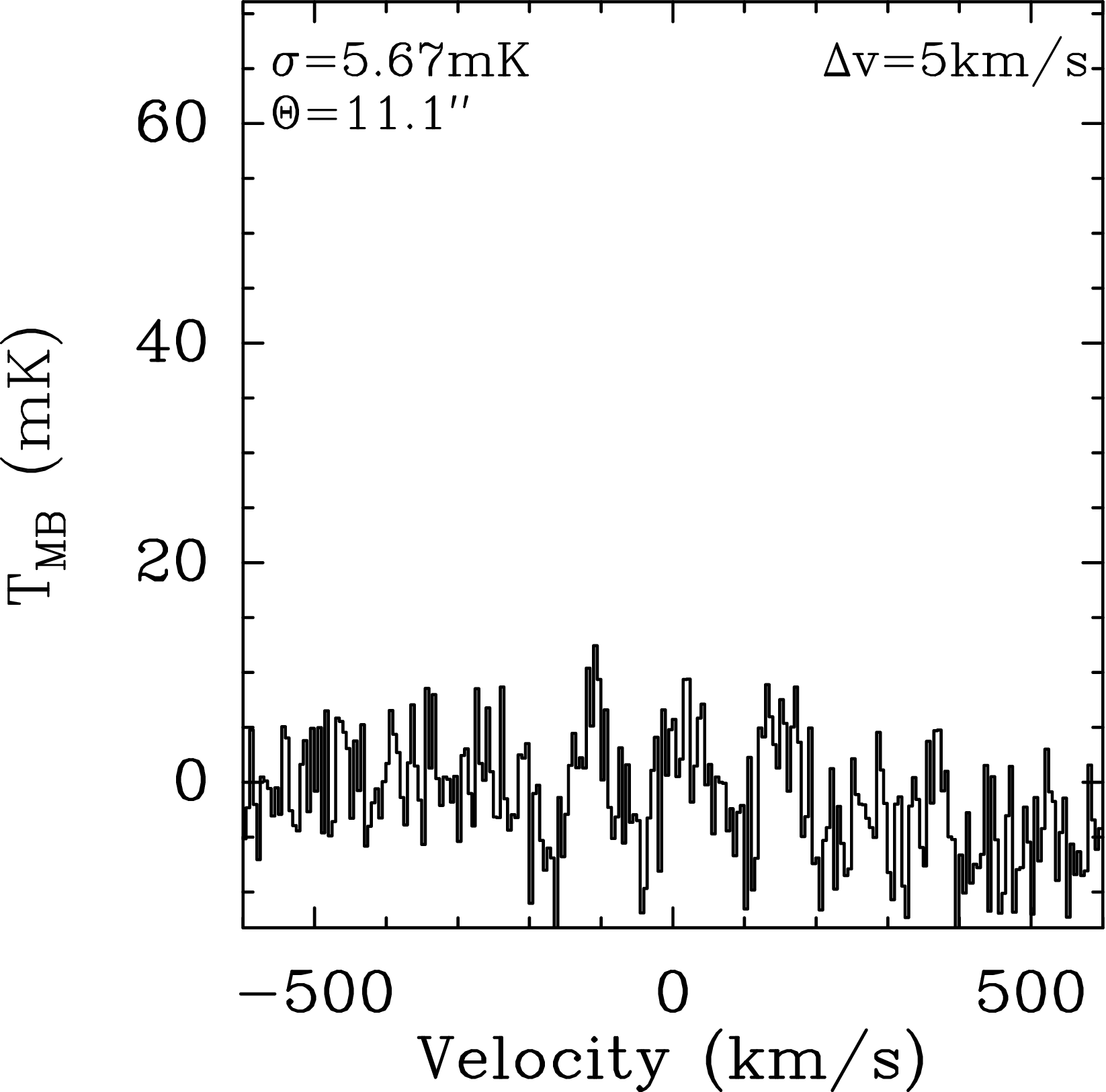}}\hspace*{\fill}

   \subfloat{\includegraphics[width=3.1cm]{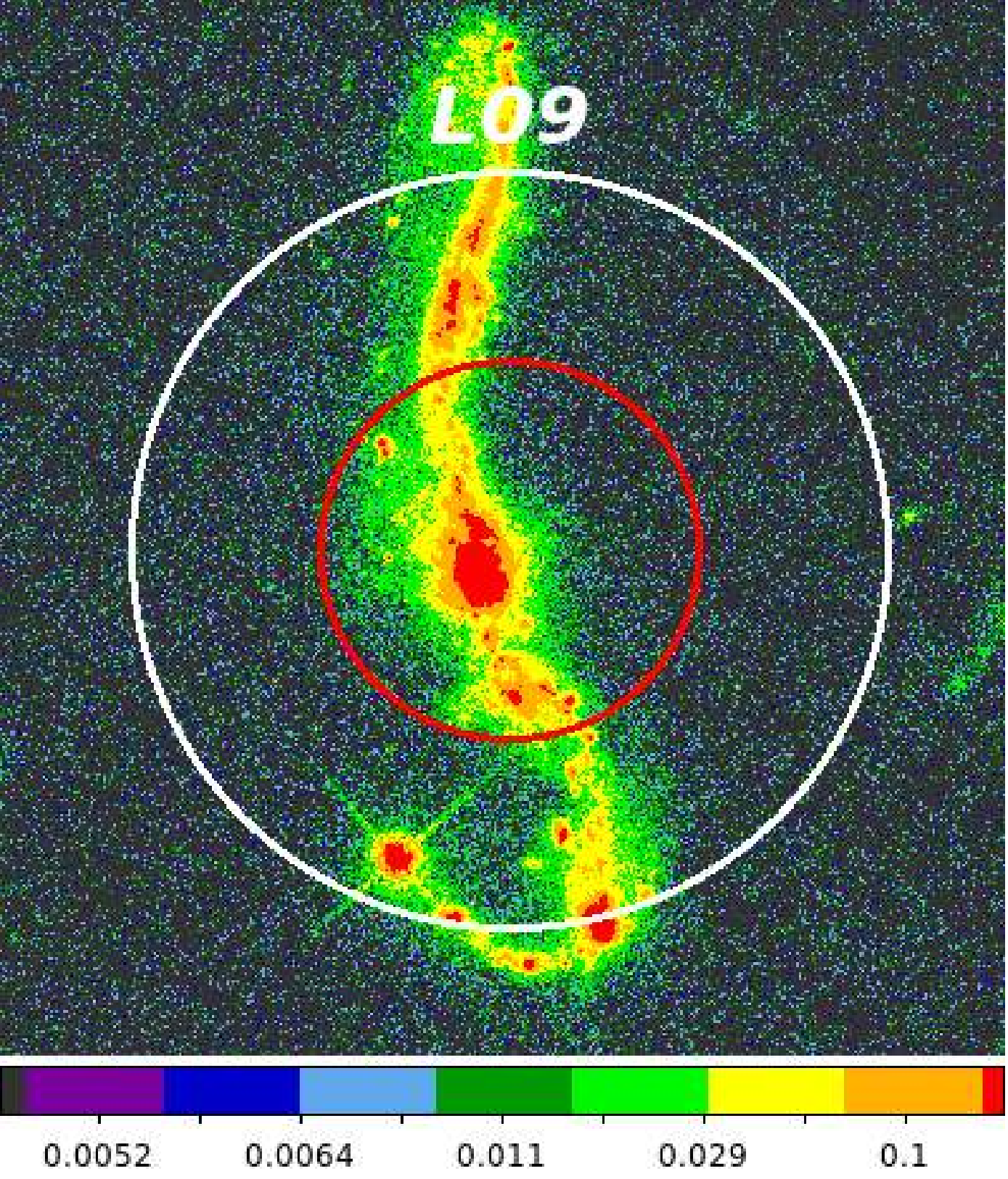}}\hspace*{\fill}
   \subfloat{\includegraphics[width=3.7cm]{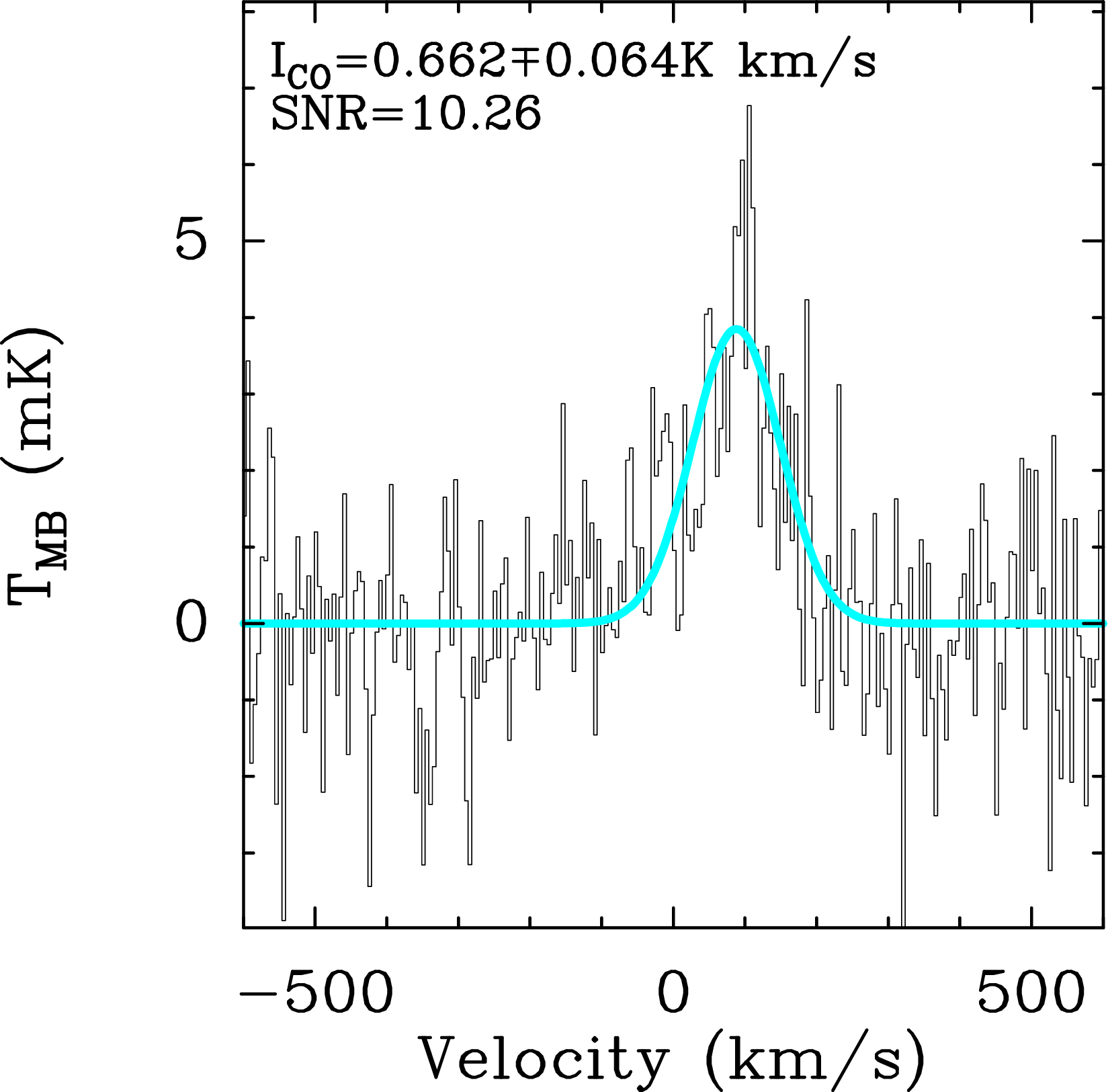}}\hspace*{\fill}
   \subfloat{\includegraphics[width=3.7cm]{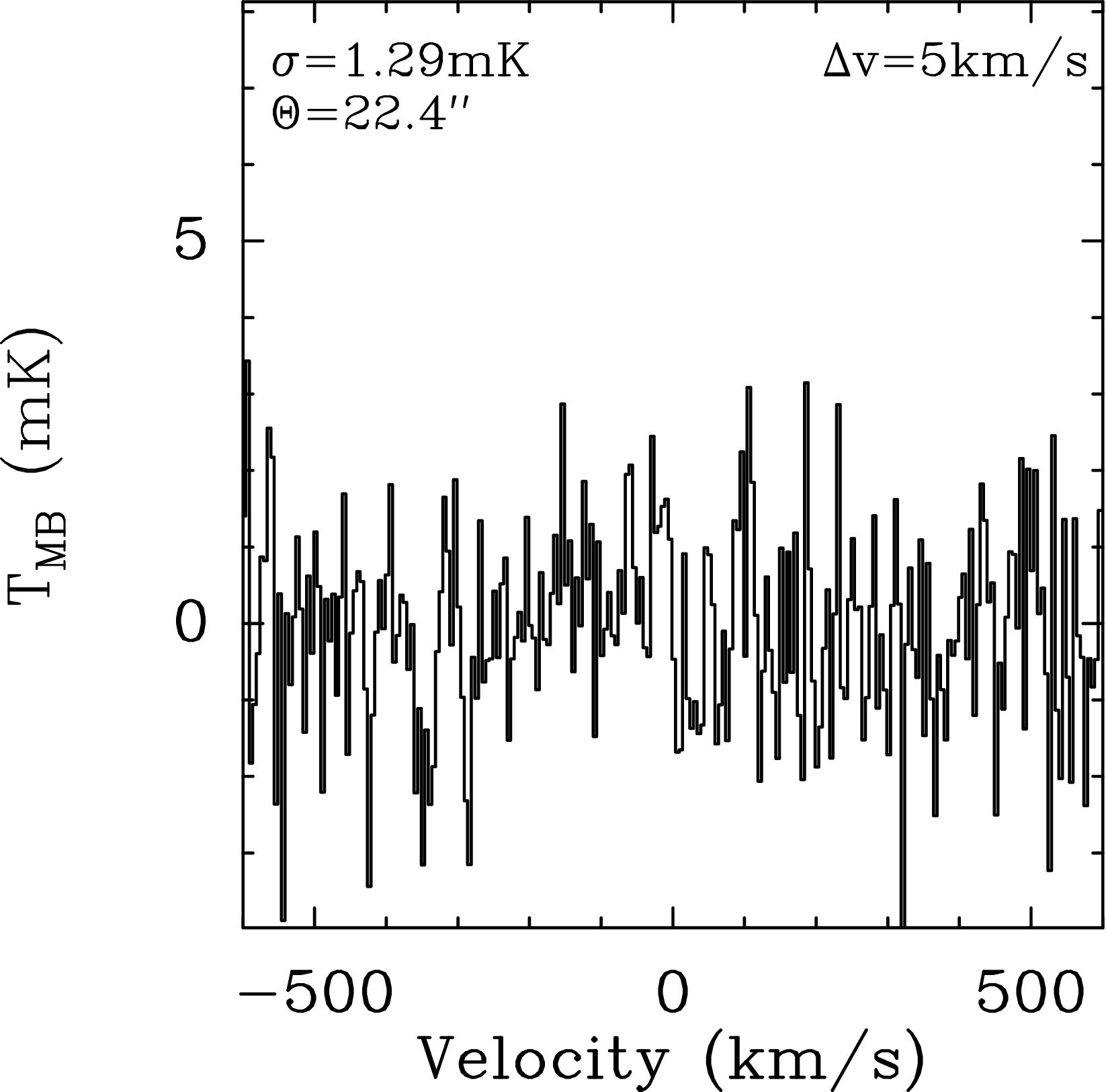}}\hspace*{\fill}
   \subfloat{\includegraphics[width=3.7cm]{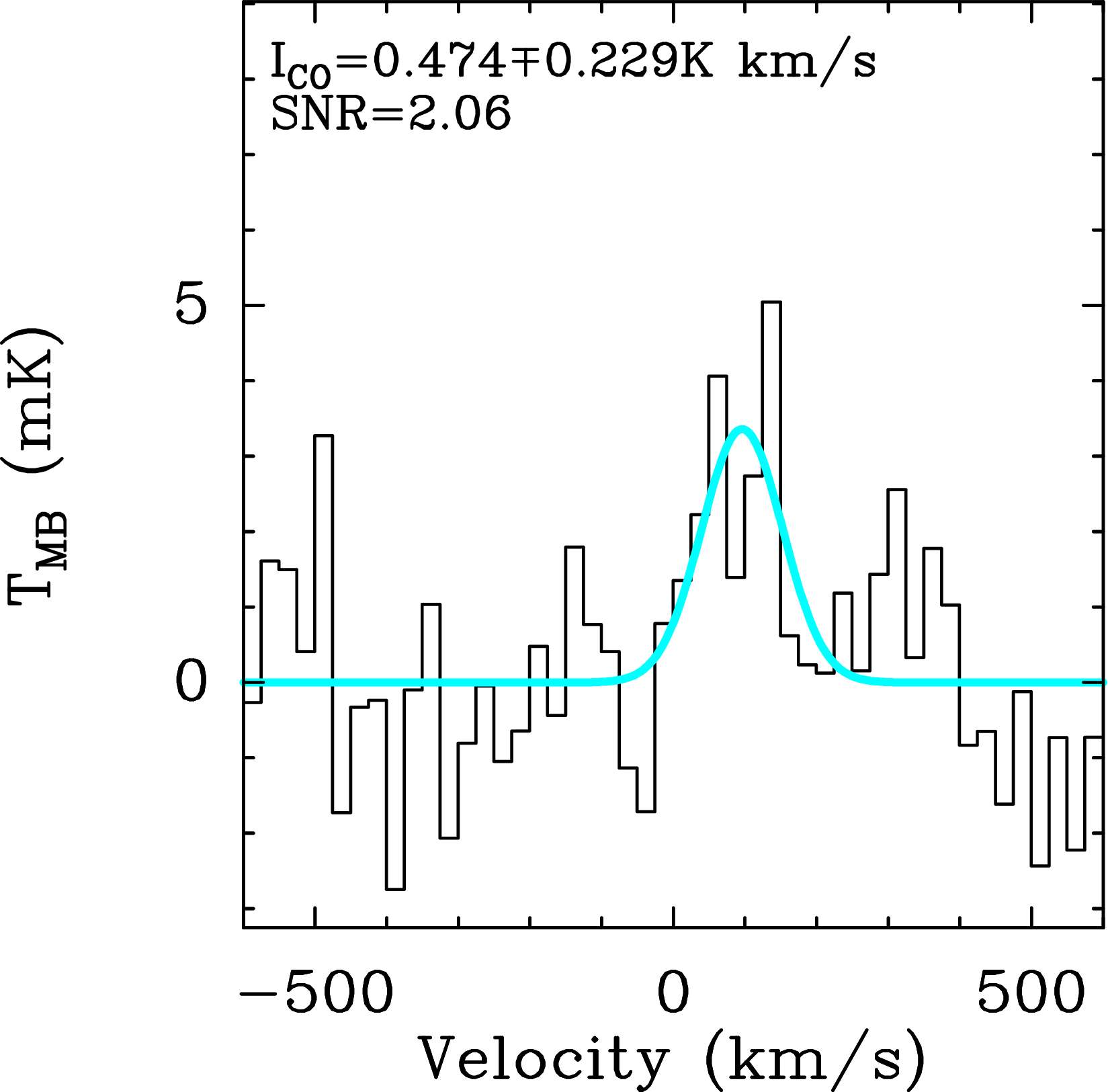}}\hspace*{\fill}
   \subfloat{\includegraphics[width=3.7cm]{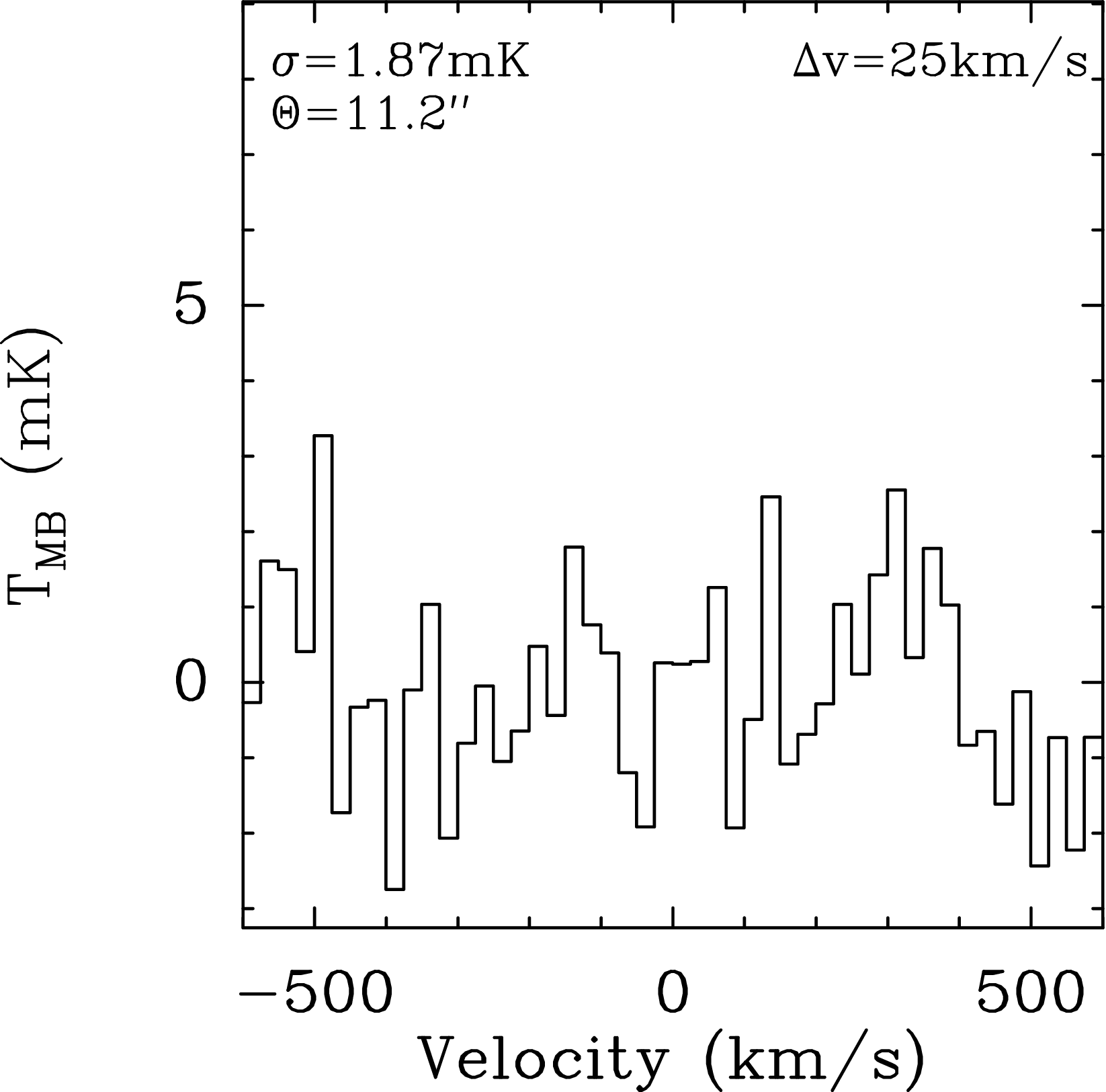}}\hspace*{\fill}

   \subfloat{\includegraphics[width=3.1cm]{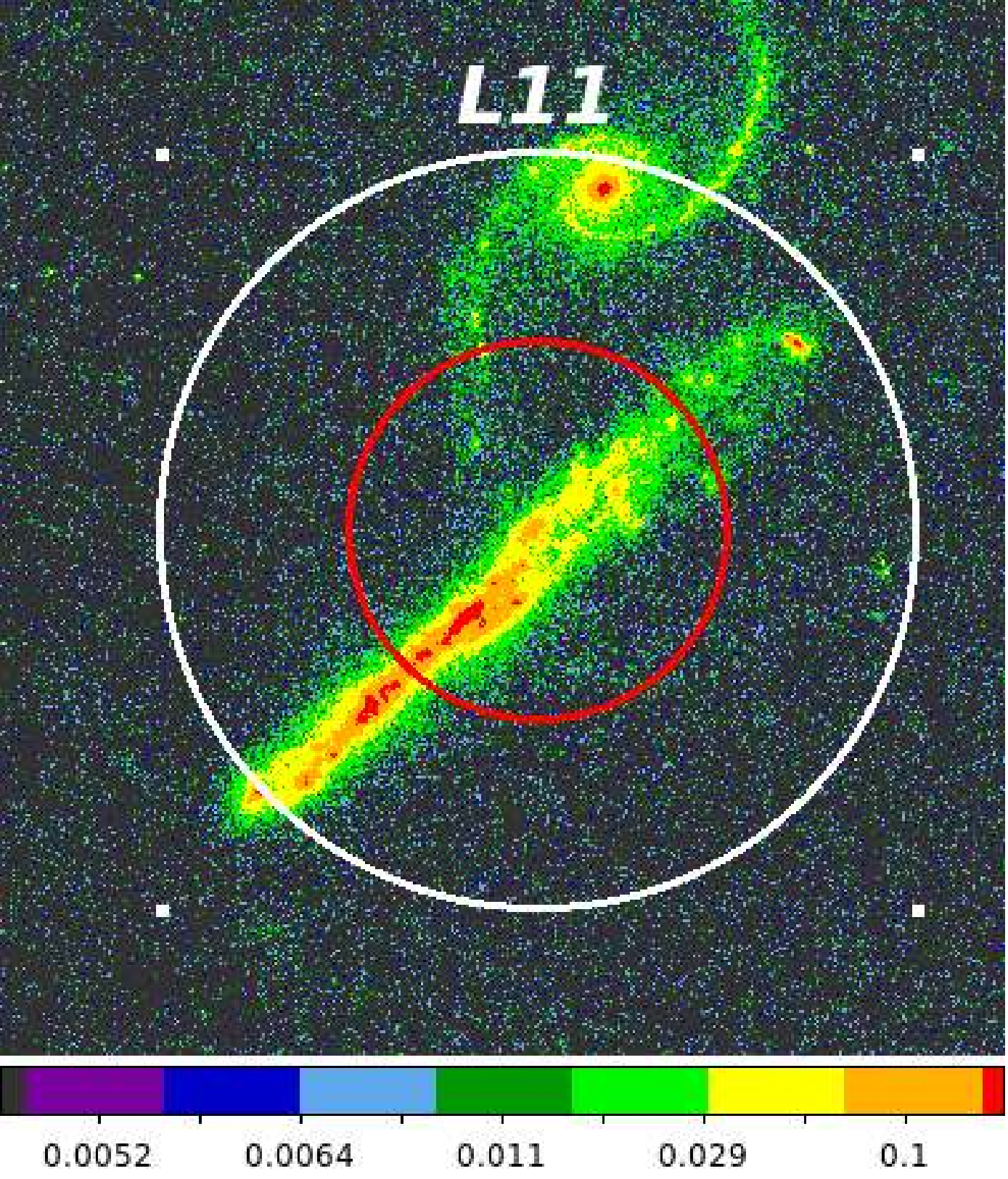}}\hspace*{\fill}
   \subfloat{\includegraphics[width=3.7cm]{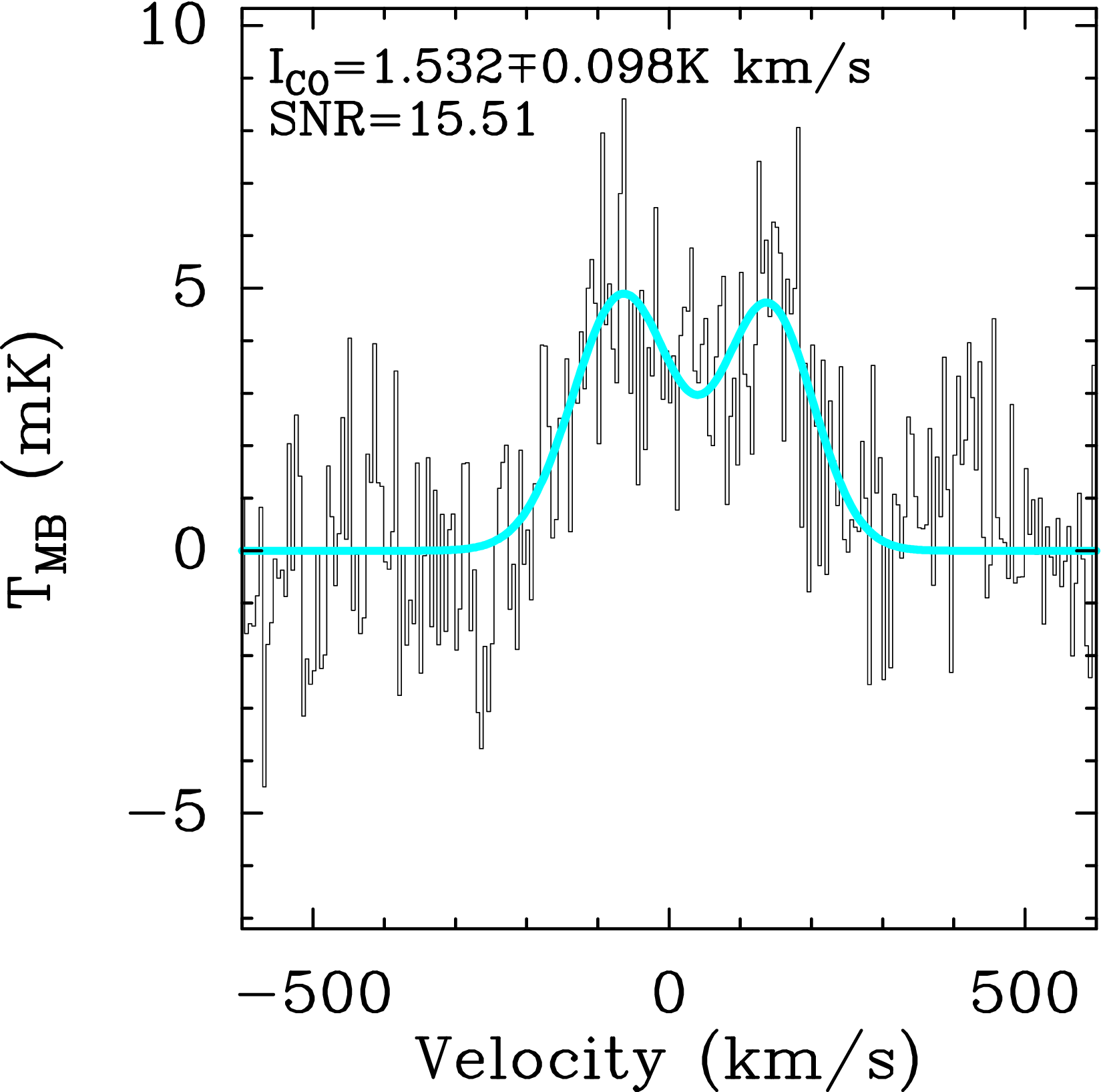}}\hspace*{\fill}
   \subfloat{\includegraphics[width=3.7cm]{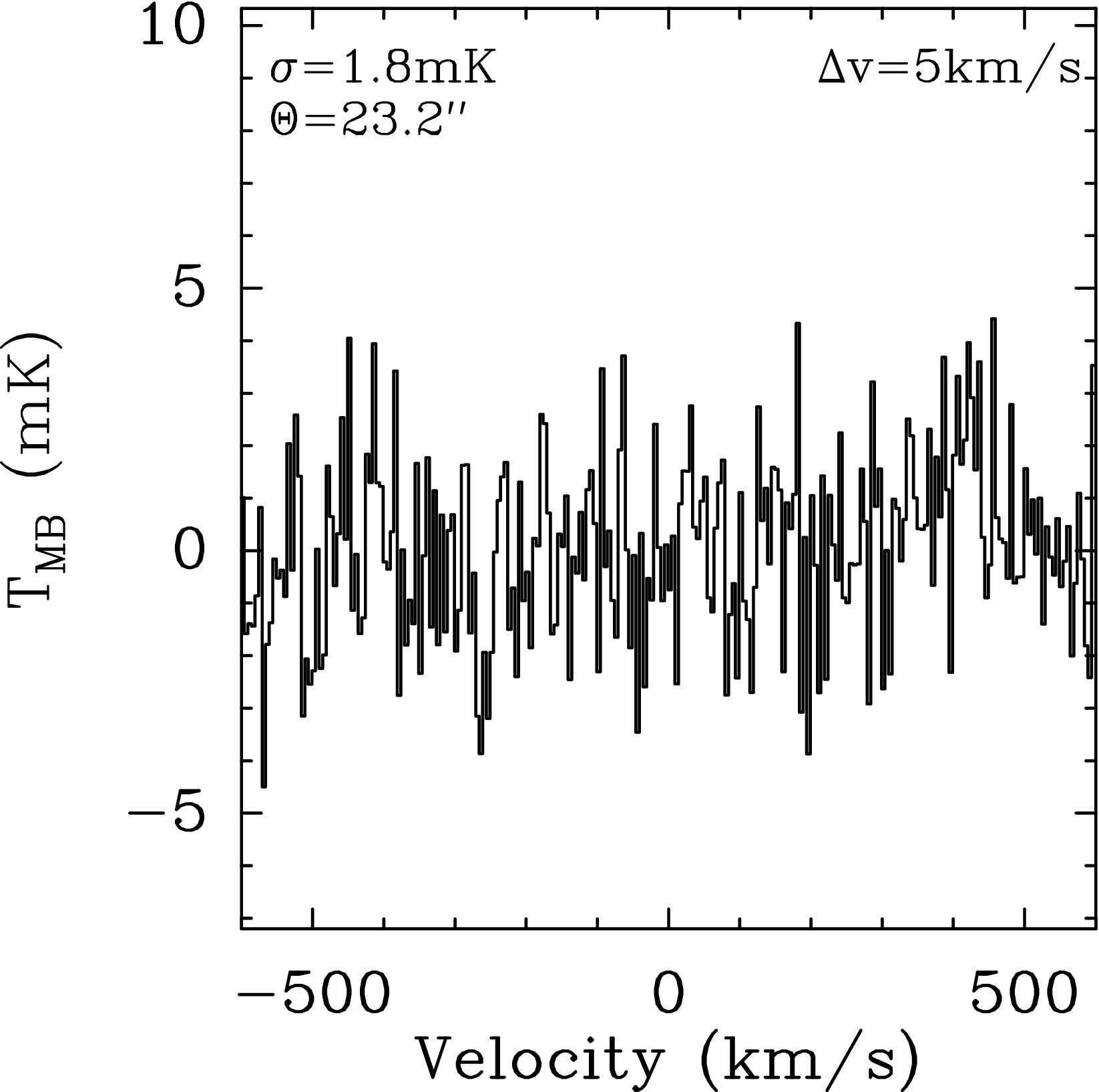}}\hspace*{\fill}
   \subfloat{\includegraphics[width=3.7cm]{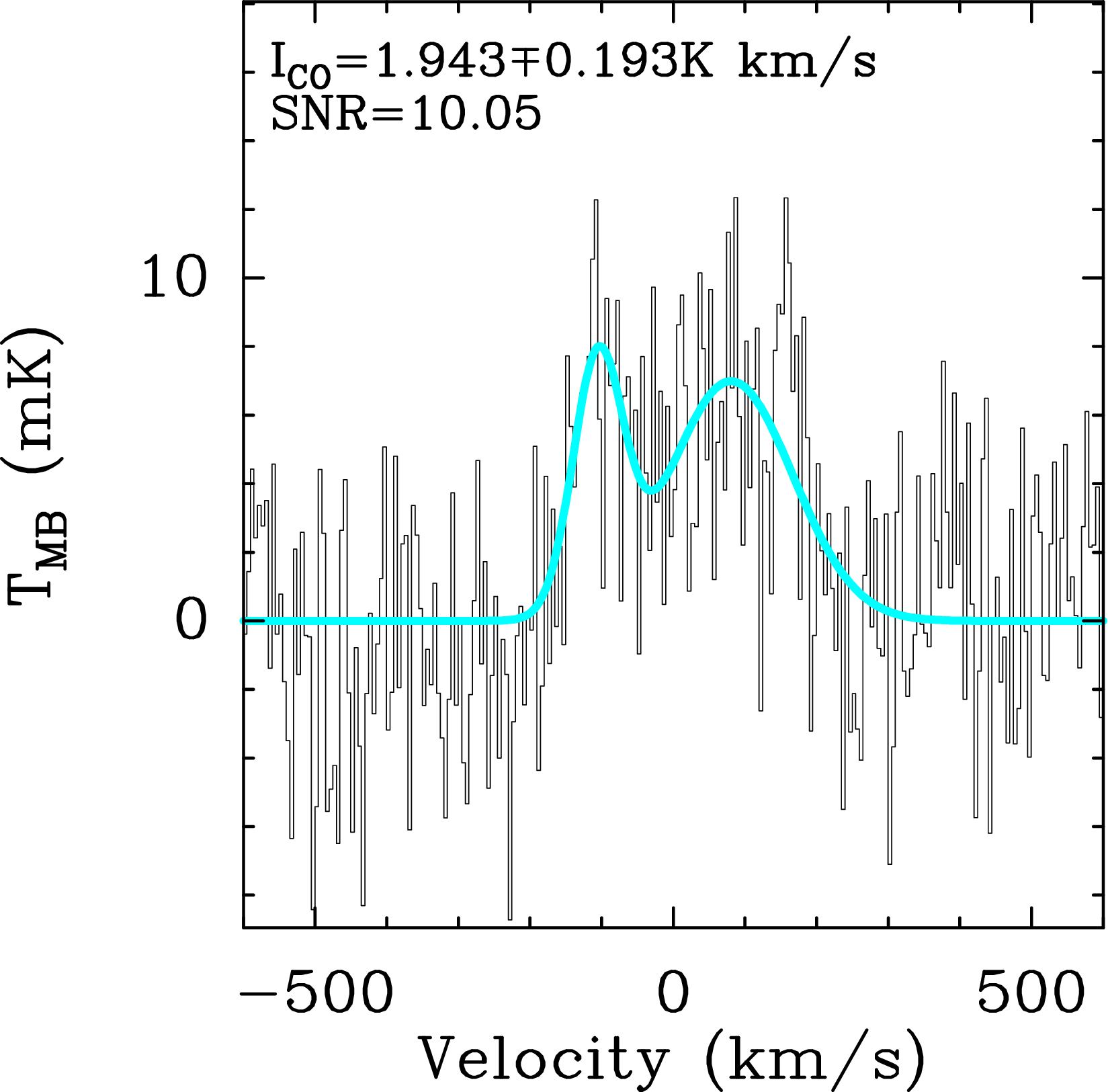}}\hspace*{\fill}
   \subfloat{\includegraphics[width=3.7cm]{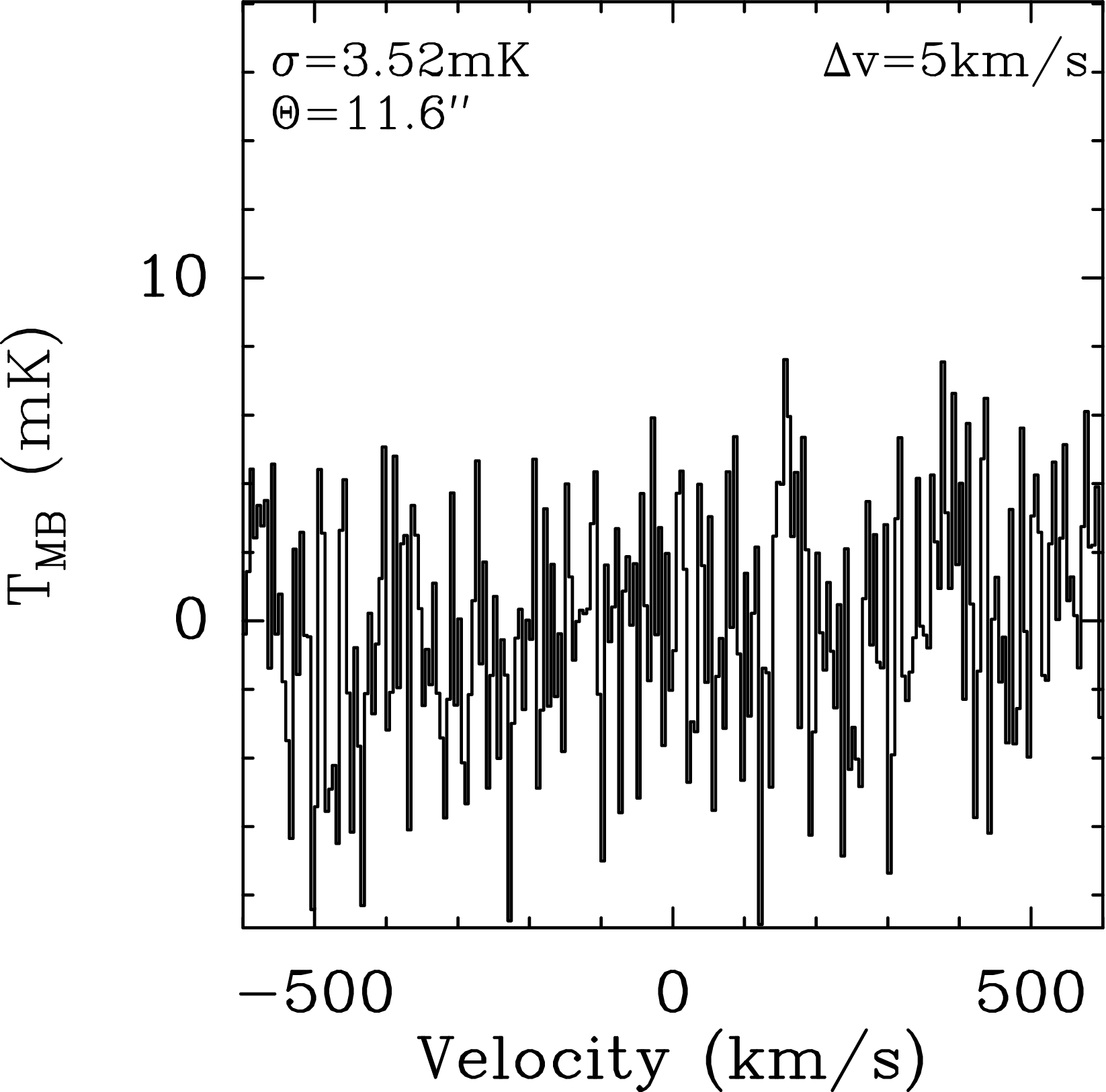}}\hspace*{\fill}

   \subfloat[(1) H$\upalpha$ \& HPBW]{\includegraphics[width=3.1cm]{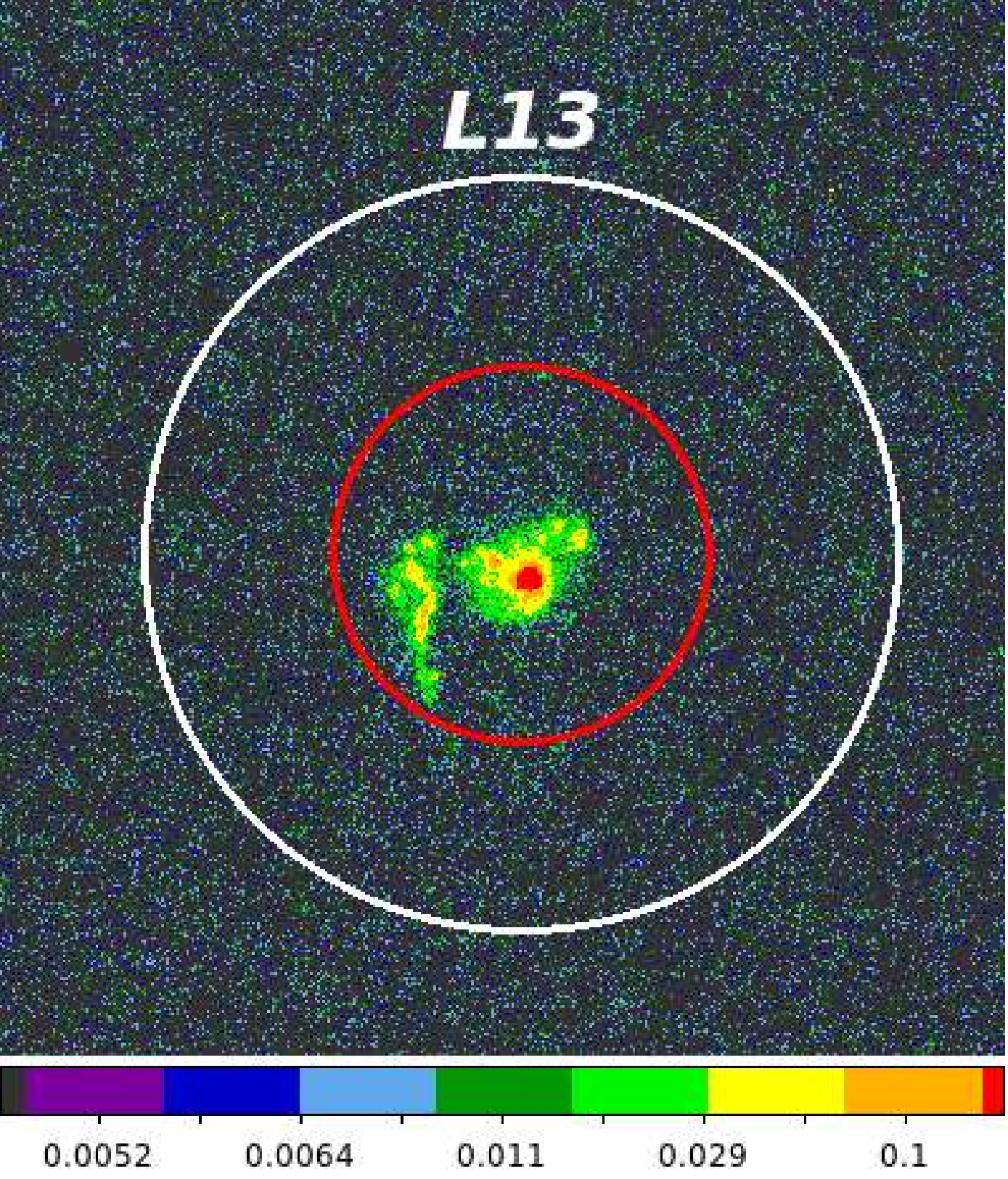}}\hspace*{\fill}
   \subfloat[\ \ \ \ \ \ \ \ \ \ \ \ \ (2) CO (1-0)]{\includegraphics[width=3.7cm]{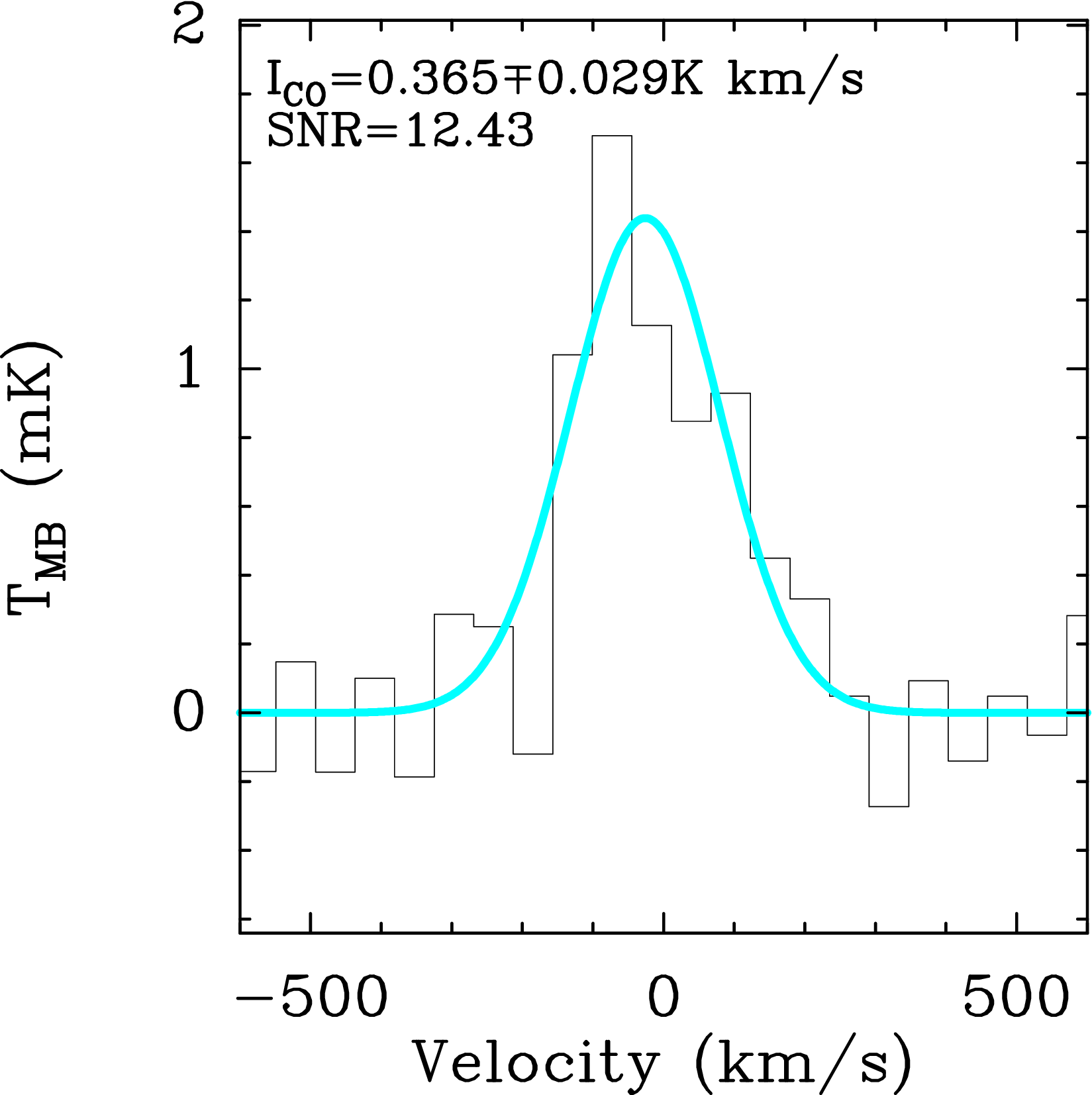}}\hspace*{\fill}
   \subfloat[\ \ \ \ \ \ \ \ \ \ \ \ \ (3) residual]{\includegraphics[width=3.7cm]{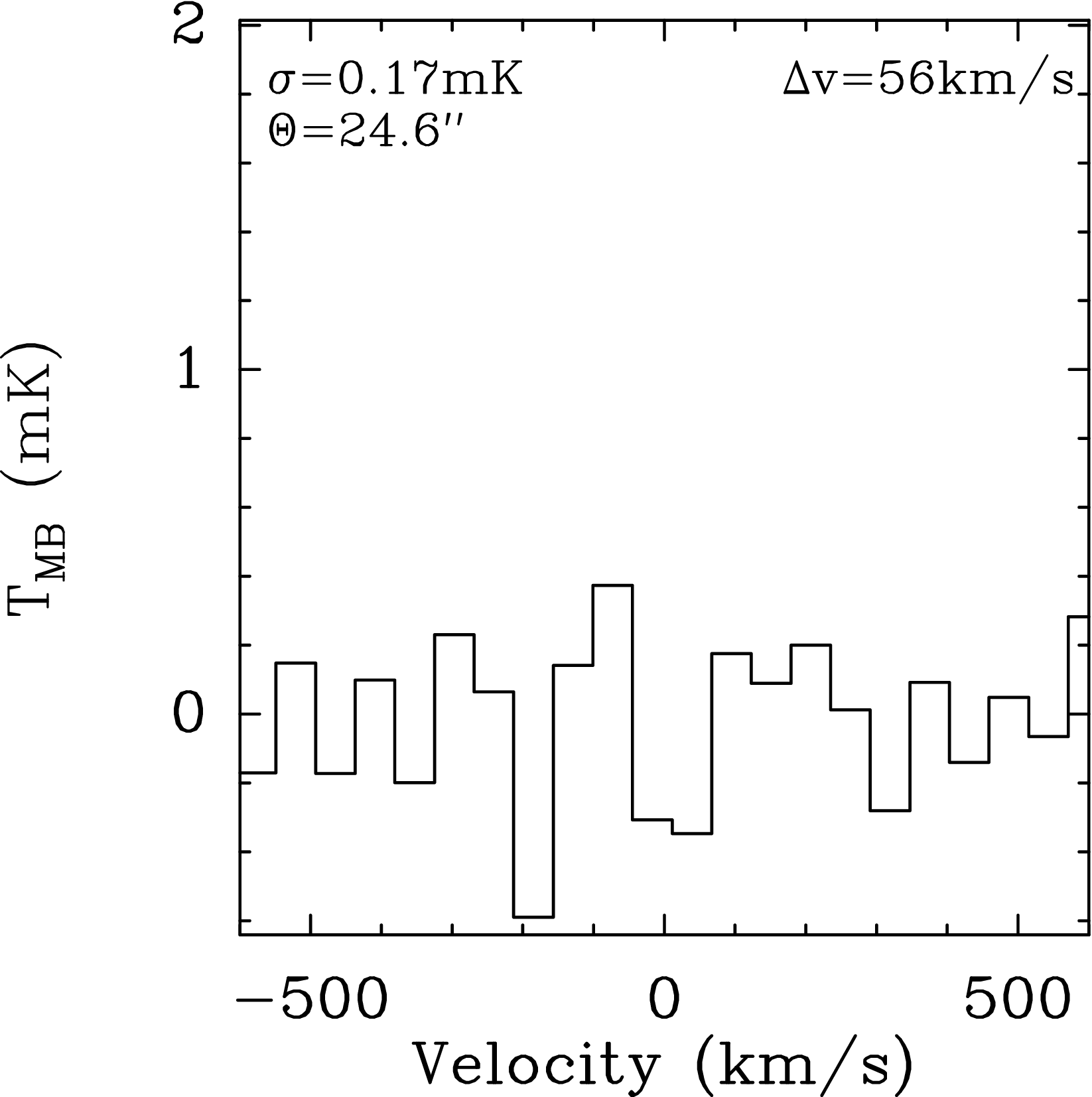}}\hspace*{\fill}
   \subfloat[\ \ \ \ \ \ \ \ \ \ \ \ \ (4) CO (2-1)]{\includegraphics[width=3.7cm]{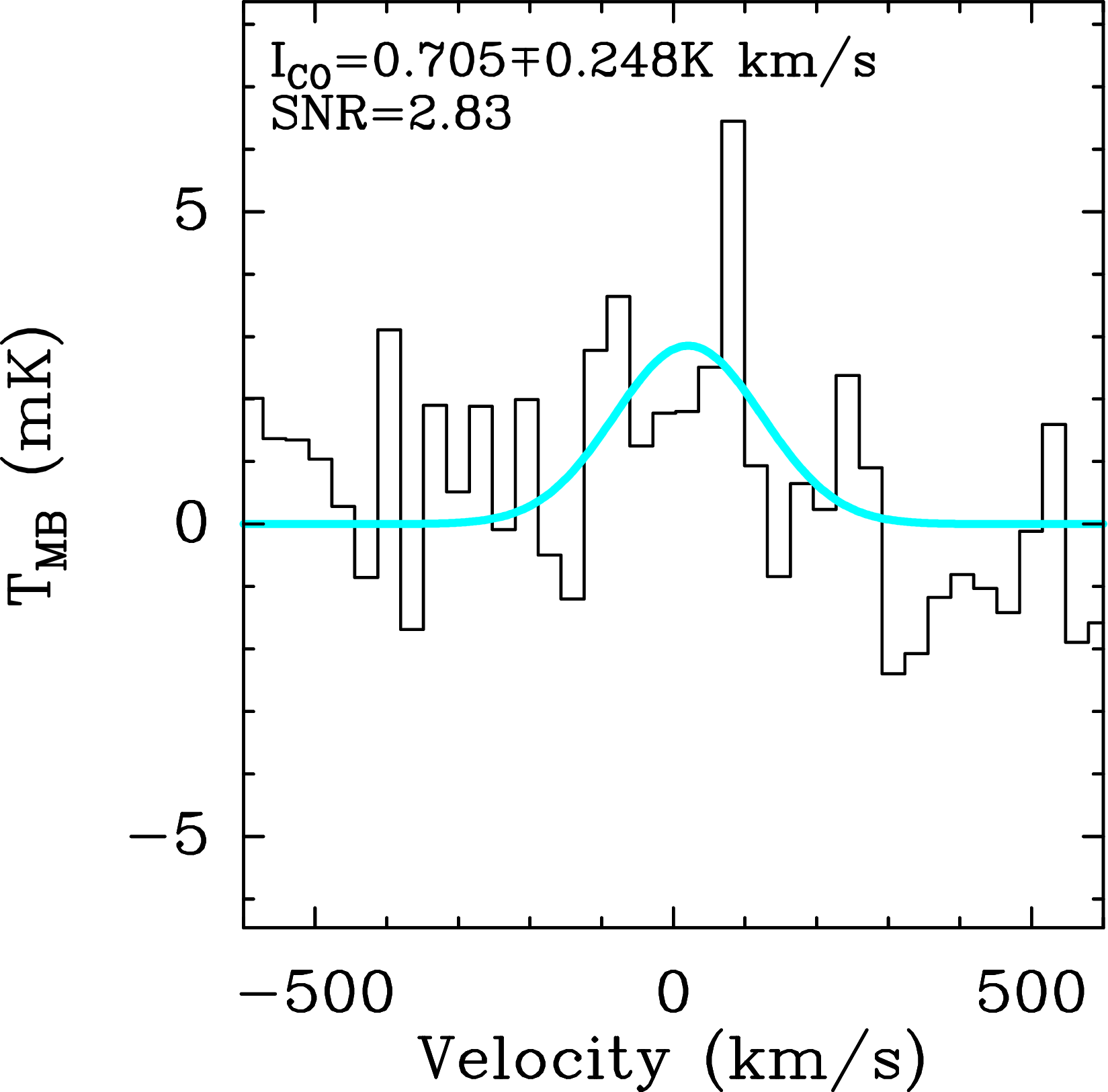}}\hspace*{\fill}
   \subfloat[\ \ \ \ \ \ \ \ \ \ \ \ \ (5) residual]{\includegraphics[width=3.7cm]{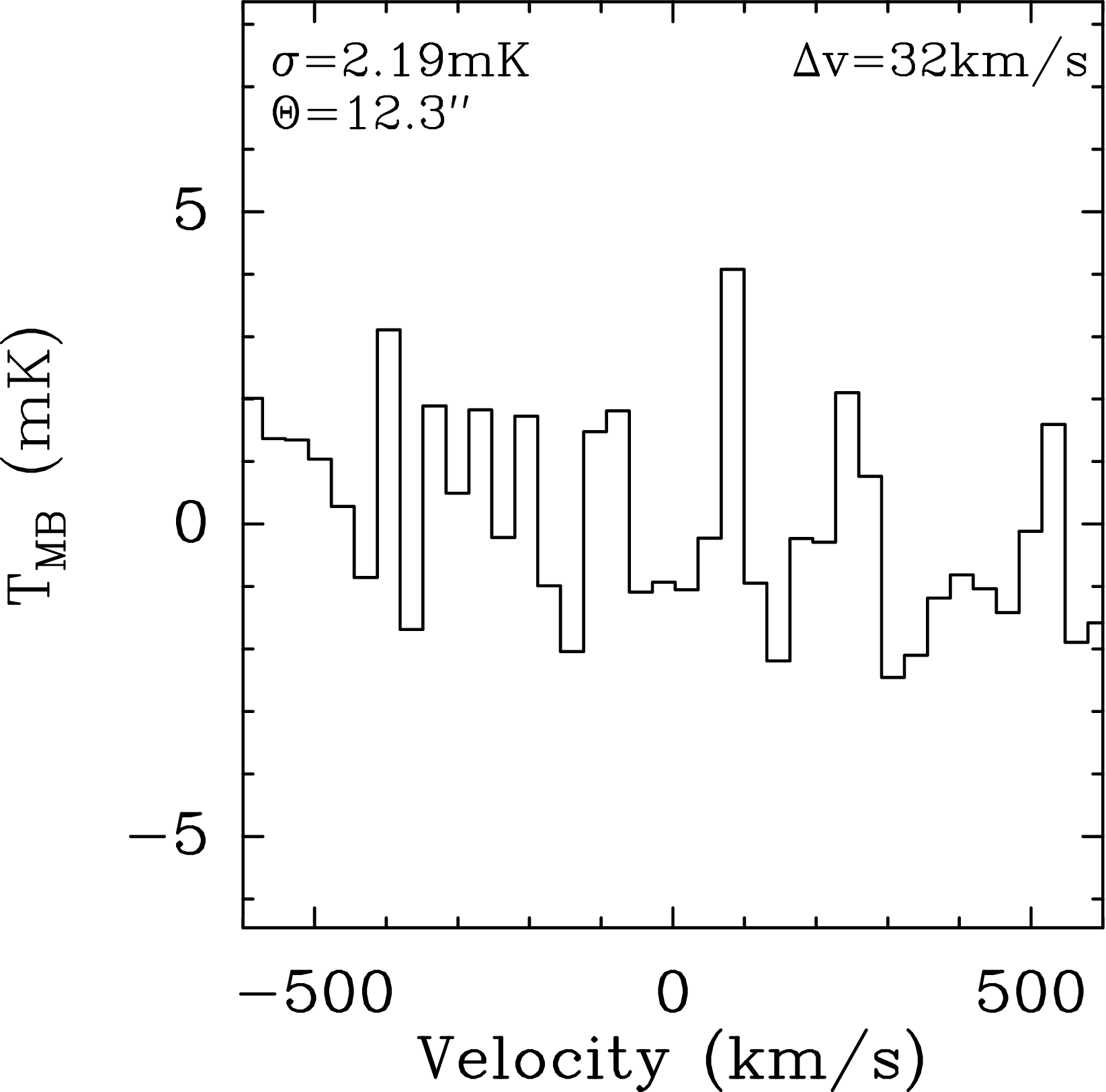}}\hspace*{\fill}
   
   \caption{CO (1-0) and (2-1) emission lines of LARS galaxies obtained with the IRAM 30m telescope, shown on the T$_{MB}$ scale in mK.
   The beamsizes for the transitions are shown in column (1)
   on top of our HST H$\upalpha$ narrowband images: \textit{white:} CO (1-0) and \textit{red:} CO (2-1). CO (1-0) and CO (2-1) spectra are shown in
   columns (2) and (4) respectively. Single- or double-Gaussian fits were performed for calculating the line flux. They are shown as cyan solid lines.
   The residuals after subtraction of the fits from the data are given in columns (3) and (5).}
   
   \label{fig:COLARS_IRAM}
\end{figure*}

Five galaxies out of our sample were
detected (simultaneously) in CO~(1-0) and CO~(2-1) using the \textit{Eight Mixer Receiver} (EMIR).
The observed CO lines are shown in Figure \ref{fig:COLARS_IRAM}
on the main beam brightness temperature scale. Beamsizes, main beam efficiencies,
Gaussian fit results and flux densities per beam are summarized in Table \ref{tab:COLARS_IRAM}.
Note that we also detected CO~(1--0) in the companion galaxy of LARS~11. As it is not part of our sample,
no analysis was performed.
Six LARS galaxies remained undetected down to a baseline r.m.s. level of $\sim$1mK@20km/s,
corresponding to a flux density of roughly 5mJy per CO~(1-0) beam (see Table \ref{tab:NONDECT_IRAM} and Figure \ref{fig:NONDECT_IRAM}).
Additionally, in the two galaxies with the strongest CO emission, LARS 3 and 8,
HCN and HCO$^+$~(1-0) were detected, however at signal-to-noise ratios of $\sim$3--6 only.

Additional CO~(3-2) observations of two galaxies were carried out with APEX,
the \textit{Atacama Pathfinder Experiment} \citep{APEX} within the Swedish time allocation
during periods P94 and P96 under project codes F-9340A-2014 and F-9329A-2015.
LARS~8 was detected in CO~(2-1) and (3-2) and LARS~13 was detected in CO~(3-2) using
the \textit{Swedish heterodyne Facility Instrument} \citep{APEX_SHFI2,APEX_SHFI1}.

All molecular scans (IRAM 30m and APEX) were obtained in wobbler-switching mode and the
calibrated data was reduced using \texttt{GILDAS/CLASS}\footnote{\url{http://www.iram.fr/IRAMFR/GILDAS}},
a collection of state-of-the-art software oriented toward (sub-)millimeter radioastronomical applications.
IRAM scans obtained with the Fast Fourier Transform Spectrometer (FTS) unit were corrected for platforming.
In our pipeline, we initially chose a wide constant line window ranging from -300 to +300~km/s around the expected zero-velocity position
of the line, extract a range of -1000 to +1000 km/s and applied a linear baseline correction on each
individual scan. In the next step, we masked spikes and then examined plots of system temperature and r.m.s. noise
for all scans, which we used to filter scans that do not follow the radiometer equation.
Scans that were obtained under conditions of high precipitable water vapor, i.e. when the atmospheric transmission
at the given frequency was lower than $\sim$33 percent, were excluded from further processing.
The remaining scans per galaxy were average stacked and re-sampled to a resolution of 1~km/s.
From that point on, each galaxy was manually processed using individual
line windows, baseline fits and smoothing to resolutions between 5 and 56~km/s per channel.
Gaussian fits were performed to reveal velocity components. Total fluxes were
derived via summation of channels along the line window. Errors on the fluxes are the r.m.s. times
the number of channels along the window.

\begin{table*}
\resizebox{\textwidth}{!}{%
\begin{threeparttable}
\caption{Summary of the detected molecular lines}
\label{tab:COLARS_IRAM}
\begin{filecontents*}{detections_iram_v2.csv}
ID,line,B$_{eff}$,$\theta$,t$_{on}$,v$_{peak}$,FWHM,SNR,I$_{mb}$,S
,,,[arcsec],[h],[km/s],[km/s],,[K km/s],[Jy km/s]
3,CO(1-0),0.79,22.10,0.6, 49.6 [17.6] -129.8 [26.4], 157.7 [29.35] 190.7 [41.43],20.0,7.13$\pm$0.36,35.6$\pm$1.8
3,CO(2-1),0.58,11.00,0.6, -127.2 [10] 53.1 [7.9], 178.3 [16.13] 152.9 [12.2],37.4,15.93$\pm$0.43,79.6$\pm$2.1
3,HCN(1-0),0.84,28.70,1.3, -40.3 [51.5], 623.7 [126.79],7.0,1.06$\pm$0.15,5.3$\pm$0.8
3,HCO+(1-0),0.84,28.60,1.3, -64.4 [56.7], 556.4 [136.97],6.1,1.08$\pm$0.18,5.4$\pm$0.9
8,CO(1-0),0.80,22.20,3.2, -75.8 [2.8] 108.4 [3.9], 127 [6.17] 144.7 [8.16],43.4,6.85$\pm$0.16,34.2$\pm$0.8
8,CO(2-1),0.59,11.10,3.4, -68 [0.2] 118.9 [2.6], 139.1 [4.04] 114.9 [5.82],37.7,11.7$\pm$0.31,58.5$\pm$1.6
8,CO(3-2),0.73,18.50,0.9, -59.8 [4.4] 124.6 [4.7], 148.7 [10.45] 118.5 [10.19],29.8,6.5$\pm$0.22,203.1$\pm$6.8
8,HCN(1-0),0.84,28.90,3.2, -121.9 [26.5], 151.8 [47.73],4.0,0.29$\pm$0.07,1.5$\pm$0.4
8,HCO+(1-0),0.84,28.70,3.2, -30.9 [36.1] 0 [0], 216.4 [98.46],3.2,0.19$\pm$0.06,0.9$\pm$0.3
9,CO(1-0),0.80,22.40,6.9, 90 [5.6], 143.1 [17.01],9.9,0.66$\pm$0.07,3.3$\pm$0.3
9,CO(2-1),0.59,11.20,4.5, 97.1 [19.5], 130.9 [40.19],3.0,0.51$\pm$0.17,2.6$\pm$0.8
11,CO(1-0),0.80,23.20,3.1, -64.9 [15.8] 139.1 [14.8], 165.6 [29.89] 149.2 [39.68],15.5,1.53$\pm$0.1,7.7$\pm$0.5
11,CO(2-1),0.60,11.60,1.4, -85 [3.6] 101.8 [9.7], 123.6 [16.35] 159.6 [16.24],10.1,1.94$\pm$0.19,9.7$\pm$1.0
13,CO(1-0),0.82,24.60,4.0, -20.2 [21.3], 256.1 [50.78],11.3,0.38$\pm$0.03,1.9$\pm$0.2
13,CO(2-1),0.63,12.30,3.3, 42.6 [33], 242.9 [79.22],6.1,0.62$\pm$0.1,3.1$\pm$0.5
13,CO(3-2),0.73,20.50,5.4, -11 [24.1], 320.8 [65.29],6.2,0.52$\pm$0.08,16.2$\pm$2.6
\end{filecontents*}
\csvreader[no head,column count=10,tabular=rlcclrrrrr,table head=\hline\hline,late after line=\\]{detections_iram_v2.csv}
  {1=\one,2=\two,3=\three,4=\four,5=\five,6=\six,7=\seven,8=\eight,9=\nine,10=\ten}
  {\one & \two & \three & \four & \five & \six & \seven & \eight & \nine & \ten}
\begin{tablenotes}
	\item For each transition, we show
	the beam efficiency B$_{eff}$ that was used to translate the antenna temperature to the T$_{mb}$ scale, the beam size $\theta$, the
	total effective on-source observing time (after processing and filtering) T$_{on}$ as well as results from Gaussian fits and fluxes.
	v$_{peak}$ is the velocity (with respect to systemic velocity of H~I) at the Gaussian peak brightness temperature with the error on
        v$_{peak}$ in square brackets (in cases with
	more than one Gaussian fit, v$_{peak}$ is given for each component), FWHM is the full-width-at-half-maximum calculated from the Gaussian
	sigma value(s) and SNR is the signal-to-noise ratio. The velocity-integrated source brightness temperatures on the main beam brightness
	temperature scale as well as velocity-integrated flux densities are indicated by $I_{mb}$ and $S$ respectively.
\end{tablenotes}
\end{threeparttable}}
\end{table*}

\begin{table}
\caption{Summary of LARS galaxies not detected in CO}
\label{tab:NONDECT_IRAM}
\begin{filecontents*}{nondetections_iram.csv}
ID,Beff,HPBW,t$_{on}$,$\sigma_{rms,20km/s}$,$\sigma_{rms,20km/s}$
,,[arcsec],[h],[mK],[mJy],,
1,0.79,22.0,3.40,1.01,5.06,,
2,0.79,22.0,1.90,1.38,6.93,,
5,0.79,22.1,3.55,0.77,3.88,,
7,0.80,22.2,5.70,0.68,3.42,,
10,0.80,22.6,1.05,1.71,8.57,,
12,0.81,23.6,2.30,0.90,4.54,,
\end{filecontents*}
\csvreader[no head,tabular=rccllrll,table head=\hline\hline,late after line=\\]{nondetections_iram.csv}
  {1=\one,2=\two,3=\three,4=\four,5=\five,6=\six}
  {\one & \two & \three & \four & \five & \six}
\end{table}


\begin{figure*}[!htbp]
   \subfloat[\ \ \ \ \ \ \ \ \ \ \ \ \ \ \ (1) LARS 08 CO (3-2)]{\includegraphics[width=4.3cm]{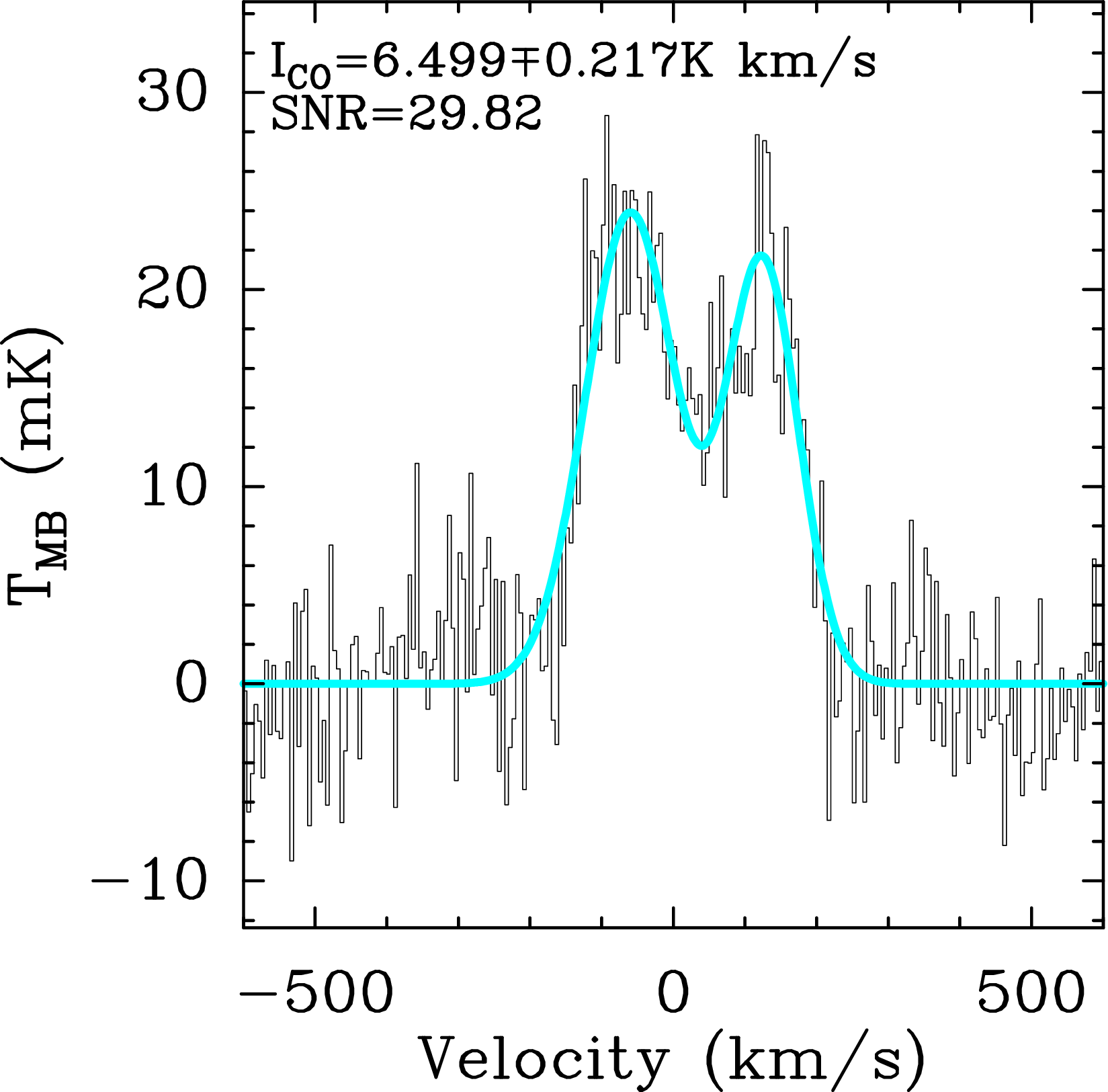}}\hspace*{\fill}
   \subfloat[\ \ \ \ \ \ \ \ \ \ \ \ \ \ \ \ (2) residual]{\includegraphics[width=4.3cm]{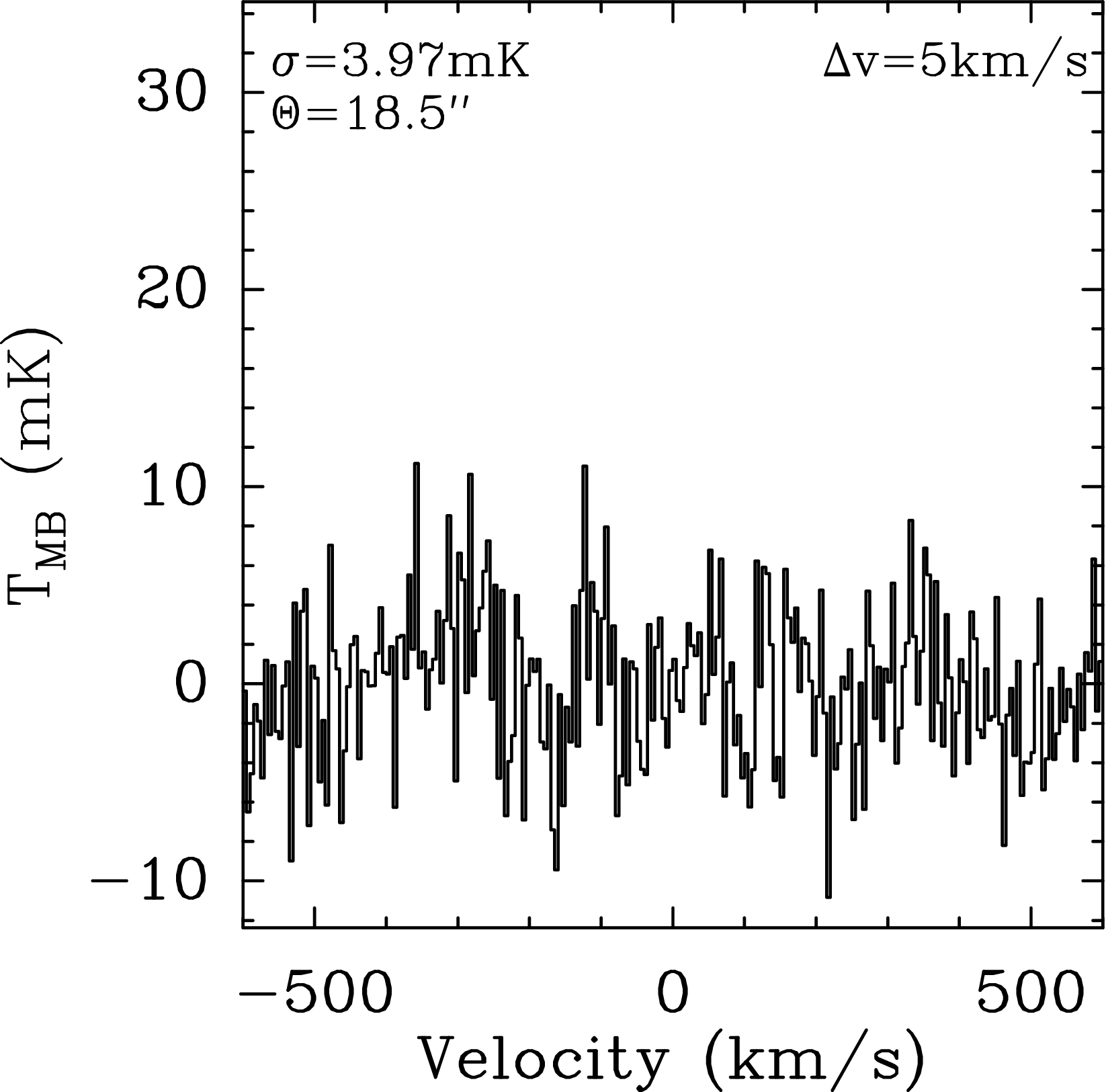}}\hspace*{\fill}
   \subfloat[\ \ \ \ \ \ \ \ \ \ \ \ \ \ \ \ \ (3) LARS 13 CO (3-2)]{\includegraphics[width=4.3cm]{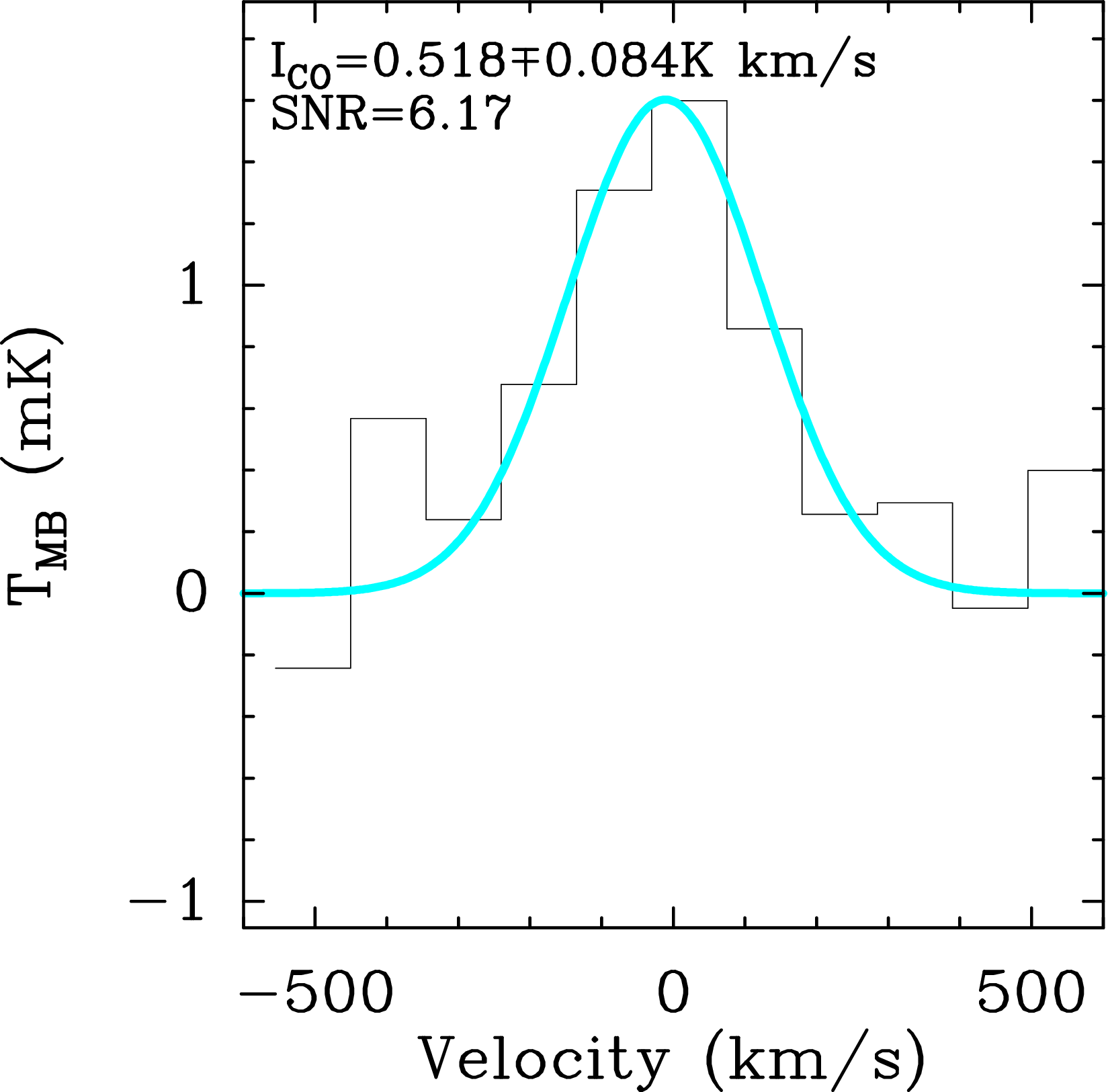}}\hspace*{\fill}
   \subfloat[\ \ \ \ \ \ \ \ \ \ \ \ \ \ \ \ (4) residual]{\includegraphics[width=4.3cm]{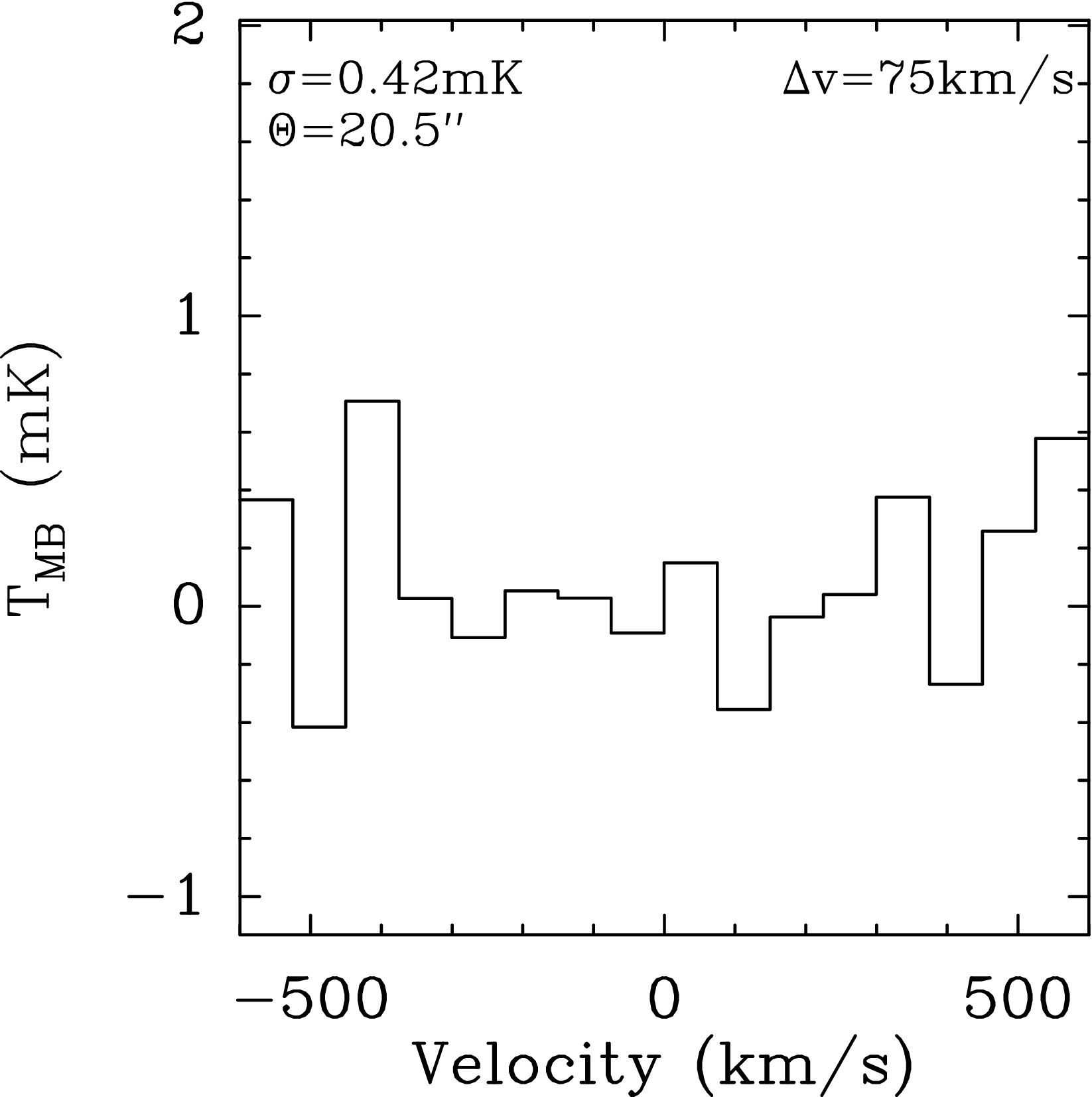}}\hspace*{\fill}

   \caption{
	Detections of CO (3-2) with APEX/SHeFI for LARS 8 (\textit{column 1}) and 13 (\textit{column 3}). Spectra are shown on the T$_{MB}$ scale in mK.
        Single- or double-Gaussian fits were performed for calculating the line flux. They are shown as cyan solid lines.
        The residuals after subtraction of the fits from the data are shown in panels (2) and (4).
   }
   
   \label{fig:COLARS_APEX}
\end{figure*}

\begin{figure*}[!htbp]
   \subfloat{\includegraphics[width=3.8cm]{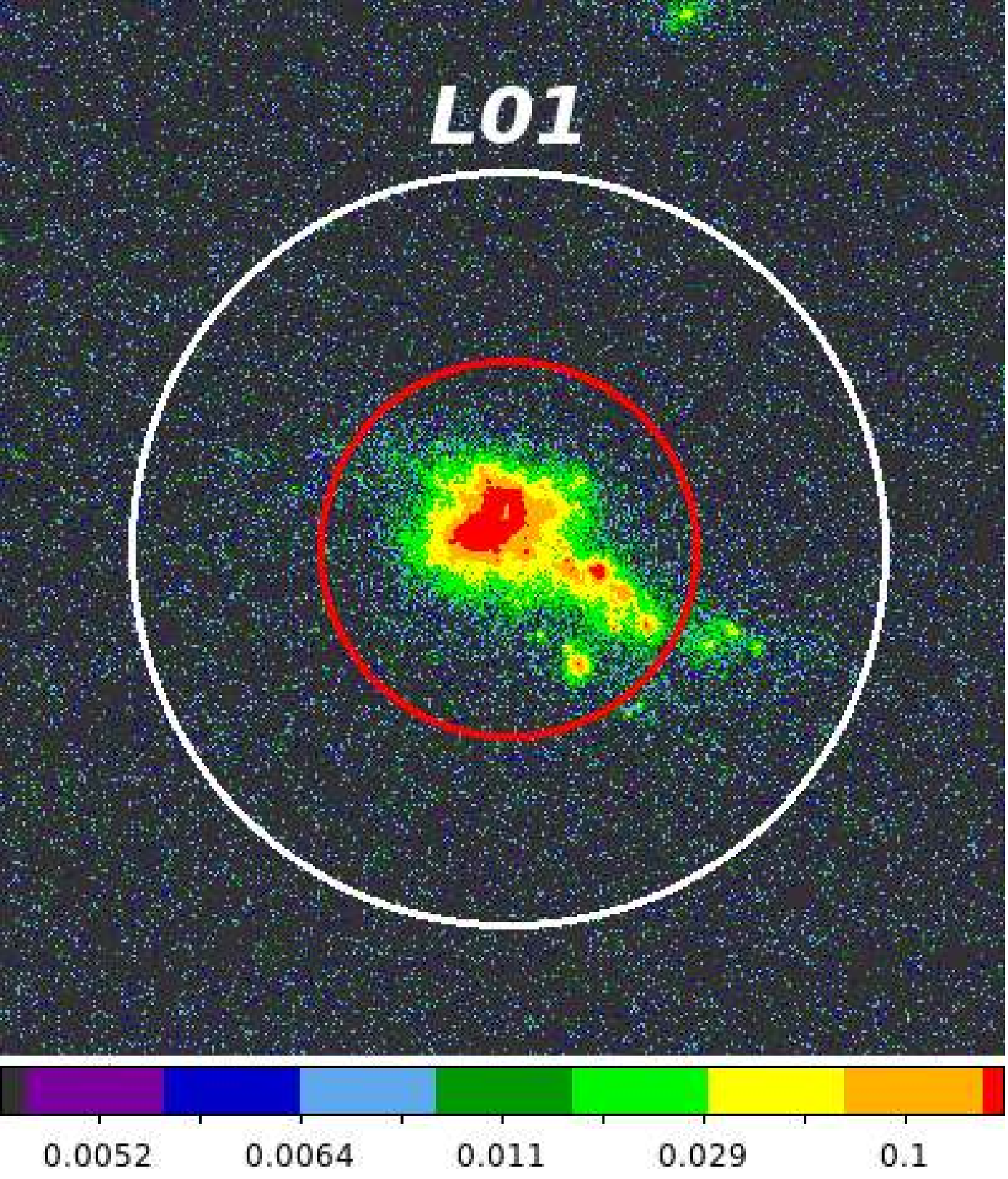}}\hspace*{\fill}
   \subfloat{\includegraphics[width=4.5cm]{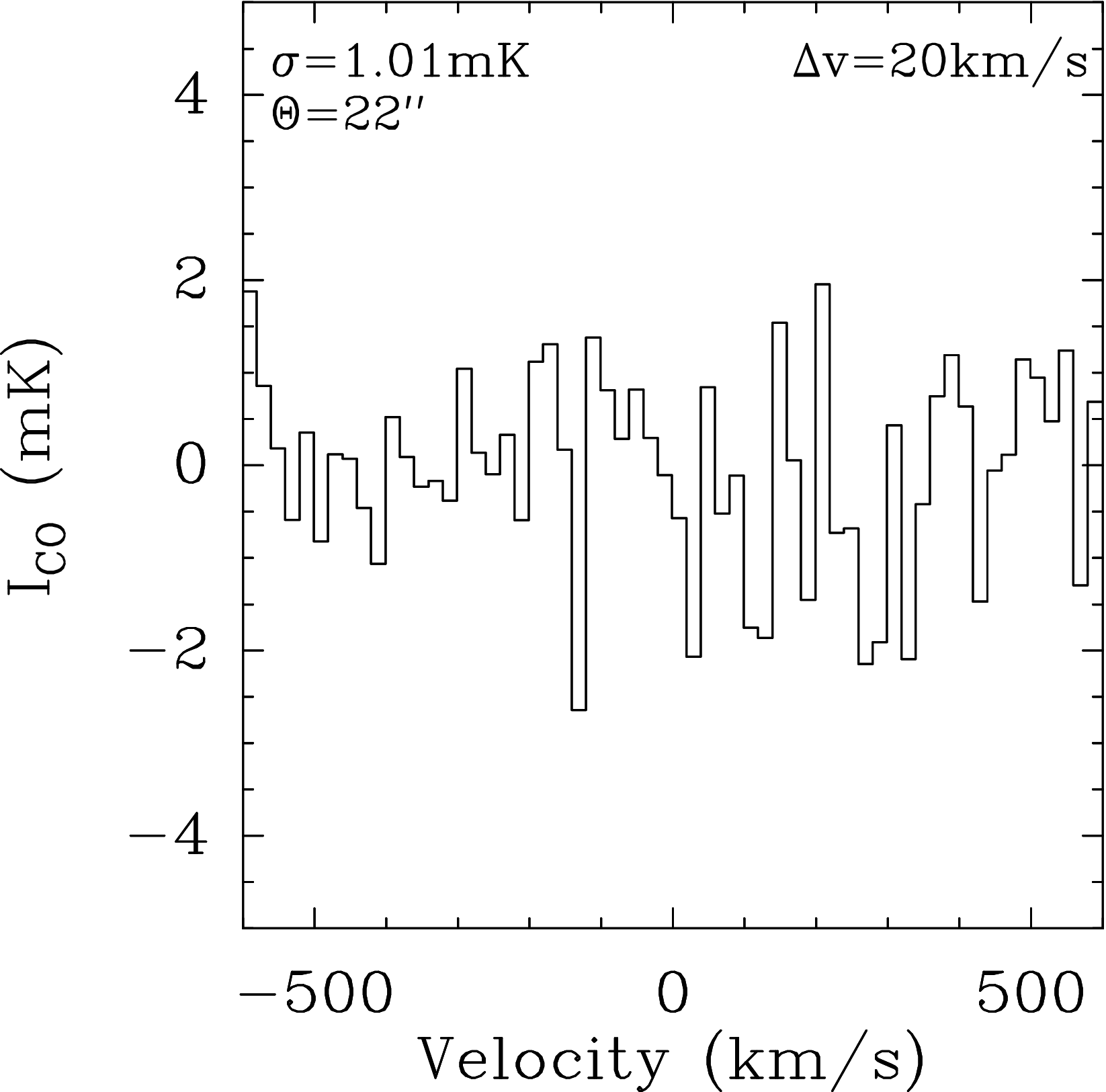}}\hspace*{\fill}
   \subfloat{\includegraphics[width=3.8cm]{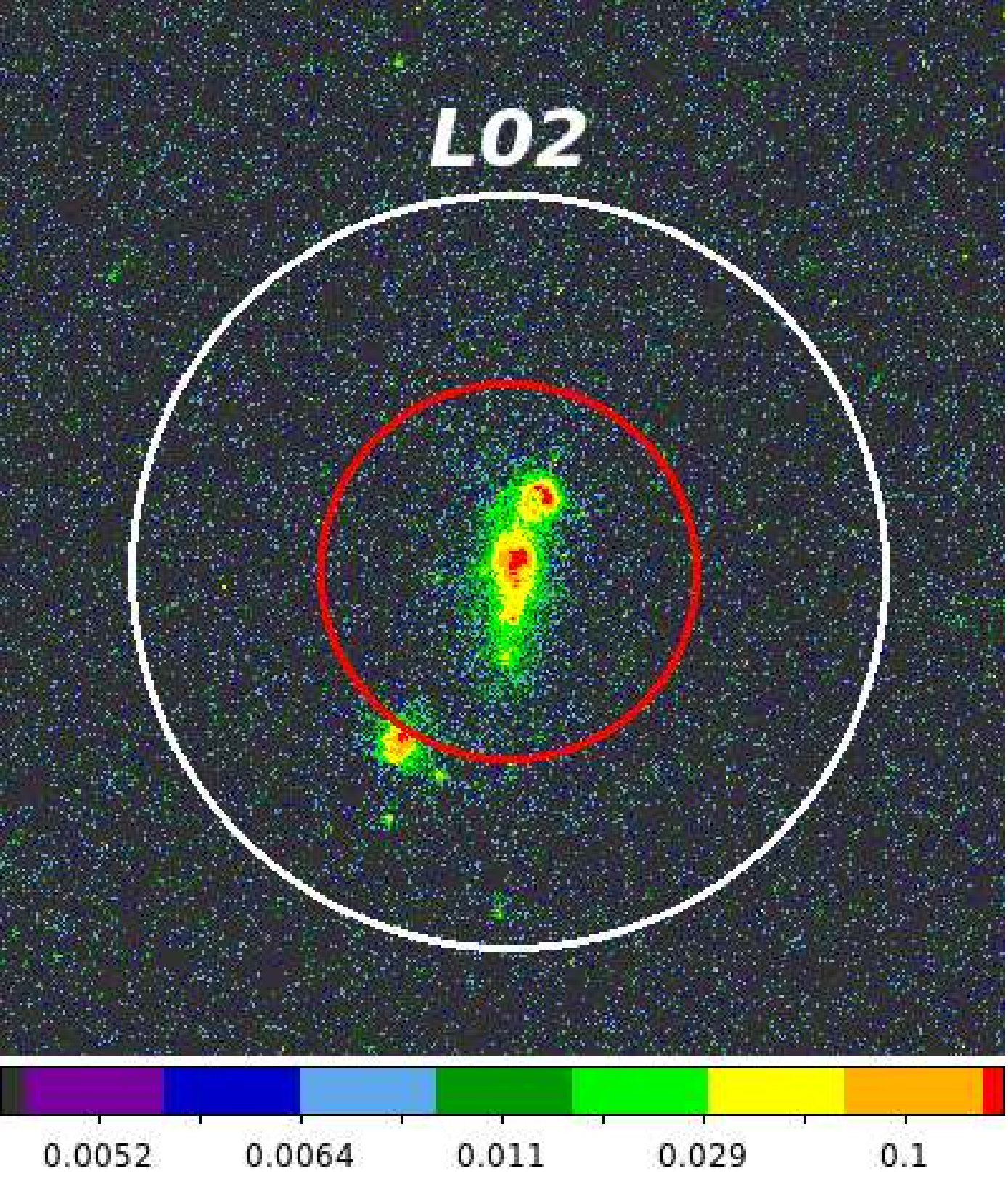}}\hspace*{\fill}
   \subfloat{\includegraphics[width=4.5cm]{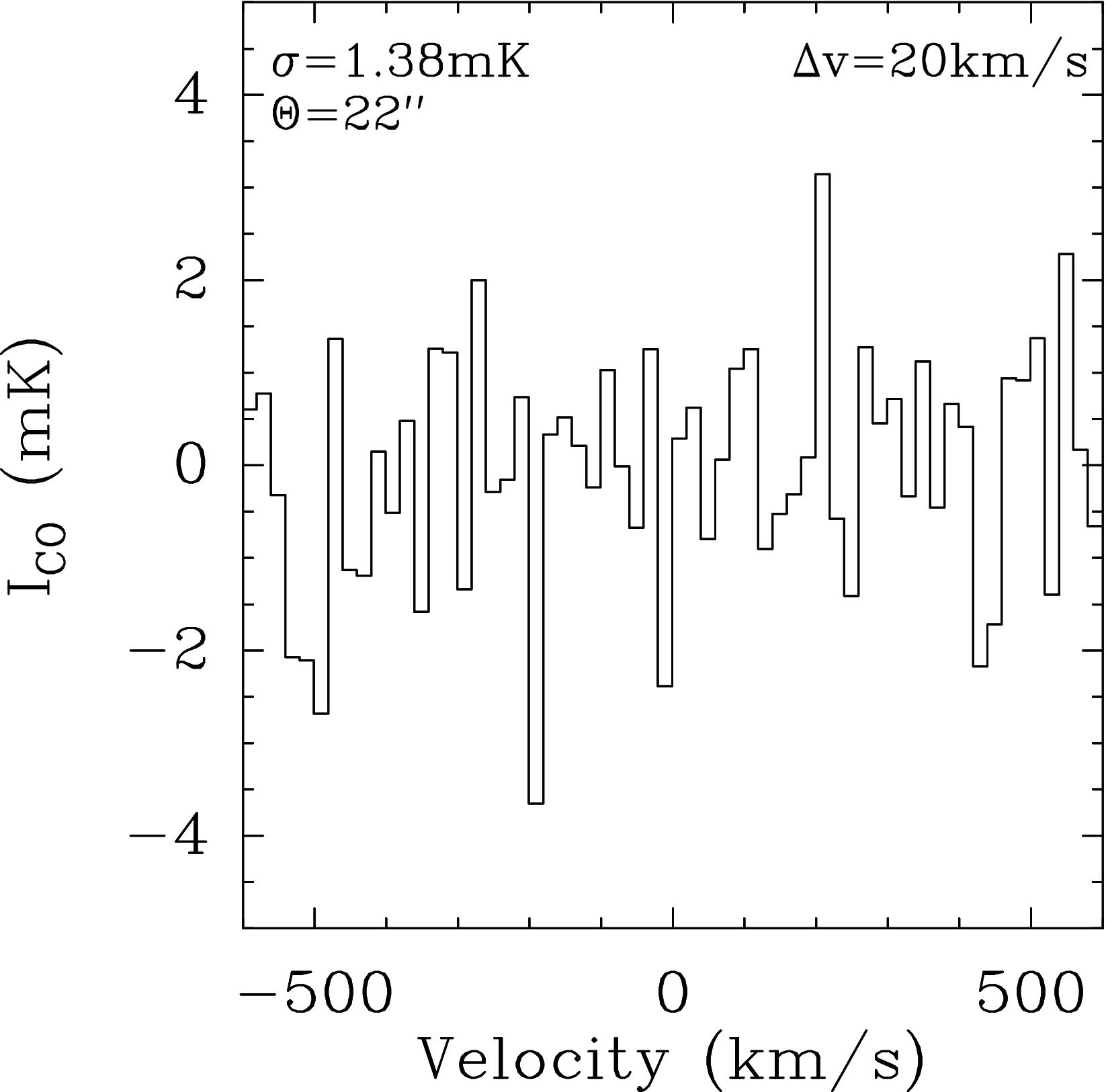}}\hspace*{\fill}

   \subfloat{\includegraphics[width=3.8cm]{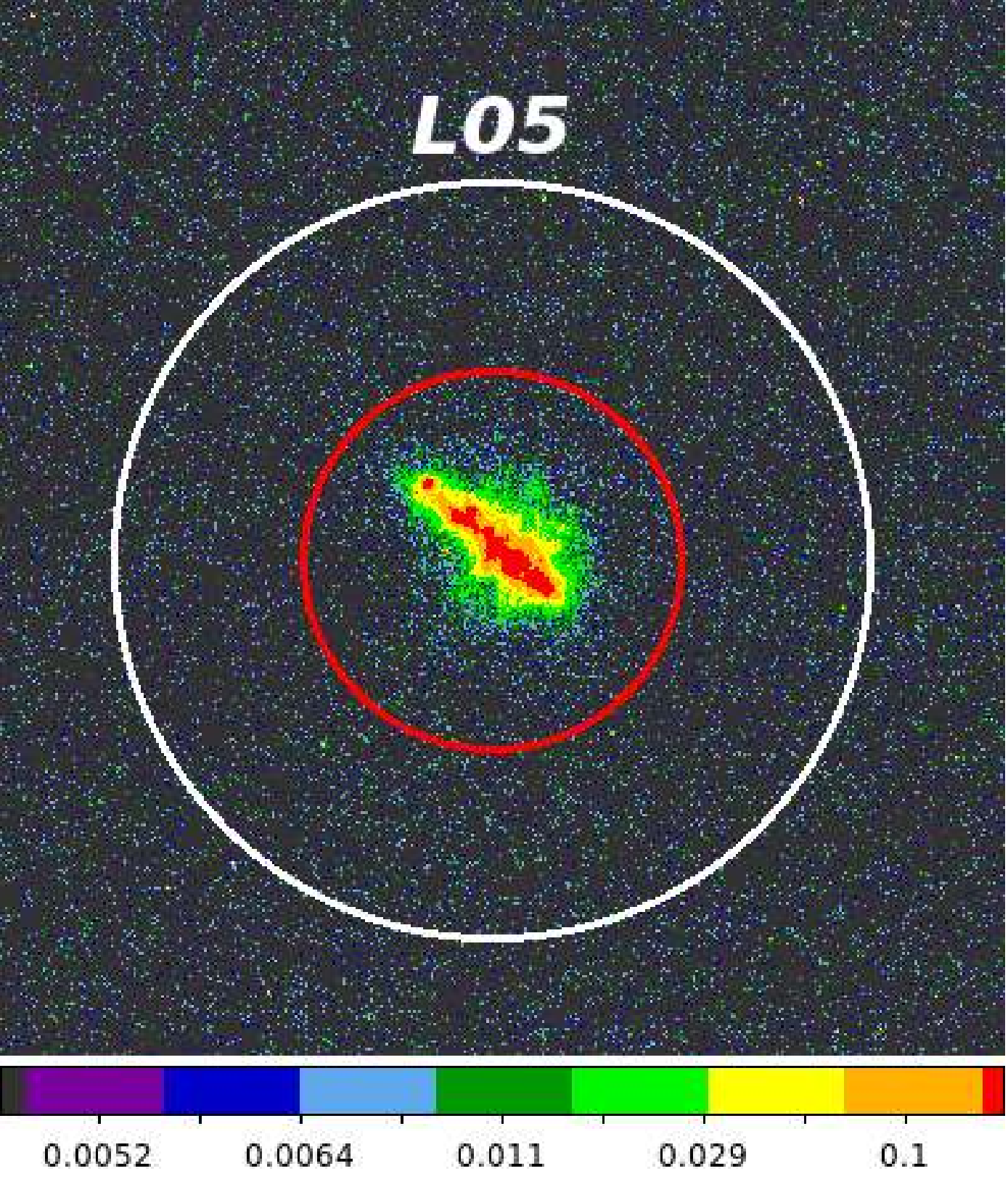}}\hspace*{\fill}
   \subfloat{\includegraphics[width=4.5cm]{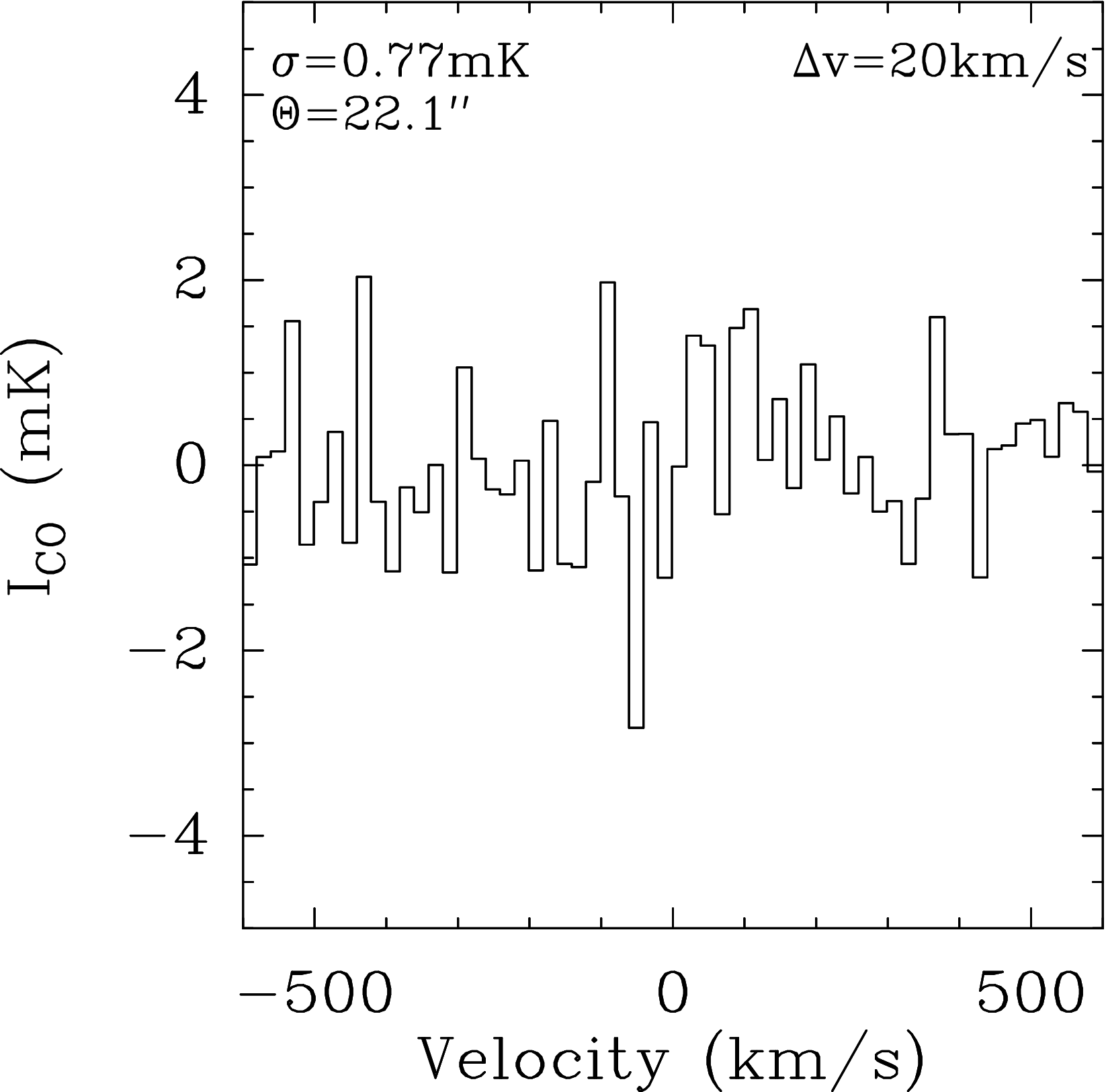}}\hspace*{\fill}
   \subfloat{\includegraphics[width=3.8cm]{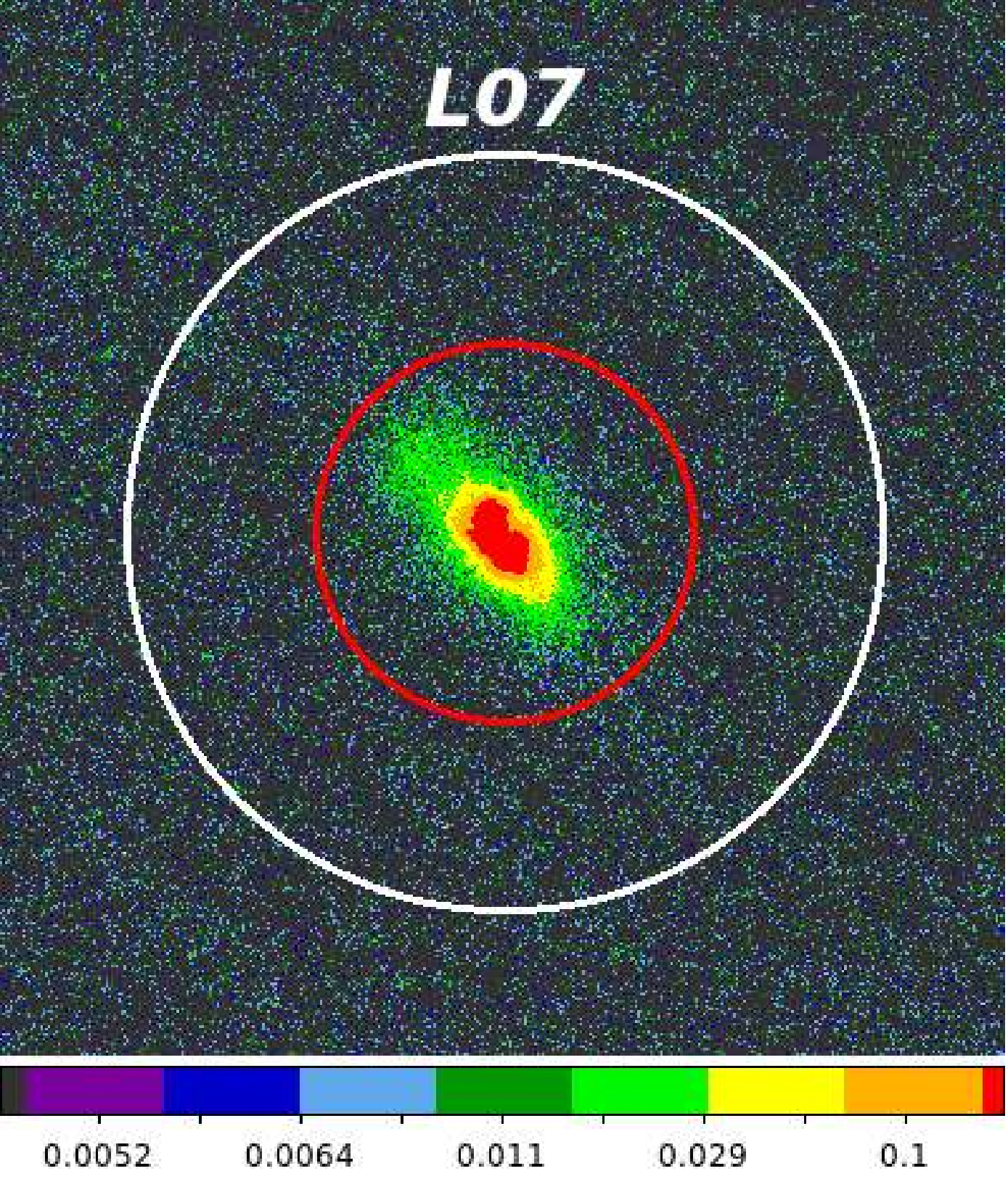}}\hspace*{\fill}
   \subfloat{\includegraphics[width=4.5cm]{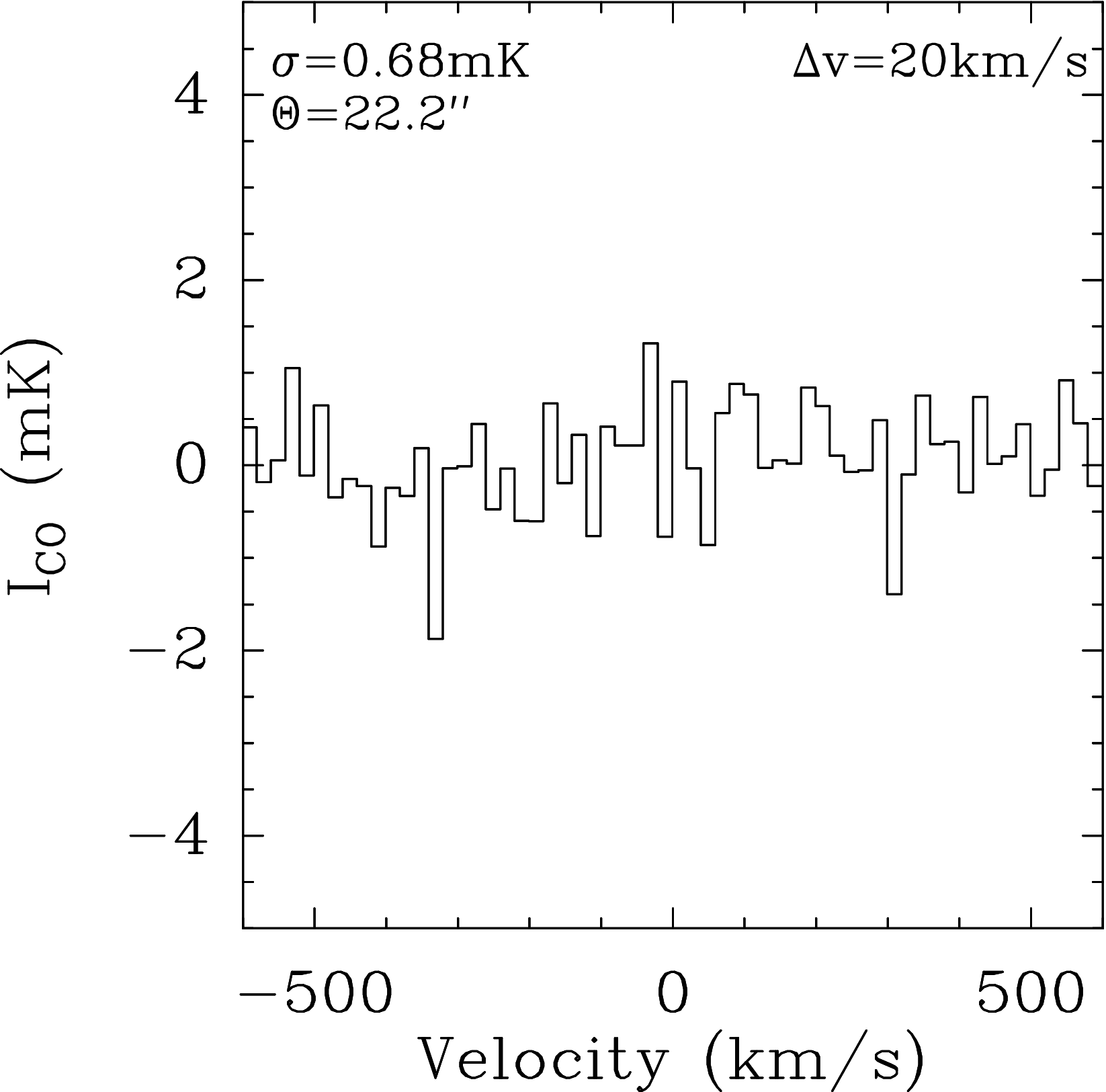}}\hspace*{\fill}

   \subfloat[(1) H$\upalpha$ \& HPBW]{\includegraphics[width=3.8cm]{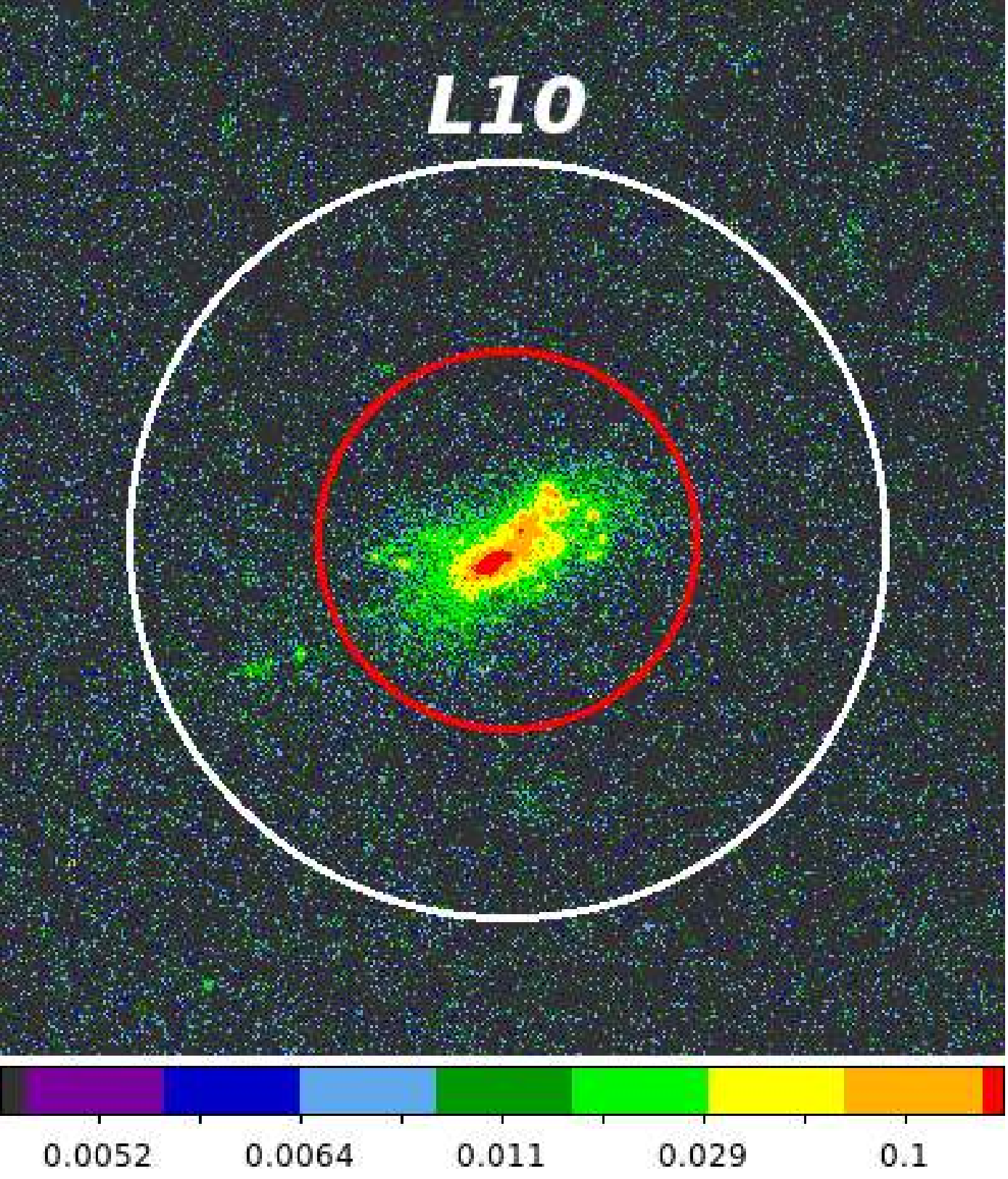}}\hspace*{\fill}
   \subfloat[\ \ \ \ \ \ \ \ \ \ \ (2) T$_{MB}$]{\includegraphics[width=4.5cm]{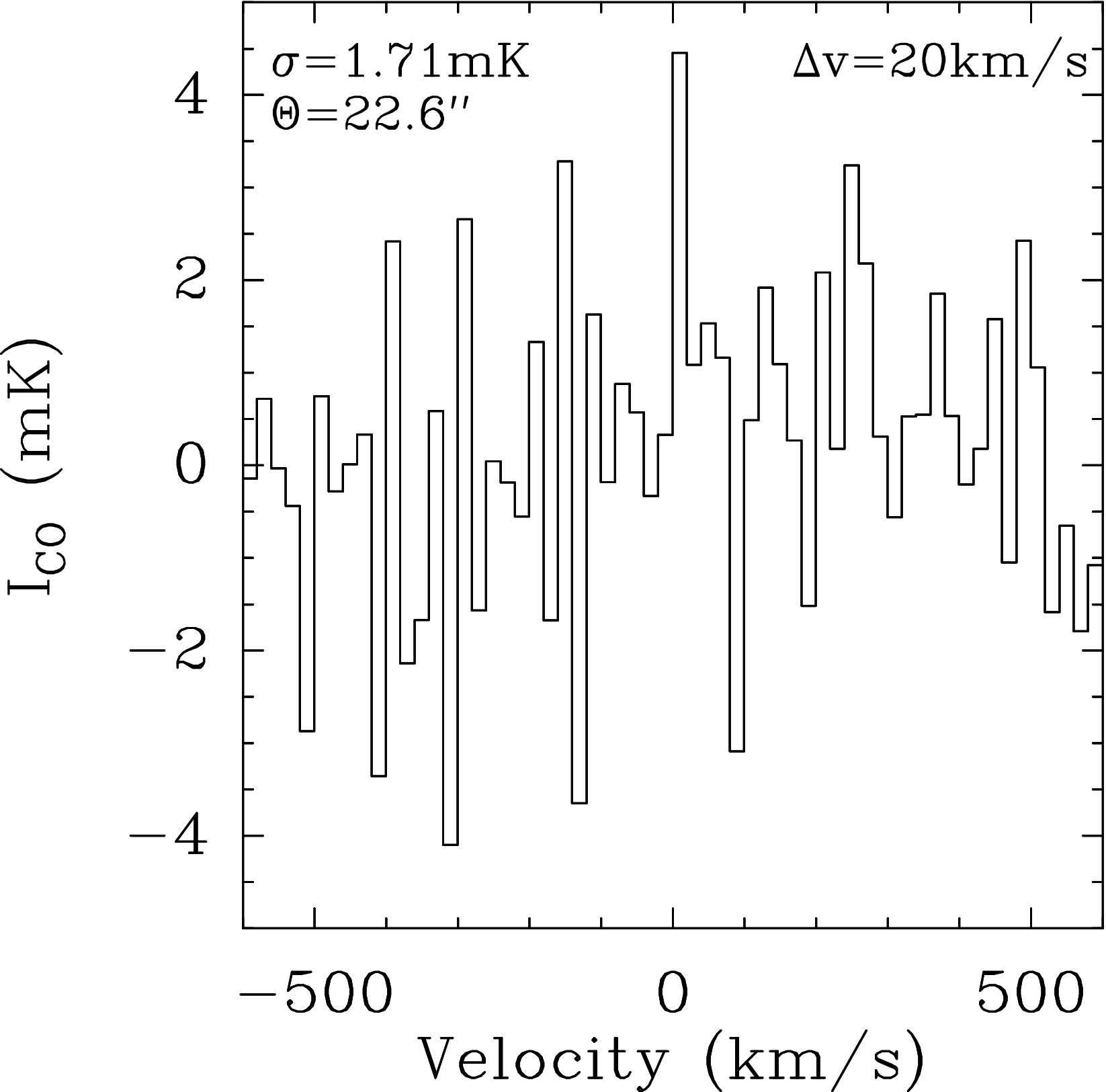}}\hspace*{\fill}
   \subfloat[\ \ \ \ \ \ \ \ \ \ \ (3) H$\upalpha$ \& HPBW]{\includegraphics[width=3.8cm]{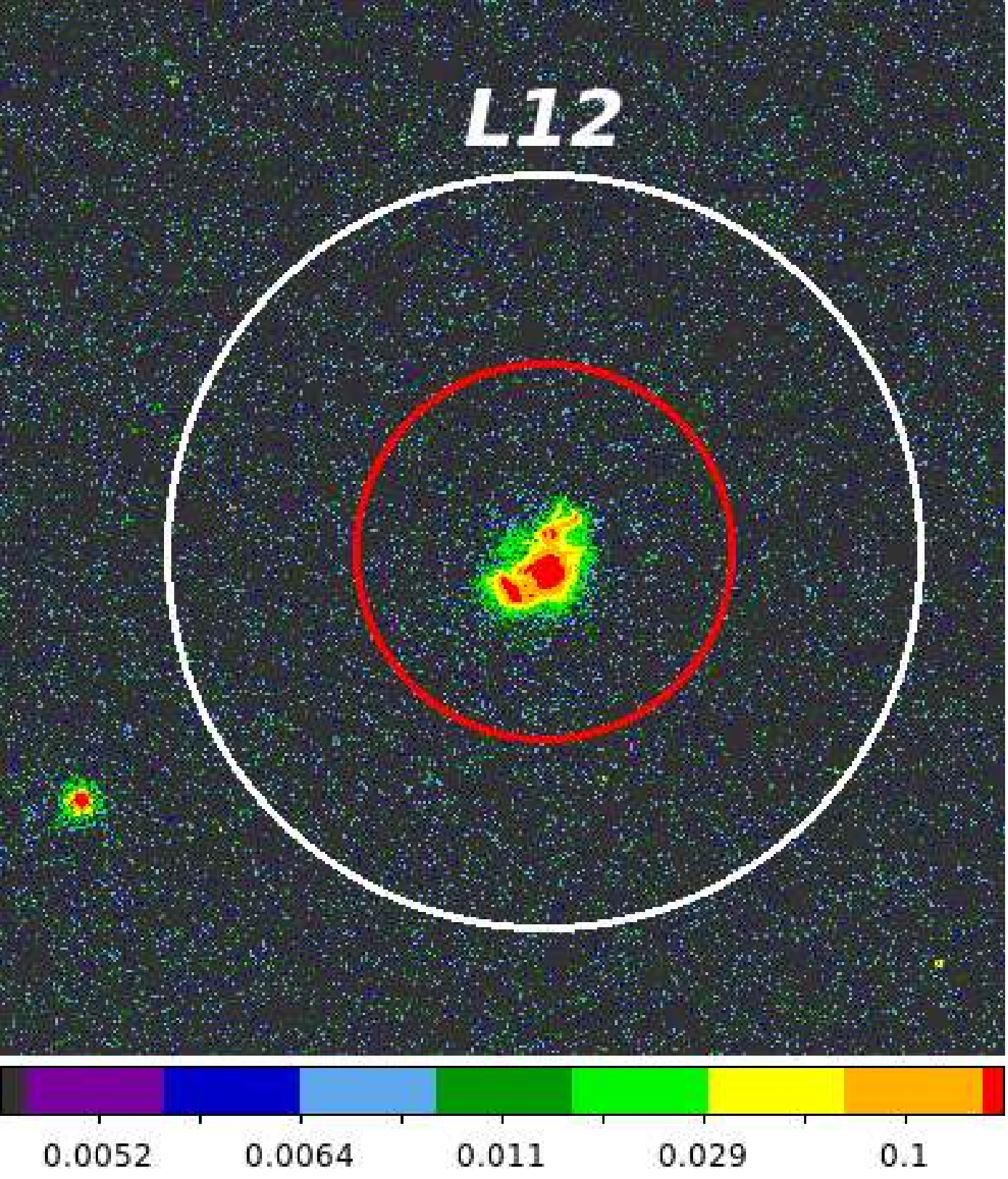}}\hspace*{\fill}
   \subfloat[\ \ \ \ \ \ \ \ \ \ \ (4) T$_{MB}$]{\includegraphics[width=4.5cm]{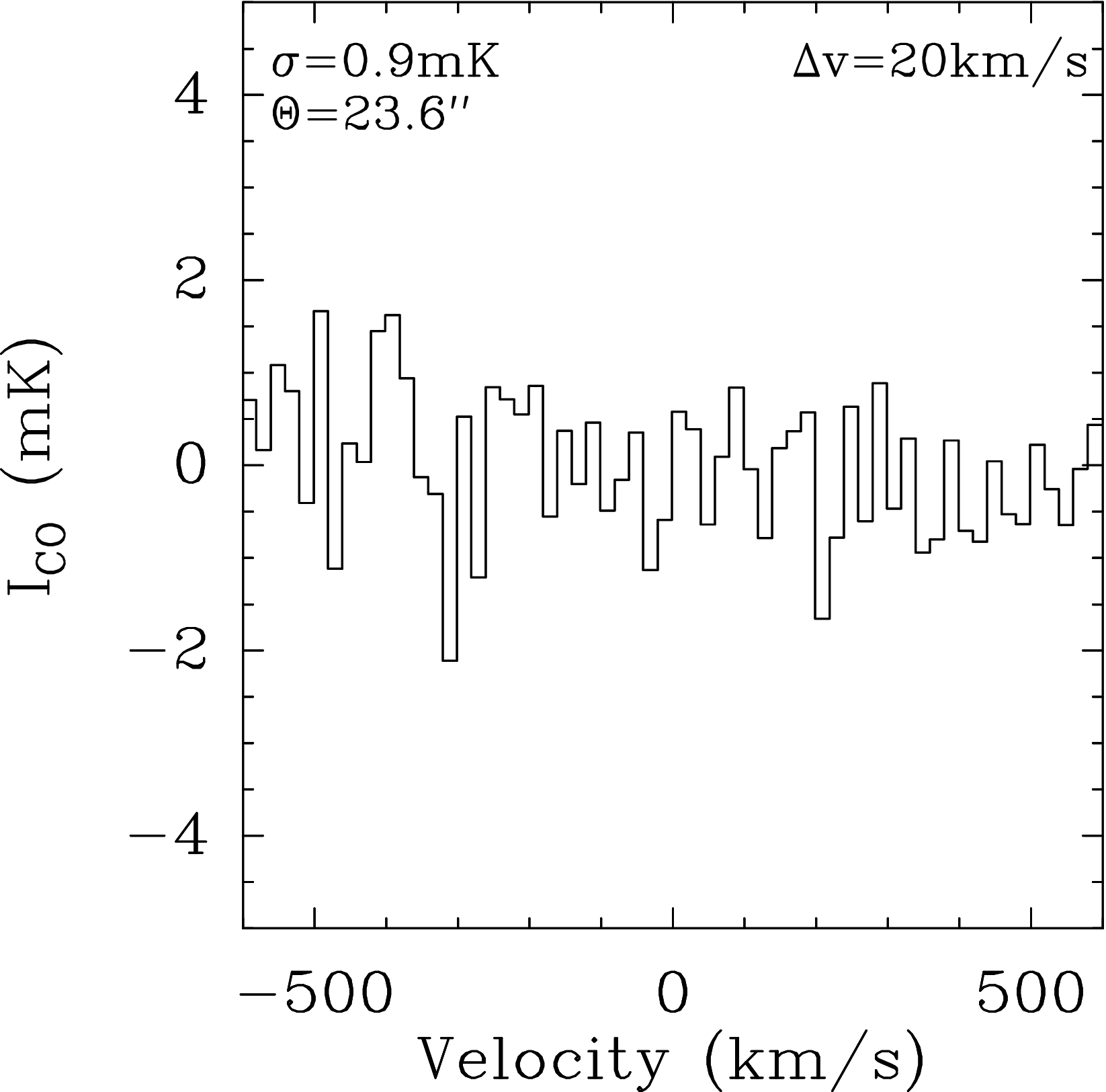}}\hspace*{\fill}
   
   \caption{CO (1-0) nondetections of LARS galaxies.
   The beamsize for the CO (1-0) transition is shown in columns (1) and (3) as white circle
   on top of our HST H$\upalpha$ narrowband images. The measured signals on the main beam
   brightness temperature scales are shown in columns (2) and (4).}
   
   \label{fig:NONDECT_IRAM}
\end{figure*}

\begin{figure*}[!htbp]
   \subfloat{\includegraphics[width=3.1cm]{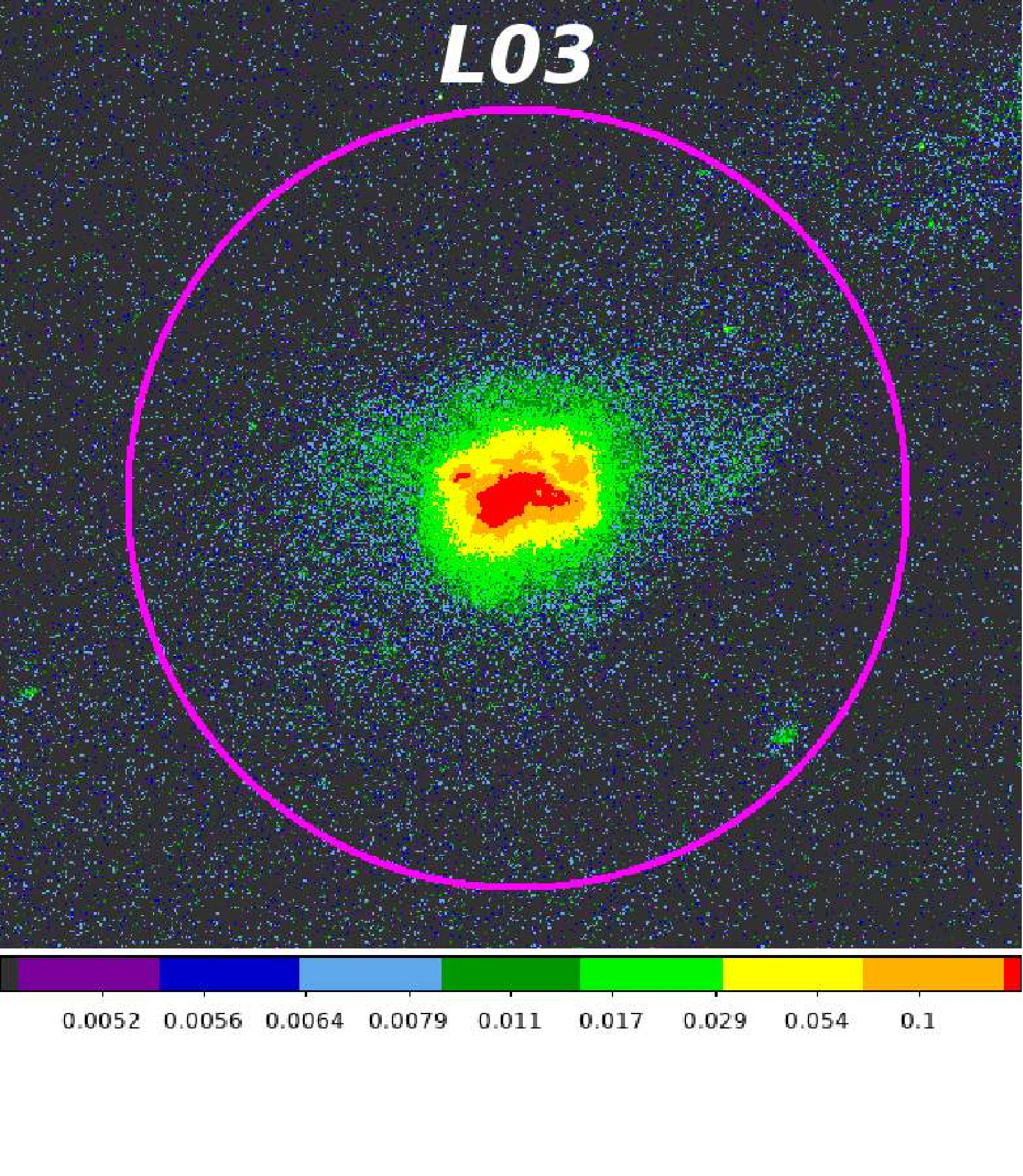}}\hspace*{\fill}
   \subfloat{\includegraphics[width=3.7cm]{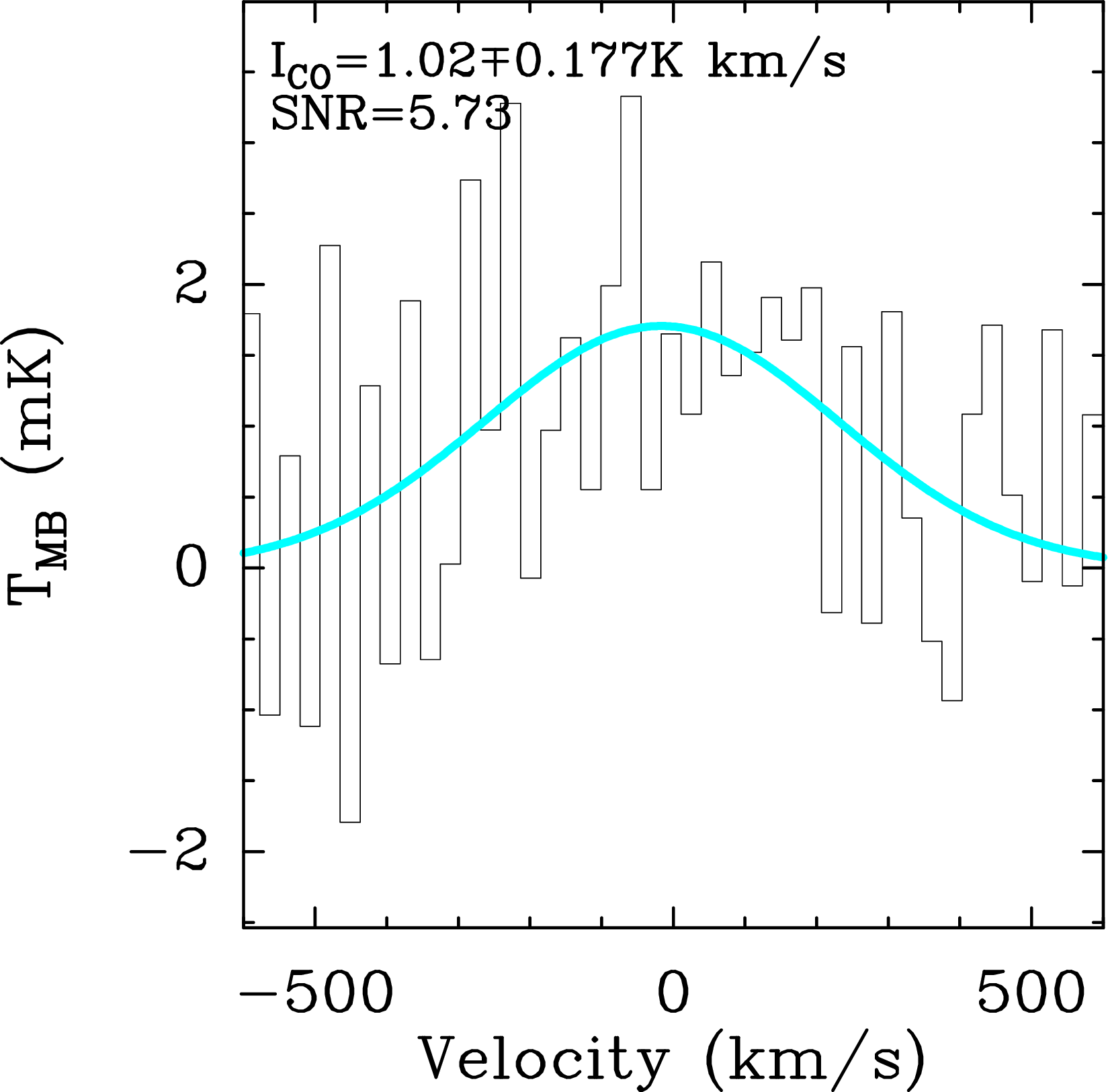}}\hspace*{\fill}
   \subfloat{\includegraphics[width=3.7cm]{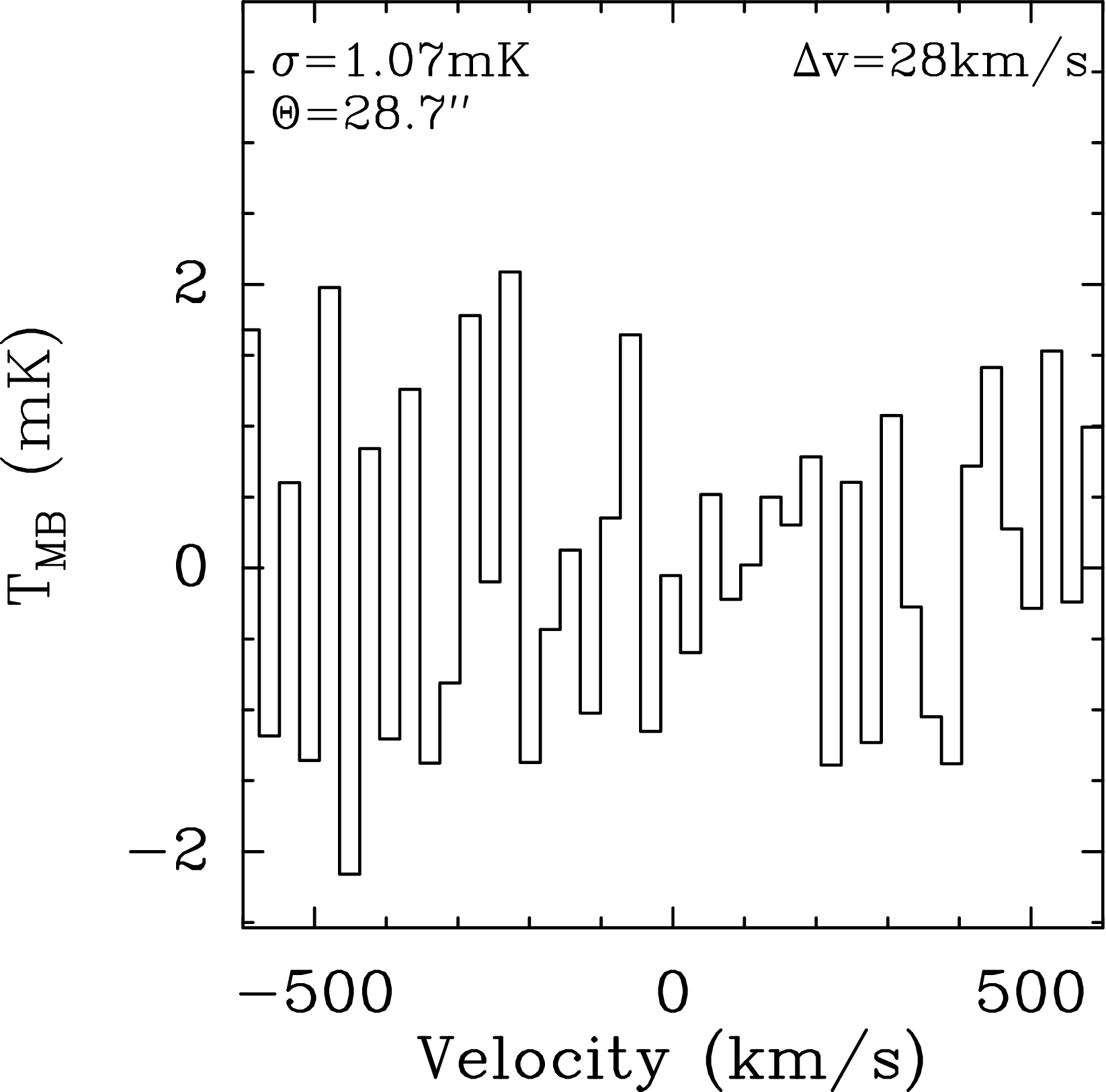}}\hspace*{\fill}
   \subfloat{\includegraphics[width=3.7cm]{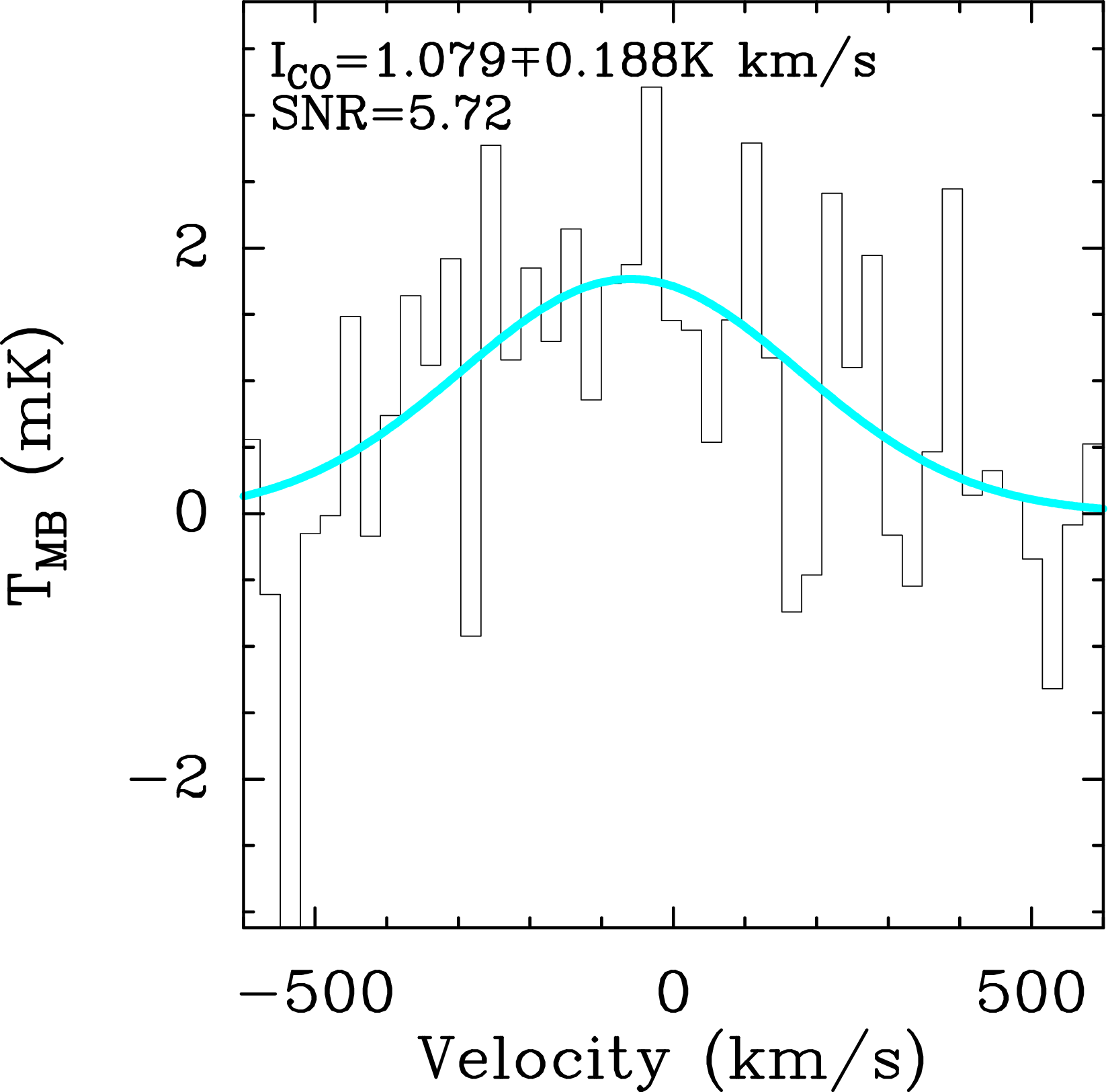}}\hspace*{\fill}
   \subfloat{\includegraphics[width=3.7cm]{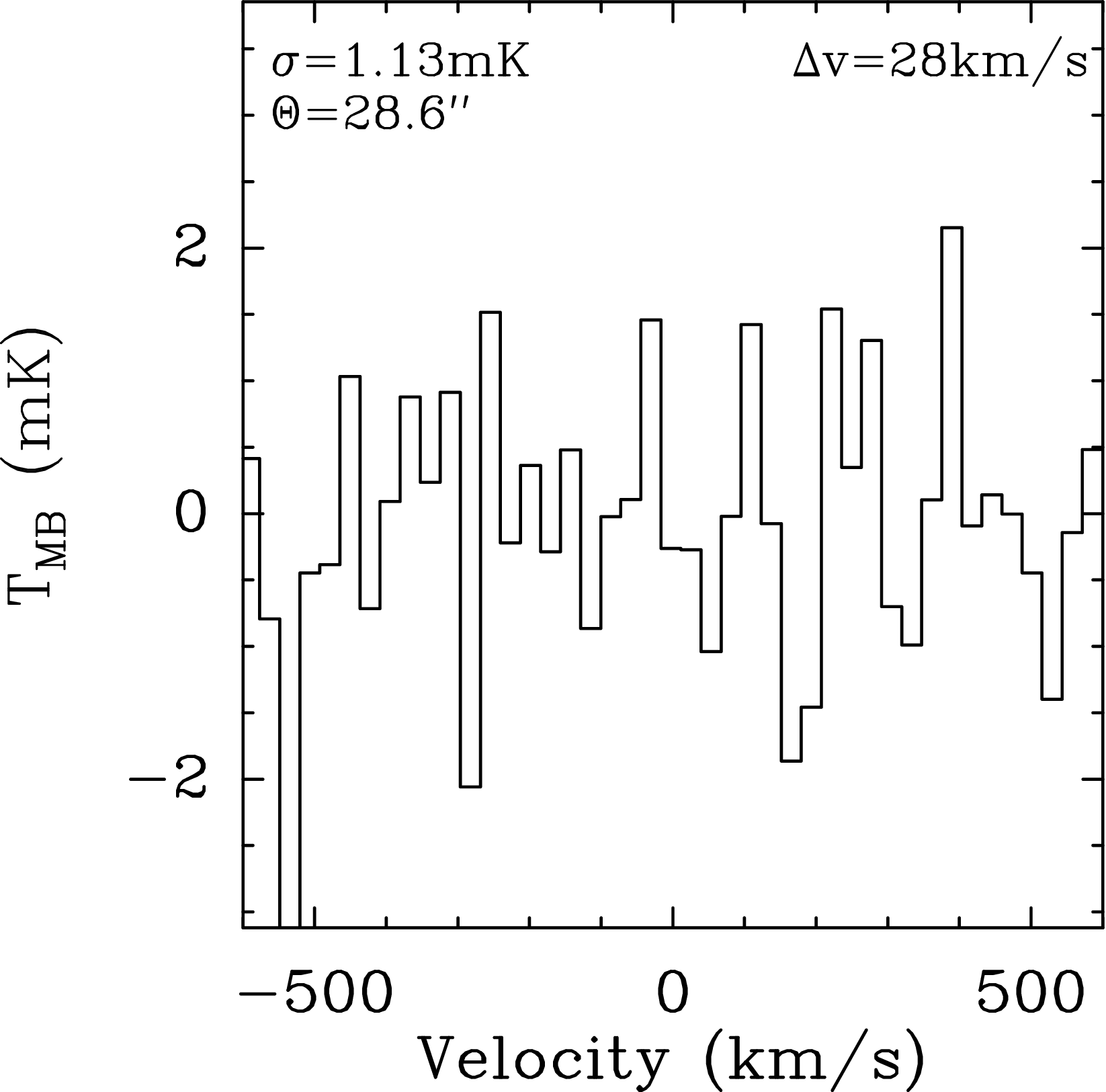}}\hspace*{\fill}

   \subfloat[(1) H$\upalpha$ \& HPBW]{\includegraphics[width=3.1cm]{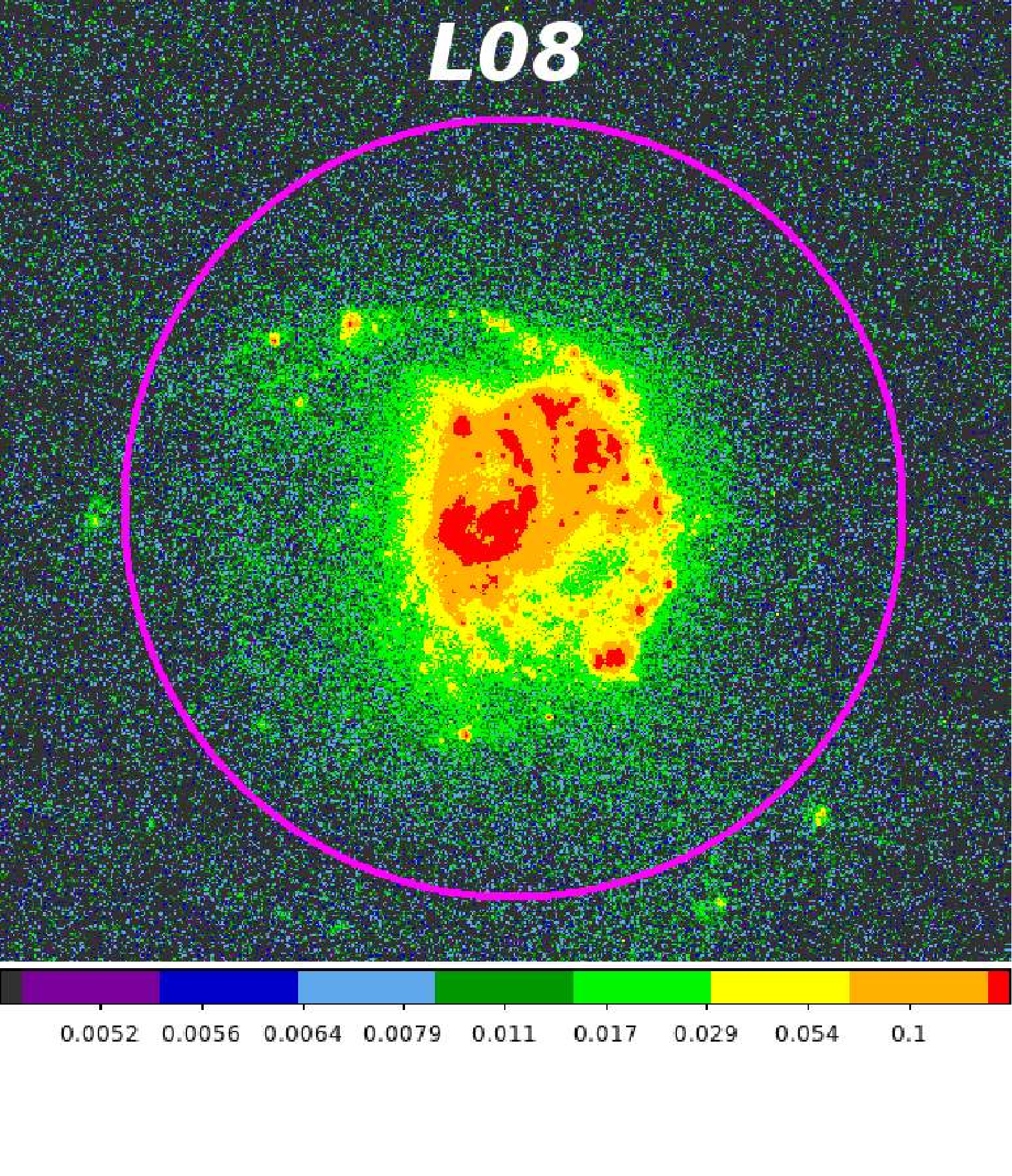}}\hspace*{\fill}
   \subfloat[\ \ \ \ \ \ \ \ \ \ \ (2) HCN (1-0)]{\includegraphics[width=3.7cm]{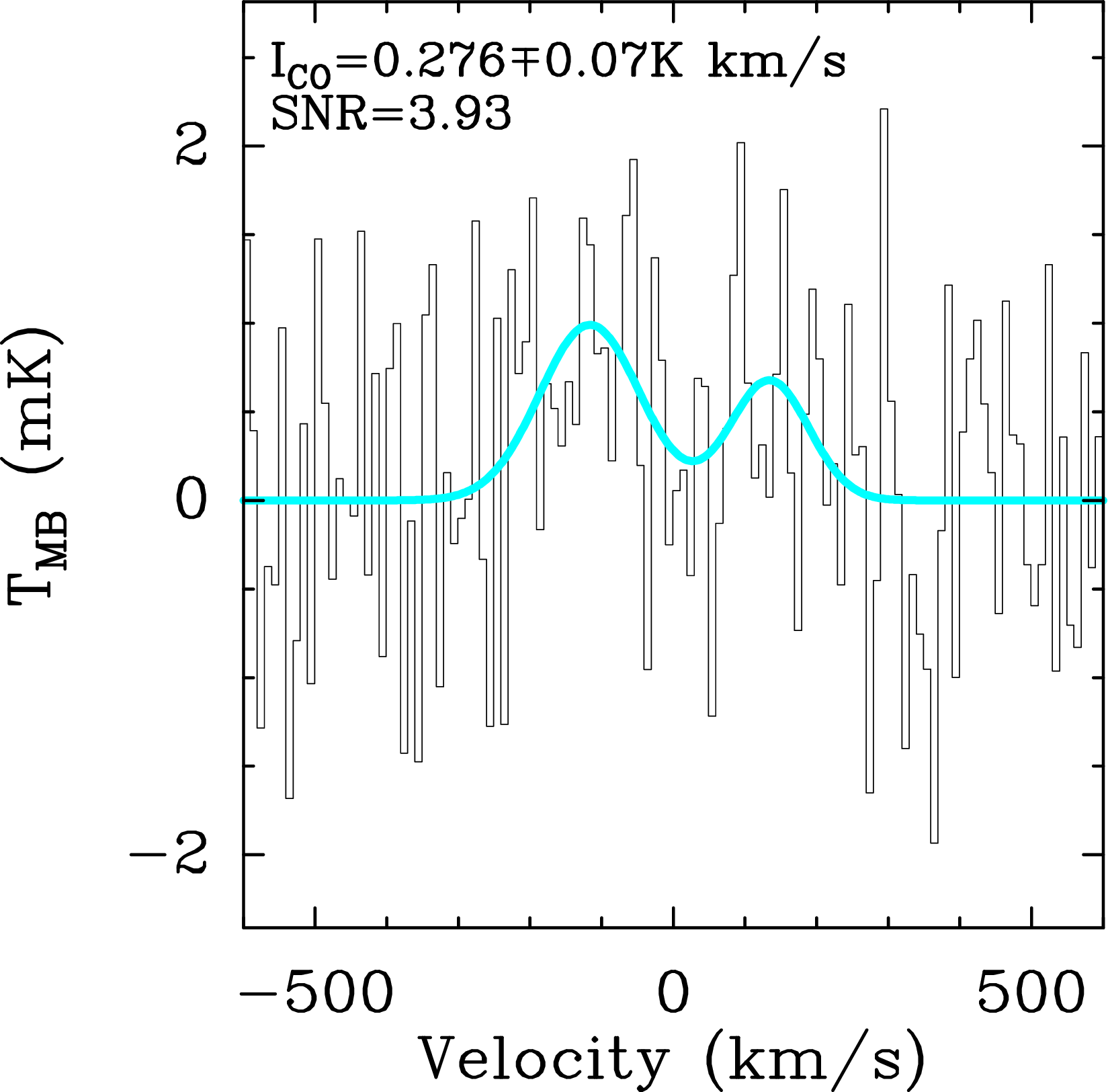}}\hspace*{\fill}
   \subfloat[\ \ \ \ \ \ \ \ \ \ \ (3) residual]{\includegraphics[width=3.7cm]{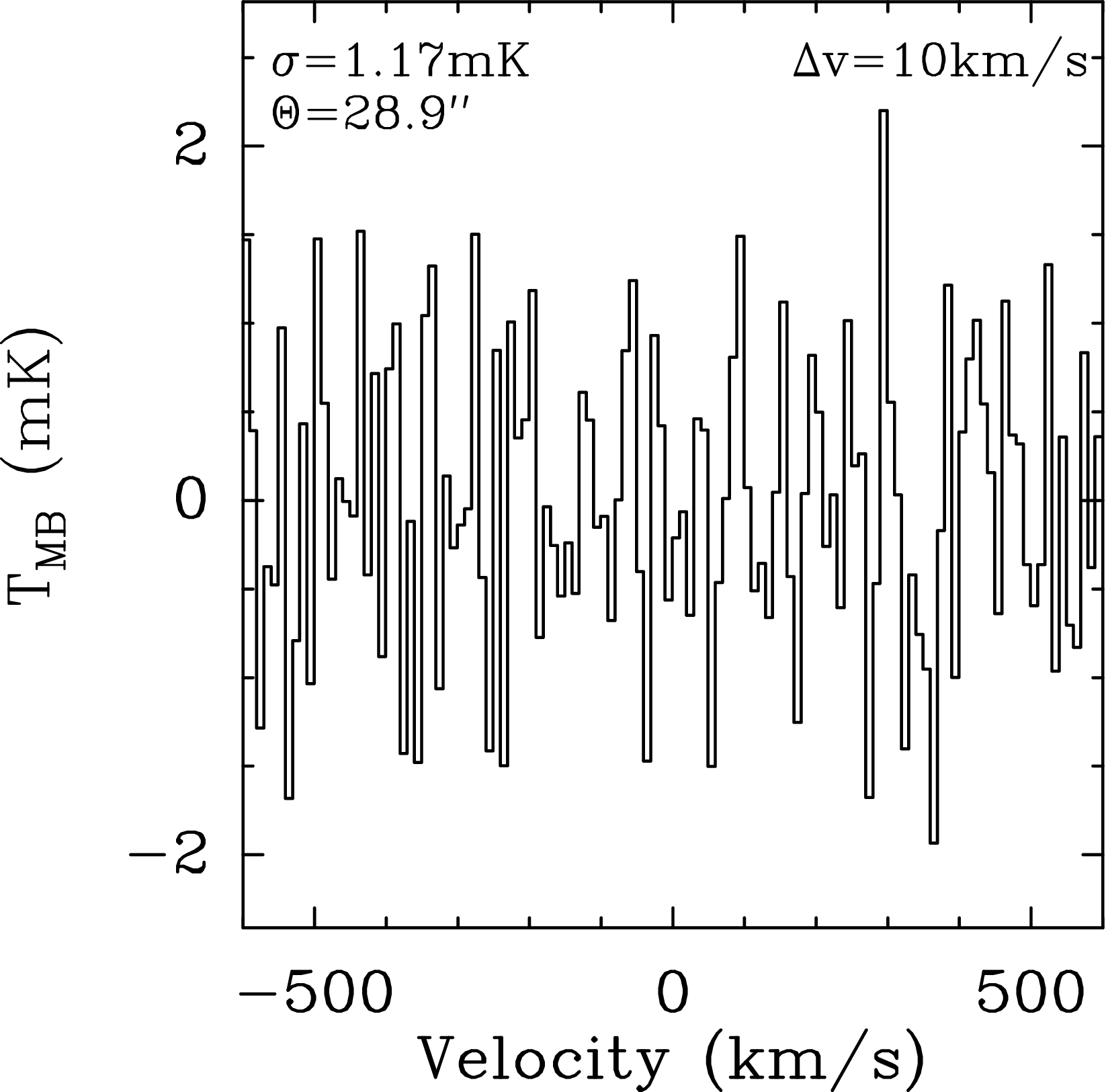}}\hspace*{\fill}
   \subfloat[\ \ \ \ \ \ \ \ \ \ \ (4) HCO$^+$ (1-0)]{\includegraphics[width=3.7cm]{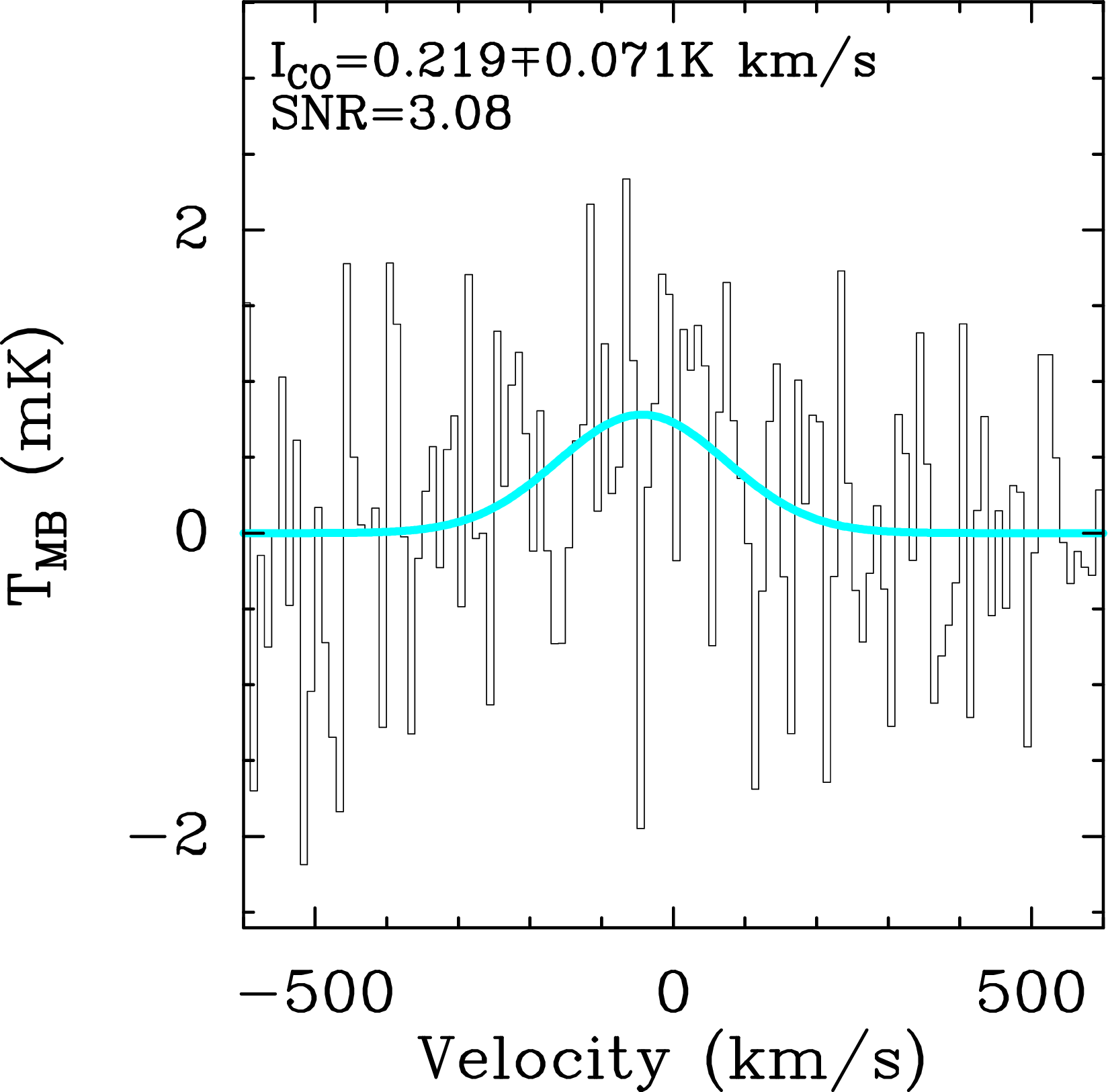}}\hspace*{\fill}
   \subfloat[\ \ \ \ \ \ \ \ \ \ \ (5) residual]{\includegraphics[width=3.7cm]{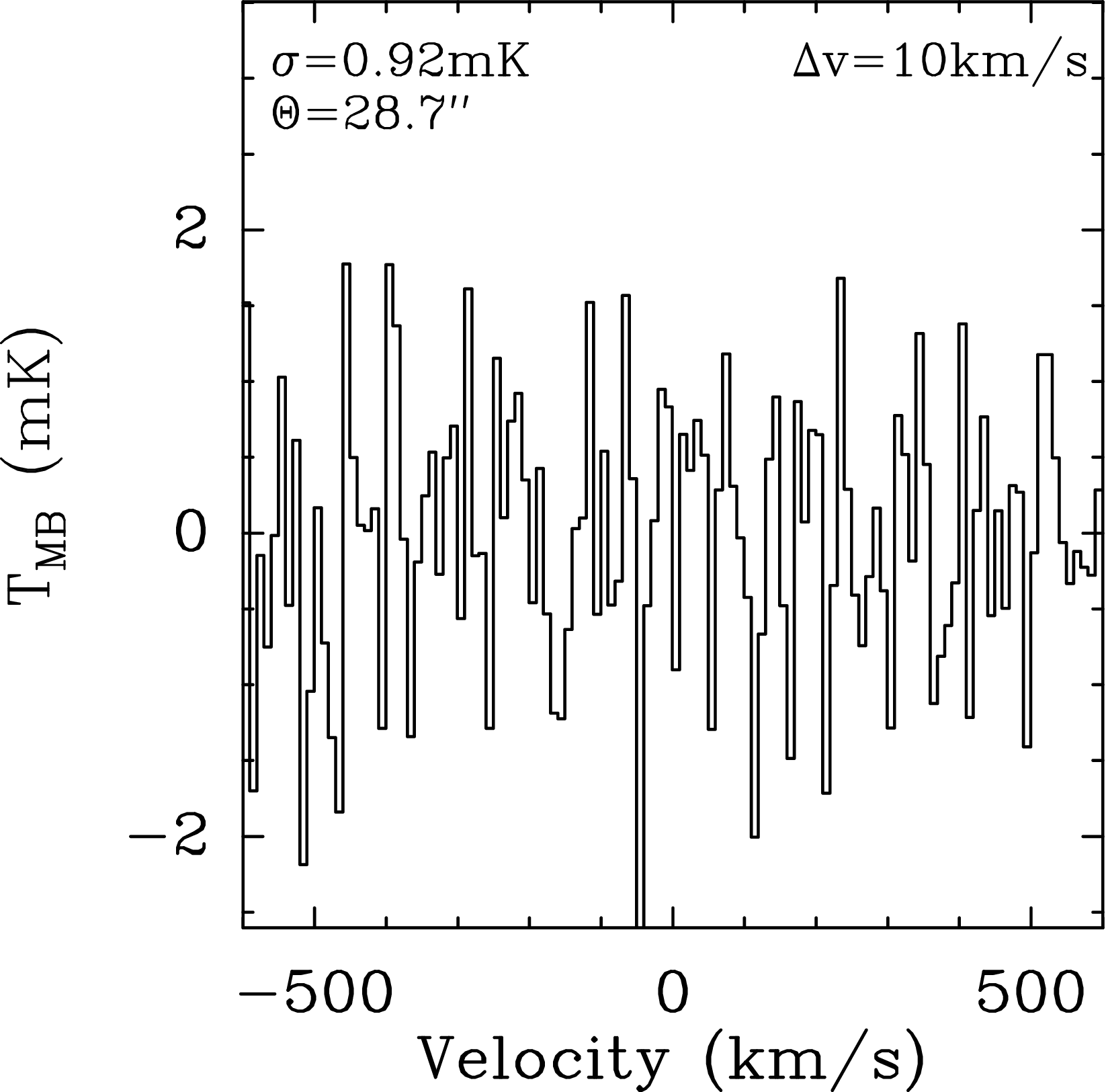}}\hspace*{\fill}
   
   \caption{Detections of dense gas tracers in LARS 3 (\textit{top row}) and 8 (\textit{bottom row}).
	The beamsizes for the transitions are shown in column (1)
	on top of our HST H$\upalpha$ narrowband images.
	HCN and HCO$^+$ (1-0) spectra are shown in columns (2) and (4).
	Spectra are given on the T$_{MB}$ scale in mK.
        Single- or double-Gaussian fits were performed for calculating the line flux. They are shown as cyan solid lines.
        The residuals after subtraction of the fits from the data are shown in columns (3) and (5).
}
   
   \label{fig:HCNLARS_IRAM}
\end{figure*}

\subsection{Infrared Observations}

\subsubsection{Herschel PACS Observations}
Far infrared observations of LARS galaxies (2, 4, 6, 8, 9) were executed with the \textit{Photoconductor Array Camera and Spectrometer} \citep[PACS][]{Poglitsch2010}
on board the \textit{Herschel Space Observatory} \citep{Pilbratt2010} under program OT2\_mhayes\_4, for a total time of 12.1 hours.
Our observations were carried out in the standard single pointing chop-nod mode.
PACS was operated in parallel mode (using the photometer and spectrometer simultaneously), targeting
photometry at 70, 100 and 160$\upmu$m as well as spectral lines of [O~I]63$\upmu$m, [O~III]88$\upmu$m and [C~II]158$\upmu$m.
The point spread function (PSF) is well described by a two dimensional Gaussian with sizes of
roughly 6x12, 7x13 and 12x16 arcsec for the blue, green and red channel respectively.

Further PACS imaging of LARS galaxies is found in the archive.
LARS~3 was
observed under the guarantee time key program KPGT\_esturm\_1 (PI: Eckhard Sturm) as part of a comprehensive far-infrared survey
of infrared bright galaxies. LARS~5 was observed in program OT2\_dhunter\_4 (PI: Deidre Hunter), as part of an investigation of
dust and star formation in local low metallicity galaxies. LARS~7 and 10 are covered by program KPOT\_seales\_01\_2, the
\textit{Herschel Thousand Degree Survey} (PI: Stephen Eales). LARS~12, 13 and 14 were observed as part of a survey of
local Lyman Break Analogues under program OT1\_roverzie\_1 (PI: Roderik Overzier).

All PACS imaging data was reduced using the \textit{Herschel Interactive Processing Environment} (HIPE; \citep{Ott2010})
version 14.0.1. To create a set of homogeneous images of our galaxies, we
re-reduce all observations starting from level 1 data products. The mapping was performed using the
MADMap maximum likelihood code for LARS 7 and 10 and JScanam otherwise. Due to the different observations available for the galaxies, our final reduction
levels range from 2.0 to 3.0, where level 2.0 indicates a fully reduced dataset, 
level 2.5 is the result of combining pairs of observations acquired in the scan and cross-scan mode,
and level 3.0 are mosaics of level 2.5 products that belong to the same sky field and observing program.
A summary of the photometric PACS observations used in this paper is shown in Table \ref{tab:pacsdata}.

In order to measure the fluxes of our galaxies, we have used the \textit{Daophot} algorithm as
implemented in the \texttt{annularSkyAperturePhotometry} task of HIPE's Jython console. We also examined
source sizes and flux measurements using the \texttt{sourceFitting} task. However, since most of our sources
are not or only marginally resolved, in particular in the red channel, we finally used the results from
the aperture photometry, with an aperture correction factor applied. Our flux measurements are summarized in
Table \ref{tab:wise_pacs}.

PACS spectroscopic data of our observing program was not re-reduced, instead we use the level 2.0
data products provided through the standard PACS spectrometer pipeline. The fluxes of the spectral lines
were derived after summation over all spaxels and subsequent fitting of the continuum and the Gaussian line.
The continuum subtracted line fluxes are then found from the area below the Gaussian fits. A summary of
the measurements is shown in Table \ref{tab:IRSPECLINES}.

\begin{table}
\caption{Summary of PACS data used in this paper}
\label{tab:pacsdata}
\begin{tabular}{r l l c}
\hline \hline
ID &  PACS   & Proposal & Reduction \\
   &  Obsid  & Code     & Level \\
2  &  1342270859                         & OT2\_mhayes\_4     & 2.0 \\
3  &  1342209025,1342209027              & KPGT\_esturm\_1    & 3.0 \\
4  &  1342247700                         & OT2\_mhayes\_4     & 2.0 \\
5  &  1342253524,1342253526              & OT2\_lhunt\_4      & 3.0 \\
6  &  1342261848,1342262223              & OT2\_mhayes\_4     & 3.0 \\
7  &  1342210558,1342210903,             & KPOT\_seales01\_2  & 3.0 \\
   &  1342210946,1342210963,             &                    &     \\
   &  1342222626                         &                    &     \\
8  &  1342248630                         & OT2\_mhayes\_4     & 2.0 \\
9  &  1342270829,1342270830              & OT2\_mhayes\_4     & 3.0 \\
10 &  1342210567,1342210963,             & KPOT\_seales01\_2  & 3.0 \\
   &  1342222626   		         &                    &     \\
12 &  1342231634,1342231636              & OT1\_roverzie\_1   & 3.0 \\
13 &  1342237833,1342223573              & OT1\_roverzie\_1,  & 2.5 \\
   &                                     & GT1\_aconturs\_1   &    \\
14 &  1342231640,1342231642              & OT1\_roverzie\_1   & 3.0 \\
\end{tabular}
\end{table}

\begin{table*}
\begin{threeparttable}
\caption{Summary of Gaussian fit results for the PACS and FIFI-LS spectral lines of [C~II]158, [O~I]63 and [O~III]88}
\label{tab:IRSPECLINES}
\begin{tabular}{rllllll}
\hline \hline
ID & line         & center & FWHM   & flux         & instrument \\
   &              & [km s$^{-1}$] & [km s$^{-1}$] & [10$^{-15}$ erg s$^{-1}$ cm$^{-2}$] &    \\
2  &  [C~II]158  &  -20  &  209  &  8.4$\pm$4.0    & PACS \\
2  &  [O~I]63    &  177  &  $<$572  &  $<$30.3  & PACS \\
3  &  [C~II]158  &  -52  &  398  &  688.0$\pm$15.6 & PACS \\
3  &  [O~I]63    &  3  &  380  &  1317.0$\pm$63.8  & PACS \\
3  &  [O~III]88  &  -43  &  440  &  843.8$\pm$54.7 & PACS \\
5  &  [C~II]158  &  -117  &  584  &  479.1$\pm$84.7 & FIFI-LS \\
5  &  [O~III]88  &  -110  &  $>$222$\dagger$  &  254.9$\pm$43.3 & FIFI-LS \\
8  &  [C~II]158  &  -20  &  497  &  1052.0$\pm$161.3 & FIFI-LS \\
9  &  [C~II]158  &  105  &  284  &  285.7$\pm$5.7 & PACS \\
9  &  [O~I]63    &  153  &  218  &  249.3$\pm$23.1 & PACS \\
11  &  [C~II]158 &  -118  &  388  &  502.3$\pm$30.5 & FIFI-LS \\
12  &  [C~II]158 &  38  &  214  &  14.5$\pm$2.1 & PACS \\
12  &  [O~I]63   &  134  &  212  &  20.1$\pm$6.0 & PACS \\
13  &  [C~II]158 &  -18  &  299  &  30.8$\pm$3.1 & PACS \\
13  &  [O~I]63   &  127  &  280  &  36.6$\pm$7.7 & PACS
\end{tabular}
\begin{tablenotes}
	\item $\dagger$ Due to strong atmospheric absorption, the red side of the line is truncated and the
	reported value is considered to be a lower limit. \\
	Also note that the spectral resolution varies between the different instruments and channels.
	PACS offers spectral resolutions of $\sim$240 km/s for the [C~II]158$\upmu$m observations
	and $\sim$100 km/s for [O~I]63$\upmu$m.
	FIFI-LS has R$\sim$1200 ($\sim$250 $km\ s^{-1}$) for the observed wavelength of [C~II]158$\upmu$m
	and R$\sim$670 ($\sim$450 $km\ s^{-1}$) for [O~III]88$\upmu$m.
\end{tablenotes}
\end{threeparttable}
\end{table*}

\subsubsection{SOFIA FIR Line Spectroscopy}
\begin{figure*}
\centering
   \subfloat{\includegraphics[width=1.0\textwidth]{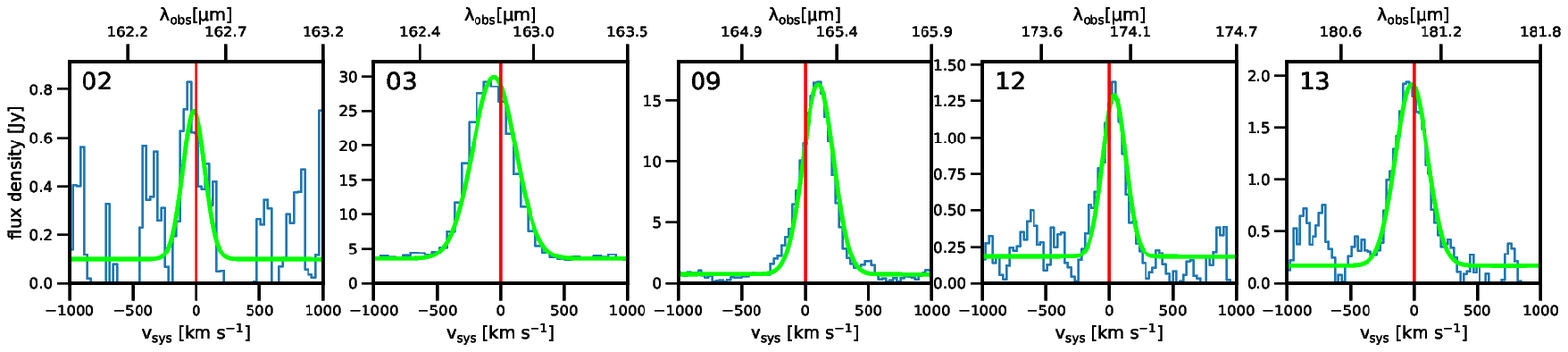}}\hspace*{\fill}

   \subfloat{\includegraphics[width=1.0\textwidth]{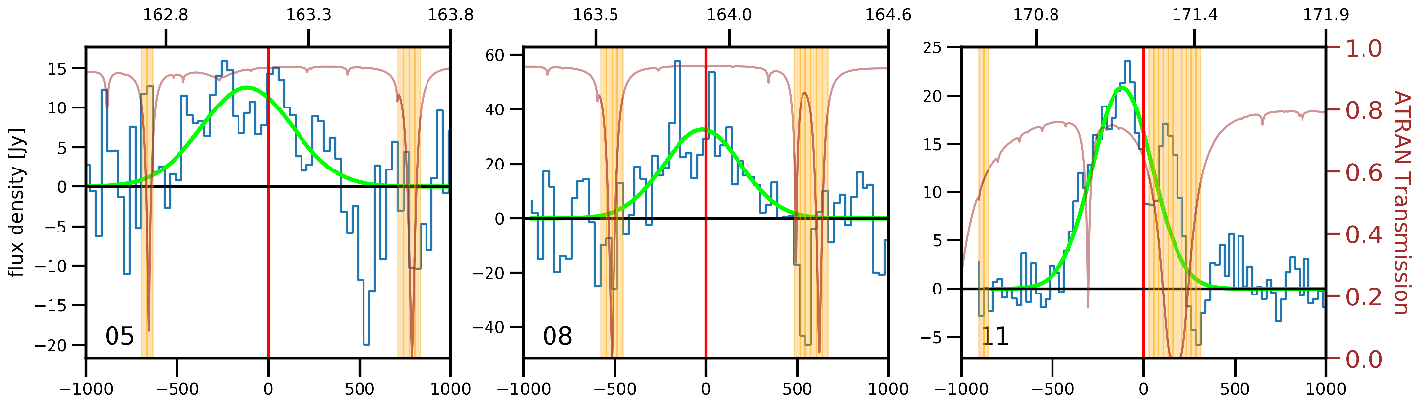}}\hspace*{\fill}
   \caption{[C~II]158$\upmu$m fine structure lines of LARS galaxies. Herschel/PACS spectra are shown in the top row, and
    SOFIA/FIFI-LS spectra in the bottom row. Total fluxes were derived using Gaussian fits (see Table \ref{tab:IRSPECLINES}). For FIFI-LS data,
    the atmospheric absorption, calculated using \texttt{ATRAN} \citep{ATRAN},
    is shown as red curve.
    Spectral ranges with strong atmospheric absorption (see orange range) were masked out for the fitting process.
    The zero-velocity (red vertical line) is related to the H~I systemic velocity from \cite{Pardy2014} for all LARS galaxies
    but LARS 12-14 (which have no H~I detection). For those, the systemic velocity is based on SDSS optical redshifts.}
   \label{fig:lars_ciispec}
\end{figure*}

\begin{figure*}
\centering
   \includegraphics[width=1.0\textwidth]{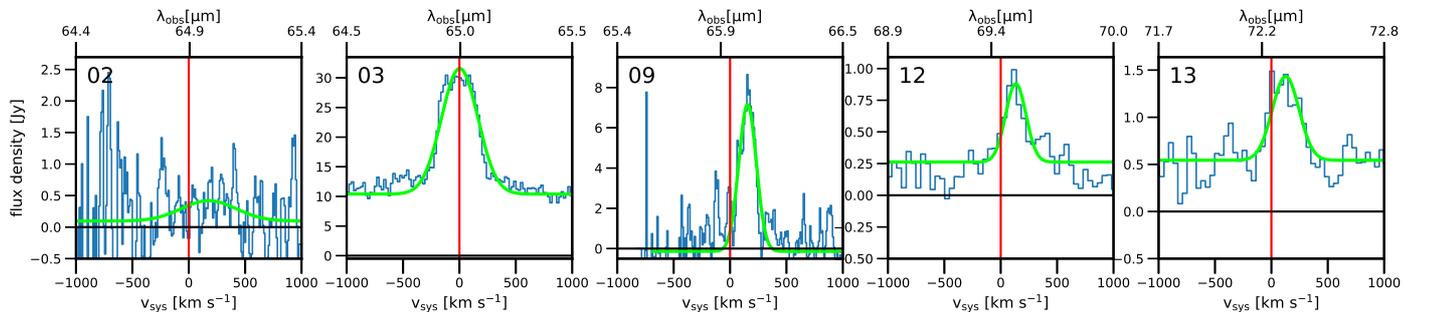}
   \caption{[O~I]63$\upmu$m fine structure lines of LARS galaxies, obtained with Herschel/PACS.
    Total fluxes were derived using Gaussian fits (see Table \ref{tab:IRSPECLINES}. The zero-velocity (red vertical line)
    is related to the H~I systemic velocity from \cite{Pardy2014} for all LARS galaxies
    but LARS 12-14 (which have no H~I detection). For those, the systemic velocity is based on SDSS optical redshifts.}
   \label{fig:lars_oispec}
\end{figure*}

\begin{figure}
\centering
   \subfloat{\includegraphics[width=0.2\columnwidth]{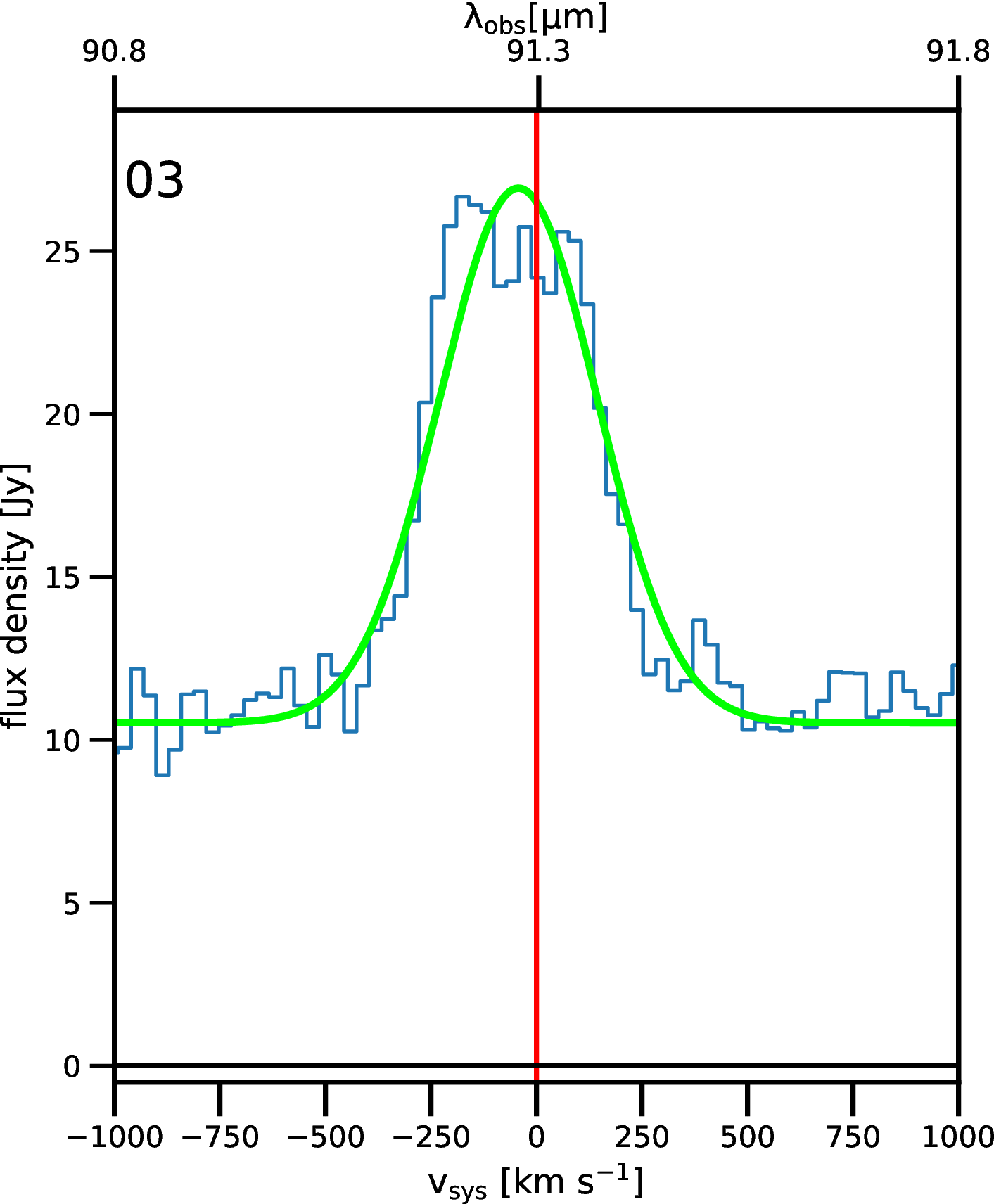}}
   \subfloat{\includegraphics[width=0.2\columnwidth]{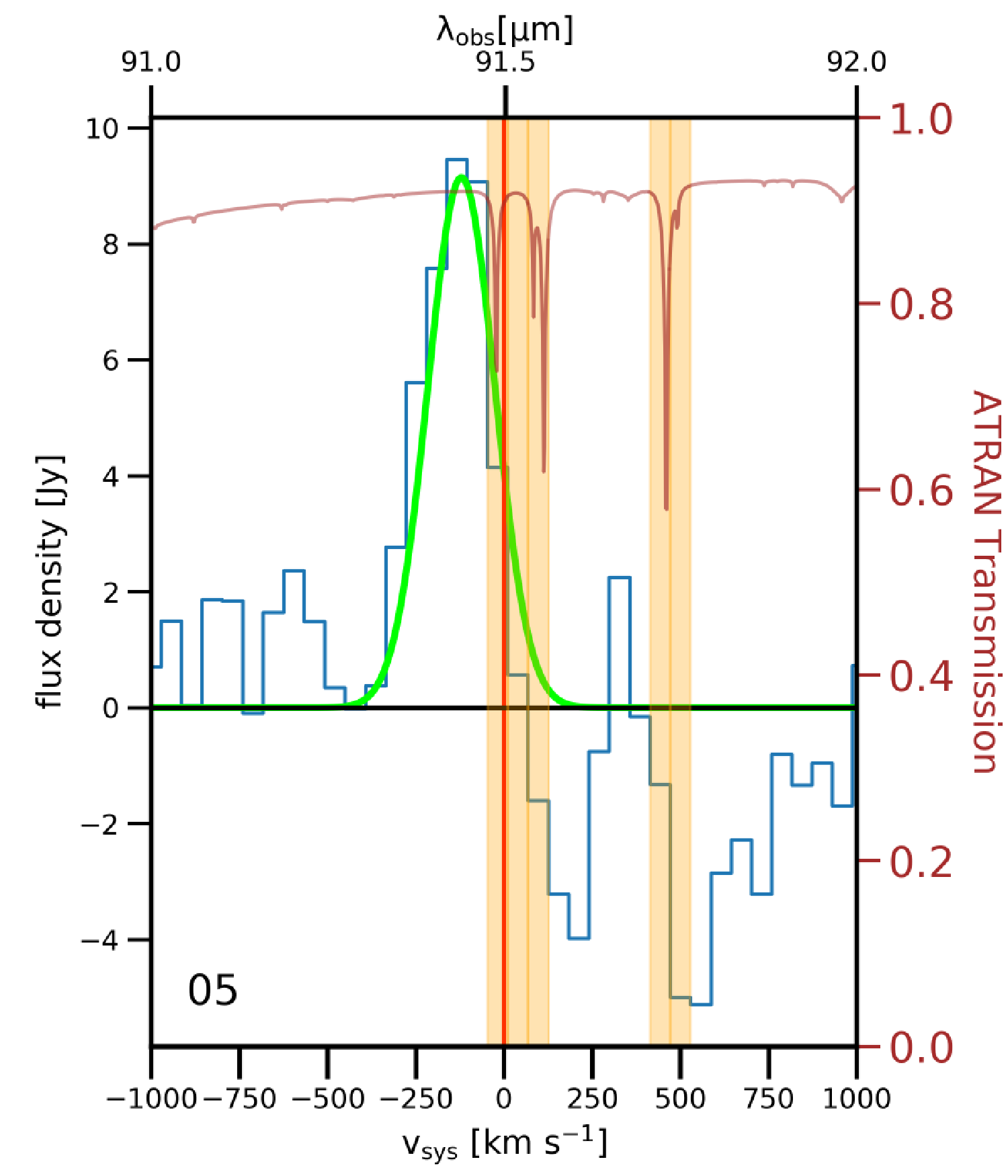}}

   \caption{[O~III]88$\upmu$m fine structure lines of LARS 3 (\textit{left}) and 5 (\textit{right}), obtained with Herschel/PACS
    and SOFIA/FIFI-LS respectively. Total fluxes were derived using Gaussian fits (see Table \ref{tab:IRSPECLINES}.
    For FIFI-LS data,
    the atmospheric absorption, calculated using \texttt{ATRAN} \citep{ATRAN},
    is shown as red curve.
    Spectral ranges with strong atmospheric absorption (see orange range) were masked out for the fitting process.
    The zero-velocity (red vertical line)
    is related to the H~I systemic velocity from \cite{Pardy2014}.}
   \label{fig:lars_oiiispec}
\end{figure}

We performed observations of LARS galaxies targeting FIR fine structure lines of [C~II]158$\upmu$m,
[O~I]63$\upmu$m and [O~III]88$\upmu$m with the \textit{Far Infrared Field-Imaging Line Spectrometer} (FIFI-LS) aboard the
Stratospheric Observatory for Infrared Astronomy (SOFIA) airplane (see Figures \ref{fig:lars_ciispec}, \ref{fig:lars_oispec} and \ref{fig:lars_oiiispec}).
Observations were carried out in Cycle 3 under project 03\_0059.

Basic data calibration was provided by the FIFI-LS team,
i.e. level 3 data products, consisting of $\sim$30 second exposures.
The data was further processed using the same method as described
in \cite{Smirnova-Pinchukova2019}, i.e. an additional selection routine and
background subtraction. Final data cubes with a sampling of 6" are
produced with the Drizzle algorithm \citep{Fruchter2002}.
Total line fluxes were calculated using Gaussian fits to the
summed spectra within an aperture of 36". Spectral ranges with
strong atmospheric absorptions were masked out for the fitting process.
A summary of the measurements is shown in Table \ref{tab:IRSPECLINES}.
Note that the spectral resolution varies between the channels,
with R$\sim$1200 ($\sim$250 $km\ s^{-1}$) for the observed wavelength of [C~II]158$\upmu$m
and R$\sim$670 ($\sim$450 $km\ s^{-1}$) for [O~III]88$\upmu$m.

\subsubsection{unWISE Source Extractions}
NASA's Wide-field Infrared Survey Explorer (WISE; \cite{Wright2010} mapped the sky at 3.4, 4.6, 12, and 22$\upmu$m
(W1, W2, W3, W4) in 2010 with an angular resolution of 6.1, 6.4, 6.5 and 12.0 arcsec respectively.
We checked the WISE all-sky source catalog for entries of our targets. Unfortunately, photometry was not available for
all LARS galaxies. Thus, we performed source extraction using \texttt{Sextractor} \citep{Bertin1996} on
unWISE\footnote{\url{http://unwise.me/}} \citep{Lang2014} images, which are unblurred coadds of the original WISE imaging.
For the extraction, we used the provided weight maps as RMS maps, applied Gaussian filters to the W3 and W4 bands in order
to smooth the irregular PSFs, and used an RMS-based absolute detection threshold. For those LARS galaxies that are unresolved,
and have a WISE catalog entry, we find that our extracted flux levels (FLUX\_AUTO) agree to a level better than 10 percent compared to
the catalog's profile fitting magnitude (mpro). We convert the derived magnitudes to flux densities
using W1-W4 zero magnitude fluxes\footnote{\url{http://wise2.ipac.caltech.edu/docs/release/allsky/expsup/sec4\_4h.html}}
of 309.54, 171.787, 31.674 and 8.363 Jy. The photometric results for the LARS galaxies in units of mJy are summarized in
Table \ref{tab:wise_pacs}.

\begin{table*}
\caption{WISE and Herschel/PACS photometric measurements of LARS galaxies}
\label{tab:wise_pacs}
\begin{tabular}{rrrrrrrr}
\hline \hline
ID  & W1           & W2           & W3            & W4              & P70             & P100            & P160            \\
    & [mJy]        & [mJy]        & [mJy]         & [mJy]           & [mJy]           & [mJy]           & [mJy]           \\ 
1 & 1.7$\pm$0.1  & 1.2$\pm$0.1  & 14.1$\pm$0.4  & 97.3$\pm$3.1    & --              & --              & --              \\
2 & 0.5$\pm$0.1  & 0.2$\pm$0.1  & 1.0$\pm$0.3   & 4.5$\pm$1.5     & 80$\pm$11   & --              & 121$\pm$14  \\
3 & 15.5$\pm$0.1 & 16.7$\pm$0.2 & 236.3$\pm$3.6 & 1586.5$\pm$24.0 & 9306$\pm$8  & 8628$\pm$7  & 4745$\pm$19 \\
4 & 2.0$\pm$0.2  & 1.4$\pm$0.1  & 12.5$\pm$0.5  & 79.9$\pm$2.5    & 474$\pm$8   & --              & 331$\pm$11  \\
5 & 0.5$\pm$0.1  & 0.5$\pm$0.1  & 4.5$\pm$0.7   & 29.1$\pm$5.5    & 113$\pm$5   & 117$\pm$4   & 40$\pm$4    \\
6 & 0.4$\pm$0.2  & 0.4$\pm$0.2  & --            & --              & 51$\pm$8    & 24$\pm$8    & 58$\pm$10    \\
7 & 0.9$\pm$0.1  & 0.8$\pm$0.1  & 8.1$\pm$0.5   & 51.9$\pm$2.1    & --              & 303$\pm$78  & 309$\pm$50  \\
8 & 16.1$\pm$0.1 & 11.3$\pm$0.1 & 108.7$\pm$1.2 & 323.5$\pm$3.3   & --              & 4322$\pm$27 & 3346$\pm$34 \\
9 & 2.3$\pm$0.5  & 4.0$\pm$0.1  & 36.2$\pm$0.6  & 171.6$\pm$4.1   & 1147$\pm$17 & 1415$\pm$20 & 1306$\pm$27 \\
10 & 1.0$\pm$0.1  & 0.8$\pm$0.2  & 3.4$\pm$0.6   & 9.3$\pm$3.2     & --              & 336$\pm$77  & 413$\pm$50  \\
11 & 4.1$\pm$0.3  & 3.1$\pm$0.1  & 27.9$\pm$0.7  & 56.7$\pm$4.9    & --              & --              & --              \\
12 & 0.3$\pm$0.2  & 0.3$\pm$0.1  & 2.8$\pm$0.6   & 17.4$\pm$3.5    & 104$\pm$4   & 113$\pm$6   & 62$\pm$4    \\
13 & 0.6$\pm$0.1  & 1.1$\pm$0.2  & 10.6$\pm$0.7  & 39.7$\pm$4.5    & 506$\pm$3   & 464$\pm$3   & 286$\pm$3   \\
14 & 0.2$\pm$0.1  & 0.1$\pm$0.1  & 0.3$\pm$0.2   & 2.8$\pm$0.7     & 16$\pm$2    & 12$\pm$1    & -- \\
\end{tabular}
\end{table*}

\subsubsection{IRIS Photometry}
Only six LARS galaxies were found in the Infrared Astronomical Satellite (IRAS) point source catalgue (PSC) and the
faint source catalogue (FSC). Therefore, we have downloaded images
of the improved reprocessing of the IRAS Survey (IRIS\footnote{\url{http://irsa.ipac.caltech.edu/data/IRIS/index\_cutouts.html}}),
and performed photometric flux measurements for all LARS galaxies using these data products.
The originally downloaded IRIS images cover fields of two square degree (NASA's \texttt{Montage} tool\footnote{\label{foot:montage}\url{http://montage.ipac.caltech.edu/}}
was used to create smaller cutouts). The full width at the half maximum of the point spread function in the
new images is 3.8, 3.8, 4.0 and 4.3 arcmin for the four IRAS bands of 12, 25, 60 and 100$\upmu$m.
All LARS galaxies are thus unresolved. After conversion from MJy/sr to mJy/px, flux densities were extracted
using \texttt{SExtractor}'s \citep{Bertin1996} FLUX\_AUTO value.
The results are shown in Table \ref{tab:iras_akari}.

\subsubsection{AKARI Photometry}
%
We also checked the archive of the AKARI \citep{Murakami2007} infrared satellite mission
for LARS galaxy data.
AKARI comprises two main instruments with 13 spectral bands in total covering a wavelength range of 1.8 to 180$\upmu$m.
For our purpose we focused on the four FIS (Far-Infrared Surveyor; \citep{Kawada2007}) bands with
central wavelengths of 65, 90, 140, and 160$\upmu$m. Only three LARS galaxies fluxes were published in the
FIS Bright Source Catalogue Ver. 2.0 (released April 2016), so we downloaded cutouts of
the AKARI Far-infrared All-Sky Survey Maps \citep{Doi2015,Takita2015}.
The originally downloaded images cover fields of six degree (NASA's \texttt{Montage} tool$^{\ref{foot:montage}}$
was used to create smaller cutouts).
The full width at the half maximum of the PSF
in these maps is roughly 60 arcsec for the 65 and the 90$\upmu$m bands and 90 arcsec for the 140 and 160$\upmu$m bands.
All LARS galaxies are thus unresolved. After conversion from MJy/sr to mJy/px, flux densities were extracted
using \texttt{SExtractor}'s \citep{Bertin1996}
FLUX\_AUTO value.
The results are summarized in Table \ref{tab:iras_akari}.

\begin{table*}
\caption{IRAS (I) and AKARI (A) photometric measurements of LARS galaxies}
\label{tab:iras_akari}
\begin{tabular}{rrrrrrrrr}
\hline \hline
ID  & I12             & I25              & I60                & I100               & A65               & A90              & A140              & A165              \\
    & [mJy]           & [mJy]            & [mJy]             & [mJy]              & [mJy]             & [mJy]            & [mJy]             & [mJy]             \\
1 & --              & --               & 810$\pm$257   & --                 & --                & 384$\pm$146  & --                & --                \\
3 & 307$\pm$108 & 1694$\pm$203 & 11461$\pm$733 & 15350$\pm$4109 & 7663$\pm$1054 & 9249$\pm$485 & 3311$\pm$200  & 1552$\pm$251  \\
4 & --              & --               & --                & --                 & --                & 321$\pm$197  & 218$\pm$148   & --                \\
6 & --              & --               & --                & --                 & --                & --               & 222$\pm$163   & --                \\
8 & --              & --               & 4096$\pm$464  & 10633$\pm$4930 & 1488$\pm$617  & 3593$\pm$689 & 3769$\pm$1786 & 3441$\pm$2531 \\
9 & --              & --               & 1085$\pm$209  & 1979$\pm$907   & --                & 1022$\pm$246 & --                & --                \\
11 & --              & --               & 505$\pm$209   & 1929$\pm$1057  & --                & 810$\pm$272  & 859$\pm$155   & --                \\
13 & --              & --               & --                & --                 & --                & 298$\pm$194  & --                & --                \\
\end{tabular}
\end{table*}

\subsection{Previously Derived Lyman Alpha Related Quantities}

\subsubsection{Lyman Alpha Escape Fraction and Scattering Length}
\cite{Hayes2014} have computed the \lya\ escape fractions of LARS galaxies in the following way.
Assuming case~B recombination theory, the observed \lya\ flux was compared to
that expected from extinction corrected H$\upalpha$.
Note that H$\upalpha$ was obtained through HST narrowband imaging, i.e. capturing all of the ionized gas.
Adopting an intrinsic value of \lya$_{intrinsic}$/H$\upalpha$=8.7, the escape fraction was calculated as
\lya$_{obs}$/\lya$_{intrinsic}$.

The \lya\ scattering lengths of LARS galaxies were published by \cite{Bridge2018}.
They are derived from modeling the \lya\ morphology based on H$\upalpha$ images, which again
when multiplied by 8.7 should give the approximate intrinsic morphology of \lya.
The scattering is then modeled using a Gaussian convolution, with an additional component
of dust attenuation (from H$\upalpha$/H$\upbeta$). Using the MCMC method,
the observed \lya\ image is re-produced and the scattering length is
found from the size of the best-fit Gaussian convolution kernel.

\section{Methods}\label{sec:methods}

\subsection{Dust Based Gas Masses}\label{sec:DL07}
Given that we could detect only one third of the LARS sample in CO (1-0),
we employ the models\footnote{\url{ftp://ftp.astro.princeton.edu/draine/dust/irem4/}}
of \cite{Draine2007model} to derive dust masses,
which finally enables us to infer the total gas masses for all LARS galaxies in a homogeneous way.
The models predict infrared spectra for wavelengths between 1$\upmu$m and 1cm.
They are based on a mixture of dust grains containing carbonaceous grains (including
PAHs) and amorphous silicates, with grain size distributions -- or extinction curves -- as found in the Milky Way (MW),
the Large Magellanic Cloud (LMC) and the Small Magellanic Cloud (SMC).
Within each distribution type (MW, LMC or SMC), a variety of PAH dust mass fractions $q_{PAH}$ is implemented,
ranging from 0.1\% to 4.6\%, with the lowest and highest $q_{PAH}$ value being
consistent with observations of the dust in the SMC and Milky Way, respectively.
The total power radiated per H-nucleon per unit frequency per steradian, $j_{\nu}$, then depends on
two radiative components: 1) the average stellar intensity found in the diffuse ISM, described by the energy density parameter $U_{min}$,
and 2) a fraction $\upgamma$ of dust mass that is exposed to a power law distribution of intensities ranging
from $U_{min}$ to $U_{max}$ with a power law slope $\upalpha=-2$. The latter component is related to
photodissociation regions (hereafter PDRs).
$U$ is a scale factor to the local Galactic interstellar radiation field estimated by \cite{Mathis1983},
and can be related to the \cite{Habing1968} field $G_0$ via $U=0.88 G_0$.
The final value of $j_{\nu}$ is given by a linear combination of the two components with
a $(1-\upgamma)$ mass fraction that is originating from dust exposed to the average, diffuse stellar intensity and
a fraction $\upgamma$ related to the currently ongoing star formation, with $0\leq\upgamma\leq1$.
Thus, the fraction of dust mass associated with PDR regions is given by the product
of $\upgamma$ and $M_{dust}$.

Another parameter included in the models is the solid angle subtended by (old) stars, $\Omega_*$,
since they also contribute to the overall emission spectrum, albeit mainly at wavelengths of few $\upmu$m,
which is at the low end of the range we are considering.
Assuming an average stellar surface temperature T of 5000K, their share is calculated through the
Planck function $B_\nu (T)$.

In the analysis that follows we keep the upper energy density cutoff $U_{max}$ fixed at a value of
$10^6$, which decreases the number of degrees of freedom by 1, while having only little influence
on the fitting result, as shown by \cite{Draine2007}.
Hence, we are left with five free parameters for the models: $q_{PAH}$, $U_{min}$, $\upgamma$, $M_{dust}$
and $\Omega_*$.

We derive those quantities for LARS galaxies using Bayesian inference, similar to the method described in
\cite{Galliano2018b}. Our logarithmic prior is set to a constant (0.0) within the following intervals for our parameters:
U$_{min}$: 0.1-25, $q_{PAH}$: 0.1-4.6\%, $M_{dust}$: 10$^6$-10$^9\ M_{\odot}$,
$\upgamma$: 10$^{-3}$-10$^{-0.1}$ and $\Omega_*$: 10$^{-10}$-10$^{-7}$ sterad. Outside these intervals, the
prior is set to negative infinity.
Using the photometric data from WISE, Herschel/PACS, AKARI and IRAS allows us to infer
the logarithmic likelihood for the whole parameter space.
Note that -- beside the measured photometric error -- a 10-percent model-based error is also included
when calculating the logarithmic likelihood of the parameter space.
The latter is explored through
the Markov-Chain Monte Carlo (MCMC) method as implemented in the \texttt{Python} package \texttt{emcee} (version 3.0.2).
The final joint posterior probability distribution of the derived parameters is
shown in Appendix \ref{sec:appendix_corner}.
The best-fit SED was found from the 50\% quantile of the logarithmic posterior distribution and the
uncertainties are derived from a quantile-based credible interval
corresponding to 16\% and 84\% (i.e. $\pm$1$\upsigma$).
A summary of the derived quantities is found in
Table \ref{tab:dl07fitresults}, where also the number of degrees of freedom ($\nu$) is denoted.

The dust masses are finally converted into total gas mass estimates ($M_{gas_{est}}$) using a metallicity-dependent
gas-to-dust ratio as described in the following section.

\begin{table*}
\caption{Results from Bayesian inference of Draine \& Li (2007) dust model parameters for LARS galaxies}
\label{tab:dl07fitresults}
\begin{tabular}{llccccccc}
\hline \hline
$ID$ & $q_{PAH}$  & $log(\Omega_*)$ & $U_{min}$    & $log(\gamma)$ & $log(M_{dust})$ & $T_{dust}$ & $\nu$ \\
     &            & $[sr]$          & $(U=0.88G_0)$   &               & $[M_{\odot}]$   & $[K]$      &     \\
\\
1  & $0.86_{-0.21}^{+0.74}$ & $-8.96_{-0.21}^{+0.21}$ & $7.37_{-4.64}^{+8.07}$  & $-0.68_{-0.27}^{+0.27}$ & $7.40_{-0.66}^{+0.58}$ & $25.11_{-3.83}^{+3.29}$ & 1  \\
\\
2  & $0.83_{-0.32}^{+0.64}$ & $-9.54_{-0.28}^{+0.13}$ & $8.45_{-3.49}^{+6.89}$  & $-2.28_{-0.80}^{+0.38}$ & $6.86_{-0.56}^{+0.29}$ & $25.69_{-2.18}^{+2.68}$ & 1  \\
\\
3  & $1.21_{-0.67}^{+0.90}$ & $-8.09_{-0.44}^{+0.16}$ & $12.73_{-2.97}^{+5.29}$ & $-0.80_{-0.14}^{+0.19}$ & $8.01_{-0.20}^{+0.26}$ & $29.15_{-1.19}^{+1.64}$ & 10 \\
\\
4  & $1.17_{-0.44}^{+0.56}$ & $-8.91_{-0.16}^{+0.15}$ & $6.67_{-1.83}^{+2.04}$  & $-0.87_{-0.17}^{+0.19}$ & $7.13_{-0.18}^{+0.51}$ & $24.70_{-1.29}^{+1.12}$ & 3  \\
\\
5  & $1.71_{-0.61}^{+0.86}$ & $-9.42_{-0.18}^{+0.21}$ & $18.14_{-8.93}^{+7.20}$ & $-1.11_{-0.33}^{+0.35}$ & $6.01_{-0.66}^{+0.93}$ & $29.18_{-3.12}^{+1.67}$ & 2  \\
\\
6  & $3.27_{-1.71}^{+1.46}$ & $-9.67_{-0.80}^{+0.35}$ & $10.74_{-4.95}^{+3.57}$ & $-1.91_{-1.23}^{+1.11}$ & $5.73_{-0.64}^{+0.61}$ & $26.74_{-2.62}^{+1.31}$ & 1  \\
\\
7  & $1.23_{-0.59}^{+1.11}$ & $-9.21_{-0.15}^{+0.16}$ & $9.34_{-5.04}^{+5.80}$  & $-0.84_{-0.22}^{+0.17}$ & $7.20_{-0.45}^{+0.50}$ & $26.12_{-3.17}^{+2.19}$ & 1  \\
\\
8  & $1.64_{-1.08}^{+0.91}$ & $-7.93_{-0.14}^{+0.18}$ & $9.09_{-3.13}^{+3.77}$  & $-1.34_{-0.11}^{+0.16}$ & $8.14_{-0.17}^{+0.41}$ & $24.24_{-1.77}^{+1.55}$ & 7  \\
\\
9  & $1.78_{-0.67}^{+0.61}$ & $-8.54_{-0.08}^{+0.10}$ & $6.58_{-2.76}^{+2.49}$  & $-1.07_{-0.16}^{+0.22}$ & $7.95_{-0.18}^{+0.56}$ & $24.64_{-2.13}^{+1.35}$ & 5  \\
\\
10 & $1.36_{-0.38}^{+0.64}$ & $-8.99_{-0.10}^{+0.09}$ & $7.15_{-4.25}^{+6.36}$  & $-2.35_{-0.81}^{+0.50}$ & $7.63_{-0.28}^{+0.38}$ & $24.98_{-3.49}^{+2.80}$ & 1  \\
\\
11 & $4.38_{-1.66}^{+0.81}$ & $-8.48_{-0.10}^{+0.09}$ & $7.30_{-5.00}^{+6.81}$  & $-1.43_{-0.29}^{+0.29}$ & $8.25_{-0.33}^{+0.96}$ & $25.07_{-4.39}^{+2.91}$ & 3  \\
\\
12 & $2.70_{-1.44}^{+1.45}$ & $-10.20_{-1.01}^{+0.64}$ & $17.70_{-5.77}^{+6.48}$ & $-0.94_{-0.24}^{+0.30}$ & $7.08_{-0.12}^{+0.17}$ & $29.06_{-1.85}^{+1.55}$ & 2  \\
\\
13 & $3.47_{-0.92}^{+1.00}$ & $-9.43_{-0.32}^{+0.23}$  & $19.22_{-4.97}^{+7.02}$ & $-1.11_{-0.24}^{+0.11}$ & $8.11_{-0.15}^{+0.15}$ & $29.46_{-1.43}^{+1.57}$ & 3  \\
\\
14 & $2.81_{-1.47}^{+1.50}$ & $-10.41_{-0.97}^{+0.52}$ & $11.34_{-7.25}^{+8.58}$ & $-0.68_{-0.75}^{+0.64}$ & $6.90_{-0.28}^{+0.64}$ & $26.98_{-4.22}^{2.66}$ & 1 
\end{tabular}
\end{table*}

\subsubsection{Metallicity-Dependent Gas-to-Dust Ratio}\label{sec:gdr}
Once the dust masses are calculated, we convert them into total gas masses
using a metallicity-dependent gas-to-dust ratio (hereafter GDR).
GDR is found to scale with Z$^{-1}$ at near-solar metallicities
\citep{Lisenfeld1998,Hirashita2002,James2002,Hunt2005,Draine2007,Engelbracht2008,
Galliano2008,MunozMateos2009,Bendo2010,Galametz2011,Magrini2011,Sandstrom2013,RemyRuyer2014,RomanDuval2014,Cortese2016,Kahre2018}.
However, this trend is not seen in low-metallicity dwarf galaxies, which
show higher GDRs than expected
\citep{Lisenfeld1998,Galliano2003,Galliano2005,Galliano2011,Galametz2011,RemyRuyer2014,Kahre2018}.
Given the generally low metal content of LARS galaxies,
ranging from $\sim$13\% solar to $\sim\sfrac{2}{3}$ solar, we adopt
a metallicity-dependent GDR as defined in \cite{RemyRuyer2014}, which is a broken power-law,
i.e. a linear scaling down to a metallicity of 12+log(O/H)=8.1,
and a power law with an index of 3.1 at even lower Z:
For Z=12+log(O/H)$>$8.10: \\
\begin{equation}
	log(GDR)=2.21+1.0 \times\ (8.69-Z),
\end{equation}
and for Z=12+log(O/H)$\le$8.10: \\
\begin{equation}
	log(GDR)=0.96+3.1 \times\ (8.69-Z).
\end{equation}
Note that several other prescriptions for GDR calculations have been published to date. We
discuss the reasoning of our choice in Appendix \ref{sec:discuss_gdr_alpha}.

\subsubsection{Far-infrared SEDs of LARS Galaxies}
SEDs fits for all LARS galaxies are shown in Appendix \ref{sec:appendix_sed}.
The agreement between the models and the data is generally very good, in particular below $\sim$100$\upmu m$.
However, due to the lack of observations at wavelengths longer than 160$\upmu m$, the uncertainty largely increases
to this end.
The fit for LARS 6 is somewhat unreliable, owed to the fact that it is the
faintest galaxy in the PACS bands of the whole sample.
An examination of the shape of the SEDs shows that while most galaxies have a rather
flat distribution in the range between 50--100 $\upmu$m, LARS 2 and 10
show significant steepening within that range. This is a result of the lack of hot dust directly associated to star formation
that would show up as excess in the mid-infrared. Consequently, they have
lower $\gamma$ values, i.e. the fraction of dust mass linked to PDR regions.

We further caution that our flux measurements were obtained using varying aperture sizes,
reflecting the capabilities of the different instruments at varying wavelengths.
In particular, a single source in IRAS or AKARI may show up as several sources in the blue WISE bands.
In most cases the latter are likely of stellar origin and thus do not impact measurements in
the far infrared. However, some might still be of extragalactic origin, contaminating our measurements
and SED fits.

\subsection{CO~(1--0) Based Molecular Masses}

\subsubsection{CO~(1--0) Luminosity}
We convert our measured CO (1-0) velocity-integrated main beam brightness temperatures I$_{CO}$
(given in units of K km/s) to CO luminosities using the definition of $L'_{CO}$ by \cite{Solomon2005},
with an aperture correction factor added:
\begin{equation}\label{eqn:LPCO}
L'_{CO} = C_{AP} * 23.5\ \Omega\ I_{CO}\ (1 + z)^{-3}\ D_L^2
\end{equation}
$L'_{CO}$ is then given in $K\,km\,s^{-1}\,pc^2$, $\Omega$\ is the solid angle of the Gaussian beam
in arcsec$^2$, $z$ the redshift and $D_L$ the luminosity distance in Mpc.
Note that in the original formula found in \cite{Solomon2005} (equation 2 therein), $\Omega$\ is the solid angle
of the source convolved with the beam. In our case, however, we perform this convolution within
the calculation of the aperture correction factor and thus must use the solid angle of
the Gaussian beam instead.

\subsubsection{Aperture Correction for the CO~(1--0) Data}
Although the majority of the galaxies in our sample is smaller than the beam size $\theta$ of the IRAM
30m telescope, some galaxies have blue diameters, D$_{25}$, much larger than $\theta$ (compare Tables \ref{tab:lars_general}, \ref{tab:COLARS_IRAM} and Figures \ref{fig:COLARS_IRAM}, \ref{fig:NONDECT_IRAM}).
In order to estimate the \textit{total} molecular masses of those, an aperture correction is needed.
We employ a similar method as described in \citet{Lisenfeld2011} and \citet{Stark2013}, which we briefly outline here.
Assuming an exponential CO source brightness distribution, the total CO flux of a
circular ($\theta=0..2\pi$) galaxy can be written in polar coordinates as:
\begin{equation}
        I_{CO,extra}=\int_{0}^{2\pi} \int_{0}^{\infty} I(r) 2\pi\ r\ dr\ d\theta
\end{equation}
Assuming I$_{CO}$(r) exponentially distributed with a central peak I$_0$ and a scale length of r$_{e,CO}$, the above equation
can be re-written:
\begin{equation}
        I_{CO,extra}=I_0 \int_{0}^{2\pi} \int_{0}^{\infty}\ exp\left[-\left(\dfrac{r}{r_{e,CO}}\right)\right]\ 2\pi\ r\ dr\ d\theta
\end{equation}
This integral can be solved by partial integration and I$_{CO,extra}$ is found to be:
\begin{equation}\label{eqn:ICOtotal}
        I_{CO,extra}=I_0 2\pi\ r_{e,CO}^2
\end{equation}
The shape of the main beam can be typically estimated as a Gaussian. In fact, it represents a weighting function, which
attenuates the source flux. Its 2D Cartesian formula for a unity peak height is:
\begin{equation}\label{eqn:Gauss2D}
        G(x,y) = exp\left[-\left(\dfrac{x^2}{\sigma_{x}^2}\ +\ \dfrac{y^2}{\sigma_{y}^2}\right)\right]
\end{equation}
The standard deviation $\sigma$ is related to the full width at half maximum or rather the half power beam width ($\theta$) by:
\begin{equation}\label{eqn:FWHMsigma}
        \theta = 2 \sqrt{2 ln(2)}\ \sigma
\end{equation}
Since the measured flux is the result of a convolution of the source brightness distribution with a Gaussian, the observed flux
within the beam can be expressed as: \\
\begin{equation}\label{eqn:GaussSourceConv}
\begin{split}
I_{CO,obs} = I_{CO,extra}\ \ast\ G(x,y) \\
= 4I_0\int_{0}^{\infty} \int_{0}^{\infty}\ exp[-\dfrac{\sqrt{x^2+y^2}}{r_{e,CO}}] \\
\times\ exp\left[-4ln(2) \left(\dfrac{x^2}{\theta^2}\ + \dfrac{(y\ cos(i))^2}{\theta^2}\right)\right]\ dx\ dy
\end{split}
\end{equation}
The inclination \textit{i} of the galaxy is also accounted for in equation \eqref{eqn:GaussSourceConv}. It may be not clear at first sight
that the effect of \textit{i} can be written within the Gaussian term, because in principle it
should be taken into account in the source and not in the beam.
This can be understood, when bringing to mind that the integration interval $dy$ on the sky corresponds to a larger space interval on the
source. This means that in the case of accounting for inclination in the source term, also
the integration interval had to be changed to $dy'\ =\ dy / cos(i)$, where $dy'$ is the interval on the source.
If one takes into account this effect, the result is the same as in the given formula, i.e.
equation \eqref{eqn:GaussSourceConv} can be derived via variable substitution from $y$ to $y'\ =\ y\ cos(i)$.
Since this cannot be solved analytically, we carry out a numerical integration to find the
aperture correction factor C$_{AP}$:
\begin{equation}\label{eqn:ApertCorrFac}
        C_{AP}\ =\ \dfrac{I_{CO,extra}}{I_{CO,obs}}
\end{equation}
For LARS galaxies with CO~(1-0) detections C$_{AP}$ ranges from 1.01--1.94, with an average correction of 30\%.
When calculating C$_{AP}$, the peak brightness temperature $I_0$, cancels out, but the scale length
of the CO distribution, $r_{e,CO}$, is needed as input. Previous studies showed that $r_{e,CO}$
can be estimated from the optical, blue $D_{25}$ diameter \citep{Young1995,Leroy2008,Lisenfeld2011,Davis2013a},
roughly independent of the Hubble type:
\begin{equation}\label{eqn:COscale}
r_{e,CO} = a\ \times\ \sfrac{1}{2} D_{25},
\end{equation}
with \textit{a} typically ranging from 0.2 to 0.25. For our work, we choose a value in-between, $a=0.23$.
The inclination of the galaxy is calculated using the relation of \citet{Hubble1926}:
\begin{equation}\label{eqn:inclsimple}
        cos^2(i)\ =\ \frac{\big(\frac{b}{a}\big)^2\ -\ q^2}{1\ -\ q^2}
\end{equation}
We adopt q=0.2, and minor-to-major axis ratios, $\sfrac{b}{a}_{M18}$, previously derived by
\cite{Micheva2018}, see Table \ref{tab:lars_general}.

\subsubsection{Metallicity Dependent CO-to-H$_2$ Conversion}
The CO-to-H$_2$ conversion factor, $\upalpha_{CO}$ is a (virialized) mass-to-light ratio
in units of $(K\ km\ s^{-1}\ pc^2)^{-1}$ that is used to translate $L'_{CO}$ into
the molecular mass (including Helium).
Given the extreme conditions in the ISM of (some) LARS galaxies,
will potentially cause large uncertainties in $\upalpha_{CO}$, e.g. due to cloud-cloud collisions.
See \cite{Bolatto2013} for a review on the topic.

However, the strongest effect on $\upalpha_{CO}$ variations may emerge due to low metallicity,
because at some (low Z) point CO is not able to shield itself from the ambient
radiation field, and is photo-dissociated, while $H_2$ survives at much larger column densities.
Thus, the fraction of so called CO-dark molecular gas
increases, an effect that needs to be accounted for by $\upalpha_{CO}$.

A variety of prescriptions for the metallicity-dependence of $\upalpha_{CO}$
exist in the literature \citep{Magdis2011,Genzel2012,Narayanan2012,Schruba2012,Hunt2015,Amorin2016,Accurso2017}.
We decided to make use of a semi-analytic relation
between the metallicity $Z'$, the O/H abundance in proportion to the solar abundance,
and the conversion factor $\upalpha_{CO}$, published by \cite{Narayanan2012}.
We discuss the reasoning of our choice in
Appendix \ref{sec:discuss_gdr_alpha}.
\begin{equation}
	\upalpha_{CO}=\frac{min[6.3,10.7\ \times\ I_{CO}^{-0.32}]}{Z'^{0.65}}
\end{equation}
Note that for the abundances of LARS galaxies given in Table \ref{tab:lars_general}, a solar reference value
of [$12+log(O/H)=8.7$] was chosen.

\subsubsection{Upper Limit Molecular Masses from CO~(1--0)}
For the six LARS galaxies that remained undetected in CO~(1--0),
we calculate upper limit CO luminosities from the
2-sigma baseline r.m.s. (measured outside the line window) multiplied by
the square-root of number of channels found within a (typical) 200km/s wide
line window.
The derived CO upper limit luminosity
is then further multiplied by the aperture correction factor and metallicity-dependent
conversion factor $\upalpha_{CO}$ exactly as described above.
However, we caution that the aperture correction
is strictly only valid for peaked CO distributions with an exponential radial decline.
For sources with more complex morphologies -- and this is the case for
few of our galaxies --  
beam dilution might lower the perceived signal
in an unforeseeable way. Moreover, the relatively low
metallicity of the galaxies, leads to hardly predictable beam filling factors,
due to large fractions of CO-dark gas \citep{Schruba2017}, which
further reduces the ability to accurately derive limits. The reported
upper limits have thus to be taken with caution.

\subsection{Star Formation Rates from SED Models, 22$\upmu$m, [C~II]158$\upmu$m and [O~I]63$\upmu$m}
Star formation rates can be derived through many different methods and tracers
\citep[see][for a review on the topic]{Kennicutt2012}.
In what follows, unless otherwise noted, we calculate SFRs from the total
infrared luminosity $L_{TIR}$ that we find from integration
of our best-fit dust model SEDs over a range of 3--1100$\upmu$m.
Using the prescription in \cite[Table 1 therein]{Kennicutt2012}, which is based on work
by \cite{Hao2011} and \cite{Murphy2011}, we translate $L_{TIR}$ into
SFR, assuming a \cite{Kroupa2003} IMF. Consequently, whenever comparing
any of our results to values in the literature, we pay attention to the
underlying IMF and make adoptions, if needed.
Thus, we convert from a \cite{Salpeter1955} to a Kroupa IMF using a constant factor
of 0.66. Given the similarity between Kroupa and \cite{Chabrier2003} IMFs,
no changes are made in that case.

Next, we briefly compare and discuss few other prescriptions
to estimate star formation rates: an extrapolation from 22$\upmu$m single-band observations
\citep{Rieke2009}, estimates from extinction corrected H$\upalpha$ and SFRs from [C~II]158$\upmu$m
and [O~I]63$\upmu$m emission lines (using prescriptions of \cite{deLooze2014}).
Taking the far-infrared SED-based SFRs as a reference,
we find that both, the 22$\upmu$m photometry and the [O~I]63$\upmu$m line,
agree very well, with only $\sim$10\% differences on average (see Table \ref{tab:sfrcomparison}).

The situation is worse for (extinction corrected) H$\upalpha$ and [C~II] that lead to
SFRs that are (on average) higher by factors of $\sim$3 and $\sim$5 compared to the infrared based estimates.
However, for the case of [C~II], the discrepancy is mainly driven by LARS 5,
a galaxy that is known to have a large scale outflow \citep{Duval2016}.
Our finding is thus in agreement with other studies \citep{Appleton2013,Appleton2018,Smirnova-Pinchukova2019} that reported on
excess [C~II]158$\upmu$m emission due to shocks connected to outflowing gas.
Excluding LARS 5, the [C~II] prescription of \cite{deLooze2014} leads to SFRs that are only $\sim$80\% higher.
It is further recognized that SFRs from (extinction corrected) H$\upalpha$ are comparable to those from [C~II],
at least once LARS 5 is excluded. In that case the ratio between SFRs from [C~II] and H$\upalpha$
is 1.7 only.

The reason for the elevated SFRs from H$\upalpha$ and [C~II] compared to the IR based results,
is likely due to the sensitivity to different timescales of the tracers.
Both, H$\upalpha$ and [C~II]158$\upmu$m are sensitive to the currently ongoing star formation
on timescales of $\sim$10Myr, because they are directly linked to the recombination within H~II regions,
driven by young, massive stars.
On the other side, for the continuum at infrared wavelengths and the TIR emission, a significant fraction of
the heating is attributed to more evolved stars. Hence, the timescale probed by TIR is longer ($\sim$100Myr).
The aforementioned discrepancy between the SFRs might thus be caused by
varying star formation histories (Melinder et al. in prep.),
given the relatively young age of LARS galaxies,
due to their selection with H$\upalpha$ equivalent widths of more than 100\AA\ \citep{Oestlin2014}.

\begin{table}
\begin{threeparttable}
\caption[SFR comparison]{Comparison of SFR estimates for LARS galaxies}
\label{tab:sfrcomparison}
\begin{tabular}{rrrrrr}
\hline \hline
ID & SFR$_{H\upalpha}$ & SFR$_{22\upmu m}$ & SFR$_{[C~II]}$ & SFR$_{[O~I]}$ & SFR$_{TIR}$ \\
(1) & (2) & (3) & (4) & (5) & (6) \\[0.6em]
1  &  6.5     & 4.6$\pm$0.12  & --           & --           & $2.13_{-0.56}^{+1.66}$ \\[0.6em]
2  &  1.4     & 0.2$\pm$0.06  & 0.4$\pm$0.2  & $<$2.3       & $0.28_{-0.11}^{+0.23}$ \\[0.6em]
3  &  26.3    & 105.7$\pm$0.3 & 35.5$\pm$0.8 & 73$\pm$3     & $65.76_{-2.41}^{+17.48}$ \\[0.6em]
4  &  6.1     & 5.2$\pm$0.15  & --           & --           & $4.00_{-0.18}^{+0.47}$ \\[0.6em]
5  &  5.8     & 2.0$\pm$0.1   & 29$\pm$5     & --           & $0.81_{-0.17}^{+0.41}$ \\[0.6em]
6  &  0.6     & --            & --           & --           & $0.27_{-0.06}^{+0.18}$ \\[0.6em]
7  &  9.3     & 4.6$\pm$0.2   & --           & --           & $2.99_{-0.15}^{+0.74}$ \\[0.6em]
8  &  36.8    & 31.0$\pm$0.32 & 81$\pm$13    & --           & $29.02_{-6.38}^{+9.25}$ \\[0.6em]
9  &  40.7    & 24.7$\pm$0.44 & 33.4$\pm$0.6 & 34$\pm$3     & $25.31_{-3.56}^{+0.40}$ \\[0.6em]
10 &  5.6     & 1.9$\pm$0.3   & --           & --           & $6.69_{-2.22}^{+0.93}$ \\[0.6em]
11 &  30.5    & 28.4$\pm$1.01 & 203$\pm$12   & --           & $38.78_{-5.96}^{+14.54}$ \\[0.6em]
12 &  97.0    & 12.7$\pm$0.72 & 8.7$\pm$1.2  & 16$\pm$4     & $9.99_{-0.12}^{+0.88}$ \\[0.6em]
13 &  64.8    & 66.5$\pm$3.02 & 40.5$\pm$4.1 & 54$\pm$10    & $106.71_{-10.07}^{+6.35}$ \\[0.6em]
14 &  24.8    & 7.0$\pm$2.13  & --           & --           & $5.79_{-0.14}^{+0.98}$ \\[0.6em]
\end{tabular}
\begin{tablenotes}
\item The SFRs are given in units of M$_{\odot}\ yr^{-1}$.
The LARS ID is shown in \textit{column 1}. Previously published SFRs from
extinction corrected H$\upalpha$ \citep{Hayes2014} are found in \textit{column 2}. New SFR estimates from
WISE 22$\upmu$m, [C~II]158$\upmu$m, [O~I]63$\upmu$m and
total infrared luminosities are given in \textit{columns 3-6}. The latter one is used for all analysis (KS plots etc.) in this paper.
\end{tablenotes}
\end{threeparttable}
\end{table}

\subsection{SFR and Gas Surface Densities}
For calculations of star formation rate surface densities, $\Sigma_{SFR}$, and gas surface densities, $\Sigma_{gas}$,
we normalize the measured integrated values to the area enclosed by the 25 mag arcsec$^{-2}$ isophote in the
optical B-band, i.e. $D_{25_{SDSS}}=\pi \times \sfrac{a}{2} \times \sfrac{b}{2}$.
D$_{25_{SDSS}}$ was calculated from the SDSS g-band, which typically gives sizes that are $\sim$1.3 times larger
than those measured in the Johnson B band \citep{Hakobyan2012}. Thus, we divided the g-band diameters by that factor
first.

\begin{sidewaystable*}%
\resizebox{\textheight}{!}{%
\begin{threeparttable}
\caption{Derived properties for LARS galaxies}
\label{tab:lars_derived_props}
\begin{tabular}{lrrrrrrrrrrrrrrrr}
\hline \hline
$ID$ & $log_{10}$       & $log_{10}$            & $GDR$  & $\upalpha_{CO}$ & $C_{AP}$ & $log_{10}$     & $log_{10}$      & $log_{10}$              & $log_{10}$        & $log_{10}$       & $log_{10}$      & $\tau_{gas}$ & $f_{gas}$ & $f_{HI}$ & $f_{mol}$ & $f_{PDR}$ \\
     &  $\Sigma_{SFR}$  & $\Sigma_{SFR}^{rcut}$ &        &               &          & $L'_{CO}$      & $M_{H_{2_{obs}}}$ & $M_{H_{2_{est}}}$     & $M_{gas_{obs}}$   & $M_{gas_{est}}$  & $\Sigma_{gas}$  & $\times10^8$ &           &          &           &  \\
(1)  & (2)              & (3)                   & (4)    & (5)           & (6)      & (7)            & (8)             & (9)                     & (10)              & (11)           & (12)          & (13)     & (14)     & (15)      & (16)  & (17) \\
1   &  -1.862  &  -0.277  &  455.0  &  11.40  &  1.07  &  $<$8.443  &   $<$9.438  &   9.816  &   $<$9.719  &  10.058  &  1.868  &   53.70  &  0.65  &   0.22  &  0.57  &  0.21  \\
2   &  -2.901  &  -0.641  &  472.1  &  10.60  &  1.15  &  $<$8.666  &   $<$9.629  &   8.779  &   $<$9.849  &   9.534  &  1.185  &  122.10  &  0.59  &   0.82  &  0.18  &  0.01  \\
3   &  -1.444  &   1.722  &  306.2  &   5.10  &  1.94  &  9.549     &  10.198     &  10.289  &  10.356     &  10.496  &  1.234  &    4.80  &  0.61  &   0.22  &  0.62  &  0.16  \\
4   &  -1.866  &  -0.348  &  511.7  &  --     &  --    &  --        &  --         &   9.054  &  --         &  10.009  &  1.541  &   25.50  &  0.44  &   0.75  &  0.11  &  0.13  \\
5   &  -2.339  &  -0.523  &  597.0  &  14.80  &  1.06  &  $<$8.482  &   $<$9.596  &  --      &   $<$9.835  &   8.786  &  0.539  &    7.50  &  0.13  &   4.75  &  --    &  0.08  \\
6   &  -3.053  &  -0.828  &  699.5  &  --     &  --    &  --        &  --         &  --      &  --         &   8.575  &  0.090  &   13.90  &  0.15  &  --     &  --    &  0.01  \\
7   &  -2.029  &   0.106  &  353.2  &  11.00  &  1.10  &  $<$8.539  &   $<$9.526  &   9.227  &   $<$9.810  &   9.748  &  1.244  &   18.70  &  0.54  &   0.55  &  0.30  &  0.14  \\
8   &  -1.846  &   0.152  &  248.3  &   4.50  &  1.36  &  9.546     &  10.145     &  10.030  &  10.556     &  10.535  &  1.226  &   11.80  &  0.27  &   0.64  &  0.31  &  0.05  \\
9   &  -0.425  &   0.157  &  342.0  &  11.70  &  1.01  &  8.573     &   9.595     &  10.173  &  10.229     &  10.484  &  2.656  &   12.00  &  0.37  &   0.43  &  0.49  &  0.09  \\
10  &  -2.250  &   0.164  &  248.3  &   6.30  &  1.13  &  $<$9.353  &  $<$10.114  &   9.781  &  $<$10.243  &  10.025  &  0.950  &   15.80  &  0.33  &   0.42  &  0.57  &  0.00  \\
11  &  -1.638  &   0.465  &  291.1  &   8.00  &  1.19  &  9.498     &  10.388     &  10.377  &  10.703     &  10.714  &  1.487  &   13.30  &  0.30  &   0.50  &  0.46  &  0.04  \\
12  &  -2.021  &  -0.135  &  361.4  &  10.20  &  1.03  &  $<$9.457  &  $<$10.465  &  --      &  --         &   9.638  &  0.617  &    4.30  &  0.37  &  --     &  --    &  0.11  \\
13  &  -1.333  &   0.543  &  249.5  &  11.30  &  1.03  &  9.273     &  10.364     &  --      &  --         &  10.507  &  1.146  &    3.00  &  0.35  &  --     &  --    &  0.08  \\
14  &  -2.474  &  -0.536  &  848.2  &  --     &  --    &  --        &  --         &  --      &  --         &   9.829  &  0.592  &   11.60  &  0.79  &  --     &  --    &  0.21
\end{tabular}
\begin{tablenotes}\small
\item Surface densities in columns (2) and (10) are normalized to the
B-band 25 mag arcsec$^{-2}$ area, while the SFR surface density in column (3) is normalized to the star forming area only.
The CO luminosity in column (7) was derived from the CO velocity-integrated main beam brightness temperature and subsequent
multiplication with the aperture correction factor in column (6). For galaxies with CO detection, M$_{H_{2_{obs}}}$ in column
(8) was calculated from the CO luminosity in column (7) via application of the conversion factor in column (5).
For galaxies with positive CO and H~I detections, M$_{gas_{obs}}$ is the sum of the atomic mass from \cite{Pardy2014} and the
molecular mass in column (8).
Total gas masses M$_{gas_{est}}$ in column (11) were derived from the dust masses using a metallicity-dependent gas-to-dust ratio shown in column (4).
The estimated molecular mass M$_{H_{2_{est}}}$ is the result of the dust based total mass M$_{gas_{est}}$ minus the observed atomic gas mass.
The total gas surface density in column (12), the total gas depletion time $\tau_{gas}$ in column (13)
as well as the total gas fraction $f_{gas}=\sfrac{M_{gas_{est}}}{(M*+M_{gas_{est}})}$ in column (14) are using the dust-based total gas mass estimate.
The atomic gas fraction $f_{HI}$ is the ratio between the observed H~I gas mass and the total dust-based gas mass.
The molecular gas fraction $f_{mol}$ is the ratio between the estimated molecular gas mass and the total dust-based gas mass.
The PDR gas mass fractions in columns (17) are the $\gamma$ values from Table \ref{tab:dl07fitresults} on a linear scale.
Note that the large discrepancies between the reported upper limits in columns (8) and (10) and the dust-based estimated quantities
are likely the result of beam dilution, i.e. the upper limits are likely misleading.
\item The units of the derived quantities are:
\item (2), (3): $M_{\odot}\ yr^{-1}\ kpc^{-2}$
\item (5): $(K\ km\ s^{-1}\ pc^2)^{-1}$
\item (7): $K\ km\ s^{-1}\ pc^2$
\item (8)-(11): $M_{\odot}$
\item (12): $M_{\odot} pc^{-2}$
\item (13): $yr$
\end{tablenotes}
\end{threeparttable}}
\end{sidewaystable*}

\section{Results}\label{sec:results}

\subsection{Scaling of Gas Mass and SFR}\label{sec:total_gas_vs_sfr}
\begin{figure}
\centering
     \includegraphics[width=0.6\columnwidth]{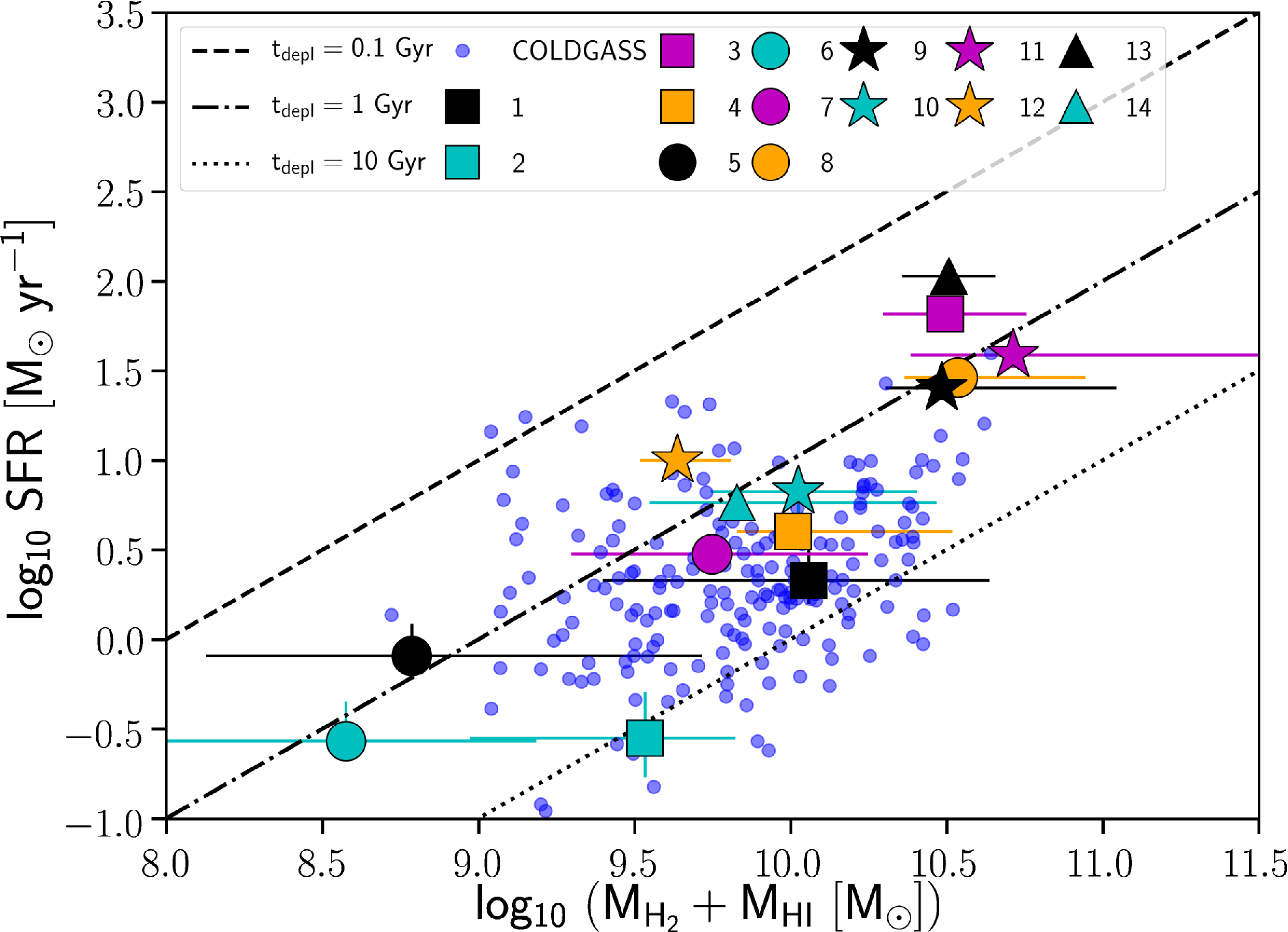}
     \caption[Gas vs. SFR]{Gas mass and SFR (from total integrated infrared luminosity) scaling for LARS galaxies and COLDGASS, a z$\sim$0 representative sample
     of galaxies.}
     \label{fig:total_gas_vs_sfr}
\end{figure}

\begin{figure}
\centering
     \includegraphics[width=0.6\columnwidth]{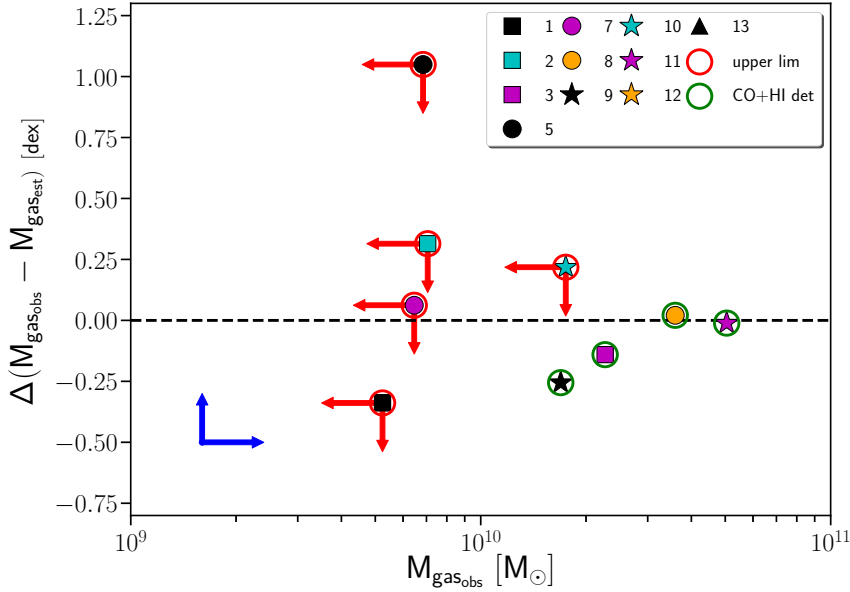}
     \caption{Total gas masses from combined CO (1-0) and H~I observations vs. the difference between observed
     total gas mass (CO+HI) and the dust-based total gas mass.
     Galaxies with detections in CO+HI are encircled in \textit{green}, while measurements with CO and/or H~I
     upper limits are shown in \textit{red}. Note that the CO upper limits are most likely unreliable due to beam
     dilution that works in the direction of the \textit{blue arrows}.}
     \label{fig:gas_obs_vs_est}
\end{figure}

The scaling relation between dust-traced total gas mass and star formation rate
in Figure \ref{fig:total_gas_vs_sfr} shows that LARS galaxies span a more than two orders of magnitude wide range
in both quantities. LARS comprises the whole parameter space of the COLDGASS survey 
\citep{Saintonge2011a,Saintonge2011b,Saintonge2012,Catinella2012},
a representative sample of z$\sim$0 galaxies with masses above $10^{10}\ M_{\odot}$.
A comparison between the two applied methods to calculate gas masses
(dust-based vs. CO+HI based) is shown in Figure \ref{fig:gas_obs_vs_est}.
For those galaxies (LARS 3, 8, 9, 11) with combined measurements of CO+HI available (LARS 13 was not detected in H~I previously),
we find an excellent match between the two methods, i.e. on average they agree on a level better
than $\sim$0.2dex. Note that the upper limit molecular masses from our CO (1-0)
observations are most likely affected by beam dilution,
leading to an increase of the limits towards higher masses.
We only use dust-based gas mass estimates for the following analysis.

\subsection{KS Relation, Gas Fractions and Depletion Times}\label{sec:KS}
\begin{figure*}
\centering
     \subfloat{\includegraphics[width=12cm]{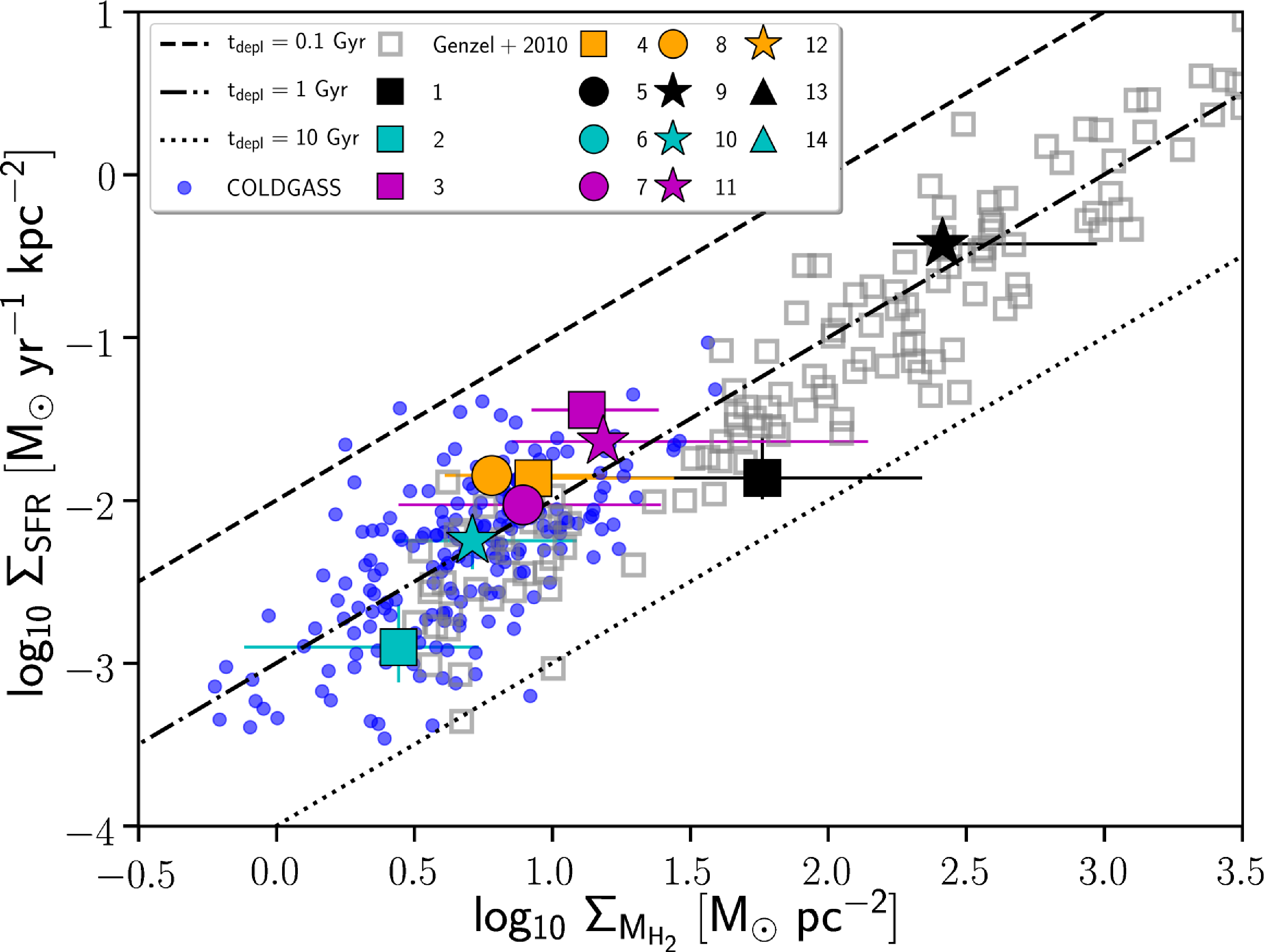}}

     \subfloat{\includegraphics[width=12cm]{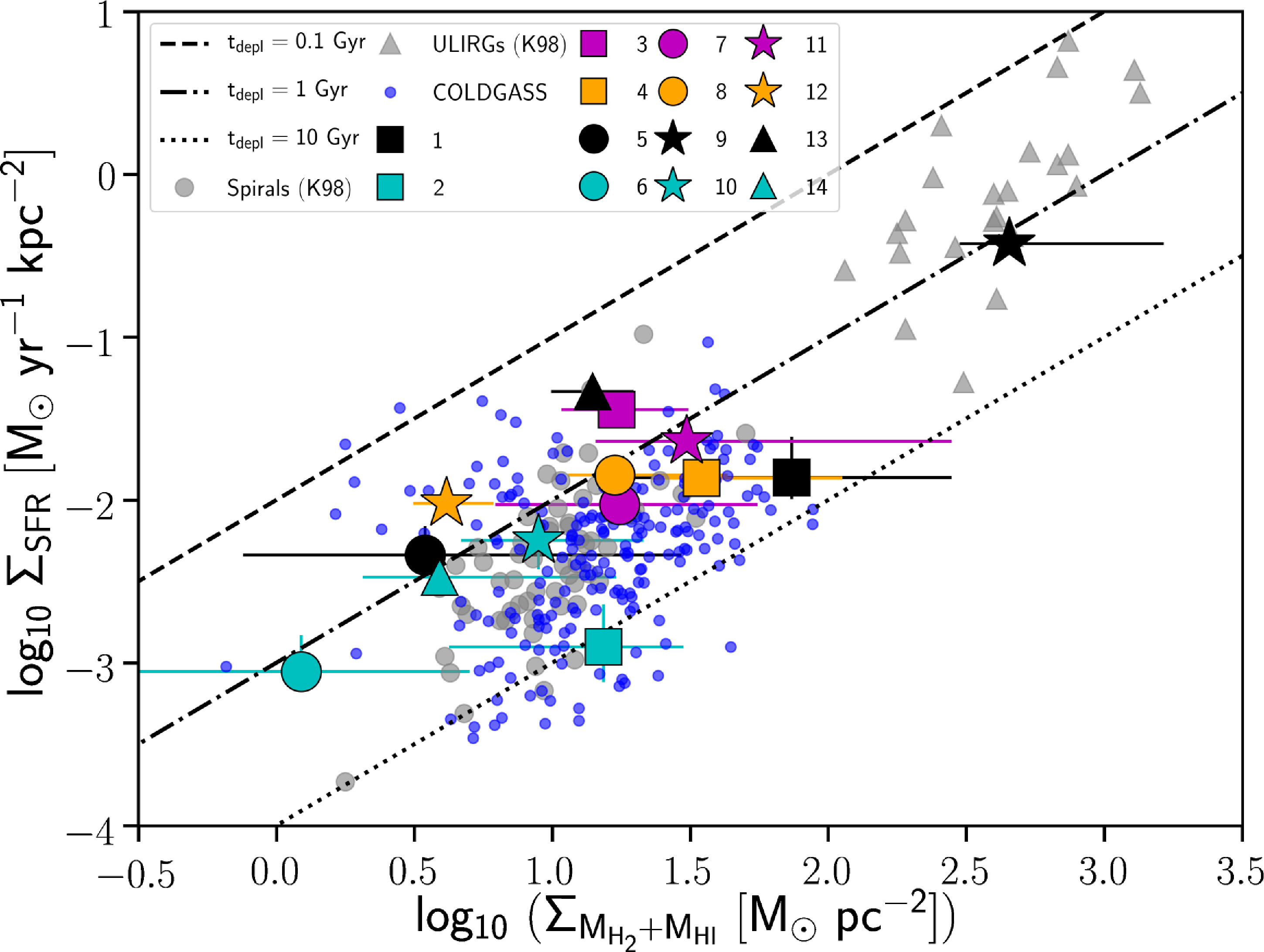}}

     \caption{\textit{Top panel}: Molecular KS relation: H$_2$ surface density (from dust-based gas mass minus atomic mass) versus star formation rate surface
     density (from total integrated L$_{TIR}$) for LARS compared to COLDGASS (\textit{blue}) and a sample of z$\sim$1-3 star forming galaxies from \cite{Genzel2010}
     (\textit{gray open squares}).
     \textit{Bottom panel}: KS relation: Total gas surface density (dust-based) vs. star formation rate surface density (from total integrated L$_{TIR}$).
     The COLDGASS galaxies (\textit{blue}) and \cite{Kennicutt1998} samples of local spirals (\textit{gray filled circles}) and ULIRGs
     \textit{gray filled triangles} are shown as reference.
     Black diagonal lines indicate (from top to bottom) constant gas depletion times of 0.1, 1 and 10 Gyr.
     }
     \label{fig:KS}
\end{figure*}

The global molecular Kennicutt-Schmidt relation for LARS is shown in the top panel of Figure \ref{fig:KS}.
We compare our sample to the COLDGASS population as well as to z$\sim$1-3 massive star forming main sequence galaxies
of \cite{Genzel2010}.

Compared to COLDGASS galaxies our sample is biased towards higher H$_2$ surface densities,
with LARS 9 at the extreme end, where typically only high-z star-forming galaxies are located.
At the lowest end, LARS 2 is found.
The molecular gas depletion time, defined as the ratio between $\Sigma_{M_{H_2}}$ and $\Sigma_{SFR}$ seems to
be rather consistent for the whole sample, with a scatter of only $\pm$1\,Gyr around the median value of
0.7\,Gyr.
The classical KS relation in the bottom panel of Figure \ref{fig:KS} is slightly different.
The total gas depletion time, i.e. the ratio between $\Sigma_{M_{H_2}+M_{HI}}$ and $\Sigma_{SFR}$,
has a 3 times larger scatter ($\pm$3\,Gyr) around the median of 1.3\,Gyr for the whole sample.

The inverse of the gas depletion time is often denoted as the
\textit{star formation efficiency} (hereafter SFE), i.e. the star formation
rate per unit gas mass.
However, it is important to understand that the SFE as defined above is not
\text{necessarily} an indication of a true efficiency or ability
to form stars per unit gas mass, because not all
the observed gas is readily available to fuel the current star formation.
The efficiency is thus degenerate with the gas fraction.
For that reason, \cite{Krumholz2012} suggest to normalize the SFE
by the cloud-scale gravitational free-fall time $\tau_{ff}$. That way, the
scatter in the classical KS law decreases. Unfortunately, $\tau_{ff}$
cannot be assessed from our observations, because
the determination of $\tau_{ff}$ involves knowledge of the volume density of the gas,
a quantity that cannot be performed based on our observations.
For that reason we will mainly make use of the inverse of the SFE
as defined above, which is the gas depletion time ($\tau_{gas}$),
i.e. the timespan star formation may be maintained at the current level given
the amount of gas available.

With previously published stellar masses \citep{Hayes2014} that are given in column 3 of Table
\ref{tab:lars_general}, the total gas fraction can be defined: $f_{gas}=\sfrac{M_{gas_{est}}}{(M*+M_{gas_{est}})}$.
For LARS the total gas fractions vary from $\sim$15--80\%.
All derived quantities discussed in this section are summarized in Table \ref{tab:lars_derived_props}.

\subsection{Compactness and Characteristic Size of Star Formation Activity}
\begin{figure}
\centering
     \includegraphics[width=0.6\columnwidth]{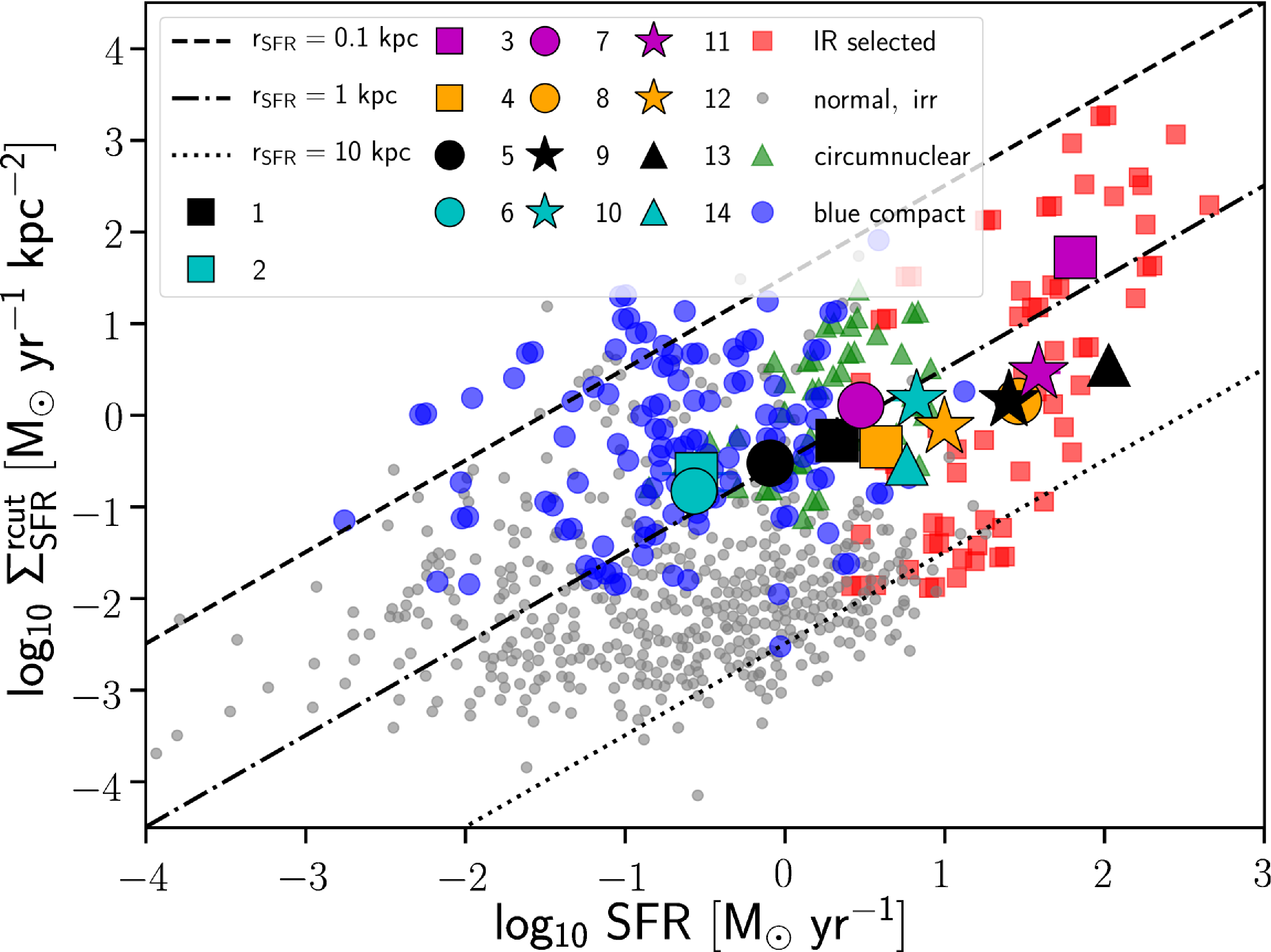}
     \caption{Total SFR versus the star formation rate density (both using SED integrated L$_{TIR}$ as tracer) normalized by the area of activity (as seen in UV).
     Overplotted are comparison samples of \cite{Kennicutt2012}.}
     \label{fig:KS_Fig9}
\end{figure}

Figure \ref{fig:KS_Fig9} shows the star formation rate versus the 
surface density of star formation, $\Sigma_{SFR}^{rcut}$ --
in this case the SFR is normalized to the area of star forming activity
(derived by \cite{Micheva2018} from FUV imaging) rather than stellar content.
It is striking that compared to the previous KS plots the scatter in the y direction
is largely decreased, and also that compared to other galaxy samples,
the size of the star forming activity in LARS galaxies is found to be
roughly constant with a median diameter of $\sim$1kpc, similar in size
to circumnuclear star formation activity. However, a weak but tight trend is seen
between $\Sigma_{SFR}^{rcut}$ and SFR, in a sense that galaxies with
high star formation rates exhibit larger and more crowded areas of star formation.
A possible explanation for the tighter spread seen in $\Sigma_{SFR}^{rcut}$
compared to $\Sigma_{SFR}$ may be found in the star formation history,
which varies from one LARS galaxy to another and has led to varying fractions and sizes
of a (older) stellar populations, while the currently ongoing
star formation density is similar.

\subsection{Grouping LARS According to $\tau_{gas}$}\label{sec:lars_groups}
\begin{figure}
\centering
   \includegraphics[width=0.6\columnwidth]{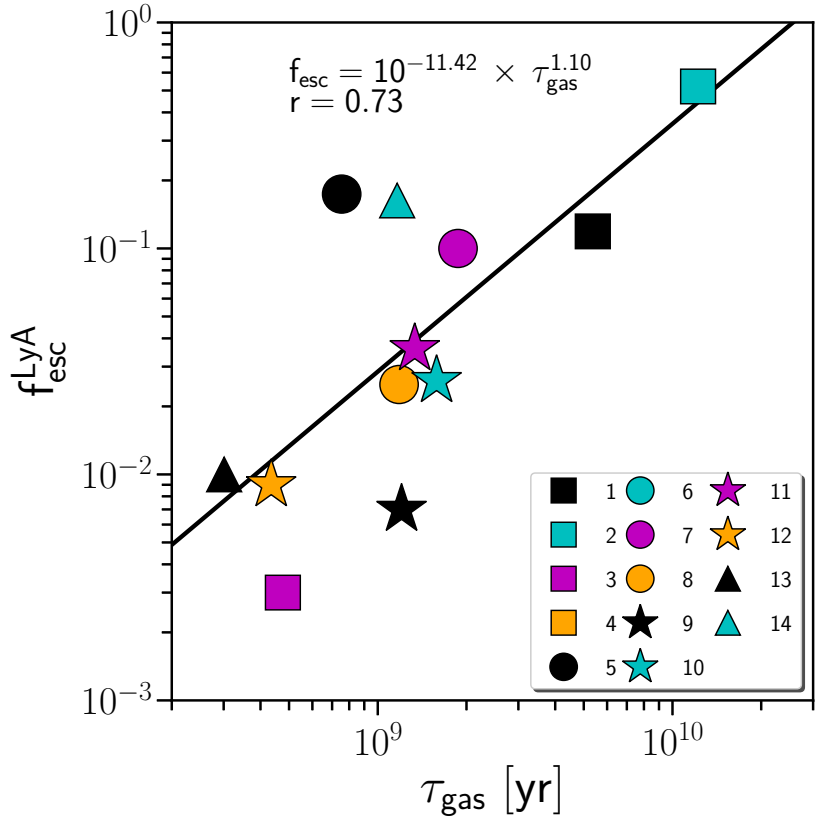}
   \caption[Gas Depletion Time vs. \lya\ Escape Fraction]{Relation between total gas depletion time and \lya\ escape fraction.
	Note that LARS 4 and 6 have a \textit{global} \lya\ escape fraction of zero and are thus omitted in the fit.
}
\label{fig:depl_vs_lya}
\end{figure}

An examination of the gas depletion times and gas fractions
(see Table \ref{tab:lars_derived_props}) reveals that the heterogeneity
of LARS galaxies that is seen in the KS law is related to \lya\ escape.
While, no trend is seen between total gas surface density and \lya\ escape,
we find a strong trend between the total gas depletion time and \lya\ escape fraction (see Figure \ref{fig:depl_vs_lya}).
This is also seen when grouping LARS galaxies into bins of increasing total gas
depletion times ($\tau_{gas}$):
\begin{itemize}
\item \textit{Group~1} ($\tau_{gas} \lesssim 1 Gyr$): \\
As seen in Figure \ref{fig:KS}, the galaxies with the shortest gas depletion times
are LARS 3, 5, 12 and 13.

\item \textit{Group~2} ($1 Gyr \lesssim \tau_{gas} \lesssim 2 Gyr$): \\
In the group of intermediate gas depletion times -- with values that are typically found in main sequence galaxies --
we find LARS 6, 8, 9, 10, 11 and 14. This group is the largest of the three and the galaxies within are quite diverse.
In fact we could break up the group into high- and low- stellar mass sub-samples. While the high-mass galaxies (LARS 8, 9, 10 and 11)
might indeed be similar to z$\sim$1--2 main sequence galaxies, a characterisation of the low-mass galaxies (6, 14) is difficult.

\item \textit{Group~3} ($\tau_{gas} \gtrsim 2 Gyr$): \\
Those LARS galaxies that have relatively high gas surface densities and low SFR surface densities (1, 2, 4, 7)
have much larger total gas fractions.
\end{itemize}

\begin{table*}
\caption[Groups of LARS galaxies]{Average and median properties of LARS galaxies grouped into bins of increasing gas depletion times. Median quantities are given in the brackets.}
\label{tab:taudeplbins}
\begin{tabular}{lcccccc}
\hline \hline
                               & $\tau_{gas}$        & f$_{esc}^{LyA}$      & D$_{scatt}$         & f$_{gas}$        & f$_{mol}$                  & U$_{min}$          \\ 
                               & [10$^8$ yr]         & [\%]                 & [kpc]               & [\%]             & [\%]                       &                    \\ 
Group~1 (3,5,12,13)            & 5 (5)               & 4.9 (1.0)            & 1.0 (0.7)           & 37 (36)          & 62 (62)                    & 17.0 (17.9)        \\ 
Group~2 (6,8,9,10,11,14)       & 13 (13)             & 5.1 (2.6)            & 0.9 (0.5)           & 37 (32)          & 46 (48)                    & 8.7 (8.2)          \\
Group~3 (1,2,4,7)              & \textbf{55 (40)}    & \textbf{24.6 (11.9)} & \textbf{0.5 (0.5)}  & \textbf{56 (57)} & \textbf{29 (24)}           & \textbf{7.9 (7.9)} \\
All LARS (incl. 6)             & 23 (13)             & 9.9 (3.1)            & 0.8 (0.5)           & 42 (37)          & 40 (46)                    & 10.8 (9.2)         \\
\end{tabular}
\end{table*}

Next, we calculate for each $\tau_{gas}$ group the average and median physical quantities
that we previously derived and relate those to the properties of
\lya\ escape. The result is shown in Table \ref{tab:taudeplbins}.
Note that the results hold-up when the median is used instead of the average.
It is eye-catching that LARS galaxies with the longest depletion times (Group~3)
also exhibit the highest \lya\ escape fractions (more than twice of the average in the sample), while
they have the shortest \lya\ scattering distances.
Moreover their average environment and ISM conditions are significantly different from the other two groups,
in a sense that their mean (FUV) energy density $U_{min}$ is the lowest
and their total gas fractions are highest (while having the lowest molecular gas fraction).
We discuss this further in Sections \ref{sec:discuss_accretion}
and \ref{sec:discuss_feedback}.

\subsection{Extreme [C~II]158$\upmu$m Line Strength in LARS 5 and Basic PDR Analysis}\label{sec:PDR}
\begin{figure*}
\centering
   \includegraphics[width=1\textwidth]{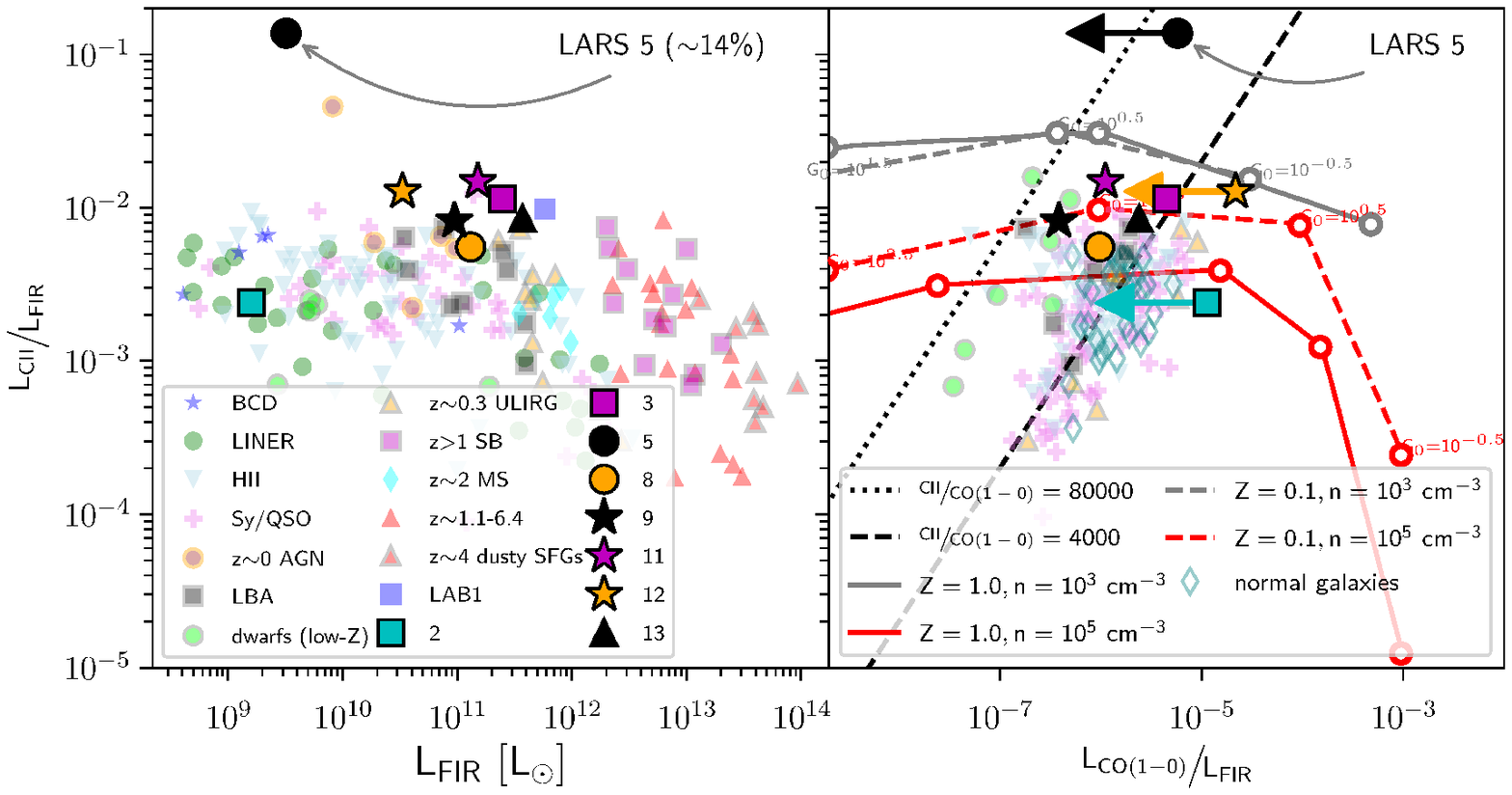}
   \caption[\[CII\] 158$\mu m$ line to FIR continuum ratio]{$L_{CII}$ 158$\mu m$ line to $L_{FIR_{n,H88}}$ ratio for LARS galaxies
   as a function of $L_{FIR_{w}}$ (\textit{left panel}) and as a function of $L_{CO10}$ line to $L_{FIR_{n,H88}}$ ratio (\textit{right panel}),
   together with reference data. For the left plot, we updated the compilation in \cite[Fig. 1]{GraciaCarpio2011},
   and added samples of Lyman Break Analogs (LBA) from \cite{Contursi2017}, local low-metallicity dwarfs from \cite{Cormier2014,Cormier2015},
   z$\sim$0 AGNs from the CARS survey \citep{Smirnova-Pinchukova2019},
   z$\sim$0.3 ULIRGs from \cite{Magdis2014}, z$>$1 starbursts (SB>1)
   taken from \cite{Carilli2013}, z$\sim$2 main-sequence (MS) galaxies \citep{Zanella2018}, the Lyman Alpha Blob 1 (LAB1) at z$\sim$3.1 from
   \cite{Umehata2017} and z$>$4 dusty star-forming galaxies
   \citep{Bothwell2017}. The right hand plot is based on \cite[Fig. 9]{Contursi2017}, which we updated by adding
   normal galaxies from \cite{deBreuck2011} (originally drawn from \cite{Stacey2010}), as well as local low-metallicity dwarfs from
   \cite{Cormier2010,Cormier2014,Cormier2015,Madden2018}. Black dotted and dashed lines in the right panel indicate
   constant [C~II]-to-CO(1-0) ratios of 80000 and 4000 respectively. Red and gray solid curves show PDR models of \cite{Kaufman1999}
   for densities of $n=10^5$ and $n=10^3\ cm^{-3}$ at solar metallictiy (Z=1.0), while dashed curves indicate low metallicity (Z=0.1).
   For the low metallicity cases, the changing FUV radiation field along the curves is given in steps of 1dex in $G_0$ from right to left.}
\label{fig:pdr_cii_fir}
\end{figure*}

In eight LARS galaxies, the [C~II]158$\upmu$m line was successfully detected, either with Herschel/PACS or SOFIA/FIFI-LS.
We first examine the relative strength of the [C~II]158$\upmu$m line compared to the total far-infrared energy.
As seen in Figure \ref{fig:pdr_cii_fir}, LARS 5 is an extreme outlier in the plot. Neither 
LINER, Seyfert or QSOs from the SHINING survey \citep{HerreraCamus2018a,HerreraCamus2018b,Zhao2016},
nor any other non-AGN galaxy observed to date (see figure caption for details about the comparison samples),
has a global [C~II]158$\upmu$m to FIR ratio on the order of
$\sim$14$\pm$3 percent
as seen in LARS 5 (see also Table \ref{tab:pdr_luminosities}).
The mechanism that drives the extreme ratio is likely connected to the large scale galactic wind
that was previously reported and studied by \cite{Duval2016}.
This is supported by the
fact that both, the [C~II]158$\upmu$m and [O~III]88$\upmu$m emission lines are blueshifted from the systemic velocity,
as seen in Figures \ref{fig:lars_ciispec} and \ref{fig:lars_oiiispec}.
Note that shock-enhanced [C~II]158$\upmu$m excess was previously observed in other galaxies
with outflows \citep{Appleton2013,Appleton2018,Smirnova-Pinchukova2019}.

While LARS 5 is an extreme outlier, most of the other
galaxies detected in [C~II]158$\upmu$m (3, 8, 9, 11, 12 and 13)
also lie on the upper end of what is typically observed in local galaxies.
On the other hand, LARS 2 has a much weaker relative [C~II] line strength, which is interesting,
considering the fact that LARS 2 is the only galaxy
with a [C~II] detection that is member of Group~3 (while LARS 7 despite having the longest
integration time remained undetected).

We perform a very basic PDR analysis based on the models of \cite{Kaufman1999,Kaufman2006}.
Using the far infrared luminosity ($L_{FIR_{w}}$) and emission lines of [C~II]158$\upmu$m and
[O~I]63$\upmu$m, we apply
\texttt{PDR Toolbox} \citep{PDRToolbox} to fit the observed line ratios
against the model predictions. The results for the best-fit model are summarized in
Table \ref{tab:result_pdrtoolbox}. 
The quality of the fits can be assessed from Figure \ref{fig:pdr_chi2}, where the
contours indicate 1, 2 and 3$\sigma$ significance regions that
are confined by $\upchi^2_{\upalpha}$=$\upchi^2_{min}$+$\updelta(\upnu,\upalpha)$ \citep{Wall1996},
with the $\upchi^2$-difference $\updelta$ being a function of degrees of freedom $\upnu$ (in our case $\upnu$=1)
and $\upalpha$ the desired significance (in our case 0.68, 0.95 and 0.99).
Note that for LARS 3 and 9, the models are not able to reproduce the observations, i.e.
$\upchi^2_{min}\gg1$. This might be related to the fact that both are merging galaxies. However, for the
remaining galaxies LARS 2, 12 and 13 we find model fits with uncertainties of $\pm$0.6,
$\pm$0.15 and $\pm$0.1 dex respectively.
It may also be recognized in Figure \ref{fig:pdr_chi2} that the solution plane shows
two valleys (bimodality), with density and radiation field strength being degenerate. However, solutions
at lower gas densities and stronger radiation field strengths G$_0$ are more pronounced.
For the second valley the solutions would lead to G$_0$<3,
i.e. relatively low given the starburst nature of our galaxies.
We caution that we apply models that are valid
only for a particular regime (PDR) to unresolved observations
that cover a multi-phase ISM. Thus, the power of such an analysis is limited.
However, a comparison of the derived PDR gas volume densities between
LARS 2 (which is member of Group~3) and LARS 12 and 13 (Group~1)
shows that the density in LARS 2 is $\sim$5 times (or 0.7 dex)
higher than the average density of LARS 12 and 13.
Assuming that the high PDR gas density holds for the whole Group~3,
this might signalize that bulk of the star forming regions are still deeply embedded in
their birth clouds that were not yet disrupted due to stellar feedback. We are thus likely witnessing
early stages of star formation. This scenario is supported by low $\Sigma_{SFR}$, low $U_{min}$ values and
high total gas fractions found in Group~3.

Assuming that the high observed PDR gas volume density in LARS 2 is reflecting
a high density of the cold, star-forming gas,
the free-fall time in LARS 2 is $\sim$3 times shorter,
suggesting that the scatter in the KS law -- once normalized to the free-fall time --
could indeed be largely reduced, given that LARS 2 is the most extreme outlier in the relation.

We also put the LARS galaxies onto a widely used diagnostic diagram as shown in the
right panel of Figure \ref{fig:pdr_cii_fir} and overplot lines of constant density
together with knot points indicating the FUV field strength G$_0$.
The shown models were calculated for two metallicities, one with solar and the other one
with one tenth solar abundances. We further consider only the case for A$_V$=1.
Note that with this diagram we mainly want to showcase the general sensitivity of the
diagnostic lines. While the [C~II]158$\upmu$m to FIR ratio mainly defines the density,
the CO~(1--0) to FIR ratio is most sensitive to the radiation field.

\begin{table*}
\resizebox{\textwidth}{!}{%
\begin{threeparttable}
\caption[FIR continuum and line luminosities]{FIR Continuum and Line Luminosities}
\label{tab:pdr_luminosities}
\begin{tabular}{rrrrrrrrr}
\hline \hline 
ID   & $L_{TIR}$   & $L_{FIR_{w}}$ &  $L_{FIR_{n,SED}}$ & $L_{FIR_{n,H88}}$ & $L_{CII}$ & $L_{OI_{63}}$ & $L_{OIII_{88}}$ & $L_{CO10}$     \\
     & $3-1100\mu m$ & $40-500\mu m$ &  $40-120\mu m$ & $40-120\mu m$          & & & & \\
     & [10$^{42}$ erg s$^{-1}$] & [10$^{42}$ erg s$^{-1}$] & [10$^{42}$ erg s$^{-1}$] & [10$^{42}$ erg s$^{-1}$] & [10$^{39}$ erg s$^{-1}$] & [10$^{39}$ erg s$^{-1}$] & [10$^{39}$ erg s$^{-1}$] & [10$^{36}$ erg s$^{-1}$] \\[0.6em]
1	&	$54.81_{-14.4}^{42.48}$		& $31.43_{-6.79}^{28.54}$		& $24.6_{-4.03}^{16.81}$	&	$19.64_{-0.04}^{14.89}$     &    --            &   --           &  --           &  <48 \\[0.6em]
2	&	$7.27_{-2.66}^{5.73}$		& $6.14_{-2.2}^{4.62}$		& $3.86_{-1.41}^{3.39}$	&	$7.08_{-3.09}^{0.01}$               &    17$\pm$8      &   62$\pm$43    &  --           &  <80 \\[0.6em]
3	&	$1690.33_{-61.83}^{449.23}$	& $950.04_{-15.56}^{135.73}$	& $796.7_{-0.08}^{147.86}$	&	$136.79_{-102.34}^{287.82}$ &    1564$\pm$36   &   2994$\pm$145 &  1918$\pm$124 &  619$\pm$31 \\[0.6em]
4	&	$102.71_{-4.48}^{12.08}$	& $58.66_{-2.2}^{6.25}$		& $47.34_{-3.84}^{3.49}$	&	$27.22_{-0.22}^{21.18}$         &    --            &   --           &  --           &  -- \\[0.6em]
5	&	$20.88_{-4.32}^{10.39}$		& $12.35_{-1.01}^{3.28}$		& $10.44_{-0.69}^{2.75}$	&	$9.15_{-0.01}^{7.07}$       &    1260$\pm$223  &   --           &  671$\pm$114  &  <53 \\[0.6em]
6	&	$6.91_{-1.67}^{4.64}$		& $4.85_{-0.44}^{1.65}$		& $3.65_{-0.12}^{1.17}$	&	$5.15_{-0.02}^{4.06}$               &    --            &   --           &  --           &  -- \\[0.6em]
7	&	$76.85_{-3.93}^{19.1}$		& $43.81_{-3.55}^{12.16}$		& $34.61_{-4.94}^{7.06}$	&	$20.62_{-0.04}^{15.66}$     &    --            &   --           &  --           &  <62\\[0.6em]
8	&	$745.8_{-164.1}^{238}$		& $502.59_{-81.5}^{171.45}$	& $369.91_{-64.07}^{145.05}$	&	$645.18_{-2.25}^{0.5}$      &    3580$\pm$549  &   --           &  --           &  629$\pm$14 \\[0.6em]
9	&	$650.46_{-91.46}^{10.24}$	& $362.57_{-25.97}^{19.57}$	& $291.66_{-38.07}^{3.35}$	&	$178.83_{-0.58}^{0.55}$         &    1471$\pm$29   &   1284$\pm$119 &  --           &  69$\pm$7 \\[0.6em]
10	&	$171.95_{-57.06}^{23.8}$	& $139.8_{-47.5}^{16.97}$		& $100.63_{-54.53}^{24.54}$	&	$83.99_{-0.14}^{0.01}$      &    --            &   --           &  --           &  <434 \\[0.6em]
11	&	$996.71_{-153.26}^{373.73}$	& $579.55_{-73.8}^{247.91}$	& $428.6_{-21.22}^{195.41}$	&	$603.07_{-0.06}^{1.56}$         &    8930$\pm$542  &   --           &  --           &  671$\pm$43 \\[0.6em]
12	&	$256.77_{-3.14}^{22.65}$	& $127.87_{-5.96}^{20.97}$	& $106.62_{-6.55}^{20.6}$	&	$29.93_{-21.21}^{60.7}$         &    383$\pm$55    &   532$\pm$158  &  --           &  <653 \\[0.6em]
13	&	$2742.85_{-258.7}^{163.12}$	& $1424.78_{-55.27}^{39.71}$	& $1225.98_{-72.1}^{8.15}$	&	$208.94_{-1.05}^{146.47}$   &    1785$\pm$182  &   2124$\pm$446 &  --           &  502$\pm$45 \\[0.6em]
14	&	$148.8_{-3.65}^{25.16}$		& $59.7_{-7.83}^{21.27}$		& $51.61_{-1.9}^{10.75}$	&	$99.89_{-0.03}^{76.77}$     &    --            &   --           &  --           &  -- \\[0.6em]
\end{tabular}
\begin{tablenotes}
\item For continnum (TIR and FIR) luminosities we integrate for each galaxy the best-fit \cite{Draine2007} SED model, using a variety of
ranges along the wavelength axis. For TIR, we use 3-1100$\mu m$, as defined in \cite{Kennicutt2012}.
In order to compare our measurements to previous studies (as done in Figure \ref{fig:pdr_cii_fir}), we further calculate several
versions of the FIR luminosity: $L_{FIR_{w}}$, i.e. integrated over $40-500\mu m$ and used on the x-axis of Figure \ref{fig:pdr_cii_fir}
as well as $L_{FIR_{n,H88}}$, i.e. the FIR luminosity valid for the range of $40-120\mu m$, but estimated from single flux densities at 60 and 100
$\mu m$ as defined by \cite{Helou1988}.
$L_{FIR_{n,H88}}$ is used for the [C~II]-to-FIR ratio that is given on the y-axis in Figure \ref{fig:pdr_cii_fir}. For comparison,
we also provide $L_{FIR_{n,SED}}$, i.e. the SED-integrated luminosity along the same range of $40-120\mu m$.
The line luminosities are derived via Gaussian fitting.
\end{tablenotes}
\end{threeparttable}}
\end{table*}

\begin{table}
\caption[PDR model results]{PDR model results.
The solution space (n,G$_0$,$\chi^2$) is shown in Figure \ref{fig:pdr_chi2}.}
\label{tab:result_pdrtoolbox}
\begin{tabular}{lrrr}
\hline \hline
ID & $\chi^2$  & $n$      & $G_0$     \\
   &           & [$cm^-3$] & \\
2 & 0.05 & 5620 & 562 \\
12 & 0.4 & 1780 & 562 \\
13 & 0.1 & 562 & 1000 \\
\end{tabular}
\end{table}

\subsection{Dense Gas Fractions in LARS 3 and 8}
For the two LARS galaxies with detections in HCN~(1-0), we calculate the
line ratios of HCN(1-0)/CO(1-0). Given the fact that the critical
density of HCN~(1-0) is almost 2 orders of magnitude higher than for CO~(1-0),
the ratio among these two lines is often used as a tracer of the \textit{dense gas
fraction} \citep{Gao2004,Bigiel2016,JimenezDonaire2017b}.
We find line ratios of 0.13 and 0.05 for LARS 3 and LARS 8 respectively. While the
latter is similar to values found e.g. in M51 \citep{Bigiel2016}, the former
ratio indicates an extremely high dense gas fraction that is on the upper
limit of what is typically found in infrared-luminous star-forming galaxies \citep[compare to e.g.][Figure 6]{Juneau2009}.
The high dense gas fraction in LARS 3 is likely the result of the ongoing
merging process that triggers a nuclear starburst in the galaxy.

\subsection{Derived Properties from Line Ratios of CO, HCN and HCO+}
The IRAM 30m telescope is capable of using two heterodyne receivers at the same time.
We made use of this feature and observed CO~(1--0) and CO~(2--1) simultaneously.
However, given the different beamsizes and thus beam filling factors,
in combination with the complex morphology of our galaxies, one must be
cautious when interpreting line ratios between CO~(2--1) and CO~(1--0).
On the other side, observations of the CO~(3--2) line with APEX
and CO~(1--0) with IRAM, give similar beam widths. Hence, the ratio among these two lines
may give insight into the average conditions of the molecular gas.
As shown in Figure \ref{fig:COLARS_APEX}, we detected CO~(3--2)
in two of our galaxies.
We find
CO(3--2)/CO(1--0)
line ratios of 0.5 and 1.0 (on the brightness temperature scale) for LARS 8 and
13 respectively. The former value indicates sub-thermal excitation, whereas the
conditions in LARS 13 are more extreme and the line ratio suggests thermalized gas
up to the CO J=3 level. This result is in agreement with our derivations of U$_{min}$,
the average energy density, which was found to be extremely high for LARS 13
and moderate in LARS 8 (see Table \ref{tab:dl07fitresults}).

\subsubsection{Line Radiative Transfer Modeling for LARS 3 and 8}
Given the variety of emission lines now available
(i.e. CO~(1--0), CO~(2--1), HCN~(1--0) and HCO+~(1--0)) for LARS 3 and 8, as well as
CO~(3--2) for the latter one, we are able
to estimate the molecular gas density using
novel line radiative transfer models, 
described in detail in a forthcoming paper by Puschnig et al. (in prep).
Note that a first release of the \texttt{Dense Gas Toolbox} \citep{DGT}
is readily available, either as stand-alone web application\footnote{\url{http://www.densegastoolbox.com}} or
source code\footnote{\url{https://doi.org/10.5281/zenodo.3686329}}.
The radiative transfer models rely on \texttt{RADEX}
\citep{vanderTak2007}, but extend the original code to take into account
that molecular emission lines may emerge from a distribution of densities
rather than from a single-zone (single-density) medium.
This approach is more realistic, in particular for unresolved or low-resolution
observations as in our case, and has strong impact on interpreting
line ratios in terms of physical quantities, as shown by \cite{Leroy2017}.
We adopt a density distribution that is log-normal
and derive the \textit{mass-weighted} mean density
of that underlying gas distribution. Making use of the previously derived
dust properties, we assume fixed temperatures,
i.e. 30 and 25\,K for LARS 3 and 8 respectively.
A summary of the line ratios used as input for the modeling is shown in Table
\ref{tab:lineratios}.
This approach further allows to derive the \textit{dense gas fraction},
defined as the fraction of gas mass with densities higher than 10$^{4.5}\ cm^{-3}$.
The results for LARS 3 and 8 are summarized in Table \ref{tab:dgt},
and the $\chi^2$ solution planes are shown in Figure \ref{fig:lars_molecular_radtrans}.

\begin{table*}
\begin{threeparttable}
\caption{Interpretation of the molecular line ratios using the \texttt{Dense Gas Toolbox} \citep{DGT}}
\label{tab:dgt}
\begin{tabular}{rrrrrr}
\hline \hline
ID     &    $<n>$     &     T     &  Width   &   f$_{dense}$  & $\chi^2$ \\
       &  [$cm^{-3}$] &    [K]    & [dex]    &    [\%]        &  \\
3      &     $\sim$8000    &     30     &  0.6     &    36         & 1.5 \\
8      &       $\sim$80    &     25     &  0.6     &    <1         & 5.1
\end{tabular}
\begin{tablenotes}
\item Assuming a log-normal distribution of densities with a fixed width of 0.6 dex and a fixed temperature
(using results from the dust models), we derive the mass-weighted mean density $<n>$,
as well as the corresponding dense gas fraction $f_{dense}$.
\end{tablenotes}
\end{threeparttable}
\end{table*}

\section{Discussion}\label{sec:discussion}

\subsection{Turbulence Driven Lyman Alpha Escape}\label{sec:discuss_accretion}
As shown in Table \ref{tab:taudeplbins}, those LARS galaxies with the longest
gas depletion times (Group~3: LARS 1, 2, 4, 7) have (on average) largest total gas fractions
(while having the lowest molecular gas fractions) and the lowest energy densities ($U_{min}$),
compared to the other two groups of LARS galaxies.
These galaxies really stand out in that they have very high \lya\
escape fractions, while their scattering distances, i.e. a
measure of the path length a \lya\ photon travels before it escapes, are short.
The combined properties of being gas-rich and having short scattering distances,
is evidence for a kinematic-driven escape in which \lya\ photons are shifted
out of resonance relatively close to their origin.
Previous high-resolution HST imaging in the optical, H$\upalpha$ and UV further reveal
a sufficiently clumpy ISM \citep{Messa2019} and disks without substantial bulge in those LARS galaxies.
Moreover, \cite{Herenz2016} studied H$\upalpha$ kinematics of LARS galaxies and revealed
that the highest \lya\ escape fractions are found in dispersion dominated systems.
Hence, we are likely witnessing early stages of disk formation,
similar to the theoretical predictions of \cite{Dekel2009} for high-z galaxies.
In such a scenario, cold gas accretion primarily causes disk instabilities that
lead to the formation of massive clumps due to shear (each containing a few percent of
the disk mass).
Encounters between the clumps in the disk further stir up velocity dispersion,
which at the same time enables \lya\ photons to escape.
The timescale of this process, however, is similar to the timescale of turbulent dissipation, so
that the dynamical friction between the massive clumps finally leads to
loss of angular momentum and migration to the center where a bulge is formed.
We thus conclude that this process significantly enhances \lya\ escape
on timescales of a few hundred Myr.
Note that this scenario is also in agreement with simulations of
molecular gas (at high redshift) by \cite{Kimm2019}, who find that \lya\ scatterings and escape are
already significant on cloud-scales.

\subsection{Stellar Feedback Driven Lyman Alpha Escape}\label{sec:discuss_feedback}
Other LARS galaxies (Group~1: 3, 5, 12, 13) have much shorter gas depletion times.
They will run out of fuel in less than $\sim$1\,Gyr.
Also, their physical properties
are distinct. They have the highest energy density ($U_{min}$) and longest \lya\ scattering distances.
From HST imaging we also recognize that these galaxies show a rather compact
and more centralized light distribution. This suggests that bulk of the gas and
star formation is concentrated in a bulge-like structure, explaining higher average
energy densities.
This is also supported by our molecular line radiative transfer modeling of LARS 3 (member of Group~1),
that indicate a very high \textit{mean} molecular gas density of $\sim10^4\ cm^{-3}$.
On top of that we observed $CO(3-2)/CO(1-0)\sim1$ in LARS 13 (also member of Group~1),
which is consistent with highly excited gas that is thermalized even up to the CO J=3 level.

Our observations thus imply strong stellar feedback where \lya\ photons may escape through
channels in the ISM.
Although the overall low gas content supports final escape,
the photons need to travel significantly larger distances until they reach
the point of last scattering. In this scenario the dust content thus plays
an important role for \lya\ to escape.

\subsection{Robustness of our Results against GDR prescriptions}\label{sec:discuss_robust}
We test whether the high gas fractions and long total gas depletion times seen in Group~3 of LARS galaxies
might be caused by the adopted power law in the gas-to-dust-ratio vs. metallicity prescription. For that purpose, we
calculate depletion times and gas fractions using a constant GDR of 500. Given the fact that GDR=500 is
slightly higher than the average GDR found from the aforementioned metallicity-dependent prescription,
we find an average increase of the gas fraction of 5\% for the whole sample. Roughly the same increase of
5\% is seen for the three groups of LARS galaxies. This is not surprising given the fact that none of
the groups are strongly biased towards lower or higher metallicities compared to the other groups.
Thus, a constant GDR has no effect on the relative change of gas fractions or
total gas depletion times among the groups of LARS galaxies and our results cannot be caused by
an over-prediction of the dependence between GDR and metallicity.

\section{Conclusion and Summary}\label{sec:summary}
LARS is a sample of 14 z$\sim$0 star-forming galaxies that
have continuum sizes, stellar masses and rest-frame absolute magnitudes
similar to 2$<$z$<$3 star-forming galaxies and massive Lyman Alpha emitters \citep{Guaita2015}.

In this paper, we present observations of LARS galaxies obtained with Herschel/PACS, SOFIA/FIFI-LS,
the IRAM 30m telescope and APEX, targeting far-infrared continuum
and emission lines of [C~II]158$\upmu$m, [O~I]63$\upmu$m and [O~III]88$\upmu$m
as well as low-J CO lines.
In combination with archival far-infrared data (WISE, AKARI and IRAS), we applied the
models of \cite{Draine2007model} and derived parameters such
as dust mass, average energy density and PDR mass using a Bayesian approach.
Total gas masses were calculated for all LARS galaxies in
a homogeneous way using a metallicity-dependent gas-to-dust ratio (GDR),
allowing us to establish the Kennicutt-Schmidt relation for all 14 LARS galaxies (see Figure \ref{fig:KS}).

For those eight galaxies with a detection in [C~II]158$\upmu$m, we
compared the relative [C~II]158$\upmu$m line strength to the total
far-infrared luminosity (see Figure \ref{fig:pdr_cii_fir}) and for five LARS galaxies -- 
with detections of at least two emission lines -- be it a finestructure
or molecular line -- a basic PDR analysis was performed, enabling us to
estimate average PDR gas density and FUV radiation field strength (see Figure \ref{fig:pdr_chi2}).

We have further applied novel radiative transfers models\footnote{\url{http://www.densegastoolbox.com}} (Puschnig et al. in prep.),
taking into account that
molecular emission lines emerge from a multi-density medium (with a lognorm density distribution) rather
than from a single density gas. Using multi-J CO, HCN~(1-0) and HCO+~(1-0) observations of LARS 3 and 8,
enabled us to derive mean mass-weighted molecular gas densities of these two galaxies.

\begin{figure}
\centering
   \includegraphics[width=0.6\columnwidth]{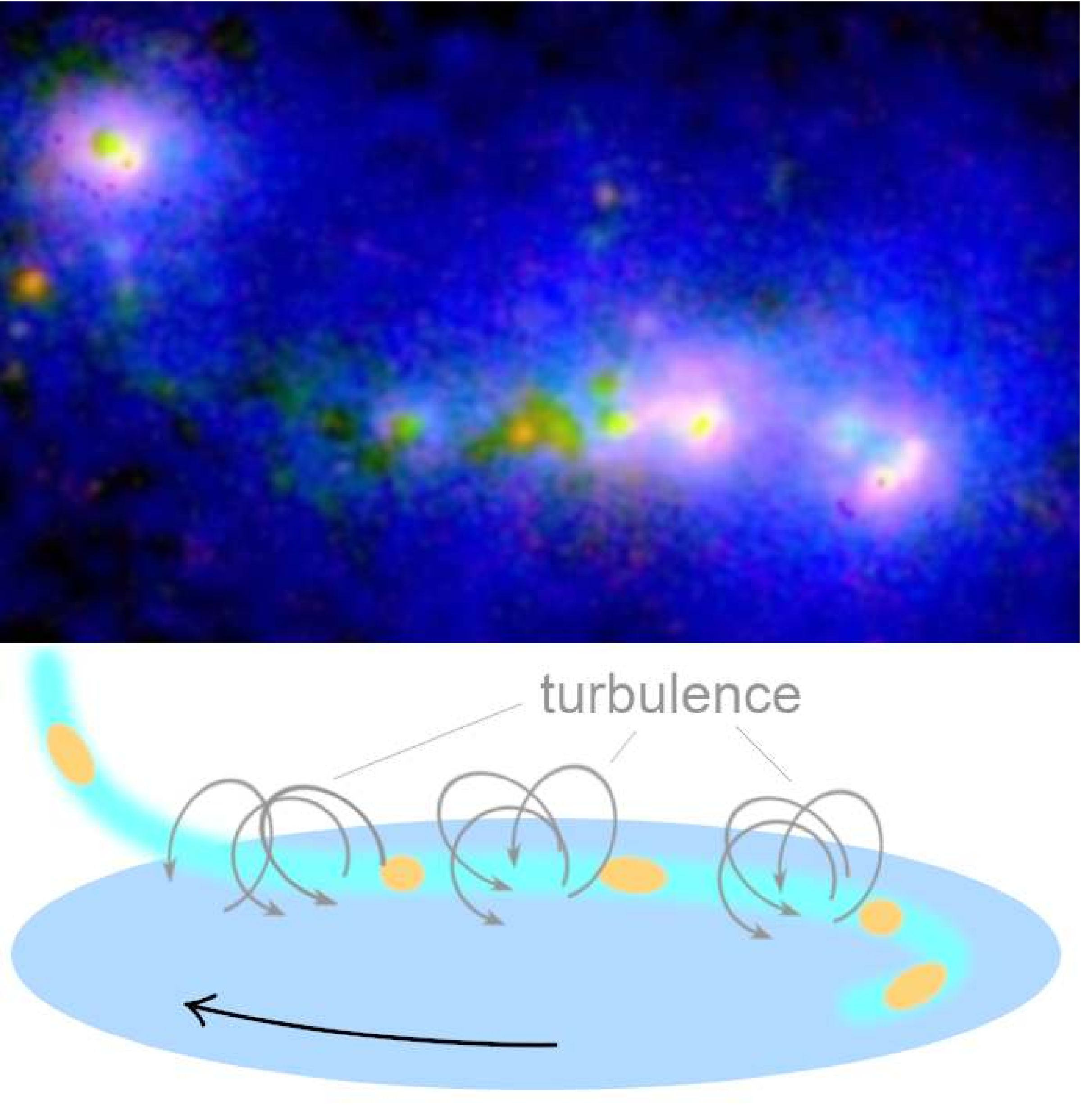}
     \caption{
     \textit{Top}: RGB image of LARS 2 with red, green and blue channels encoding H$\upalpha$, UV continuum and \lya\ (scaled to show details).
     \textit{Bottom}: Sketch of turbulence driven enhanced \lya\ escape during disk formation.}
     \label{fig:turbulence_lars2}
\end{figure}

From the analysis of our data we conclude the following:

\begin{itemize}

\item LARS covers a wide dynamic range in the derived properties, with FIR-based star formation rates
from $\sim$0.5--100 $M_{\odot}\ yr^{-1}$, gas fractions between $\sim$15--80\% and
gas depletion times ranging from a few hundred Myr up to more than 10~Gyr.

\item The distribution of LARS galaxies 
in the $\Sigma_{gas}$ vs. $\Sigma_{SFR}$ is quite heterogeneous. However, after defining 
three groups of galaxies according to their
gas depletion times, we observe that the group (LARS 1, 2, 4, 7) with the longest gas depletion
times, i.e. relatively high gas surface densities ($\Sigma_{gas}$) and low
star formation rate densities ($\Sigma_{SFR}$), has (by far) the highest \lya\ escape fraction.
A relatively strong $\sim$linear trend is found between \lya\ escape fraction and
total gas depletion time (see Figure \ref{fig:depl_vs_lya}).
We argue that the \lya\ escape in those galaxies is driven by 
accretion-induced turbulence in the star-forming gas that shifts the \lya\ photons out of resonance
close to the places where they originate (see Figure \ref{fig:turbulence_lars2}).
This scenario is supported by several other findings:
From previously published optical, H$\upalpha$ and UV imaging,
we recognize that these galaxies are very
clumpy and most importantly do not show any form of strong bulge,
in agreement with our finding of a low average energy density.
We speculate that the clumps are the result of recent or ongoing cold gas accretion,
which 1) triggered the clump formation and 2) injected turbulence that has
not yet dissipated, and thus facilitates \lya\ escape.

\item
Another grouping of LARS galaxies is found in the KS plot. They have relatively low gas surface
densities $\Sigma_{gas}$ that are more similar to observations in normal spirals or main
sequence galaxies. However, their extreme star formation rate densities suggest very
high star formation efficiencies (LARS 3, 5, 12, 13), corresponding to gas depletion times of a few 100Myr only.
We argue that the \lya\ escape in those galaxies (which is on the order of a few percent only) is
facilitated by an environment that is radiation-dominated and highly ionized. This is
supported by our observations of very high average energy densities (U$_{min}$).
HST imaging of these galaxies further shows a compact, centralized light distribution
in those galaxies. In such an environment \lya\ most likely escapes through channels that are a product of strong stellar feedback.
This scenario is also supported by high-J CO observations of LARS 13 that suggest highly excited gas that is
thermalized at least up to the J=3 level. Further evidence is found from molecular gas radiative transfer modeling
of LARS 3 that suggest very high \textit{mean} molecular gas densities of $\sim10^4\ cm^{-3}$.
However, it seems that this scenario is less efficient in driving up \lya\ escape fractions (compared to the previously described
turbulence scenario found in other LARS galaxies), because the photons still undergo scatterings farther out in the halo
(suggested by the longer scattering distances) and are thus more prone to dust absorption.

\item We further report on an extreme [C~II]158$\upmu$m excess in LARS 5, the highest
[C~II]-to-FIR ratio observed in a non-AGN galaxy to date.
LARS 5 is known to have an extreme stellar driven outflow of gas. We find that the extreme [C~II]158$\upmu$m
line strength
(corresponding to $\sim$14$\pm3$\%
of the FIR) must be related to the outflow as well. This is
supported by the fact that the [C~II]158$\upmu$m line is blueshifted compared to the systemic velocity.

\end{itemize}

\section*{Acknowledgements}
\small
JP ows special thanks to all IRAM Granada staff. In particular JP thanks Claudia Marka
for numerous fruitful discussions, help with the heterodyne receiver setup and data reduction;
Nicolas Billot for assigning additional telescope time when observing time was lost due to
scheduling constraints or weather; Israel Hermelo for helping to create OTF scripts;
Sandra Trevino for support during the first heterodyne pool week in September 2014;
and Carsten Kramer for discussions about our observing strategy and absolute calibration accuracy.
This work has benefited from research funding from the European Community's Seventh Framework Programme,
which covered JP's travelling expenses for IRAM 30m observations connected to programs 082-14 and 064-15.
M.H. acknowledges the support of the Swedish Research Council,
Vetenskapsr{\aa}det and the Swedish National Space Board (SNSB), and is Fellow
of the Knut and Alice Wallenberg Foundation.
D.K. is supported by the Centre National d'\'Etudes Spatiales (CNES)/Centre National de la Recherche Scientifique (CNRS); convention no 131425.
JP thanks all APEX staff in Chile, in particular Karl Torstensson for giving great support by
enlighting new features of the data reduction environment GILDAS/CLASS.
GILDAS is a collection of state-of-the-art softwares oriented toward (sub-)millimeter radioastronomical
applications (either single-dish or interferometer). JP thanks J\'{e}r\^{o}me Pety and S\'{e}bastien Bardeau
for help with GILDAS/CLASS related questions.
This work also benefits from data obtained with the Herschel spacecraft, which was designed,
built, tested, and launched under a contract to ESA managed by the Herschel/Planck Project
team by an industrial consortium under the overall responsibility of the prime contractor
Thales Alenia Space (Cannes), and including Astrium (Friedrichshafen) responsible for the
payload module and for system testing at spacecraft level, Thales Alenia Space (Turin) responsible
for the service module, and Astrium (Toulouse) responsible for the telescope, with in excess of
a hundred subcontractors.
This publication makes use of data products from the Wide-field Infrared Survey Explorer, which is a
joint project of the University of California, Los Angeles, and the Jet Propulsion Laboratory/California
Institute of Technology, funded by the National Aeronautics and Space Administration.
This research is based on observations with AKARI, a JAXA project with the
participation of ESA.
This research has made use of the NASA/ IPAC Infrared Science Archive, which
is operated by the Jet Propulsion Laboratory, California Institute of
Technology, under contract with the National Aeronautics and Space
Administration.
Funding for the Sloan Digital Sky Survey (SDSS) has been provided by the Alfred P. Sloan Foundation,
the Participating Institutions, the National Aeronautics and Space Administration, the National Science
Foundation, the U.S. Department of Energy, the Japanese Monbukagakusho, and the Max Planck Society.
The SDSS Web site is http://www.sdss.org/.
The SDSS is managed by the Astrophysical Research Consortium (ARC) for the Participating Institutions.
The Participating Institutions are The University of Chicago, Fermilab, the Institute for Advanced Study,
the Japan Participation Group, The Johns Hopkins University, Los Alamos National Laboratory,
the Max-Planck-Institute for Astronomy (MPIA), the Max-Planck-Institute for Astrophysics (MPA),
New Mexico State University, University of Pittsburgh, Princeton University, the United
States Naval Observatory, and the University of Washington.
This research made use of Montage, funded by the National Aeronautics and Space Administration's
Earth Science Technology Office, Computational Technnologies Project, under
Cooperative Agreement Number NCC5-626 between NASA and the California Institute of Technology.
The code is maintained by the NASA/IPAC Infrared Science Archive.
\normalsize

\bibliographystyle{unsrtnat}
\bibliography{copapers} 

\begin{thebibliography}{189}
\providecommand{\natexlab}[1]{#1}
\providecommand{\url}[1]{\texttt{#1}}
\expandafter\ifx\csname urlstyle\endcsname\relax
  \providecommand{\doi}[1]{doi: #1}\else
  \providecommand{\doi}{doi: \begingroup \urlstyle{rm}\Url}\fi

\bibitem[{Partridge} and {Peebles}(1967)]{Partridge1967}
R.~B. {Partridge} and P.~J.~E. {Peebles}.
\newblock {Are Young Galaxies Visible?}
\newblock \emph{\apj}, 147:\penalty0 868, March 1967.
\newblock \doi{10.1086/149079}.

\bibitem[{Meier} and {Terlevich}(1981)]{Meier1981}
D.~L. {Meier} and R.~{Terlevich}.
\newblock {Extragalactic H II regions in the UV - Implications for primeval
  galaxies}.
\newblock \emph{\apjl}, 246:\penalty0 L109--L113, June 1981.
\newblock \doi{10.1086/183565}.

\bibitem[{Cowie} and {Hu}(1998)]{Cowie1998}
L.~L. {Cowie} and E.~M. {Hu}.
\newblock {High-z Lyalpha Emitters. I. A Blank-Field Search for Objects near
  Redshift Z = 3.4 in and around the Hubble Deep Field and the Hawaii Deep
  Field SSA 22}.
\newblock \emph{\aj}, 115:\penalty0 1319--1328, April 1998.
\newblock \doi{10.1086/300309}.

\bibitem[{Ouchi} et~al.(2008){Ouchi}, {Shimasaku}, {Akiyama}, {Simpson},
  {Saito}, {Ueda}, {Furusawa}, {Sekiguchi}, {Yamada}, {Kodama}, {Kashikawa},
  {Okamura}, {Iye}, {Takata}, {Yoshida}, and {Yoshida}]{Ouchi2008}
M.~{Ouchi}, K.~{Shimasaku}, M.~{Akiyama}, C.~{Simpson}, T.~{Saito}, Y.~{Ueda},
  H.~{Furusawa}, K.~{Sekiguchi}, T.~{Yamada}, T.~{Kodama}, N.~{Kashikawa},
  S.~{Okamura}, M.~{Iye}, T.~{Takata}, M.~{Yoshida}, and M.~{Yoshida}.
\newblock {The Subaru/XMM-Newton Deep Survey (SXDS). IV. Evolution of
  Ly{$\alpha$} Emitters from z = 3.1 to 5.7 in the 1 deg$^{2}$ Field:
  Luminosity Functions and AGN}.
\newblock \emph{\apjs}, 176:\penalty0 301--330, June 2008.
\newblock \doi{10.1086/527673}.

\bibitem[{Ouchi} et~al.(2010){Ouchi}, {Shimasaku}, {Furusawa}, {Saito},
  {Yoshida}, {Akiyama}, {Ono}, {Yamada}, {Ota}, {Kashikawa}, {Iye}, {Kodama},
  {Okamura}, {Simpson}, and {Yoshida}]{Ouchi2010}
M.~{Ouchi}, K.~{Shimasaku}, H.~{Furusawa}, T.~{Saito}, M.~{Yoshida},
  M.~{Akiyama}, Y.~{Ono}, T.~{Yamada}, K.~{Ota}, N.~{Kashikawa}, M.~{Iye},
  T.~{Kodama}, S.~{Okamura}, C.~{Simpson}, and M.~{Yoshida}.
\newblock {Statistics of 207 Ly{$\alpha$} Emitters at a Redshift Near 7:
  Constraints on Reionization and Galaxy Formation Models}.
\newblock \emph{\apj}, 723:\penalty0 869--894, November 2010.
\newblock \doi{10.1088/0004-637X/723/1/869}.

\bibitem[{Adams} et~al.(2011){Adams}, {Uson}, {Hill}, and
  {MacQueen}]{Adams2011}
J.~J. {Adams}, J.~M. {Uson}, G.~J. {Hill}, and P.~J. {MacQueen}.
\newblock {A New z = 0 Metagalactic Ultraviolet Background Limit}.
\newblock \emph{\apj}, 728:\penalty0 107, February 2011.
\newblock \doi{10.1088/0004-637X/728/2/107}.

\bibitem[{Matthee} et~al.(2015){Matthee}, {Sobral}, {Santos}, {R{\"o}ttgering},
  {Darvish}, and {Mobasher}]{Matthee2015}
J.~{Matthee}, D.~{Sobral}, S.~{Santos}, H.~{R{\"o}ttgering}, B.~{Darvish}, and
  B.~{Mobasher}.
\newblock {Identification of the brightest Ly{$\alpha$} emitters at z = 6.6:
  implications for the evolution of the luminosity function in the reionization
  era}.
\newblock \emph{\mnras}, 451:\penalty0 400--417, July 2015.
\newblock \doi{10.1093/mnras/stv947}.

\bibitem[{Santos} et~al.(2016){Santos}, {Sobral}, and {Matthee}]{Santos2016}
S.~{Santos}, D.~{Sobral}, and J.~{Matthee}.
\newblock {The Ly{$\alpha$} luminosity function at z = 5.7 - 6.6 and the steep
  drop of the faint end: implications for reionization}.
\newblock \emph{\mnras}, 463:\penalty0 1678--1691, December 2016.
\newblock \doi{10.1093/mnras/stw2076}.

\bibitem[{Herenz} et~al.(2017){Herenz}, {Urrutia}, {Wisotzki}, {Kerutt},
  {Saust}, {Werhahn}, {Schmidt}, {Caruana}, {Diener}, {Bacon}, {Brinchmann},
  {Schaye}, {Maseda}, and {Weilbacher}]{Herenz2017}
E.~C. {Herenz}, T.~{Urrutia}, L.~{Wisotzki}, J.~{Kerutt}, R.~{Saust},
  M.~{Werhahn}, K.~B. {Schmidt}, J.~{Caruana}, C.~{Diener}, R.~{Bacon},
  J.~{Brinchmann}, J.~{Schaye}, M.~{Maseda}, and P.~M. {Weilbacher}.
\newblock {The MUSE-Wide survey: A first catalogue of 831 emission line
  galaxies}.
\newblock \emph{\aap}, 606:\penalty0 A12, September 2017.
\newblock \doi{10.1051/0004-6361/201731055}.

\bibitem[{Sobral} et~al.(2018){Sobral}, {Santos}, {Matthee}, {Paulino-Afonso},
  {Ribeiro}, {Calhau}, and {Khostovan}]{Sobral2018}
D.~{Sobral}, S.~{Santos}, J.~{Matthee}, A.~{Paulino-Afonso}, B.~{Ribeiro},
  J.~{Calhau}, and A.~A. {Khostovan}.
\newblock {Slicing COSMOS with SC4K: the evolution of typical Ly $\alpha$
  emitters and the Ly $\alpha$ escape fraction from z $\sim$ 2 to 6}.
\newblock \emph{\mnras}, 476:\penalty0 4725--4752, June 2018.
\newblock \doi{10.1093/mnras/sty378}.

\bibitem[{Urrutia} et~al.(2019){Urrutia}, {Wisotzki}, {Kerutt}, {Schmidt},
  {Herenz}, {Klar}, {Saust}, {Werhahn}, {Diener}, {Caruana}, {Krajnovi{\'c}},
  {Bacon}, {Boogaard}, {Brinchmann}, {Enke}, {Maseda}, {Nanayakkara},
  {Richard}, {Steinmetz}, and {Weilbacher}]{Urrutia2019}
T.~{Urrutia}, L.~{Wisotzki}, J.~{Kerutt}, K.~B. {Schmidt}, E.~C. {Herenz},
  J.~{Klar}, R.~{Saust}, M.~{Werhahn}, C.~{Diener}, J.~{Caruana},
  D.~{Krajnovi{\'c}}, R.~{Bacon}, L.~{Boogaard}, J.~{Brinchmann}, H.~{Enke},
  M.~{Maseda}, T.~{Nanayakkara}, J.~{Richard}, M.~{Steinmetz}, and P.~M.
  {Weilbacher}.
\newblock {The MUSE-Wide Survey: survey description and first data release}.
\newblock \emph{\aap}, 624:\penalty0 A141, Apr 2019.
\newblock \doi{10.1051/0004-6361/201834656}.

\bibitem[{Neufeld}(1990)]{Neufeld1990}
D.~A. {Neufeld}.
\newblock {The transfer of resonance-line radiation in static astrophysical
  media}.
\newblock \emph{\apj}, 350:\penalty0 216--241, February 1990.
\newblock \doi{10.1086/168375}.

\bibitem[{Ahn} et~al.(2001){Ahn}, {Lee}, and {Lee}]{Ahn2001}
S.-H. {Ahn}, H.-W. {Lee}, and H.~M. {Lee}.
\newblock {Ly{$\alpha$} Line Formation in Starbursting Galaxies. I. Moderately
  Thick, Dustless, and Static H I Media}.
\newblock \emph{\apj}, 554:\penalty0 604--614, June 2001.
\newblock \doi{10.1086/321374}.

\bibitem[{Ahn} et~al.(2002){Ahn}, {Lee}, and {Lee}]{Ahn2002}
S.-H. {Ahn}, H.-W. {Lee}, and H.~M. {Lee}.
\newblock {Ly{$\alpha$} Line Formation in Starbursting Galaxies. II. Extremely
  Thick, Dustless, and Static H I Media}.
\newblock \emph{\apj}, 567:\penalty0 922--930, March 2002.
\newblock \doi{10.1086/338497}.

\bibitem[{Dijkstra} et~al.(2006{\natexlab{a}}){Dijkstra}, {Haiman}, and
  {Spaans}]{Dijkstra2006a}
M.~{Dijkstra}, Z.~{Haiman}, and M.~{Spaans}.
\newblock {Ly{$\alpha$} Radiation from Collapsing Protogalaxies. I.
  Characteristics of the Emergent Spectrum}.
\newblock \emph{\apj}, 649:\penalty0 14--36, September 2006{\natexlab{a}}.
\newblock \doi{10.1086/506243}.

\bibitem[{Dijkstra} et~al.(2006{\natexlab{b}}){Dijkstra}, {Haiman}, and
  {Spaans}]{Dijkstra2006b}
M.~{Dijkstra}, Z.~{Haiman}, and M.~{Spaans}.
\newblock {Ly{$\alpha$} Radiation from Collapsing Protogalaxies. II.
  Observational Evidence for Gas Infall}.
\newblock \emph{\apj}, 649:\penalty0 37--47, September 2006{\natexlab{b}}.
\newblock \doi{10.1086/506244}.

\bibitem[{Verhamme} et~al.(2006){Verhamme}, {Schaerer}, and
  {Maselli}]{Verhamme2006}
A.~{Verhamme}, D.~{Schaerer}, and A.~{Maselli}.
\newblock {3D Ly{$\alpha$} radiation transfer. I. Understanding Ly{$\alpha$}
  line profile morphologies}.
\newblock \emph{\aap}, 460:\penalty0 397--413, December 2006.
\newblock \doi{10.1051/0004-6361:20065554}.

\bibitem[{Schaerer} and {Verhamme}(2008)]{Schaerer2008}
D.~{Schaerer} and A.~{Verhamme}.
\newblock {3D Ly{$\alpha$} radiation transfer. II. Fitting the Lyman break
  galaxy MS 1512-cB58 and implications for Ly{$\alpha$} emission in high-z
  starbursts}.
\newblock \emph{\aap}, 480:\penalty0 369--377, March 2008.
\newblock \doi{10.1051/0004-6361:20078913}.

\bibitem[{Verhamme} et~al.(2008){Verhamme}, {Schaerer}, {Atek}, and
  {Tapken}]{Verhamme2008}
A.~{Verhamme}, D.~{Schaerer}, H.~{Atek}, and C.~{Tapken}.
\newblock {3D Ly{$\alpha$} radiation transfer. III. Constraints on gas and
  stellar properties of z \~{} 3 Lyman break galaxies (LBG) and implications
  for high-z LBGs and Ly{$\alpha$} emitters}.
\newblock \emph{\aap}, 491:\penalty0 89--111, November 2008.
\newblock \doi{10.1051/0004-6361:200809648}.

\bibitem[{Hayes} et~al.(2010){Hayes}, {{\"O}stlin}, {Schaerer}, {Mas-Hesse},
  {Leitherer}, {Atek}, {Kunth}, {Verhamme}, {de Barros}, and
  {Melinder}]{Hayes2010}
M.~{Hayes}, G.~{{\"O}stlin}, D.~{Schaerer}, J.~M. {Mas-Hesse}, C.~{Leitherer},
  H.~{Atek}, D.~{Kunth}, A.~{Verhamme}, S.~{de Barros}, and J.~{Melinder}.
\newblock {Escape of about five per cent of Lyman-{$\alpha$} photons from
  high-redshift star-forming galaxies}.
\newblock \emph{\nat}, 464:\penalty0 562--565, March 2010.
\newblock \doi{10.1038/nature08881}.

\bibitem[{Scarlata} et~al.(2009){Scarlata}, {Colbert}, {Teplitz}, {Panagia},
  {Hayes}, {Siana}, {Rau}, {Francis}, {Caon}, {Pizzella}, and
  {Bridge}]{Scarlata2009}
C.~{Scarlata}, J.~{Colbert}, H.~I. {Teplitz}, N.~{Panagia}, M.~{Hayes},
  B.~{Siana}, A.~{Rau}, P.~{Francis}, A.~{Caon}, A.~{Pizzella}, and
  C.~{Bridge}.
\newblock {The Effect of Dust Geometry on the Ly{$\alpha$} Output of Galaxies}.
\newblock \emph{\apjl}, 704:\penalty0 L98--L102, October 2009.
\newblock \doi{10.1088/0004-637X/704/2/L98}.

\bibitem[{Zheng} et~al.(2017){Zheng}, {Wang}, {Rhoads}, {Infante}, {Malhotra},
  {Hu}, {Walker}, {Jiang}, {Jiang}, {Hibon}, {Gonzalez}, {Kong}, {Zheng},
  {Galaz}, and {Barrientos}]{Zheng2017}
Zhen-Ya {Zheng}, Junxian {Wang}, James {Rhoads}, Leopoldo {Infante}, Sangeeta
  {Malhotra}, Weida {Hu}, Alistair~R. {Walker}, Linhua {Jiang}, Chunyan
  {Jiang}, Pascale {Hibon}, Alicia {Gonzalez}, Xu~{Kong}, XianZhong {Zheng},
  Gaspar {Galaz}, and L.~Felipe {Barrientos}.
\newblock {First Results from the Lyman Alpha Galaxies in the Epoch of
  Reionization (LAGER) Survey: Cosmological Reionization at z $\sim$ 7}.
\newblock \emph{\apjl}, 842\penalty0 (2):\penalty0 L22, Jun 2017.
\newblock \doi{10.3847/2041-8213/aa794f}.

\bibitem[{Kunth} et~al.(1998){Kunth}, {Mas-Hesse}, {Terlevich}, {Terlevich},
  {Lequeux}, and {Fall}]{Kunth1998}
Daniel {Kunth}, J.~M. {Mas-Hesse}, E.~{Terlevich}, R.~{Terlevich},
  J.~{Lequeux}, and S.~Michael {Fall}.
\newblock {HST study of Lyman-alpha emission in star-forming galaxies: the
  effect of neutral gas flows}.
\newblock \emph{\aap}, 334:\penalty0 11--20, Jun 1998.

\bibitem[{Cannon} et~al.(2004){Cannon}, {McClure-Griffiths}, {Skillman}, and
  {C{\^o}t{\'e}}]{Cannon2004}
J.~M. {Cannon}, N.~M. {McClure-Griffiths}, E.~D. {Skillman}, and
  S.~{C{\^o}t{\'e}}.
\newblock {The Complex Neutral Gas Dynamics of the Dwarf Starburst Galaxy NGC
  625}.
\newblock \emph{\apj}, 607:\penalty0 274--284, May 2004.
\newblock \doi{10.1086/383408}.

\bibitem[{Wofford} et~al.(2013){Wofford}, {Leitherer}, and
  {Salzer}]{Wofford2013}
Aida {Wofford}, Claus {Leitherer}, and John {Salzer}.
\newblock {Ly{\ensuremath{\alpha}} Escape from z
  \raisebox{-0.5ex}\textasciitilde 0.03 Star-forming Galaxies: The Dominant
  Role of Outflows}.
\newblock \emph{\apj}, 765\penalty0 (2):\penalty0 118, Mar 2013.
\newblock \doi{10.1088/0004-637X/765/2/118}.

\bibitem[{Pardy} et~al.(2014){Pardy}, {Cannon}, {{\"O}stlin}, {Hayes},
  {Rivera-Thorsen}, {Sandberg}, {Adamo}, {Freeland}, {Herenz}, {Guaita},
  {Kunth}, {Laursen}, {Mas-Hesse}, {Melinder}, {Orlitov{\'a}},
  {Ot{\'{\i}}-Floranes}, {Puschnig}, {Schaerer}, and {Verhamme}]{Pardy2014}
S.~A. {Pardy}, J.~M. {Cannon}, G.~{{\"O}stlin}, M.~{Hayes},
  T.~{Rivera-Thorsen}, A.~{Sandberg}, A.~{Adamo}, E.~{Freeland}, E.~C.
  {Herenz}, L.~{Guaita}, D.~{Kunth}, P.~{Laursen}, J.~M. {Mas-Hesse},
  J.~{Melinder}, I.~{Orlitov{\'a}}, H.~{Ot{\'{\i}}-Floranes}, J.~{Puschnig},
  D.~{Schaerer}, and A.~{Verhamme}.
\newblock {The Lyman Alpha Reference Sample. III. Properties of the Neutral ISM
  from GBT and VLA Observations}.
\newblock \emph{\apj}, 794:\penalty0 101, October 2014.
\newblock \doi{10.1088/0004-637X/794/2/101}.

\bibitem[{Tenorio-Tagle} et~al.(1999){Tenorio-Tagle}, {Silich}, {Kunth},
  {Terlevich}, and {Terlevich}]{TenorioTagle1999}
G.~{Tenorio-Tagle}, S.~A. {Silich}, D.~{Kunth}, E.~{Terlevich}, and
  R.~{Terlevich}.
\newblock {The evolution of superbubbles and the detection of Ly{$\alpha$} in
  star-forming galaxies}.
\newblock \emph{\mnras}, 309:\penalty0 332--342, October 1999.
\newblock \doi{10.1046/j.1365-8711.1999.02809.x}.

\bibitem[{Mas-Hesse} et~al.(2003){Mas-Hesse}, {Kunth}, {Tenorio-Tagle},
  {Leitherer}, {Terlevich}, and {Terlevich}]{MasHesse2003}
J.~M. {Mas-Hesse}, D.~{Kunth}, G.~{Tenorio-Tagle}, C.~{Leitherer}, R.~J.
  {Terlevich}, and E.~{Terlevich}.
\newblock {Ly{$\alpha$} Emission in Starbursts: Implications for Galaxies at
  High Redshift}.
\newblock \emph{\apj}, 598:\penalty0 858--877, December 2003.
\newblock \doi{10.1086/379116}.

\bibitem[{Duval} et~al.(2014){Duval}, {Schaerer}, {{\"O}stlin}, and
  {Laursen}]{Duval2014}
F.~{Duval}, D.~{Schaerer}, G.~{{\"O}stlin}, and P.~{Laursen}.
\newblock {Lyman {$\alpha$} line and continuum radiative transfer in a clumpy
  interstellar medium}.
\newblock \emph{\aap}, 562:\penalty0 A52, February 2014.
\newblock \doi{10.1051/0004-6361/201220455}.

\bibitem[{Jaskot} and {Oey}(2014)]{Jaskot2014}
A.~E. {Jaskot} and M.~S. {Oey}.
\newblock {Linking Ly$\alpha$ and Low-ionization Transitions at Low Optical
  Depth}.
\newblock \emph{\apjl}, 791\penalty0 (2):\penalty0 L19, Aug 2014.
\newblock \doi{10.1088/2041-8205/791/2/L19}.

\bibitem[{Hayes}(2015)]{Hayes2015}
M.~{Hayes}.
\newblock {Lyman Alpha Emitting Galaxies in the Nearby Universe}.
\newblock \emph{\pasa}, 32:\penalty0 e027, July 2015.
\newblock \doi{10.1017/pasa.2015.25}.

\bibitem[{Bridge} et~al.(2018){Bridge}, {Hayes}, {Melinder}, {{\"O}stlin},
  {Gronwall}, {Ciardullo}, {Atek}, {Cannon}, {Gronke}, {Guaita}, {Hagen},
  {Herenz}, {Kunth}, {Laursen}, {Mas-Hesse}, and {Pardy}]{Bridge2018}
J.~S. {Bridge}, M.~{Hayes}, J.~{Melinder}, G.~{{\"O}stlin}, C.~{Gronwall},
  R.~{Ciardullo}, H.~{Atek}, J.~M. {Cannon}, M.~{Gronke}, L.~{Guaita},
  A.~{Hagen}, E.~C. {Herenz}, D.~{Kunth}, P.~{Laursen}, J.~M. {Mas-Hesse}, and
  S.~A. {Pardy}.
\newblock {The Ly{$\alpha$} Reference Sample. VIII. Characterizing Ly{$\alpha$}
  Scattering in Nearby Galaxies}.
\newblock \emph{\apj}, 852:\penalty0 9, January 2018.
\newblock \doi{10.3847/1538-4357/aa9932}.

\bibitem[{Verhamme} et~al.(2015){Verhamme}, {Orlitov{\'a}}, {Schaerer}, and
  {Hayes}]{Verhamme2015}
A.~{Verhamme}, I.~{Orlitov{\'a}}, D.~{Schaerer}, and M.~{Hayes}.
\newblock {Using Lyman-{$\alpha$} to detect galaxies that leak Lyman
  continuum}.
\newblock \emph{\aap}, 578:\penalty0 A7, June 2015.
\newblock \doi{10.1051/0004-6361/201423978}.

\bibitem[{Puschnig} et~al.(2017){Puschnig}, {Hayes}, {{\"O}stlin},
  {Rivera-Thorsen}, {Melinder}, {Cannon}, {Menacho}, {Zackrisson}, {Bergvall},
  and {Leitet}]{Puschnig2017}
J.~{Puschnig}, M.~{Hayes}, G.~{{\"O}stlin}, T.~E. {Rivera-Thorsen},
  J.~{Melinder}, J.~M. {Cannon}, V.~{Menacho}, E.~{Zackrisson}, N.~{Bergvall},
  and E.~{Leitet}.
\newblock {The Lyman continuum escape and ISM properties in Tololo 1247-232 -
  new insights from HST and VLA}.
\newblock \emph{\mnras}, 469:\penalty0 3252--3269, August 2017.
\newblock \doi{10.1093/mnras/stx951}.

\bibitem[{Trainor} et~al.(2019){Trainor}, {Strom}, {Steidel}, {Rudie}, {Chen},
  and {Theios}]{Trainor2019}
Ryan~F. {Trainor}, Allison~L. {Strom}, Charles~C. {Steidel}, Gwen~C. {Rudie},
  Yuguang {Chen}, and Rachel~L. {Theios}.
\newblock {Predicting Ly{\ensuremath{\alpha}} Emission from Galaxies via
  Empirical Markers of Production and Escape in the KBSS}.
\newblock \emph{\apj}, 887\penalty0 (1):\penalty0 85, December 2019.
\newblock \doi{10.3847/1538-4357/ab4993}.

\bibitem[{Sobral} and {Matthee}(2019)]{Sobral2019}
David {Sobral} and Jorryt {Matthee}.
\newblock {Predicting Ly{\ensuremath{\alpha}} escape fractions with a simple
  observable. Ly{\ensuremath{\alpha}} in emission as an empirically calibrated
  star formation rate indicator}.
\newblock \emph{\aap}, 623:\penalty0 A157, Mar 2019.
\newblock \doi{10.1051/0004-6361/201833075}.

\bibitem[{Runnholm} et~al.(2020){Runnholm}, {Hayes}, {Melinder},
  {Rivera-Thorsen}, {{\"O}stlin}, {Cannon}, and {Kunth}]{Runnholm2020}
Axel {Runnholm}, Matthew {Hayes}, Jens {Melinder}, Emil {Rivera-Thorsen},
  G{\"o}ran {{\"O}stlin}, John {Cannon}, and Daniel {Kunth}.
\newblock {Lyman Alpha Reference Sample: X. Predicting Lyman alpha output from
  starforming galaxies using multivariate regression}.
\newblock \emph{arXiv e-prints}, art. arXiv:2002.12378, February 2020.

\bibitem[{Hayes} et~al.(2013){Hayes}, {{\"O}stlin}, {Schaerer}, {Verhamme},
  {Mas-Hesse}, {Adamo}, {Atek}, {Cannon}, {Duval}, {Guaita}, {Herenz}, {Kunth},
  {Laursen}, {Melinder}, {Orlitov{\'a}}, {Ot{\'{\i}}-Floranes}, and
  {Sandberg}]{Hayes2013}
M.~{Hayes}, G.~{{\"O}stlin}, D.~{Schaerer}, A.~{Verhamme}, J.~M. {Mas-Hesse},
  A.~{Adamo}, H.~{Atek}, J.~M. {Cannon}, F.~{Duval}, L.~{Guaita}, E.~C.
  {Herenz}, D.~{Kunth}, P.~{Laursen}, J.~{Melinder}, I.~{Orlitov{\'a}},
  H.~{Ot{\'{\i}}-Floranes}, and A.~{Sandberg}.
\newblock {The Lyman Alpha Reference Sample: Extended Lyman Alpha Halos
  Produced at Low Dust Content}.
\newblock \emph{\apjl}, 765:\penalty0 L27, March 2013.
\newblock \doi{10.1088/2041-8205/765/2/L27}.

\bibitem[{Hayes} et~al.(2014){Hayes}, {{\"O}stlin}, {Duval}, {Sandberg},
  {Guaita}, {Melinder}, {Adamo}, {Schaerer}, {Verhamme}, {Orlitov{\'a}},
  {Mas-Hesse}, {Cannon}, {Atek}, {Kunth}, {Laursen}, {Ot{\'{\i}}-Floranes},
  {Pardy}, {Rivera-Thorsen}, and {Herenz}]{Hayes2014}
M.~{Hayes}, G.~{{\"O}stlin}, F.~{Duval}, A.~{Sandberg}, L.~{Guaita},
  J.~{Melinder}, A.~{Adamo}, D.~{Schaerer}, A.~{Verhamme}, I.~{Orlitov{\'a}},
  J.~M. {Mas-Hesse}, J.~M. {Cannon}, H.~{Atek}, D.~{Kunth}, P.~{Laursen},
  H.~{Ot{\'{\i}}-Floranes}, S.~{Pardy}, T.~{Rivera-Thorsen}, and E.~C.
  {Herenz}.
\newblock {The Lyman Alpha Reference Sample. II. Hubble Space Telescope Imaging
  Results, Integrated Properties, and Trends}.
\newblock \emph{\apj}, 782:\penalty0 6, February 2014.
\newblock \doi{10.1088/0004-637X/782/1/6}.

\bibitem[{{\"O}stlin} et~al.(2014){{\"O}stlin}, {Hayes}, {Duval}, {Sandberg},
  {Rivera-Thorsen}, {Marquart}, {Orlitov{\'a}}, {Adamo}, {Melinder}, {Guaita},
  {Atek}, {Cannon}, {Gruyters}, {Herenz}, {Kunth}, {Laursen}, {Mas-Hesse},
  {Micheva}, {Ot{\'{\i}}-Floranes}, {Pardy}, {Roth}, {Schaerer}, and
  {Verhamme}]{Oestlin2014}
G.~{{\"O}stlin}, M.~{Hayes}, F.~{Duval}, A.~{Sandberg}, T.~{Rivera-Thorsen},
  T.~{Marquart}, I.~{Orlitov{\'a}}, A.~{Adamo}, J.~{Melinder}, L.~{Guaita},
  H.~{Atek}, J.~M. {Cannon}, P.~{Gruyters}, E.~C. {Herenz}, D.~{Kunth},
  P.~{Laursen}, J.~M. {Mas-Hesse}, G.~{Micheva}, H.~{Ot{\'{\i}}-Floranes},
  S.~A. {Pardy}, M.~M. {Roth}, D.~{Schaerer}, and A.~{Verhamme}.
\newblock {The Ly{$\alpha$} Reference Sample. I. Survey Outline and First
  Results for Markarian 259}.
\newblock \emph{\apj}, 797:\penalty0 11, December 2014.
\newblock \doi{10.1088/0004-637X/797/1/11}.

\bibitem[{Cowie} et~al.(2011){Cowie}, {Barger}, and {Hu}]{Cowie2011}
Lennox~L. {Cowie}, Amy~J. {Barger}, and Esther~M. {Hu}.
\newblock {Ly{\ensuremath{\alpha}} Emitting Galaxies as Early Stages in Galaxy
  Formation}.
\newblock \emph{\apj}, 738\penalty0 (2):\penalty0 136, Sep 2011.
\newblock \doi{10.1088/0004-637X/738/2/136}.

\bibitem[{Rivera-Thorsen} et~al.(2015){Rivera-Thorsen}, {Hayes}, {{\"O}stlin},
  {Duval}, {Orlitov{\'a}}, {Verhamme}, {Mas-Hesse}, {Schaerer}, {Cannon},
  {Ot{\'{\i}}-Floranes}, {Sandberg}, {Guaita}, {Adamo}, {Atek}, {Herenz},
  {Kunth}, {Laursen}, and {Melinder}]{RiveraThorsen2015}
T.~E. {Rivera-Thorsen}, M.~{Hayes}, G.~{{\"O}stlin}, F.~{Duval},
  I.~{Orlitov{\'a}}, A.~{Verhamme}, J.~M. {Mas-Hesse}, D.~{Schaerer}, J.~M.
  {Cannon}, H.~{Ot{\'{\i}}-Floranes}, A.~{Sandberg}, L.~{Guaita}, A.~{Adamo},
  H.~{Atek}, E.~C. {Herenz}, D.~{Kunth}, P.~{Laursen}, and J.~{Melinder}.
\newblock {The Lyman Alpha Reference Sample. V. The Impact of Neutral ISM
  Kinematics and Geometry on Ly{$\alpha$} Escape}.
\newblock \emph{\apj}, 805:\penalty0 14, May 2015.
\newblock \doi{10.1088/0004-637X/805/1/14}.

\bibitem[{Micheva} et~al.(2018){Micheva}, {{\"O}stlin}, {Zackrisson}, {Hayes},
  {Melinder}, {Guaita}, {Cannon}, {Bridge}, {Kunth}, and
  {Sandberg}]{Micheva2018}
G.~{Micheva}, G.~{{\"O}stlin}, E.~{Zackrisson}, M.~{Hayes}, J.~{Melinder},
  L.~{Guaita}, J.~M. {Cannon}, J.~S. {Bridge}, D.~{Kunth}, and A.~{Sandberg}.
\newblock {The Lyman Alpha Reference Sample. IX. Revelations from deep surface
  photometry}.
\newblock \emph{\aap}, 615:\penalty0 A46, July 2018.
\newblock \doi{10.1051/0004-6361/201832584}.

\bibitem[{Herenz} et~al.(2016){Herenz}, {Gruyters}, {Orlitova}, {Hayes},
  {{\"O}stlin}, {Cannon}, {Roth}, {Bik}, {Pardy}, {Ot{\'{\i}}-Floranes},
  {Mas-Hesse}, {Adamo}, {Atek}, {Duval}, {Guaita}, {Kunth}, {Laursen},
  {Melinder}, {Puschnig}, {Rivera-Thorsen}, {Schaerer}, and
  {Verhamme}]{Herenz2016}
E.~C. {Herenz}, P.~{Gruyters}, I.~{Orlitova}, M.~{Hayes}, G.~{{\"O}stlin},
  J.~M. {Cannon}, M.~M. {Roth}, A.~{Bik}, S.~{Pardy}, H.~{Ot{\'{\i}}-Floranes},
  J.~M. {Mas-Hesse}, A.~{Adamo}, H.~{Atek}, F.~{Duval}, L.~{Guaita},
  D.~{Kunth}, P.~{Laursen}, J.~{Melinder}, J.~{Puschnig}, T.~E.
  {Rivera-Thorsen}, D.~{Schaerer}, and A.~{Verhamme}.
\newblock {The Lyman alpha reference sample. VII. Spatially resolved
  H{$\alpha$} kinematics}.
\newblock \emph{\aap}, 587:\penalty0 A78, March 2016.
\newblock \doi{10.1051/0004-6361/201527373}.

\bibitem[{Schmidt}(1959)]{Schmidt1959}
M.~{Schmidt}.
\newblock {The Rate of Star Formation.}
\newblock \emph{\apj}, 129:\penalty0 243, March 1959.
\newblock \doi{10.1086/146614}.

\bibitem[{Kennicutt}(1998)]{Kennicutt1998}
R.~C. {Kennicutt}, Jr.
\newblock {The Global Schmidt Law in Star-forming Galaxies}.
\newblock \emph{\apj}, 498:\penalty0 541--552, May 1998.
\newblock \doi{10.1086/305588}.

\bibitem[{Wong} and {Blitz}(2002)]{Wong2002}
T.~{Wong} and L.~{Blitz}.
\newblock {The Relationship between Gas Content and Star Formation in
  Molecule-rich Spiral Galaxies}.
\newblock \emph{\apj}, 569:\penalty0 157--183, April 2002.
\newblock \doi{10.1086/339287}.

\bibitem[{Bigiel} et~al.(2008){Bigiel}, {Leroy}, {Walter}, {Brinks}, {de Blok},
  {Madore}, and {Thornley}]{Bigiel2008}
F.~{Bigiel}, A.~{Leroy}, F.~{Walter}, E.~{Brinks}, W.~J.~G. {de Blok},
  B.~{Madore}, and M.~D. {Thornley}.
\newblock {The Star Formation Law in Nearby Galaxies on Sub-Kpc Scales}.
\newblock \emph{\aj}, 136:\penalty0 2846--2871, December 2008.
\newblock \doi{10.1088/0004-6256/136/6/2846}.

\bibitem[{Bigiel} et~al.(2011){Bigiel}, {Leroy}, {Walter}, {Brinks}, {de Blok},
  {Kramer}, {Rix}, {Schruba}, {Schuster}, {Usero}, and
  {Wiesemeyer}]{Bigiel2011}
F.~{Bigiel}, A.~K. {Leroy}, F.~{Walter}, E.~{Brinks}, W.~J.~G. {de Blok},
  C.~{Kramer}, H.~W. {Rix}, A.~{Schruba}, K.-F. {Schuster}, A.~{Usero}, and
  H.~W. {Wiesemeyer}.
\newblock {A Constant Molecular Gas Depletion Time in Nearby Disk Galaxies}.
\newblock \emph{\apjl}, 730:\penalty0 L13, April 2011.
\newblock \doi{10.1088/2041-8205/730/2/L13}.

\bibitem[{Leroy} et~al.(2008){Leroy}, {Walter}, {Brinks}, {Bigiel}, {de Blok},
  {Madore}, and {Thornley}]{Leroy2008}
A.~K. {Leroy}, F.~{Walter}, E.~{Brinks}, F.~{Bigiel}, W.~J.~G. {de Blok},
  B.~{Madore}, and M.~D. {Thornley}.
\newblock {The Star Formation Efficiency in Nearby Galaxies: Measuring Where
  Gas Forms Stars Effectively}.
\newblock \emph{\aj}, 136:\penalty0 2782--2845, December 2008.
\newblock \doi{10.1088/0004-6256/136/6/2782}.

\bibitem[{Leroy} et~al.(2013){Leroy}, {Walter}, {Sandstrom}, {Schruba},
  {Munoz-Mateos}, {Bigiel}, {Bolatto}, {Brinks}, {de Blok}, {Meidt}, {Rix},
  {Rosolowsky}, {Schinnerer}, {Schuster}, and {Usero}]{Leroy2013}
A.~K. {Leroy}, F.~{Walter}, K.~{Sandstrom}, A.~{Schruba}, J.-C. {Munoz-Mateos},
  F.~{Bigiel}, A.~{Bolatto}, E.~{Brinks}, W.~J.~G. {de Blok}, S.~{Meidt}, H.-W.
  {Rix}, E.~{Rosolowsky}, E.~{Schinnerer}, K.-F. {Schuster}, and A.~{Usero}.
\newblock {Molecular Gas and Star Formation in nearby Disk Galaxies}.
\newblock \emph{\aj}, 146:\penalty0 19, August 2013.
\newblock \doi{10.1088/0004-6256/146/2/19}.

\bibitem[{Bolatto} et~al.(2017){Bolatto}, {Wong}, {Utomo}, {Blitz}, {Vogel},
  {S{\'a}nchez}, {Barrera-Ballesteros}, {Cao}, {Colombo}, {Dannerbauer},
  {Garc{\'{\i}}a-Benito}, {Herrera-Camus}, {Husemann}, {Kalinova}, {Leroy},
  {Leung}, {Levy}, {Mast}, {Ostriker}, {Rosolowsky}, {Sandstrom}, {Teuben},
  {van de Ven}, and {Walter}]{Bolatto2017}
A.~D. {Bolatto}, T.~{Wong}, D.~{Utomo}, L.~{Blitz}, S.~N. {Vogel}, S.~F.
  {S{\'a}nchez}, J.~{Barrera-Ballesteros}, Y.~{Cao}, D.~{Colombo},
  H.~{Dannerbauer}, R.~{Garc{\'{\i}}a-Benito}, R.~{Herrera-Camus},
  B.~{Husemann}, V.~{Kalinova}, A.~K. {Leroy}, G.~{Leung}, R.~C. {Levy},
  D.~{Mast}, E.~{Ostriker}, E.~{Rosolowsky}, K.~M. {Sandstrom}, P.~{Teuben},
  G.~{van de Ven}, and F.~{Walter}.
\newblock {The EDGE-CALIFA Survey: Interferometric Observations of 126 Galaxies
  with CARMA}.
\newblock \emph{\apj}, 846:\penalty0 159, September 2017.
\newblock \doi{10.3847/1538-4357/aa86aa}.

\bibitem[{Utomo} et~al.(2018){Utomo}, {Sun}, {Leroy}, {Kruijssen},
  {Schinnerer}, {Schruba}, {Bigiel}, {Blanc}, {Chevance}, {Emsellem},
  {Herrera}, {Hygate}, {Kreckel}, {Ostriker}, {Pety}, {Querejeta},
  {Rosolowsky}, {Sandstrom}, and {Usero}]{Utomo2018}
D.~{Utomo}, J.~{Sun}, A.~K. {Leroy}, J.~M.~D. {Kruijssen}, E.~{Schinnerer},
  A.~{Schruba}, F.~{Bigiel}, G.~A. {Blanc}, M.~{Chevance}, E.~{Emsellem},
  C.~{Herrera}, A.~P.~S. {Hygate}, K.~{Kreckel}, E.~C. {Ostriker}, J.~{Pety},
  M.~{Querejeta}, E.~{Rosolowsky}, K.~M. {Sandstrom}, and A.~{Usero}.
\newblock {Star Formation Efficiency per Free-fall Time in nearby Galaxies}.
\newblock \emph{\apjl}, 861:\penalty0 L18, July 2018.
\newblock \doi{10.3847/2041-8213/aacf8f}.

\bibitem[{Colombo} et~al.(2018){Colombo}, {Kalinova}, {Utomo}, {Rosolowsky},
  {Bolatto}, {Levy}, {Wong}, {Sanchez}, {Leroy}, {Ostriker}, {Blitz}, {Vogel},
  {Mast}, {Garc{\'{\i}}a-Benito}, {Husemann}, {Dannerbauer}, {Ellmeier}, and
  {Cao}]{Colombo2018}
D.~{Colombo}, V.~{Kalinova}, D.~{Utomo}, E.~{Rosolowsky}, A.~D. {Bolatto},
  R.~C. {Levy}, T.~{Wong}, S.~F. {Sanchez}, A.~K. {Leroy}, E.~{Ostriker},
  L.~{Blitz}, S.~{Vogel}, D.~{Mast}, R.~{Garc{\'{\i}}a-Benito}, B.~{Husemann},
  H.~{Dannerbauer}, L.~{Ellmeier}, and Y.~{Cao}.
\newblock {The EDGE-CALIFA survey: the influence of galactic rotation on the
  molecular depletion time across the Hubble sequence}.
\newblock \emph{\mnras}, 475:\penalty0 1791--1808, April 2018.
\newblock \doi{10.1093/mnras/stx3233}.

\bibitem[{Wu} et~al.(2005){Wu}, {Evans}, {Gao}, {Solomon}, {Shirley}, and
  {Vanden Bout}]{Wu2005}
J.~{Wu}, N.~J. {Evans}, II, Y.~{Gao}, P.~M. {Solomon}, Y.~L. {Shirley}, and
  P.~A. {Vanden Bout}.
\newblock {Connecting Dense Gas Tracers of Star Formation in our Galaxy to
  High-z Star Formation}.
\newblock \emph{\apjl}, 635:\penalty0 L173--L176, December 2005.
\newblock \doi{10.1086/499623}.

\bibitem[{Schruba} et~al.(2011){Schruba}, {Leroy}, {Walter}, {Bigiel},
  {Brinks}, {de Blok}, {Dumas}, {Kramer}, {Rosolowsky}, {Sandstrom},
  {Schuster}, {Usero}, {Weiss}, and {Wiesemeyer}]{Schruba2011}
A.~{Schruba}, A.~K. {Leroy}, F.~{Walter}, F.~{Bigiel}, E.~{Brinks}, W.~J.~G.
  {de Blok}, G.~{Dumas}, C.~{Kramer}, E.~{Rosolowsky}, K.~{Sandstrom},
  K.~{Schuster}, A.~{Usero}, A.~{Weiss}, and H.~{Wiesemeyer}.
\newblock {A Molecular Star Formation Law in the Atomic-gas-dominated Regime in
  Nearby Galaxies}.
\newblock \emph{\aj}, 142:\penalty0 37, August 2011.
\newblock \doi{10.1088/0004-6256/142/2/37}.

\bibitem[{Bolatto} et~al.(2011){Bolatto}, {Leroy}, {Jameson}, {Ostriker},
  {Gordon}, {Lawton}, {Stanimirovi{\'c}}, {Israel}, {Madden}, {Hony},
  {Sandstrom}, {Bot}, {Rubio}, {Winkler}, {Roman-Duval}, {van Loon},
  {Oliveira}, and {Indebetouw}]{Bolatto2011}
A.~D. {Bolatto}, A.~K. {Leroy}, K.~{Jameson}, E.~{Ostriker}, K.~{Gordon},
  B.~{Lawton}, S.~{Stanimirovi{\'c}}, F.~P. {Israel}, S.~C. {Madden},
  S.~{Hony}, K.~M. {Sandstrom}, C.~{Bot}, M.~{Rubio}, P.~F. {Winkler},
  J.~{Roman-Duval}, J.~T. {van Loon}, J.~M. {Oliveira}, and R.~{Indebetouw}.
\newblock {The State of the Gas and the Relation between Gas and Star Formation
  at Low Metallicity: The Small Magellanic Cloud}.
\newblock \emph{\apj}, 741:\penalty0 12, November 2011.
\newblock \doi{10.1088/0004-637X/741/1/12}.

\bibitem[{Jameson} et~al.(2016){Jameson}, {Bolatto}, {Leroy}, {Meixner},
  {Roman-Duval}, {Gordon}, {Hughes}, {Israel}, {Rubio}, {Indebetouw}, {Madden},
  {Bot}, {Hony}, {Cormier}, {Pellegrini}, {Galametz}, and
  {Sonneborn}]{Jameson2016}
K.~E. {Jameson}, A.~D. {Bolatto}, A.~K. {Leroy}, M.~{Meixner},
  J.~{Roman-Duval}, K.~{Gordon}, A.~{Hughes}, F.~P. {Israel}, M.~{Rubio},
  R.~{Indebetouw}, S.~C. {Madden}, C.~{Bot}, S.~{Hony}, D.~{Cormier}, E.~W.
  {Pellegrini}, M.~{Galametz}, and G.~{Sonneborn}.
\newblock {The Relationship Between Molecular Gas, H I, and Star Formation in
  the Low-mass, Low-metallicity Magellanic Clouds}.
\newblock \emph{\apj}, 825:\penalty0 12, July 2016.
\newblock \doi{10.3847/0004-637X/825/1/12}.

\bibitem[{Semenov} et~al.(2019){Semenov}, {Kravtsov}, and
  {Gnedin}]{Semenov2019}
V.~A. {Semenov}, A.~V. {Kravtsov}, and N.~Y. {Gnedin}.
\newblock {What Sets the Slope of the Molecular Kennicutt-Schmidt Relation?}
\newblock \emph{\apj}, 870:\penalty0 79, January 2019.
\newblock \doi{10.3847/1538-4357/aaf163}.

\bibitem[{Evans} et~al.(2009){Evans}, {Dunham}, {J{\o}rgensen}, {Enoch},
  {Mer{\'{\i}}n}, {van Dishoeck}, {Alcal{\'a}}, {Myers}, {Stapelfeldt},
  {Huard}, {Allen}, {Harvey}, {van Kempen}, {Blake}, {Koerner}, {Mundy},
  {Padgett}, and {Sargent}]{Evans2009}
N.~J. {Evans}, II, M.~M. {Dunham}, J.~K. {J{\o}rgensen}, M.~L. {Enoch},
  B.~{Mer{\'{\i}}n}, E.~F. {van Dishoeck}, J.~M. {Alcal{\'a}}, P.~C. {Myers},
  K.~R. {Stapelfeldt}, T.~L. {Huard}, L.~E. {Allen}, P.~M. {Harvey}, T.~{van
  Kempen}, G.~A. {Blake}, D.~W. {Koerner}, L.~G. {Mundy}, D.~L. {Padgett}, and
  A.~I. {Sargent}.
\newblock {The Spitzer c2d Legacy Results: Star-Formation Rates and
  Efficiencies; Evolution and Lifetimes}.
\newblock \emph{\apjs}, 181:\penalty0 321--350, April 2009.
\newblock \doi{10.1088/0067-0049/181/2/321}.

\bibitem[{Evans} et~al.(2014){Evans}, {Heiderman}, and
  {Vutisalchavakul}]{Evans2014}
N.~J. {Evans}, II, A.~{Heiderman}, and N.~{Vutisalchavakul}.
\newblock {Star Formation Relations in Nearby Molecular Clouds}.
\newblock \emph{\apj}, 782:\penalty0 114, February 2014.
\newblock \doi{10.1088/0004-637X/782/2/114}.

\bibitem[{Heiderman} et~al.(2010){Heiderman}, {Evans}, {Allen}, {Huard}, and
  {Heyer}]{Heiderman2010}
A.~{Heiderman}, N.~J. {Evans}, II, L.~E. {Allen}, T.~{Huard}, and M.~{Heyer}.
\newblock {The Star Formation Rate and Gas Surface Density Relation in the
  Milky Way: Implications for Extragalactic Studies}.
\newblock \emph{\apj}, 723:\penalty0 1019--1037, November 2010.
\newblock \doi{10.1088/0004-637X/723/2/1019}.

\bibitem[{Gutermuth} et~al.(2011){Gutermuth}, {Pipher}, {Megeath}, {Myers},
  {Allen}, and {Allen}]{Gutermuth2011}
R.~A. {Gutermuth}, J.~L. {Pipher}, S.~T. {Megeath}, P.~C. {Myers}, L.~E.
  {Allen}, and T.~S. {Allen}.
\newblock {A Correlation between Surface Densities of Young Stellar Objects and
  Gas in Eight Nearby Molecular Clouds}.
\newblock \emph{\apj}, 739:\penalty0 84, October 2011.
\newblock \doi{10.1088/0004-637X/739/2/84}.

\bibitem[{Genzel} et~al.(2010){Genzel}, {Tacconi}, {Gracia-Carpio},
  {Sternberg}, {Cooper}, {Shapiro}, {Bolatto}, {Bouch{\'e}}, {Bournaud},
  {Burkert}, {Combes}, {Comerford}, {Cox}, {Davis}, {Schreiber},
  {Garcia-Burillo}, {Lutz}, {Naab}, {Neri}, {Omont}, {Shapley}, and
  {Weiner}]{Genzel2010}
R.~{Genzel}, L.~J. {Tacconi}, J.~{Gracia-Carpio}, A.~{Sternberg}, M.~C.
  {Cooper}, K.~{Shapiro}, A.~{Bolatto}, N.~{Bouch{\'e}}, F.~{Bournaud},
  A.~{Burkert}, F.~{Combes}, J.~{Comerford}, P.~{Cox}, M.~{Davis}, N.~M.~F.
  {Schreiber}, S.~{Garcia-Burillo}, D.~{Lutz}, T.~{Naab}, R.~{Neri},
  A.~{Omont}, A.~{Shapley}, and B.~{Weiner}.
\newblock {A study of the gas-star formation relation over cosmic time}.
\newblock \emph{\mnras}, 407:\penalty0 2091--2108, October 2010.
\newblock \doi{10.1111/j.1365-2966.2010.16969.x}.

\bibitem[{Genzel} et~al.(2015){Genzel}, {Tacconi}, {Lutz}, {Saintonge},
  {Berta}, {Magnelli}, {Combes}, {Garc{\'{\i}}a-Burillo}, {Neri}, {Bolatto},
  {Contini}, {Lilly}, {Boissier}, {Boone}, {Bouch{\'e}}, {Bournaud}, {Burkert},
  {Carollo}, {Colina}, {Cooper}, {Cox}, {Feruglio}, {F{\"o}rster Schreiber},
  {Freundlich}, {Gracia-Carpio}, {Juneau}, {Kovac}, {Lippa}, {Naab}, {Salome},
  {Renzini}, {Sternberg}, {Walter}, {Weiner}, {Weiss}, and {Wuyts}]{Genzel2015}
R.~{Genzel}, L.~J. {Tacconi}, D.~{Lutz}, A.~{Saintonge}, S.~{Berta},
  B.~{Magnelli}, F.~{Combes}, S.~{Garc{\'{\i}}a-Burillo}, R.~{Neri},
  A.~{Bolatto}, T.~{Contini}, S.~{Lilly}, J.~{Boissier}, F.~{Boone},
  N.~{Bouch{\'e}}, F.~{Bournaud}, A.~{Burkert}, M.~{Carollo}, L.~{Colina},
  M.~C. {Cooper}, P.~{Cox}, C.~{Feruglio}, N.~M. {F{\"o}rster Schreiber},
  J.~{Freundlich}, J.~{Gracia-Carpio}, S.~{Juneau}, K.~{Kovac}, M.~{Lippa},
  T.~{Naab}, P.~{Salome}, A.~{Renzini}, A.~{Sternberg}, F.~{Walter},
  B.~{Weiner}, A.~{Weiss}, and S.~{Wuyts}.
\newblock {Combined CO and Dust Scaling Relations of Depletion Time and
  Molecular Gas Fractions with Cosmic Time, Specific Star-formation Rate, and
  Stellar Mass}.
\newblock \emph{\apj}, 800:\penalty0 20, February 2015.
\newblock \doi{10.1088/0004-637X/800/1/20}.

\bibitem[{Calzetti} et~al.(2018){Calzetti}, {Wilson}, {Draine}, {Roussel},
  {Johnson}, {Heyer}, {Wall}, {Grasha}, {Battisti}, {Andrews}, {Kirkpatrick},
  {Rosa Gonz{\'a}lez}, {Vega}, {Puschnig}, {Yun}, {{\"O}stlin}, {Evans},
  {Tang}, {Lowenthal}, and {S{\'a}nchez-Arguelles}]{Calzetti2018}
D.~{Calzetti}, G.~W. {Wilson}, B.~T. {Draine}, H.~{Roussel}, K.~E. {Johnson},
  M.~H. {Heyer}, W.~F. {Wall}, K.~{Grasha}, A.~{Battisti}, J.~E. {Andrews},
  A.~{Kirkpatrick}, D.~{Rosa Gonz{\'a}lez}, O.~{Vega}, J.~{Puschnig}, M.~{Yun},
  G.~{{\"O}stlin}, A.~S. {Evans}, Y.~{Tang}, J.~{Lowenthal}, and
  D.~{S{\'a}nchez-Arguelles}.
\newblock {Spatially Resolved Dust, Gas, and Star Formation in the Dwarf
  Magellanic Irregular NGC 4449}.
\newblock \emph{\apj}, 852:\penalty0 106, January 2018.
\newblock \doi{10.3847/1538-4357/aaa1e2}.

\bibitem[{Gao} and {Solomon}(2004{\natexlab{a}})]{Gao2004b}
Y.~{Gao} and P.~M. {Solomon}.
\newblock {HCN Survey of Normal Spiral, Infrared-luminous, and Ultraluminous
  Galaxies}.
\newblock \emph{\apjs}, 152:\penalty0 63--80, May 2004{\natexlab{a}}.
\newblock \doi{10.1086/383003}.

\bibitem[{Gao} and {Solomon}(2004{\natexlab{b}})]{Gao2004a}
Y.~{Gao} and P.~M. {Solomon}.
\newblock {The Star Formation Rate and Dense Molecular Gas in Galaxies}.
\newblock \emph{\apj}, 606:\penalty0 271--290, May 2004{\natexlab{b}}.
\newblock \doi{10.1086/382999}.

\bibitem[{Lada} et~al.(2010){Lada}, {Lombardi}, and {Alves}]{Lada2010}
Charles~J. {Lada}, Marco {Lombardi}, and Jo{\~a}o~F. {Alves}.
\newblock {On the Star Formation Rates in Molecular Clouds}.
\newblock \emph{\apj}, 724\penalty0 (1):\penalty0 687--693, Nov 2010.
\newblock \doi{10.1088/0004-637X/724/1/687}.

\bibitem[{Lada} et~al.(2012){Lada}, {Forbrich}, {Lombardi}, and
  {Alves}]{Lada2012}
Charles~J. {Lada}, Jan {Forbrich}, Marco {Lombardi}, and Jo{\~a}o~F. {Alves}.
\newblock {Star Formation Rates in Molecular Clouds and the Nature of the
  Extragalactic Scaling Relations}.
\newblock \emph{\apj}, 745\penalty0 (2):\penalty0 190, Feb 2012.
\newblock \doi{10.1088/0004-637X/745/2/190}.

\bibitem[{Bigiel} et~al.(2016){Bigiel}, {Leroy}, {Jim{\'e}nez-Donaire}, {Pety},
  {Usero}, {Cormier}, {Bolatto}, {Garcia-Burillo}, {Colombo},
  {Gonz{\'a}lez-Garc{\'{\i}}a}, {Hughes}, {Kepley}, {Kramer}, {Sandstrom},
  {Schinnerer}, {Schruba}, {Schuster}, {Tomicic}, and
  {Zschaechner}]{Bigiel2016}
F.~{Bigiel}, A.~K. {Leroy}, M.~J. {Jim{\'e}nez-Donaire}, J.~{Pety}, A.~{Usero},
  D.~{Cormier}, A.~{Bolatto}, S.~{Garcia-Burillo}, D.~{Colombo},
  M.~{Gonz{\'a}lez-Garc{\'{\i}}a}, A.~{Hughes}, A.~A. {Kepley}, C.~{Kramer},
  K.~{Sandstrom}, E.~{Schinnerer}, A.~{Schruba}, K.~{Schuster}, N.~{Tomicic},
  and L.~{Zschaechner}.
\newblock {The EMPIRE Survey: Systematic Variations in the Dense Gas Fraction
  and Star Formation Efficiency from Full-disk Mapping of M51}.
\newblock \emph{\apjl}, 822:\penalty0 L26, May 2016.
\newblock \doi{10.3847/2041-8205/822/2/L26}.

\bibitem[{Jim{\'e}nez-Donaire} et~al.(2019){Jim{\'e}nez-Donaire}, {Bigiel},
  {Leroy}, {Usero}, {Cormier}, {Puschnig}, {Gallagher}, {Kepley}, {Bolatto},
  {Garc{\'\i}a-Burillo}, {Hughes}, {Kramer}, {Pety}, {Schinnerer}, {Schruba},
  {Schuster}, and {Walter}]{JimenezDonaire2019}
Mar{\'\i}a~J. {Jim{\'e}nez-Donaire}, F.~{Bigiel}, A.~K. {Leroy}, A.~{Usero},
  D.~{Cormier}, J.~{Puschnig}, M.~{Gallagher}, A.~{Kepley}, A.~{Bolatto},
  S.~{Garc{\'\i}a-Burillo}, A.~{Hughes}, C.~{Kramer}, J.~{Pety},
  E.~{Schinnerer}, A.~{Schruba}, K.~{Schuster}, and F.~{Walter}.
\newblock {EMPIRE: The IRAM 30 m Dense Gas Survey of Nearby Galaxies}.
\newblock \emph{\apj}, 880\penalty0 (2):\penalty0 127, Aug 2019.
\newblock \doi{10.3847/1538-4357/ab2b95}.

\bibitem[{Kruijssen} et~al.(2019){Kruijssen}, {Schruba}, {Chevance},
  {Longmore}, {Hygate}, {Haydon}, {McLeod}, {Dalcanton}, {Tacconi}, and {van
  Dishoeck}]{Kruijssen2019}
J.~M.~Diederik {Kruijssen}, Andreas {Schruba}, M{\'e}lanie {Chevance},
  Steven~N. {Longmore}, Alexand er P.~S. {Hygate}, Daniel~T. {Haydon}, Anna~F.
  {McLeod}, Julianne~J. {Dalcanton}, Linda~J. {Tacconi}, and Ewine~F. {van
  Dishoeck}.
\newblock {Fast and inefficient star formation due to short-lived molecular
  clouds and rapid feedback}.
\newblock \emph{\nat}, 569\penalty0 (7757):\penalty0 519--522, May 2019.
\newblock \doi{10.1038/s41586-019-1194-3}.

\bibitem[{Popesso} et~al.(2019){Popesso}, {Concas}, {Morselli}, {Schreiber},
  {Rodighiero}, {Cresci}, {Belli}, {Erfanianfar}, {Mancini}, {Inami},
  {Dickinson}, {Ilbert}, {Pannella}, and {Elbaz}]{Popesso2019}
P.~{Popesso}, A.~{Concas}, L.~{Morselli}, C.~{Schreiber}, G.~{Rodighiero},
  G.~{Cresci}, S.~{Belli}, G.~{Erfanianfar}, C.~{Mancini}, H.~{Inami},
  M.~{Dickinson}, O.~{Ilbert}, M.~{Pannella}, and D.~{Elbaz}.
\newblock {The main sequence of star-forming galaxies - I. The local relation
  and its bending}.
\newblock \emph{\mnras}, 483:\penalty0 3213--3226, March 2019.
\newblock \doi{10.1093/mnras/sty3210}.

\bibitem[{Brinchmann} et~al.(2004){Brinchmann}, {Charlot}, {White}, {Tremonti},
  {Kauffmann}, {Heckman}, and {Brinkmann}]{Brinchmann2004}
J.~{Brinchmann}, S.~{Charlot}, S.~D.~M. {White}, C.~{Tremonti}, G.~{Kauffmann},
  T.~{Heckman}, and J.~{Brinkmann}.
\newblock {The physical properties of star-forming galaxies in the low-redshift
  Universe}.
\newblock \emph{\mnras}, 351:\penalty0 1151--1179, July 2004.
\newblock \doi{10.1111/j.1365-2966.2004.07881.x}.

\bibitem[{Noeske} et~al.(2007){Noeske}, {Faber}, {Weiner}, {Koo}, {Primack},
  {Dekel}, {Papovich}, {Conselice}, {Le Floc'h}, {Rieke}, {Coil}, {Lotz},
  {Somerville}, and {Bundy}]{Noeske2007}
K.~G. {Noeske}, S.~M. {Faber}, B.~J. {Weiner}, D.~C. {Koo}, J.~R. {Primack},
  A.~{Dekel}, C.~{Papovich}, C.~J. {Conselice}, E.~{Le Floc'h}, G.~H. {Rieke},
  A.~L. {Coil}, J.~M. {Lotz}, R.~S. {Somerville}, and K.~{Bundy}.
\newblock {Star Formation in AEGIS Field Galaxies since z=1.1: Staged Galaxy
  Formation and a Model of Mass-dependent Gas Exhaustion}.
\newblock \emph{\apjl}, 660:\penalty0 L47--L50, May 2007.
\newblock \doi{10.1086/517927}.

\bibitem[{Daddi} et~al.(2007){Daddi}, {Dickinson}, {Morrison}, {Chary},
  {Cimatti}, {Elbaz}, {Frayer}, {Renzini}, {Pope}, {Alexander}, {Bauer},
  {Giavalisco}, {Huynh}, {Kurk}, and {Mignoli}]{Daddi2007}
E.~{Daddi}, M.~{Dickinson}, G.~{Morrison}, R.~{Chary}, A.~{Cimatti},
  D.~{Elbaz}, D.~{Frayer}, A.~{Renzini}, A.~{Pope}, D.~M. {Alexander}, F.~E.
  {Bauer}, M.~{Giavalisco}, M.~{Huynh}, J.~{Kurk}, and M.~{Mignoli}.
\newblock {Multiwavelength Study of Massive Galaxies at z\~{}2. I. Star
  Formation and Galaxy Growth}.
\newblock \emph{\apj}, 670:\penalty0 156--172, November 2007.
\newblock \doi{10.1086/521818}.

\bibitem[{Elbaz} et~al.(2007){Elbaz}, {Daddi}, {Le Borgne}, {Dickinson},
  {Alexander}, {Chary}, {Starck}, {Brandt}, {Kitzbichler}, {MacDonald},
  {Nonino}, {Popesso}, {Stern}, and {Vanzella}]{Elbaz2007}
D.~{Elbaz}, E.~{Daddi}, D.~{Le Borgne}, M.~{Dickinson}, D.~M. {Alexander},
  R.-R. {Chary}, J.-L. {Starck}, W.~N. {Brandt}, M.~{Kitzbichler},
  E.~{MacDonald}, M.~{Nonino}, P.~{Popesso}, D.~{Stern}, and E.~{Vanzella}.
\newblock {The reversal of the star formation-density relation in the distant
  universe}.
\newblock \emph{\aap}, 468:\penalty0 33--48, June 2007.
\newblock \doi{10.1051/0004-6361:20077525}.

\bibitem[{Peng} et~al.(2010){Peng}, {Lilly}, {Kova{\v c}}, {Bolzonella},
  {Pozzetti}, {Renzini}, {Zamorani}, {Ilbert}, {Knobel}, {Iovino}, {Maier},
  {Cucciati}, {Tasca}, {Carollo}, {Silverman}, {Kampczyk}, {de Ravel},
  {Sanders}, {Scoville}, {Contini}, {Mainieri}, {Scodeggio}, {Kneib}, {Le
  F{\`e}vre}, {Bardelli}, {Bongiorno}, {Caputi}, {Coppa}, {de la Torre},
  {Franzetti}, {Garilli}, {Lamareille}, {Le Borgne}, {Le Brun}, {Mignoli},
  {Perez Montero}, {Pello}, {Ricciardelli}, {Tanaka}, {Tresse}, {Vergani},
  {Welikala}, {Zucca}, {Oesch}, {Abbas}, {Barnes}, {Bordoloi}, {Bottini},
  {Cappi}, {Cassata}, {Cimatti}, {Fumana}, {Hasinger}, {Koekemoer},
  {Leauthaud}, {Maccagni}, {Marinoni}, {McCracken}, {Memeo}, {Meneux}, {Nair},
  {Porciani}, {Presotto}, and {Scaramella}]{Peng2010}
Y.-j. {Peng}, S.~J. {Lilly}, K.~{Kova{\v c}}, M.~{Bolzonella}, L.~{Pozzetti},
  A.~{Renzini}, G.~{Zamorani}, O.~{Ilbert}, C.~{Knobel}, A.~{Iovino},
  C.~{Maier}, O.~{Cucciati}, L.~{Tasca}, C.~M. {Carollo}, J.~{Silverman},
  P.~{Kampczyk}, L.~{de Ravel}, D.~{Sanders}, N.~{Scoville}, T.~{Contini},
  V.~{Mainieri}, M.~{Scodeggio}, J.-P. {Kneib}, O.~{Le F{\`e}vre},
  S.~{Bardelli}, A.~{Bongiorno}, K.~{Caputi}, G.~{Coppa}, S.~{de la Torre},
  P.~{Franzetti}, B.~{Garilli}, F.~{Lamareille}, J.-F. {Le Borgne}, V.~{Le
  Brun}, M.~{Mignoli}, E.~{Perez Montero}, R.~{Pello}, E.~{Ricciardelli},
  M.~{Tanaka}, L.~{Tresse}, D.~{Vergani}, N.~{Welikala}, E.~{Zucca},
  P.~{Oesch}, U.~{Abbas}, L.~{Barnes}, R.~{Bordoloi}, D.~{Bottini}, A.~{Cappi},
  P.~{Cassata}, A.~{Cimatti}, M.~{Fumana}, G.~{Hasinger}, A.~{Koekemoer},
  A.~{Leauthaud}, D.~{Maccagni}, C.~{Marinoni}, H.~{McCracken}, P.~{Memeo},
  B.~{Meneux}, P.~{Nair}, C.~{Porciani}, V.~{Presotto}, and R.~{Scaramella}.
\newblock {Mass and Environment as Drivers of Galaxy Evolution in SDSS and
  zCOSMOS and the Origin of the Schechter Function}.
\newblock \emph{\apj}, 721:\penalty0 193--221, September 2010.
\newblock \doi{10.1088/0004-637X/721/1/193}.

\bibitem[{Wuyts} et~al.(2011){Wuyts}, {F{\"o}rster Schreiber}, {van der Wel},
  {Magnelli}, {Guo}, {Genzel}, {Lutz}, {Aussel}, {Barro}, {Berta}, {Cava},
  {Graci{\'a}-Carpio}, {Hathi}, {Huang}, {Kocevski}, {Koekemoer}, {Lee}, {Le
  Floc'h}, {McGrath}, {Nordon}, {Popesso}, {Pozzi}, {Riguccini}, {Rodighiero},
  {Saintonge}, and {Tacconi}]{Wuyts2011}
S.~{Wuyts}, N.~M. {F{\"o}rster Schreiber}, A.~{van der Wel}, B.~{Magnelli},
  Y.~{Guo}, R.~{Genzel}, D.~{Lutz}, H.~{Aussel}, G.~{Barro}, S.~{Berta},
  A.~{Cava}, J.~{Graci{\'a}-Carpio}, N.~P. {Hathi}, K.-H. {Huang}, D.~D.
  {Kocevski}, A.~M. {Koekemoer}, K.-S. {Lee}, E.~{Le Floc'h}, E.~J. {McGrath},
  R.~{Nordon}, P.~{Popesso}, F.~{Pozzi}, L.~{Riguccini}, G.~{Rodighiero},
  A.~{Saintonge}, and L.~{Tacconi}.
\newblock {Galaxy Structure and Mode of Star Formation in the SFR-Mass Plane
  from z \~{} 2.5 to z \~{} 0.1}.
\newblock \emph{\apj}, 742:\penalty0 96, December 2011.
\newblock \doi{10.1088/0004-637X/742/2/96}.

\bibitem[{Whitaker} et~al.(2012){Whitaker}, {van Dokkum}, {Brammer}, and
  {Franx}]{Whitaker2012}
K.~E. {Whitaker}, P.~G. {van Dokkum}, G.~{Brammer}, and M.~{Franx}.
\newblock {The Star Formation Mass Sequence Out to z = 2.5}.
\newblock \emph{\apjl}, 754:\penalty0 L29, August 2012.
\newblock \doi{10.1088/2041-8205/754/2/L29}.

\bibitem[{Whitaker} et~al.(2014){Whitaker}, {Franx}, {Leja}, {van Dokkum},
  {Henry}, {Skelton}, {Fumagalli}, {Momcheva}, {Brammer}, {Labb{\'e}},
  {Nelson}, and {Rigby}]{Whitaker2014}
K.~E. {Whitaker}, M.~{Franx}, J.~{Leja}, P.~G. {van Dokkum}, A.~{Henry}, R.~E.
  {Skelton}, M.~{Fumagalli}, I.~G. {Momcheva}, G.~B. {Brammer}, I.~{Labb{\'e}},
  E.~J. {Nelson}, and J.~R. {Rigby}.
\newblock {Constraining the Low-mass Slope of the Star Formation Sequence at
  0.5 $<$ z $<$ 2.5}.
\newblock \emph{\apj}, 795:\penalty0 104, November 2014.
\newblock \doi{10.1088/0004-637X/795/2/104}.

\bibitem[{Tomczak} et~al.(2016){Tomczak}, {Quadri}, {Tran}, {Labb{\'e}},
  {Straatman}, {Papovich}, {Glazebrook}, {Allen}, {Brammer}, {Cowley},
  {Dickinson}, {Elbaz}, {Inami}, {Kacprzak}, {Morrison}, {Nanayakkara},
  {Persson}, {Rees}, {Salmon}, {Schreiber}, {Spitler}, and
  {Whitaker}]{Tomczak2016}
A.~R. {Tomczak}, R.~F. {Quadri}, K.-V.~H. {Tran}, I.~{Labb{\'e}}, C.~M.~S.
  {Straatman}, C.~{Papovich}, K.~{Glazebrook}, R.~{Allen}, G.~B. {Brammer},
  M.~{Cowley}, M.~{Dickinson}, D.~{Elbaz}, H.~{Inami}, G.~G. {Kacprzak}, G.~E.
  {Morrison}, T.~{Nanayakkara}, S.~E. {Persson}, G.~A. {Rees}, B.~{Salmon},
  C.~{Schreiber}, L.~R. {Spitler}, and K.~E. {Whitaker}.
\newblock {The SFR-M* Relation and Empirical Star-Formation Histories from
  ZFOURGE* at 0.5 $<$ z $<$ 4}.
\newblock \emph{\apj}, 817:\penalty0 118, February 2016.
\newblock \doi{10.3847/0004-637X/817/2/118}.

\bibitem[{Rodighiero} et~al.(2011){Rodighiero}, {Daddi}, {Baronchelli},
  {Cimatti}, {Renzini}, {Aussel}, {Popesso}, {Lutz}, {Andreani}, {Berta},
  {Cava}, {Elbaz}, {Feltre}, {Fontana}, {F{\"o}rster Schreiber},
  {Franceschini}, {Genzel}, {Grazian}, {Gruppioni}, {Ilbert}, {Le Floch},
  {Magdis}, {Magliocchetti}, {Magnelli}, {Maiolino}, {McCracken}, {Nordon},
  {Poglitsch}, {Santini}, {Pozzi}, {Riguccini}, {Tacconi}, {Wuyts}, and
  {Zamorani}]{Rodighiero2011}
G.~{Rodighiero}, E.~{Daddi}, I.~{Baronchelli}, A.~{Cimatti}, A.~{Renzini},
  H.~{Aussel}, P.~{Popesso}, D.~{Lutz}, P.~{Andreani}, S.~{Berta}, A.~{Cava},
  D.~{Elbaz}, A.~{Feltre}, A.~{Fontana}, N.~M. {F{\"o}rster Schreiber},
  A.~{Franceschini}, R.~{Genzel}, A.~{Grazian}, C.~{Gruppioni}, O.~{Ilbert},
  E.~{Le Floch}, G.~{Magdis}, M.~{Magliocchetti}, B.~{Magnelli}, R.~{Maiolino},
  H.~{McCracken}, R.~{Nordon}, A.~{Poglitsch}, P.~{Santini}, F.~{Pozzi},
  L.~{Riguccini}, L.~J. {Tacconi}, S.~{Wuyts}, and G.~{Zamorani}.
\newblock {The Lesser Role of Starbursts in Star Formation at z = 2}.
\newblock \emph{\apjl}, 739\penalty0 (2):\penalty0 L40, Oct 2011.
\newblock \doi{10.1088/2041-8205/739/2/L40}.

\bibitem[{Tacconi} et~al.(2013){Tacconi}, {Neri}, {Genzel}, {Combes},
  {Bolatto}, {Cooper}, {Wuyts}, {Bournaud}, {Burkert}, {Comerford}, {Cox},
  {Davis}, {F{\"o}rster Schreiber}, {Garc{\'{\i}}a-Burillo}, {Gracia-Carpio},
  {Lutz}, {Naab}, {Newman}, {Omont}, {Saintonge}, {Shapiro Griffin}, {Shapley},
  {Sternberg}, and {Weiner}]{Tacconi2013}
L.~J. {Tacconi}, R.~{Neri}, R.~{Genzel}, F.~{Combes}, A.~{Bolatto}, M.~C.
  {Cooper}, S.~{Wuyts}, F.~{Bournaud}, A.~{Burkert}, J.~{Comerford}, P.~{Cox},
  M.~{Davis}, N.~M. {F{\"o}rster Schreiber}, S.~{Garc{\'{\i}}a-Burillo},
  J.~{Gracia-Carpio}, D.~{Lutz}, T.~{Naab}, S.~{Newman}, A.~{Omont},
  A.~{Saintonge}, K.~{Shapiro Griffin}, A.~{Shapley}, A.~{Sternberg}, and
  B.~{Weiner}.
\newblock {Phibss: Molecular Gas Content and Scaling Relations in z \~{} 1-3
  Massive, Main-sequence Star-forming Galaxies}.
\newblock \emph{\apj}, 768:\penalty0 74, May 2013.
\newblock \doi{10.1088/0004-637X/768/1/74}.

\bibitem[{Scoville} et~al.(2017){Scoville}, {Lee}, {Vanden Bout},
  {Diaz-Santos}, {Sanders}, {Darvish}, {Bongiorno}, {Casey}, {Murchikova},
  {Koda}, {Capak}, {Vlahakis}, {Ilbert}, {Sheth}, {Morokuma-Matsui}, {Ivison},
  {Aussel}, {Laigle}, {McCracken}, {Armus}, {Pope}, {Toft}, and
  {Masters}]{Scoville2017}
N.~{Scoville}, N.~{Lee}, P.~{Vanden Bout}, T.~{Diaz-Santos}, D.~{Sanders},
  B.~{Darvish}, A.~{Bongiorno}, C.~M. {Casey}, L.~{Murchikova}, J.~{Koda},
  P.~{Capak}, C.~{Vlahakis}, O.~{Ilbert}, K.~{Sheth}, K.~{Morokuma-Matsui},
  R.~J. {Ivison}, H.~{Aussel}, C.~{Laigle}, H.~J. {McCracken}, L.~{Armus},
  A.~{Pope}, S.~{Toft}, and D.~{Masters}.
\newblock {Evolution of Interstellar Medium, Star Formation, and Accretion at
  High Redshift}.
\newblock \emph{\apj}, 837:\penalty0 150, March 2017.
\newblock \doi{10.3847/1538-4357/aa61a0}.

\bibitem[{Hakobyan} et~al.(2012){Hakobyan}, {Adibekyan}, {Aramyan},
  {Petrosian}, {Gomes}, {Mamon}, {Kunth}, and {Turatto}]{Hakobyan2012}
A.~A. {Hakobyan}, V.~Z. {Adibekyan}, L.~S. {Aramyan}, A.~R. {Petrosian}, J.~M.
  {Gomes}, G.~A. {Mamon}, D.~{Kunth}, and M.~{Turatto}.
\newblock {Supernovae and their host galaxies. I. The SDSS DR8 database and
  statistics}.
\newblock \emph{\aap}, 544:\penalty0 A81, August 2012.
\newblock \doi{10.1051/0004-6361/201219541}.

\bibitem[{Duval} et~al.(2016){Duval}, {{\"O}stlin}, {Hayes}, {Zackrisson},
  {Verhamme}, {Orlitova}, {Adamo}, {Guaita}, {Melinder}, {Cannon}, {Laursen},
  {Rivera-Thorsen}, {Herenz}, {Gruyters}, {Mas-Hesse}, {Kunth}, {Sandberg},
  {Schaerer}, and {M{\aa}nsson}]{Duval2016}
F.~{Duval}, G.~{{\"O}stlin}, M.~{Hayes}, E.~{Zackrisson}, A.~{Verhamme},
  I.~{Orlitova}, A.~{Adamo}, L.~{Guaita}, J.~{Melinder}, J.~M. {Cannon},
  P.~{Laursen}, T.~{Rivera-Thorsen}, E.~C. {Herenz}, P.~{Gruyters}, J.~M.
  {Mas-Hesse}, D.~{Kunth}, A.~{Sandberg}, D.~{Schaerer}, and T.~{M{\aa}nsson}.
\newblock {The Lyman alpha reference sample. VI. Lyman alpha escape from the
  edge-on disk galaxy Mrk 1486}.
\newblock \emph{\aap}, 587:\penalty0 A77, March 2016.
\newblock \doi{10.1051/0004-6361/201526876}.

\bibitem[{G{\"u}sten} et~al.(2006){G{\"u}sten}, {Nyman}, {Schilke}, {Menten},
  {Cesarsky}, and {Booth}]{APEX}
R.~{G{\"u}sten}, L.~{\AA}. {Nyman}, P.~{Schilke}, K.~{Menten}, C.~{Cesarsky},
  and R.~{Booth}.
\newblock {The Atacama Pathfinder EXperiment (APEX) - a new submillimeter
  facility for southern skies -}.
\newblock \emph{\aap}, 454:\penalty0 L13--L16, August 2006.
\newblock \doi{10.1051/0004-6361:20065420}.

\bibitem[{Belitsky} et~al.(2006){Belitsky}, {Lapkin}, {Monje}, {Vassilev},
  {Risacher}, {Pavolotsky}, {Meledin}, {Olberg}, {Pantaleev}, and
  {Booth}]{APEX_SHFI2}
V.~{Belitsky}, I.~{Lapkin}, R.~{Monje}, V.~{Vassilev}, C.~{Risacher},
  A.~{Pavolotsky}, D.~{Meledin}, M.~{Olberg}, M.~{Pantaleev}, and R.~{Booth}.
\newblock {Heterodyne single-pixel facility instrumentation for the APEX
  Telescope}.
\newblock In \emph{Society of Photo-Optical Instrumentation Engineers (SPIE)
  Conference Series}, volume 6275 of \emph{\procspie}, page 62750G, June 2006.
\newblock \doi{10.1117/12.671383}.

\bibitem[{Vassilev} et~al.(2008){Vassilev}, {Meledin}, {Lapkin}, {Belitsky},
  {Nystr{\"o}m}, {Henke}, {Pavolotsky}, {Monje}, {Risacher}, {Olberg},
  {Strandberg}, {Sundin}, {Fredrixon}, {Ferm}, {Desmaris}, {Dochev},
  {Pantaleev}, {Bergman}, and {Olofsson}]{APEX_SHFI1}
V.~{Vassilev}, D.~{Meledin}, I.~{Lapkin}, V.~{Belitsky}, O.~{Nystr{\"o}m},
  D.~{Henke}, A.~{Pavolotsky}, R.~{Monje}, C.~{Risacher}, M.~{Olberg},
  M.~{Strandberg}, E.~{Sundin}, M.~{Fredrixon}, S.-E. {Ferm}, V.~{Desmaris},
  D.~{Dochev}, M.~{Pantaleev}, P.~{Bergman}, and H.~{Olofsson}.
\newblock {A Swedish heterodyne facility instrument for the APEX telescope}.
\newblock \emph{\aap}, 490:\penalty0 1157--1163, November 2008.
\newblock \doi{10.1051/0004-6361:200810459}.

\bibitem[{Poglitsch} et~al.(2010){Poglitsch}, {Waelkens}, {Geis},
  {Feuchtgruber}, {Vandenbussche}, {Rodriguez}, {Krause}, {Renotte}, {van
  Hoof}, {Saraceno}, {Cepa}, {Kerschbaum}, {Agn{\`e}se}, {Ali}, {Altieri},
  {Andreani}, {Augueres}, {Balog}, {Barl}, {Bauer}, {Belbachir}, {Benedettini},
  {Billot}, {Boulade}, {Bischof}, {Blommaert}, {Callut}, {Cara}, {Cerulli},
  {Cesarsky}, {Contursi}, {Creten}, {De Meester}, {Doublier}, {Doumayrou},
  {Duband}, {Exter}, {Genzel}, {Gillis}, {Gr{\"o}zinger}, {Henning},
  {Herreros}, {Huygen}, {Inguscio}, {Jakob}, {Jamar}, {Jean}, {de Jong},
  {Katterloher}, {Kiss}, {Klaas}, {Lemke}, {Lutz}, {Madden}, {Marquet},
  {Martignac}, {Mazy}, {Merken}, {Montfort}, {Morbidelli}, {M{\"u}ller},
  {Nielbock}, {Okumura}, {Orfei}, {Ottensamer}, {Pezzuto}, {Popesso},
  {Putzeys}, {Regibo}, {Reveret}, {Royer}, {Sauvage}, {Schreiber}, {Stegmaier},
  {Schmitt}, {Schubert}, {Sturm}, {Thiel}, {Tofani}, {Vavrek}, {Wetzstein},
  {Wieprecht}, and {Wiezorrek}]{Poglitsch2010}
A.~{Poglitsch}, C.~{Waelkens}, N.~{Geis}, H.~{Feuchtgruber},
  B.~{Vandenbussche}, L.~{Rodriguez}, O.~{Krause}, E.~{Renotte}, C.~{van Hoof},
  P.~{Saraceno}, J.~{Cepa}, F.~{Kerschbaum}, P.~{Agn{\`e}se}, B.~{Ali},
  B.~{Altieri}, P.~{Andreani}, J.-L. {Augueres}, Z.~{Balog}, L.~{Barl}, O.~H.
  {Bauer}, N.~{Belbachir}, M.~{Benedettini}, N.~{Billot}, O.~{Boulade},
  H.~{Bischof}, J.~{Blommaert}, E.~{Callut}, C.~{Cara}, R.~{Cerulli},
  D.~{Cesarsky}, A.~{Contursi}, Y.~{Creten}, W.~{De Meester}, V.~{Doublier},
  E.~{Doumayrou}, L.~{Duband}, K.~{Exter}, R.~{Genzel}, J.-M. {Gillis},
  U.~{Gr{\"o}zinger}, T.~{Henning}, J.~{Herreros}, R.~{Huygen}, M.~{Inguscio},
  G.~{Jakob}, C.~{Jamar}, C.~{Jean}, J.~{de Jong}, R.~{Katterloher}, C.~{Kiss},
  U.~{Klaas}, D.~{Lemke}, D.~{Lutz}, S.~{Madden}, B.~{Marquet}, J.~{Martignac},
  A.~{Mazy}, P.~{Merken}, F.~{Montfort}, L.~{Morbidelli}, T.~{M{\"u}ller},
  M.~{Nielbock}, K.~{Okumura}, R.~{Orfei}, R.~{Ottensamer}, S.~{Pezzuto},
  P.~{Popesso}, J.~{Putzeys}, S.~{Regibo}, V.~{Reveret}, P.~{Royer},
  M.~{Sauvage}, J.~{Schreiber}, J.~{Stegmaier}, D.~{Schmitt}, J.~{Schubert},
  E.~{Sturm}, M.~{Thiel}, G.~{Tofani}, R.~{Vavrek}, M.~{Wetzstein},
  E.~{Wieprecht}, and E.~{Wiezorrek}.
\newblock {The Photodetector Array Camera and Spectrometer (PACS) on the
  Herschel Space Observatory}.
\newblock \emph{\aap}, 518:\penalty0 L2, July 2010.
\newblock \doi{10.1051/0004-6361/201014535}.

\bibitem[{Pilbratt} et~al.(2010){Pilbratt}, {Riedinger}, {Passvogel}, {Crone},
  {Doyle}, {Gageur}, {Heras}, {Jewell}, {Metcalfe}, {Ott}, and
  {Schmidt}]{Pilbratt2010}
G.~L. {Pilbratt}, J.~R. {Riedinger}, T.~{Passvogel}, G.~{Crone}, D.~{Doyle},
  U.~{Gageur}, A.~M. {Heras}, C.~{Jewell}, L.~{Metcalfe}, S.~{Ott}, and
  M.~{Schmidt}.
\newblock {Herschel Space Observatory. An ESA facility for far-infrared and
  submillimetre astronomy}.
\newblock \emph{\aap}, 518:\penalty0 L1, July 2010.
\newblock \doi{10.1051/0004-6361/201014759}.

\bibitem[{Ott}(2010)]{Ott2010}
S.~{Ott}.
\newblock {The Herschel Data Processing System - HIPE and Pipelines - Up and
  Running Since the Start of the Mission}.
\newblock In Y.~{Mizumoto}, K.-I. {Morita}, and M.~{Ohishi}, editors,
  \emph{Astronomical Data Analysis Software and Systems XIX}, volume 434 of
  \emph{Astronomical Society of the Pacific Conference Series}, page 139,
  December 2010.

\bibitem[{Smirnova-Pinchukova} et~al.(2019){Smirnova-Pinchukova}, {Husemann},
  {Busch}, {Appleton}, {Bethermin}, {Combes}, {Croom}, {Davis}, {Fischer},
  {Gaspari}, {Groves}, {Klein}, {O'Dea}, {P{\'e}rez-Torres},
  {Scharw{\"a}chter}, {Singha}, {Tremblay}, and
  {Urrutia}]{Smirnova-Pinchukova2019}
I.~{Smirnova-Pinchukova}, B.~{Husemann}, G.~{Busch}, P.~{Appleton},
  M.~{Bethermin}, F.~{Combes}, S.~{Croom}, T.~A. {Davis}, C.~{Fischer},
  M.~{Gaspari}, B.~{Groves}, R.~{Klein}, C.~P. {O'Dea}, M.~{P{\'e}rez-Torres},
  J.~{Scharw{\"a}chter}, M.~{Singha}, G.~R. {Tremblay}, and T.~{Urrutia}.
\newblock {The Close AGN Reference Survey (CARS). Discovery of a global [C II]
  158 {\ensuremath{\mu}}m line excess in AGN HE 1353-1917}.
\newblock \emph{\aap}, 626:\penalty0 L3, Jun 2019.
\newblock \doi{10.1051/0004-6361/201935577}.

\bibitem[Fruchter and Hook(2002)]{Fruchter2002}
A.~S. Fruchter and R.~N. Hook.
\newblock Drizzle: A method for the linear reconstruction of undersampled
  images.
\newblock \emph{Publications of the Astronomical Society of the Pacific},
  114\penalty0 (792):\penalty0 144--152, feb 2002.
\newblock \doi{10.1086/338393}.
\newblock URL \url{https://doi.org/10.1086\%2F338393}.

\bibitem[Lord(1992)]{ATRAN}
S.~D. Lord.
\newblock {NASA Technical Memorandum 103957}.
\newblock Technical report, NASA, 1992.
\newblock URL \url{https://atran.arc.nasa.gov/cgi-bin/atran/atran.cgi}.

\bibitem[{Wright} et~al.(2010){Wright}, {Eisenhardt}, {Mainzer}, {Ressler},
  {Cutri}, {Jarrett}, {Kirkpatrick}, {Padgett}, {McMillan}, {Skrutskie},
  {Stanford}, {Cohen}, {Walker}, {Mather}, {Leisawitz}, {Gautier}, {McLean},
  {Benford}, {Lonsdale}, {Blain}, {Mendez}, {Irace}, {Duval}, {Liu}, {Royer},
  {Heinrichsen}, {Howard}, {Shannon}, {Kendall}, {Walsh}, {Larsen}, {Cardon},
  {Schick}, {Schwalm}, {Abid}, {Fabinsky}, {Naes}, and {Tsai}]{Wright2010}
E.~L. {Wright}, P.~R.~M. {Eisenhardt}, A.~K. {Mainzer}, M.~E. {Ressler}, R.~M.
  {Cutri}, T.~{Jarrett}, J.~D. {Kirkpatrick}, D.~{Padgett}, R.~S. {McMillan},
  M.~{Skrutskie}, S.~A. {Stanford}, M.~{Cohen}, R.~G. {Walker}, J.~C. {Mather},
  D.~{Leisawitz}, T.~N. {Gautier}, III, I.~{McLean}, D.~{Benford}, C.~J.
  {Lonsdale}, A.~{Blain}, B.~{Mendez}, W.~R. {Irace}, V.~{Duval}, F.~{Liu},
  D.~{Royer}, I.~{Heinrichsen}, J.~{Howard}, M.~{Shannon}, M.~{Kendall}, A.~L.
  {Walsh}, M.~{Larsen}, J.~G. {Cardon}, S.~{Schick}, M.~{Schwalm}, M.~{Abid},
  B.~{Fabinsky}, L.~{Naes}, and C.-W. {Tsai}.
\newblock {The Wide-field Infrared Survey Explorer (WISE): Mission Description
  and Initial On-orbit Performance}.
\newblock \emph{\aj}, 140:\penalty0 1868-1881, December 2010.
\newblock \doi{10.1088/0004-6256/140/6/1868}.

\bibitem[{Bertin} and {Arnouts}(1996)]{Bertin1996}
E.~{Bertin} and S.~{Arnouts}.
\newblock {SExtractor: Software for source extraction.}
\newblock \emph{\aaps}, 117:\penalty0 393--404, June 1996.
\newblock \doi{10.1051/aas:1996164}.

\bibitem[{Lang}(2014)]{Lang2014}
D.~{Lang}.
\newblock {unWISE: Unblurred Coadds of the WISE Imaging}.
\newblock \emph{\aj}, 147:\penalty0 108, May 2014.
\newblock \doi{10.1088/0004-6256/147/5/108}.

\bibitem[{Murakami} et~al.(2007){Murakami}, {Baba}, {Barthel}, {Clements},
  {Cohen}, {Doi}, {Enya}, {Figueredo}, {Fujishiro}, {Fujiwara}, {Fujiwara},
  {Garcia-Lario}, {Goto}, {Hasegawa}, {Hibi}, {Hirao}, {Hiromoto}, {Hong},
  {Imai}, {Ishigaki}, {Ishiguro}, {Ishihara}, {Ita}, {Jeong}, {Jeong},
  {Kaneda}, {Kataza}, {Kawada}, {Kawai}, {Kawamura}, {Kessler}, {Kester},
  {Kii}, {Kim}, {Kim}, {Kobayashi}, {Koo}, {Kwon}, {Lee}, {Lorente}, {Makiuti},
  {Matsuhara}, {Matsumoto}, {Matsuo}, {Matsuura}, {M{\"U}ller}, {Murakami},
  {Nagata}, {Nakagawa}, {Naoi}, {Narita}, {Noda}, {Oh}, {Ohnishi}, {Ohyama},
  {Okada}, {Okuda}, {Oliver}, {Onaka}, {Ootsubo}, {Oyabu}, {Pak}, {Park},
  {Pearson}, {Rowan-Robinson}, {Saito}, {Sakon}, {Salama}, {Sato}, {Savage},
  {Serjeant}, {Shibai}, {Shirahata}, {Sohn}, {Suzuki}, {Takagi}, {Takahashi},
  {Tanab{\'E}}, {Takeuchi}, {Takita}, {Thomson}, {Uemizu}, {Ueno}, {Usui},
  {Verdugo}, {Wada}, {Wang}, {Watabe}, {Watarai}, {White}, {Yamamura},
  {Yamauchi}, and {Yasuda}]{Murakami2007}
H.~{Murakami}, H.~{Baba}, P.~{Barthel}, D.~L. {Clements}, M.~{Cohen}, Y.~{Doi},
  K.~{Enya}, E.~{Figueredo}, N.~{Fujishiro}, H.~{Fujiwara}, M.~{Fujiwara},
  P.~{Garcia-Lario}, T.~{Goto}, S.~{Hasegawa}, Y.~{Hibi}, T.~{Hirao},
  N.~{Hiromoto}, S.~S. {Hong}, K.~{Imai}, M.~{Ishigaki}, M.~{Ishiguro},
  D.~{Ishihara}, Y.~{Ita}, W.-S. {Jeong}, K.~S. {Jeong}, H.~{Kaneda},
  H.~{Kataza}, M.~{Kawada}, T.~{Kawai}, A.~{Kawamura}, M.~F. {Kessler},
  D.~{Kester}, T.~{Kii}, D.~C. {Kim}, W.~{Kim}, H.~{Kobayashi}, B.~C. {Koo},
  S.~M. {Kwon}, H.~M. {Lee}, R.~{Lorente}, S.~{Makiuti}, H.~{Matsuhara},
  T.~{Matsumoto}, H.~{Matsuo}, S.~{Matsuura}, T.~G. {M{\"U}ller},
  N.~{Murakami}, H.~{Nagata}, T.~{Nakagawa}, T.~{Naoi}, M.~{Narita}, M.~{Noda},
  S.~H. {Oh}, A.~{Ohnishi}, Y.~{Ohyama}, Y.~{Okada}, H.~{Okuda}, S.~{Oliver},
  T.~{Onaka}, T.~{Ootsubo}, S.~{Oyabu}, S.~{Pak}, Y.-S. {Park}, C.~P.
  {Pearson}, M.~{Rowan-Robinson}, T.~{Saito}, I.~{Sakon}, A.~{Salama},
  S.~{Sato}, R.~S. {Savage}, S.~{Serjeant}, H.~{Shibai}, M.~{Shirahata},
  J.~{Sohn}, T.~{Suzuki}, T.~{Takagi}, H.~{Takahashi}, T.~{Tanab{\'E}}, T.~T.
  {Takeuchi}, S.~{Takita}, M.~{Thomson}, K.~{Uemizu}, M.~{Ueno}, F.~{Usui},
  E.~{Verdugo}, T.~{Wada}, L.~{Wang}, T.~{Watabe}, H.~{Watarai}, G.~J. {White},
  I.~{Yamamura}, C.~{Yamauchi}, and A.~{Yasuda}.
\newblock {The Infrared Astronomical Mission AKARI*}.
\newblock \emph{\pasj}, 59:\penalty0 S369--S376, October 2007.
\newblock \doi{10.1093/pasj/59.sp2.S369}.

\bibitem[{Kawada} et~al.(2007){Kawada}, {Baba}, {Barthel}, {Clements}, {Cohen},
  {Doi}, {Figueredo}, {Fujiwara}, {Goto}, {Hasegawa}, {Hibi}, {Hirao},
  {Hiromoto}, {Jeong}, {Kaneda}, {Kawai}, {Kawamura}, {Kester}, {Kii},
  {Kobayashi}, {Kwon}, {Lee}, {Makiuti}, {Matsuo}, {Matsuura}, {M{\"u}ller},
  {Murakami}, {Nagata}, {Nakagawa}, {Narita}, {Noda}, {Oh}, {Okada}, {Okuda},
  {Oliver}, {Ootsubo}, {Pak}, {Park}, {Pearson}, {Rowan-Robinson}, {Saito},
  {Salama}, {Sato}, {Savage}, {Serjeant}, {Shibai}, {Shirahata}, {Sohn},
  {Suzuki}, {Takagi}, {Takahashi}, {Thomson}, {Usui}, {Verdugo}, {Watabe},
  {White}, {Wang}, {Yamamura}, {Yamauchi}, and {Yasuda}]{Kawada2007}
M.~{Kawada}, H.~{Baba}, P.~D. {Barthel}, D.~{Clements}, M.~{Cohen}, Y.~{Doi},
  E.~{Figueredo}, M.~{Fujiwara}, T.~{Goto}, S.~{Hasegawa}, Y.~{Hibi},
  T.~{Hirao}, N.~{Hiromoto}, W.-S. {Jeong}, H.~{Kaneda}, T.~{Kawai},
  A.~{Kawamura}, D.~{Kester}, T.~{Kii}, H.~{Kobayashi}, S.~M. {Kwon}, H.~M.
  {Lee}, S.~{Makiuti}, H.~{Matsuo}, S.~{Matsuura}, T.~G. {M{\"u}ller},
  N.~{Murakami}, H.~{Nagata}, T.~{Nakagawa}, M.~{Narita}, M.~{Noda}, S.~H.
  {Oh}, Y.~{Okada}, H.~{Okuda}, S.~{Oliver}, T.~{Ootsubo}, S.~{Pak}, Y.-S.
  {Park}, C.~P. {Pearson}, M.~{Rowan-Robinson}, T.~{Saito}, A.~{Salama},
  S.~{Sato}, R.~S. {Savage}, S.~{Serjeant}, H.~{Shibai}, M.~{Shirahata},
  J.~{Sohn}, T.~{Suzuki}, T.~{Takagi}, H.~{Takahashi}, M.~{Thomson}, F.~{Usui},
  E.~{Verdugo}, T.~{Watabe}, G.~J. {White}, L.~{Wang}, I.~{Yamamura},
  C.~{Yamauchi}, and A.~{Yasuda}.
\newblock {The Far-Infrared Surveyor (FIS) for AKARI}.
\newblock \emph{\pasj}, 59:\penalty0 S389, October 2007.
\newblock \doi{10.1093/pasj/59.sp2.S389}.

\bibitem[{Doi} et~al.(2015){Doi}, {Takita}, {Ootsubo}, {Arimatsu}, {Tanaka},
  {Kitamura}, {Kawada}, {Matsuura}, {Nakagawa}, {Morishima}, {Hattori},
  {Komugi}, {White}, {Ikeda}, {Kato}, {Chinone}, {Etxaluze}, and
  {Cypriano}]{Doi2015}
Y.~{Doi}, S.~{Takita}, T.~{Ootsubo}, K.~{Arimatsu}, M.~{Tanaka}, Y.~{Kitamura},
  M.~{Kawada}, S.~{Matsuura}, T.~{Nakagawa}, T.~{Morishima}, M.~{Hattori},
  S.~{Komugi}, G.~J. {White}, N.~{Ikeda}, D.~{Kato}, Y.~{Chinone},
  M.~{Etxaluze}, and E.~F. {Cypriano}.
\newblock {The AKARI far-infrared all-sky survey maps}.
\newblock \emph{\pasj}, 67:\penalty0 50, June 2015.
\newblock \doi{10.1093/pasj/psv022}.

\bibitem[{Takita} et~al.(2015){Takita}, {Doi}, {Ootsubo}, {Arimatsu}, {Ikeda},
  {Kawada}, {Kitamura}, {Matsuura}, {Nakagawa}, {Hattori}, {Morishima},
  {Tanaka}, and {Komugi}]{Takita2015}
S.~{Takita}, Y.~{Doi}, T.~{Ootsubo}, K.~{Arimatsu}, N.~{Ikeda}, M.~{Kawada},
  Y.~{Kitamura}, S.~{Matsuura}, T.~{Nakagawa}, M.~{Hattori}, T.~{Morishima},
  M.~{Tanaka}, and S.~{Komugi}.
\newblock {Calibration of the AKARI far-infrared all-sky survey maps}.
\newblock \emph{\pasj}, 67:\penalty0 51, June 2015.
\newblock \doi{10.1093/pasj/psv033}.

\bibitem[{Draine} and {Li}(2007)]{Draine2007model}
B.~T. {Draine} and A.~{Li}.
\newblock {Infrared Emission from Interstellar Dust. IV. The
  Silicate-Graphite-PAH Model in the Post-Spitzer Era}.
\newblock \emph{\apj}, 657:\penalty0 810--837, March 2007.
\newblock \doi{10.1086/511055}.

\bibitem[{Mathis} et~al.(1983){Mathis}, {Mezger}, and {Panagia}]{Mathis1983}
J.~S. {Mathis}, P.~G. {Mezger}, and N.~{Panagia}.
\newblock {Interstellar radiation field and dust temperatures in the diffuse
  interstellar matter and in giant molecular clouds}.
\newblock \emph{\aap}, 128:\penalty0 212--229, November 1983.

\bibitem[{Habing}(1968)]{Habing1968}
H.~J. {Habing}.
\newblock {The interstellar radiation density between 912 A and 2400 A}.
\newblock \emph{\bain}, 19:\penalty0 421, January 1968.

\bibitem[{Draine} et~al.(2007){Draine}, {Dale}, {Bendo}, {Gordon}, {Smith},
  {Armus}, {Engelbracht}, {Helou}, {Kennicutt}, {Li}, {Roussel}, {Walter},
  {Calzetti}, {Moustakas}, {Murphy}, {Rieke}, {Bot}, {Hollenbach}, {Sheth}, and
  {Teplitz}]{Draine2007}
B.~T. {Draine}, D.~A. {Dale}, G.~{Bendo}, K.~D. {Gordon}, J.~D.~T. {Smith},
  L.~{Armus}, C.~W. {Engelbracht}, G.~{Helou}, R.~C. {Kennicutt}, Jr., A.~{Li},
  H.~{Roussel}, F.~{Walter}, D.~{Calzetti}, J.~{Moustakas}, E.~J. {Murphy},
  G.~H. {Rieke}, C.~{Bot}, D.~J. {Hollenbach}, K.~{Sheth}, and H.~I. {Teplitz}.
\newblock {Dust Masses, PAH Abundances, and Starlight Intensities in the SINGS
  Galaxy Sample}.
\newblock \emph{\apj}, 663:\penalty0 866--894, July 2007.
\newblock \doi{10.1086/518306}.

\bibitem[Galliano(2018)]{Galliano2018b}
Frédéric Galliano.
\newblock {A dust spectral energy distribution model with hierarchical Bayesian
  inference – I. Formalism and benchmarking}.
\newblock \emph{Monthly Notices of the Royal Astronomical Society},
  476\penalty0 (2):\penalty0 1445--1469, 01 2018.
\newblock ISSN 0035-8711.
\newblock \doi{10.1093/mnras/sty189}.
\newblock URL \url{https://doi.org/10.1093/mnras/sty189}.

\bibitem[{Lisenfeld} and {Ferrara}(1998)]{Lisenfeld1998}
U.~{Lisenfeld} and A.~{Ferrara}.
\newblock {Dust-to-Gas Ratio and Metal Abundance in Dwarf Galaxies}.
\newblock \emph{\apj}, 496:\penalty0 145--154, March 1998.
\newblock \doi{10.1086/305354}.

\bibitem[{Hirashita} et~al.(2002){Hirashita}, {Tajiri}, and
  {Kamaya}]{Hirashita2002}
H.~{Hirashita}, Y.~Y. {Tajiri}, and H.~{Kamaya}.
\newblock {Dust-to-gas ratio and star formation history of blue compact dwarf
  galaxies}.
\newblock \emph{\aap}, 388:\penalty0 439--445, June 2002.
\newblock \doi{10.1051/0004-6361:20020605}.

\bibitem[{James} et~al.(2002){James}, {Dunne}, {Eales}, and
  {Edmunds}]{James2002}
A.~{James}, L.~{Dunne}, S.~{Eales}, and M.~G. {Edmunds}.
\newblock {SCUBA observations of galaxies with metallicity measurements: a new
  method for determining the relation between submillimetre luminosity and dust
  mass}.
\newblock \emph{\mnras}, 335:\penalty0 753--761, September 2002.
\newblock \doi{10.1046/j.1365-8711.2002.05660.x}.

\bibitem[{Hunt} et~al.(2005){Hunt}, {Bianchi}, and {Maiolino}]{Hunt2005}
L.~{Hunt}, S.~{Bianchi}, and R.~{Maiolino}.
\newblock {The optical-to-radio spectral energy distributions of
  low-metallicity blue compact dwarf galaxies}.
\newblock \emph{\aap}, 434:\penalty0 849--866, May 2005.
\newblock \doi{10.1051/0004-6361:20042157}.

\bibitem[{Engelbracht} et~al.(2008){Engelbracht}, {Rieke}, {Gordon}, {Smith},
  {Werner}, {Moustakas}, {Willmer}, and {Vanzi}]{Engelbracht2008}
C.~W. {Engelbracht}, G.~H. {Rieke}, K.~D. {Gordon}, J.-D.~T. {Smith}, M.~W.
  {Werner}, J.~{Moustakas}, C.~N.~A. {Willmer}, and L.~{Vanzi}.
\newblock {Metallicity Effects on Dust Properties in Starbursting Galaxies}.
\newblock \emph{\apj}, 678:\penalty0 804--827, May 2008.
\newblock \doi{10.1086/529513}.

\bibitem[{Galliano} et~al.(2008){Galliano}, {Dwek}, and
  {Chanial}]{Galliano2008}
F.~{Galliano}, E.~{Dwek}, and P.~{Chanial}.
\newblock {Stellar Evolutionary Effects on the Abundances of Polycyclic
  Aromatic Hydrocarbons and Supernova-Condensed Dust in Galaxies}.
\newblock \emph{\apj}, 672:\penalty0 214--243, January 2008.
\newblock \doi{10.1086/523621}.

\bibitem[{Mu{\~n}oz-Mateos} et~al.(2009){Mu{\~n}oz-Mateos}, {Gil de Paz},
  {Boissier}, {Zamorano}, {Dale}, {P{\'e}rez-Gonz{\'a}lez}, {Gallego},
  {Madore}, {Bendo}, {Thornley}, {Draine}, {Boselli}, {Buat}, {Calzetti},
  {Moustakas}, and {Kennicutt}]{MunozMateos2009}
J.~C. {Mu{\~n}oz-Mateos}, A.~{Gil de Paz}, S.~{Boissier}, J.~{Zamorano}, D.~A.
  {Dale}, P.~G. {P{\'e}rez-Gonz{\'a}lez}, J.~{Gallego}, B.~F. {Madore},
  G.~{Bendo}, M.~D. {Thornley}, B.~T. {Draine}, A.~{Boselli}, V.~{Buat},
  D.~{Calzetti}, J.~{Moustakas}, and R.~C. {Kennicutt}, Jr.
\newblock {Radial Distribution of Stars, Gas, and Dust in Sings Galaxies. II.
  Derived Dust Properties}.
\newblock \emph{\apj}, 701:\penalty0 1965--1991, August 2009.
\newblock \doi{10.1088/0004-637X/701/2/1965}.

\bibitem[{Bendo} et~al.(2010){Bendo}, {Wilson}, {Warren}, {Brinks}, {Butner},
  {Chanial}, {Clements}, {Courteau}, {Irwin}, {Israel}, {Knapen}, {Leech},
  {Matthews}, {M{\"u}hle}, {Petitpas}, {Serjeant}, {Tan}, {Tilanus}, {Usero},
  {Vaccari}, {van der Werf}, {Vlahakis}, {Wiegert}, and {Zhu}]{Bendo2010}
G.~J. {Bendo}, C.~D. {Wilson}, B.~E. {Warren}, E.~{Brinks}, H.~M. {Butner},
  P.~{Chanial}, D.~L. {Clements}, S.~{Courteau}, J.~{Irwin}, F.~P. {Israel},
  J.~H. {Knapen}, J.~{Leech}, H.~E. {Matthews}, S.~{M{\"u}hle}, G.~{Petitpas},
  S.~{Serjeant}, B.~K. {Tan}, R.~P.~J. {Tilanus}, A.~{Usero}, M.~{Vaccari},
  P.~{van der Werf}, C.~{Vlahakis}, T.~{Wiegert}, and M.~{Zhu}.
\newblock {The JCMT Nearby Galaxies Legacy Survey - III. Comparisons of cold
  dust, polycyclic aromatic hydrocarbons, molecular gas and atomic gas in NGC
  2403}.
\newblock \emph{\mnras}, 402:\penalty0 1409--1425, March 2010.
\newblock \doi{10.1111/j.1365-2966.2009.16043.x}.

\bibitem[{Galametz} et~al.(2011){Galametz}, {Madden}, {Galliano}, {Hony},
  {Bendo}, and {Sauvage}]{Galametz2011}
M.~{Galametz}, S.~C. {Madden}, F.~{Galliano}, S.~{Hony}, G.~J. {Bendo}, and
  M.~{Sauvage}.
\newblock {Probing the dust properties of galaxies up to submillimetre
  wavelengths. II. Dust-to-gas mass ratio trends with metallicity and the submm
  excess in dwarf galaxies}.
\newblock \emph{\aap}, 532:\penalty0 A56, August 2011.
\newblock \doi{10.1051/0004-6361/201014904}.

\bibitem[{Magrini} et~al.(2011){Magrini}, {Bianchi}, {Corbelli}, {Cortese},
  {Hunt}, {Smith}, {Vlahakis}, {Davies}, {Bendo}, {Baes}, {Boselli}, {Clemens},
  {Casasola}, {De Looze}, {Fritz}, {Giovanardi}, {Grossi}, {Hughes}, {Madden},
  {Pappalardo}, {Pohlen}, {di Serego Alighieri}, and {Verstappen}]{Magrini2011}
L.~{Magrini}, S.~{Bianchi}, E.~{Corbelli}, L.~{Cortese}, L.~{Hunt}, M.~{Smith},
  C.~{Vlahakis}, J.~{Davies}, G.~J. {Bendo}, M.~{Baes}, A.~{Boselli},
  M.~{Clemens}, V.~{Casasola}, I.~{De Looze}, J.~{Fritz}, C.~{Giovanardi},
  M.~{Grossi}, T.~{Hughes}, S.~{Madden}, C.~{Pappalardo}, M.~{Pohlen}, S.~{di
  Serego Alighieri}, and J.~{Verstappen}.
\newblock {The Herschel Virgo Cluster Survey. IX. Dust-to-gas mass ratio and
  metallicity gradients in four Virgo spiral galaxies}.
\newblock \emph{\aap}, 535:\penalty0 A13, November 2011.
\newblock \doi{10.1051/0004-6361/201116872}.

\bibitem[{Sandstrom} et~al.(2013){Sandstrom}, {Leroy}, {Walter}, {Bolatto},
  {Croxall}, {Draine}, {Wilson}, {Wolfire}, {Calzetti}, {Kennicutt}, {Aniano},
  {Donovan Meyer}, {Usero}, {Bigiel}, {Brinks}, {de Blok}, {Crocker}, {Dale},
  {Engelbracht}, {Galametz}, {Groves}, {Hunt}, {Koda}, {Kreckel}, {Linz},
  {Meidt}, {Pellegrini}, {Rix}, {Roussel}, {Schinnerer}, {Schruba}, {Schuster},
  {Skibba}, {van der Laan}, {Appleton}, {Armus}, {Brandl}, {Gordon}, {Hinz},
  {Krause}, {Montiel}, {Sauvage}, {Schmiedeke}, {Smith}, and
  {Vigroux}]{Sandstrom2013}
K.~M. {Sandstrom}, A.~K. {Leroy}, F.~{Walter}, A.~D. {Bolatto}, K.~V.
  {Croxall}, B.~T. {Draine}, C.~D. {Wilson}, M.~{Wolfire}, D.~{Calzetti}, R.~C.
  {Kennicutt}, G.~{Aniano}, J.~{Donovan Meyer}, A.~{Usero}, F.~{Bigiel},
  E.~{Brinks}, W.~J.~G. {de Blok}, A.~{Crocker}, D.~{Dale}, C.~W.
  {Engelbracht}, M.~{Galametz}, B.~{Groves}, L.~K. {Hunt}, J.~{Koda},
  K.~{Kreckel}, H.~{Linz}, S.~{Meidt}, E.~{Pellegrini}, H.-W. {Rix},
  H.~{Roussel}, E.~{Schinnerer}, A.~{Schruba}, K.-F. {Schuster}, R.~{Skibba},
  T.~{van der Laan}, P.~{Appleton}, L.~{Armus}, B.~{Brandl}, K.~{Gordon},
  J.~{Hinz}, O.~{Krause}, E.~{Montiel}, M.~{Sauvage}, A.~{Schmiedeke}, J.~D.~T.
  {Smith}, and L.~{Vigroux}.
\newblock {The CO-to-H$_{2}$ Conversion Factor and Dust-to-gas Ratio on
  Kiloparsec Scales in Nearby Galaxies}.
\newblock \emph{\apj}, 777:\penalty0 5, November 2013.
\newblock \doi{10.1088/0004-637X/777/1/5}.

\bibitem[{R{\'e}my-Ruyer} et~al.(2014){R{\'e}my-Ruyer}, {Madden}, {Galliano},
  {Galametz}, {Takeuchi}, {Asano}, {Zhukovska}, {Lebouteiller}, {Cormier},
  {Jones}, {Bocchio}, {Baes}, {Bendo}, {Boquien}, {Boselli}, {DeLooze},
  {Doublier-Pritchard}, {Hughes}, {Karczewski}, and {Spinoglio}]{RemyRuyer2014}
A.~{R{\'e}my-Ruyer}, S.~C. {Madden}, F.~{Galliano}, M.~{Galametz}, T.~T.
  {Takeuchi}, R.~S. {Asano}, S.~{Zhukovska}, V.~{Lebouteiller}, D.~{Cormier},
  A.~{Jones}, M.~{Bocchio}, M.~{Baes}, G.~J. {Bendo}, M.~{Boquien},
  A.~{Boselli}, I.~{DeLooze}, V.~{Doublier-Pritchard}, T.~{Hughes}, O.~{\L}.
  {Karczewski}, and L.~{Spinoglio}.
\newblock {Gas-to-dust mass ratios in local galaxies over a 2 dex metallicity
  range}.
\newblock \emph{\aap}, 563:\penalty0 A31, March 2014.
\newblock \doi{10.1051/0004-6361/201322803}.

\bibitem[{Roman-Duval} et~al.(2014){Roman-Duval}, {Gordon}, {Meixner}, {Bot},
  {Bolatto}, {Hughes}, {Wong}, {Babler}, {Bernard}, {Clayton}, {Fukui},
  {Galametz}, {Galliano}, {Glover}, {Hony}, {Israel}, {Jameson},
  {Lebouteiller}, {Lee}, {Li}, {Madden}, {Misselt}, {Montiel}, {Okumura},
  {Onishi}, {Panuzzo}, {Reach}, {Remy-Ruyer}, {Robitaille}, {Rubio}, {Sauvage},
  {Seale}, {Sewilo}, {Staveley-Smith}, and {Zhukovska}]{RomanDuval2014}
J.~{Roman-Duval}, K.~D. {Gordon}, M.~{Meixner}, C.~{Bot}, A.~{Bolatto},
  A.~{Hughes}, T.~{Wong}, B.~{Babler}, J.-P. {Bernard}, G.~C. {Clayton},
  Y.~{Fukui}, M.~{Galametz}, F.~{Galliano}, S.~{Glover}, S.~{Hony},
  F.~{Israel}, K.~{Jameson}, V.~{Lebouteiller}, M.-Y. {Lee}, A.~{Li},
  S.~{Madden}, K.~{Misselt}, E.~{Montiel}, K.~{Okumura}, T.~{Onishi},
  P.~{Panuzzo}, W.~{Reach}, A.~{Remy-Ruyer}, T.~{Robitaille}, M.~{Rubio},
  M.~{Sauvage}, J.~{Seale}, M.~{Sewilo}, L.~{Staveley-Smith}, and
  S.~{Zhukovska}.
\newblock {Dust and Gas in the Magellanic Clouds from the HERITAGE Herschel Key
  Project. II. Gas-to-dust Ratio Variations across Interstellar Medium Phases}.
\newblock \emph{\apj}, 797:\penalty0 86, December 2014.
\newblock \doi{10.1088/0004-637X/797/2/86}.

\bibitem[{Cortese} et~al.(2016){Cortese}, {Bekki}, {Boselli}, {Catinella},
  {Ciesla}, {Hughes}, {Baes}, {Bendo}, {Boquien}, {de Looze}, {Smith},
  {Spinoglio}, and {Viaene}]{Cortese2016}
L.~{Cortese}, K.~{Bekki}, A.~{Boselli}, B.~{Catinella}, L.~{Ciesla}, T.~M.
  {Hughes}, M.~{Baes}, G.~J. {Bendo}, M.~{Boquien}, I.~{de Looze}, M.~W.~L.
  {Smith}, L.~{Spinoglio}, and S.~{Viaene}.
\newblock {The selective effect of environment on the atomic and molecular
  gas-to-dust ratio of nearby galaxies in the Herschel Reference Survey}.
\newblock \emph{\mnras}, 459:\penalty0 3574--3584, July 2016.
\newblock \doi{10.1093/mnras/stw801}.

\bibitem[{Kahre} et~al.(2018){Kahre}, {Walterbos}, {Kim}, {Thilker},
  {Calzetti}, {Lee}, {Sabbi}, {Ubeda}, {Aloisi}, {Cignoni}, {Cook}, {Dale},
  {Elmegreen}, {Elmegreen}, {Fumagalli}, {Gallagher}, {Gouliermis}, {Grasha},
  {Grebel}, {Hunter}, {Sacchi}, {Smith}, {Tosi}, {Adamo}, {Andrews},
  {Ashworth}, {Bright}, {Brown}, {Chandar}, {Christian}, {de Mink}, {Dobbs},
  {Evans}, {Herrero}, {Johnson}, {Kennicutt}, {Krumholz}, {Messa}, {Nair},
  {Nota}, {Pellerin}, {Ryon}, {Schaerer}, {Shabani}, {Van Dyk}, {Whitmore}, and
  {Wofford}]{Kahre2018}
L.~{Kahre}, R.~A. {Walterbos}, H.~{Kim}, D.~{Thilker}, D.~{Calzetti}, J.~C.
  {Lee}, E.~{Sabbi}, L.~{Ubeda}, A.~{Aloisi}, M.~{Cignoni}, D.~O. {Cook}, D.~A.
  {Dale}, B.~G. {Elmegreen}, D.~M. {Elmegreen}, M.~{Fumagalli}, J.~S.
  {Gallagher}, III, D.~A. {Gouliermis}, K.~{Grasha}, E.~K. {Grebel}, D.~A.
  {Hunter}, E.~{Sacchi}, L.~J. {Smith}, M.~{Tosi}, A.~{Adamo}, J.~E. {Andrews},
  G.~{Ashworth}, S.~N. {Bright}, T.~M. {Brown}, R.~{Chandar}, C.~{Christian},
  S.~E. {de Mink}, C.~{Dobbs}, A.~S. {Evans}, A.~{Herrero}, K.~E. {Johnson},
  R.~C. {Kennicutt}, M.~R. {Krumholz}, M.~{Messa}, P.~{Nair}, A.~{Nota},
  A.~{Pellerin}, J.~E. {Ryon}, D.~{Schaerer}, F.~{Shabani}, S.~D. {Van Dyk},
  B.~C. {Whitmore}, and A.~{Wofford}.
\newblock {Extinction Maps and Dust-to-gas Ratios in Nearby Galaxies with
  LEGUS}.
\newblock \emph{\apj}, 855:\penalty0 133, March 2018.
\newblock \doi{10.3847/1538-4357/aab101}.

\bibitem[{Galliano} et~al.(2003){Galliano}, {Madden}, {Jones}, {Wilson},
  {Bernard}, and {Le Peintre}]{Galliano2003}
F.~{Galliano}, S.~C. {Madden}, A.~P. {Jones}, C.~D. {Wilson}, J.-P. {Bernard},
  and F.~{Le Peintre}.
\newblock {ISM properties in low-metallicity environments. II. The dust
  spectral energy distribution of NGC 1569}.
\newblock \emph{\aap}, 407:\penalty0 159--176, August 2003.
\newblock \doi{10.1051/0004-6361:20030814}.

\bibitem[{Galliano} et~al.(2005){Galliano}, {Madden}, {Jones}, {Wilson}, and
  {Bernard}]{Galliano2005}
F.~{Galliano}, S.~C. {Madden}, A.~P. {Jones}, C.~D. {Wilson}, and J.-P.
  {Bernard}.
\newblock {ISM properties in low-metallicity environments. III. The spectral
  energy distributions of II Zw 40, He 2-10 and NGC 1140}.
\newblock \emph{\aap}, 434:\penalty0 867--885, May 2005.
\newblock \doi{10.1051/0004-6361:20042369}.

\bibitem[{Galliano} et~al.(2011){Galliano}, {Hony}, {Bernard}, {Bot}, {Madden},
  {Roman-Duval}, {Galametz}, {Li}, {Meixner}, {Engelbracht}, {Lebouteiller},
  {Misselt}, {Montiel}, {Panuzzo}, {Reach}, and {Skibba}]{Galliano2011}
F.~{Galliano}, S.~{Hony}, J.-P. {Bernard}, C.~{Bot}, S.~C. {Madden},
  J.~{Roman-Duval}, M.~{Galametz}, A.~{Li}, M.~{Meixner}, C.~W. {Engelbracht},
  V.~{Lebouteiller}, K.~{Misselt}, E.~{Montiel}, P.~{Panuzzo}, W.~T. {Reach},
  and R.~{Skibba}.
\newblock {Non-standard grain properties, dark gas reservoir, and extended
  submillimeter excess, probed by Herschel in the Large Magellanic Cloud}.
\newblock \emph{\aap}, 536:\penalty0 A88, December 2011.
\newblock \doi{10.1051/0004-6361/201117952}.

\bibitem[{Solomon} and {Vanden Bout}(2005)]{Solomon2005}
P.~M. {Solomon} and P.~A. {Vanden Bout}.
\newblock {Molecular Gas at High Redshift}.
\newblock \emph{\araa}, 43:\penalty0 677--725, September 2005.
\newblock \doi{10.1146/annurev.astro.43.051804.102221}.

\bibitem[{Lisenfeld} et~al.(2011){Lisenfeld}, {Espada}, {Verdes-Montenegro},
  {Kuno}, {Leon}, {Sabater}, {Sato}, {Sulentic}, {Verley}, and
  {Yun}]{Lisenfeld2011}
U.~{Lisenfeld}, D.~{Espada}, L.~{Verdes-Montenegro}, N.~{Kuno}, S.~{Leon},
  J.~{Sabater}, N.~{Sato}, J.~{Sulentic}, S.~{Verley}, and M.~S. {Yun}.
\newblock {The AMIGA sample of isolated galaxies. IX. Molecular gas
  properties}.
\newblock \emph{\aap}, 534:\penalty0 A102, October 2011.
\newblock \doi{10.1051/0004-6361/201117056}.

\bibitem[{Stark} et~al.(2013){Stark}, {Kannappan}, {Wei}, {Baker}, {Leroy},
  {Eckert}, and {Vogel}]{Stark2013}
D.~V. {Stark}, S.~J. {Kannappan}, L.~H. {Wei}, A.~J. {Baker}, A.~K. {Leroy},
  K.~D. {Eckert}, and S.~N. {Vogel}.
\newblock {The Fueling Diagram: Linking Galaxy Molecular-to-atomic Gas Ratios
  to Interactions and Accretion}.
\newblock \emph{\apj}, 769:\penalty0 82, May 2013.
\newblock \doi{10.1088/0004-637X/769/1/82}.

\bibitem[{Young} et~al.(1995){Young}, {Xie}, {Tacconi}, {Knezek}, {Viscuso},
  {Tacconi-Garman}, {Scoville}, {Schneider}, {Schloerb}, {Lord}, {Lesser},
  {Kenney}, {Huang}, {Devereux}, {Claussen}, {Case}, {Carpenter}, {Berry}, and
  {Allen}]{Young1995}
J.~S. {Young}, S.~{Xie}, L.~{Tacconi}, P.~{Knezek}, P.~{Viscuso},
  L.~{Tacconi-Garman}, N.~{Scoville}, S.~{Schneider}, F.~P. {Schloerb},
  S.~{Lord}, A.~{Lesser}, J.~{Kenney}, Y.-L. {Huang}, N.~{Devereux},
  M.~{Claussen}, J.~{Case}, J.~{Carpenter}, M.~{Berry}, and L.~{Allen}.
\newblock {The FCRAO Extragalactic CO Survey. I. The Data}.
\newblock \emph{\apjs}, 98:\penalty0 219, May 1995.
\newblock \doi{10.1086/192159}.

\bibitem[{Davis} et~al.(2013){Davis}, {Alatalo}, {Bureau}, {Cappellari},
  {Scott}, {Young}, {Blitz}, {Crocker}, {Bayet}, {Bois}, {Bournaud}, {Davies},
  {de Zeeuw}, {Duc}, {Emsellem}, {Khochfar}, {Krajnovi{\'c}}, {Kuntschner},
  {Lablanche}, {McDermid}, {Morganti}, {Naab}, {Oosterloo}, {Sarzi}, {Serra},
  and {Weijmans}]{Davis2013a}
T.~A. {Davis}, K.~{Alatalo}, M.~{Bureau}, M.~{Cappellari}, N.~{Scott}, L.~M.
  {Young}, L.~{Blitz}, A.~{Crocker}, E.~{Bayet}, M.~{Bois}, F.~{Bournaud},
  R.~L. {Davies}, P.~T. {de Zeeuw}, P.-A. {Duc}, E.~{Emsellem}, S.~{Khochfar},
  D.~{Krajnovi{\'c}}, H.~{Kuntschner}, P.-Y. {Lablanche}, R.~M. {McDermid},
  R.~{Morganti}, T.~{Naab}, T.~{Oosterloo}, M.~{Sarzi}, P.~{Serra}, and A.-M.
  {Weijmans}.
\newblock {The ATLAS$^{3D}$ Project - XIV. The extent and kinematics of the
  molecular gas in early-type galaxies}.
\newblock \emph{\mnras}, 429:\penalty0 534--555, February 2013.
\newblock \doi{10.1093/mnras/sts353}.

\bibitem[{Hubble}(1926)]{Hubble1926}
E.~P. {Hubble}.
\newblock {Extragalactic nebulae.}
\newblock \emph{\apj}, 64, December 1926.
\newblock \doi{10.1086/143018}.

\bibitem[{Bolatto} et~al.(2013){Bolatto}, {Wolfire}, and {Leroy}]{Bolatto2013}
A.~D. {Bolatto}, M.~{Wolfire}, and A.~K. {Leroy}.
\newblock {The CO-to-H$_{2}$ Conversion Factor}.
\newblock \emph{\araa}, 51:\penalty0 207--268, August 2013.
\newblock \doi{10.1146/annurev-astro-082812-140944}.

\bibitem[{Magdis} et~al.(2011){Magdis}, {Daddi}, {Elbaz}, {Sargent},
  {Dickinson}, {Dannerbauer}, {Aussel}, {Walter}, {Hwang}, {Charmandaris},
  {Hodge}, {Riechers}, {Rigopoulou}, {Carilli}, {Pannella}, {Mullaney},
  {Leiton}, and {Scott}]{Magdis2011}
G.~E. {Magdis}, E.~{Daddi}, D.~{Elbaz}, M.~{Sargent}, M.~{Dickinson},
  H.~{Dannerbauer}, H.~{Aussel}, F.~{Walter}, H.~S. {Hwang}, V.~{Charmandaris},
  J.~{Hodge}, D.~{Riechers}, D.~{Rigopoulou}, C.~{Carilli}, M.~{Pannella},
  J.~{Mullaney}, R.~{Leiton}, and D.~{Scott}.
\newblock {GOODS-Herschel: Gas-to-dust Mass Ratios and CO-to-H$_{2}$ Conversion
  Factors in Normal and Starbursting Galaxies at High-z}.
\newblock \emph{\apjl}, 740:\penalty0 L15, October 2011.
\newblock \doi{10.1088/2041-8205/740/1/L15}.

\bibitem[{Genzel} et~al.(2012){Genzel}, {Tacconi}, {Combes}, {Bolatto}, {Neri},
  {Sternberg}, {Cooper}, {Bouch{\'e}}, {Bournaud}, {Burkert}, {Comerford},
  {Cox}, {Davis}, {F{\"o}rster Schreiber}, {Garcia-Burillo}, {Gracia-Carpio},
  {Lutz}, {Naab}, {Newman}, {Saintonge}, {Shapiro}, {Shapley}, and
  {Weiner}]{Genzel2012}
R.~{Genzel}, L.~J. {Tacconi}, F.~{Combes}, A.~{Bolatto}, R.~{Neri},
  A.~{Sternberg}, M.~C. {Cooper}, N.~{Bouch{\'e}}, F.~{Bournaud}, A.~{Burkert},
  J.~{Comerford}, P.~{Cox}, M.~{Davis}, N.~M. {F{\"o}rster Schreiber},
  S.~{Garcia-Burillo}, J.~{Gracia-Carpio}, D.~{Lutz}, T.~{Naab}, S.~{Newman},
  A.~{Saintonge}, K.~{Shapiro}, A.~{Shapley}, and B.~{Weiner}.
\newblock {The Metallicity Dependence of the CO-H$_{2}$ Conversion Factor in z
  $>$= 1 Star-forming Galaxies}.
\newblock \emph{\apj}, 746:\penalty0 69, February 2012.
\newblock \doi{10.1088/0004-637X/746/1/69}.

\bibitem[{Narayanan} et~al.(2012){Narayanan}, {Krumholz}, {Ostriker}, and
  {Hernquist}]{Narayanan2012}
D.~{Narayanan}, M.~R. {Krumholz}, E.~C. {Ostriker}, and L.~{Hernquist}.
\newblock {A general model for the CO-H$_{2}$ conversion factor in galaxies
  with applications to the star formation law}.
\newblock \emph{\mnras}, 421:\penalty0 3127--3146, April 2012.
\newblock \doi{10.1111/j.1365-2966.2012.20536.x}.

\bibitem[{Schruba} et~al.(2012){Schruba}, {Leroy}, {Walter}, {Bigiel},
  {Brinks}, {de Blok}, {Kramer}, {Rosolowsky}, {Sandstrom}, {Schuster},
  {Usero}, {Weiss}, and {Wiesemeyer}]{Schruba2012}
A.~{Schruba}, A.~K. {Leroy}, F.~{Walter}, F.~{Bigiel}, E.~{Brinks}, W.~J.~G.
  {de Blok}, C.~{Kramer}, E.~{Rosolowsky}, K.~{Sandstrom}, K.~{Schuster},
  A.~{Usero}, A.~{Weiss}, and H.~{Wiesemeyer}.
\newblock {Low CO Luminosities in Dwarf Galaxies}.
\newblock \emph{\aj}, 143:\penalty0 138, June 2012.
\newblock \doi{10.1088/0004-6256/143/6/138}.

\bibitem[{Hunt} et~al.(2015){Hunt}, {Garc{\'{\i}}a-Burillo}, {Casasola},
  {Caselli}, {Combes}, {Henkel}, {Lundgren}, {Maiolino}, {Menten}, {Testi}, and
  {Weiss}]{Hunt2015}
L.~K. {Hunt}, S.~{Garc{\'{\i}}a-Burillo}, V.~{Casasola}, P.~{Caselli},
  F.~{Combes}, C.~{Henkel}, A.~{Lundgren}, R.~{Maiolino}, K.~M. {Menten},
  L.~{Testi}, and A.~{Weiss}.
\newblock {Molecular depletion times and the CO-to-H$_{2}$ conversion factor in
  metal-poor galaxies}.
\newblock \emph{\aap}, 583:\penalty0 A114, November 2015.
\newblock \doi{10.1051/0004-6361/201526553}.

\bibitem[{Amor{\'{\i}}n} et~al.(2016){Amor{\'{\i}}n},
  {Mu{\~n}oz-Tu{\~n}{\'o}n}, {Aguerri}, and {Planesas}]{Amorin2016}
R.~{Amor{\'{\i}}n}, C.~{Mu{\~n}oz-Tu{\~n}{\'o}n}, J.~A.~L. {Aguerri}, and
  P.~{Planesas}.
\newblock {Molecular gas in low-metallicity starburst galaxies:. Scaling
  relations and the CO-to-H$_{2}$ conversion factor}.
\newblock \emph{\aap}, 588:\penalty0 A23, April 2016.
\newblock \doi{10.1051/0004-6361/201526397}.

\bibitem[{Accurso} et~al.(2017){Accurso}, {Saintonge}, {Catinella}, {Cortese},
  {Dav{\'e}}, {Dunsheath}, {Genzel}, {Gracia-Carpio}, {Heckman}, {Jimmy},
  {Kramer}, {Li}, {Lutz}, {Schiminovich}, {Schuster}, {Sternberg}, {Sturm},
  {Tacconi}, {Tran}, and {Wang}]{Accurso2017}
G.~{Accurso}, A.~{Saintonge}, B.~{Catinella}, L.~{Cortese}, R.~{Dav{\'e}},
  S.~H. {Dunsheath}, R.~{Genzel}, J.~{Gracia-Carpio}, T.~M. {Heckman}, {Jimmy},
  C.~{Kramer}, C.~{Li}, K.~{Lutz}, D.~{Schiminovich}, K.~{Schuster},
  A.~{Sternberg}, E.~{Sturm}, L.~J. {Tacconi}, K.~V. {Tran}, and J.~{Wang}.
\newblock {Deriving a multivariate {$\alpha$}$_{CO}$ conversion function using
  the [C II]/CO (1-0) ratio and its application to molecular gas scaling
  relations}.
\newblock \emph{\mnras}, 470:\penalty0 4750--4766, October 2017.
\newblock \doi{10.1093/mnras/stx1556}.

\bibitem[Schruba et~al.(2017)Schruba, Leroy, Kruijssen, Bigiel, Bolatto,
  de~Blok, Tacconi, van Dishoeck, and Walter]{Schruba2017}
Andreas Schruba, Adam~K. Leroy, J.~M.~Diederik Kruijssen, Frank Bigiel,
  Alberto~D. Bolatto, W.~J.~G. de~Blok, Linda Tacconi, Ewine~F. van Dishoeck,
  and Fabian Walter.
\newblock Physical properties of molecular clouds at 2 pc resolution in the
  low-metallicity dwarf galaxy {NGC} 6822 and the milky way.
\newblock \emph{The Astrophysical Journal}, 835\penalty0 (2):\penalty0 278, feb
  2017.
\newblock \doi{10.3847/1538-4357/835/2/278}.
\newblock URL \url{https://doi.org/10.3847%2F1538-4357%2F835%2F2%2F278}.

\bibitem[{Kennicutt} and {Evans}(2012)]{Kennicutt2012}
R.~C. {Kennicutt} and N.~J. {Evans}.
\newblock {Star Formation in the Milky Way and Nearby Galaxies}.
\newblock \emph{\araa}, 50:\penalty0 531--608, September 2012.
\newblock \doi{10.1146/annurev-astro-081811-125610}.

\bibitem[{Hao} et~al.(2011){Hao}, {Kennicutt}, {Johnson}, {Calzetti}, {Dale},
  and {Moustakas}]{Hao2011}
C.-N. {Hao}, R.~C. {Kennicutt}, B.~D. {Johnson}, D.~{Calzetti}, D.~A. {Dale},
  and J.~{Moustakas}.
\newblock {Dust-corrected Star Formation Rates of Galaxies. II. Combinations of
  Ultraviolet and Infrared Tracers}.
\newblock \emph{\apj}, 741:\penalty0 124, November 2011.
\newblock \doi{10.1088/0004-637X/741/2/124}.

\bibitem[{Murphy} et~al.(2011){Murphy}, {Condon}, {Schinnerer}, {Kennicutt},
  {Calzetti}, {Armus}, {Helou}, {Turner}, {Aniano}, {Beir{\~a}o}, {Bolatto},
  {Brandl}, {Croxall}, {Dale}, {Donovan Meyer}, {Draine}, {Engelbracht},
  {Hunt}, {Hao}, {Koda}, {Roussel}, {Skibba}, and {Smith}]{Murphy2011}
E.~J. {Murphy}, J.~J. {Condon}, E.~{Schinnerer}, R.~C. {Kennicutt},
  D.~{Calzetti}, L.~{Armus}, G.~{Helou}, J.~L. {Turner}, G.~{Aniano},
  P.~{Beir{\~a}o}, A.~D. {Bolatto}, B.~R. {Brandl}, K.~V. {Croxall}, D.~A.
  {Dale}, J.~L. {Donovan Meyer}, B.~T. {Draine}, C.~{Engelbracht}, L.~K.
  {Hunt}, C.-N. {Hao}, J.~{Koda}, H.~{Roussel}, R.~{Skibba}, and J.-D.~T.
  {Smith}.
\newblock {Calibrating Extinction-free Star Formation Rate Diagnostics with 33
  GHz Free-free Emission in NGC 6946}.
\newblock \emph{\apj}, 737:\penalty0 67, August 2011.
\newblock \doi{10.1088/0004-637X/737/2/67}.

\bibitem[{Kroupa} and {Weidner}(2003)]{Kroupa2003}
P.~{Kroupa} and C.~{Weidner}.
\newblock {Galactic-Field Initial Mass Functions of Massive Stars}.
\newblock \emph{\apj}, 598:\penalty0 1076--1078, December 2003.
\newblock \doi{10.1086/379105}.

\bibitem[{Salpeter}(1955)]{Salpeter1955}
E.~E. {Salpeter}.
\newblock {The Luminosity Function and Stellar Evolution.}
\newblock \emph{\apj}, 121:\penalty0 161, January 1955.
\newblock \doi{10.1086/145971}.

\bibitem[{Chabrier}(2003)]{Chabrier2003}
G.~{Chabrier}.
\newblock {Galactic Stellar and Substellar Initial Mass Function}.
\newblock \emph{\pasp}, 115:\penalty0 763--795, July 2003.
\newblock \doi{10.1086/376392}.

\bibitem[{Rieke} et~al.(2009){Rieke}, {Alonso-Herrero}, {Weiner},
  {P{\'e}rez-Gonz{\'a}lez}, {Blaylock}, {Donley}, and {Marcillac}]{Rieke2009}
G.~H. {Rieke}, A.~{Alonso-Herrero}, B.~J. {Weiner}, P.~G.
  {P{\'e}rez-Gonz{\'a}lez}, M.~{Blaylock}, J.~L. {Donley}, and D.~{Marcillac}.
\newblock {Determining Star Formation Rates for Infrared Galaxies}.
\newblock \emph{\apj}, 692:\penalty0 556--573, February 2009.
\newblock \doi{10.1088/0004-637X/692/1/556}.

\bibitem[{De Looze} et~al.(2014){De Looze}, {Cormier}, {Lebouteiller},
  {Madden}, {Baes}, {Bendo}, {Boquien}, {Boselli}, {Clements}, {Cortese},
  {Cooray}, {Galametz}, {Galliano}, {Graci{\'a}-Carpio}, {Isaak}, {Karczewski},
  {Parkin}, {Pellegrini}, {R{\'e}my-Ruyer}, {Spinoglio}, {Smith}, and
  {Sturm}]{deLooze2014}
Ilse {De Looze}, Diane {Cormier}, Vianney {Lebouteiller}, Suzanne {Madden},
  Maarten {Baes}, George~J. {Bendo}, M{\'e}d{\'e}ric {Boquien}, Alessandro
  {Boselli}, David~L. {Clements}, Luca {Cortese}, Asantha {Cooray}, Maud
  {Galametz}, Fr{\'e}d{\'e}ric {Galliano}, Javier {Graci{\'a}-Carpio}, Kate
  {Isaak}, Oskar~{\L}. {Karczewski}, Tara~J. {Parkin}, Eric~W. {Pellegrini},
  Aur{\'e}lie {R{\'e}my-Ruyer}, Luigi {Spinoglio}, Matthew W.~L. {Smith}, and
  Eckhard {Sturm}.
\newblock {The applicability of far-infrared fine-structure lines as star
  formation rate tracers over wide ranges of metallicities and galaxy types}.
\newblock \emph{\aap}, 568:\penalty0 A62, Aug 2014.
\newblock \doi{10.1051/0004-6361/201322489}.

\bibitem[{Appleton} et~al.(2013){Appleton}, {Guillard}, {Boulanger}, {Cluver},
  {Ogle}, {Falgarone}, {Pineau des For{\^e}ts}, {O'Sullivan}, {Duc},
  {Gallagher}, {Gao}, {Jarrett}, {Konstantopoulos}, {Lisenfeld}, {Lord}, {Lu},
  {Peterson}, {Struck}, {Sturm}, {Tuffs}, {Valchanov}, {van der Werf}, and
  {Xu}]{Appleton2013}
P.~N. {Appleton}, P.~{Guillard}, F.~{Boulanger}, M.~E. {Cluver}, P.~{Ogle},
  E.~{Falgarone}, G.~{Pineau des For{\^e}ts}, E.~{O'Sullivan}, P.~A. {Duc},
  S.~{Gallagher}, Y.~{Gao}, T.~{Jarrett}, I.~{Konstantopoulos}, U.~{Lisenfeld},
  S.~{Lord}, N.~{Lu}, B.~W. {Peterson}, C.~{Struck}, E.~{Sturm}, R.~{Tuffs},
  I.~{Valchanov}, P.~{van der Werf}, and K.~C. {Xu}.
\newblock {Shock-enhanced C$^{+}$ Emission and the Detection of H$_{2}$O from
  the Stephan's Quintet Group-wide Shock Using Herschel}.
\newblock \emph{\apj}, 777\penalty0 (1):\penalty0 66, Nov 2013.
\newblock \doi{10.1088/0004-637X/777/1/66}.

\bibitem[{Appleton} et~al.(2018){Appleton}, {Diaz-Santos}, {Fadda}, {Ogle},
  {Togi}, {Lanz}, {Alatalo}, {Fischer}, {Rich}, and {Guillard}]{Appleton2018}
P.~N. {Appleton}, T.~{Diaz-Santos}, D.~{Fadda}, P.~{Ogle}, A.~{Togi},
  L.~{Lanz}, K.~{Alatalo}, C.~{Fischer}, J.~{Rich}, and P.~{Guillard}.
\newblock {Jet-related Excitation of the [C II] Emission in the Active Galaxy
  NGC 4258 with SOFIA}.
\newblock \emph{\apj}, 869\penalty0 (1):\penalty0 61, Dec 2018.
\newblock \doi{10.3847/1538-4357/aaed2a}.

\bibitem[{Saintonge} et~al.(2011{\natexlab{a}}){Saintonge}, {Kauffmann},
  {Kramer}, {Tacconi}, {Buchbender}, {Catinella}, {Fabello},
  {Graci{\'a}-Carpio}, {Wang}, {Cortese}, {Fu}, {Genzel}, {Giovanelli}, {Guo},
  {Haynes}, {Heckman}, {Krumholz}, {Lemonias}, {Li}, {Moran},
  {Rodriguez-Fernandez}, {Schiminovich}, {Schuster}, and
  {Sievers}]{Saintonge2011a}
A.~{Saintonge}, G.~{Kauffmann}, C.~{Kramer}, L.~J. {Tacconi}, C.~{Buchbender},
  B.~{Catinella}, S.~{Fabello}, J.~{Graci{\'a}-Carpio}, J.~{Wang},
  L.~{Cortese}, J.~{Fu}, R.~{Genzel}, R.~{Giovanelli}, Q.~{Guo}, M.~P.
  {Haynes}, T.~M. {Heckman}, M.~R. {Krumholz}, J.~{Lemonias}, C.~{Li},
  S.~{Moran}, N.~{Rodriguez-Fernandez}, D.~{Schiminovich}, K.~{Schuster}, and
  A.~{Sievers}.
\newblock {COLD GASS, an IRAM legacy survey of molecular gas in massive
  galaxies - I. Relations between H$_{2}$, H I, stellar content and structural
  properties}.
\newblock \emph{\mnras}, 415:\penalty0 32--60, July 2011{\natexlab{a}}.
\newblock \doi{10.1111/j.1365-2966.2011.18677.x}.

\bibitem[{Saintonge} et~al.(2011{\natexlab{b}}){Saintonge}, {Kauffmann},
  {Wang}, {Kramer}, {Tacconi}, {Buchbender}, {Catinella}, {Graci{\'a}-Carpio},
  {Cortese}, {Fabello}, {Fu}, {Genzel}, {Giovanelli}, {Guo}, {Haynes},
  {Heckman}, {Krumholz}, {Lemonias}, {Li}, {Moran}, {Rodriguez-Fernandez},
  {Schiminovich}, {Schuster}, and {Sievers}]{Saintonge2011b}
A.~{Saintonge}, G.~{Kauffmann}, J.~{Wang}, C.~{Kramer}, L.~J. {Tacconi},
  C.~{Buchbender}, B.~{Catinella}, J.~{Graci{\'a}-Carpio}, L.~{Cortese},
  S.~{Fabello}, J.~{Fu}, R.~{Genzel}, R.~{Giovanelli}, Q.~{Guo}, M.~P.
  {Haynes}, T.~M. {Heckman}, M.~R. {Krumholz}, J.~{Lemonias}, C.~{Li},
  S.~{Moran}, N.~{Rodriguez-Fernandez}, D.~{Schiminovich}, K.~{Schuster}, and
  A.~{Sievers}.
\newblock {COLD GASS, an IRAM legacy survey of molecular gas in massive
  galaxies - II. The non-universality of the molecular gas depletion
  time-scale}.
\newblock \emph{\mnras}, 415:\penalty0 61--76, July 2011{\natexlab{b}}.
\newblock \doi{10.1111/j.1365-2966.2011.18823.x}.

\bibitem[{Saintonge} et~al.(2012){Saintonge}, {Tacconi}, {Fabello}, {Wang},
  {Catinella}, {Genzel}, {Graci{\'a}-Carpio}, {Kramer}, {Moran}, {Heckman},
  {Schiminovich}, {Schuster}, and {Wuyts}]{Saintonge2012}
A.~{Saintonge}, L.~J. {Tacconi}, S.~{Fabello}, J.~{Wang}, B.~{Catinella},
  R.~{Genzel}, J.~{Graci{\'a}-Carpio}, C.~{Kramer}, S.~{Moran}, T.~M.
  {Heckman}, D.~{Schiminovich}, K.~{Schuster}, and S.~{Wuyts}.
\newblock {The Impact of Interactions, Bars, Bulges, and Active Galactic Nuclei
  on Star Formation Efficiency in Local Massive Galaxies}.
\newblock \emph{\apj}, 758:\penalty0 73, October 2012.
\newblock \doi{10.1088/0004-637X/758/2/73}.

\bibitem[{Catinella} et~al.(2012){Catinella}, {Schiminovich}, {Kauffmann},
  {Fabello}, {Hummels}, {Lemonias}, {Moran}, {Wu}, {Cooper}, and
  {Wang}]{Catinella2012}
B.~{Catinella}, D.~{Schiminovich}, G.~{Kauffmann}, S.~{Fabello}, C.~{Hummels},
  J.~{Lemonias}, S.~M. {Moran}, R.~{Wu}, A.~{Cooper}, and J.~{Wang}.
\newblock {The GALEX Arecibo SDSS Survey. VI. Second data release and updated
  gas fraction scaling relations}.
\newblock \emph{\aap}, 544:\penalty0 A65, August 2012.
\newblock \doi{10.1051/0004-6361/201219261}.

\bibitem[{Krumholz} et~al.(2012){Krumholz}, {Dekel}, and {McKee}]{Krumholz2012}
M.~R. {Krumholz}, A.~{Dekel}, and C.~F. {McKee}.
\newblock {A Universal, Local Star Formation Law in Galactic Clouds, nearby
  Galaxies, High-redshift Disks, and Starbursts}.
\newblock \emph{\apj}, 745:\penalty0 69, January 2012.
\newblock \doi{10.1088/0004-637X/745/1/69}.

\bibitem[{Herrera-Camus} et~al.(2018{\natexlab{a}}){Herrera-Camus}, {Sturm},
  {Graci{\'a}-Carpio}, {Lutz}, {Contursi}, {Veilleux}, {Fischer},
  {Gonz{\'a}lez-Alfonso}, {Poglitsch}, {Tacconi}, {Genzel}, {Maiolino},
  {Sternberg}, {Davies}, and {Verma}]{HerreraCamus2018a}
R.~{Herrera-Camus}, E.~{Sturm}, J.~{Graci{\'a}-Carpio}, D.~{Lutz},
  A.~{Contursi}, S.~{Veilleux}, J.~{Fischer}, E.~{Gonz{\'a}lez-Alfonso},
  A.~{Poglitsch}, L.~{Tacconi}, R.~{Genzel}, R.~{Maiolino}, A.~{Sternberg},
  R.~{Davies}, and A.~{Verma}.
\newblock {SHINING, A Survey of Far-infrared Lines in Nearby Galaxies. I.
  Survey Description, Observational Trends, and Line Diagnostics}.
\newblock \emph{\apj}, 861:\penalty0 94, July 2018{\natexlab{a}}.
\newblock \doi{10.3847/1538-4357/aac0f6}.

\bibitem[{Herrera-Camus} et~al.(2018{\natexlab{b}}){Herrera-Camus}, {Sturm},
  {Graci{\'a}-Carpio}, {Lutz}, {Contursi}, {Veilleux}, {Fischer},
  {Gonz{\'a}lez-Alfonso}, {Poglitsch}, {Tacconi}, {Genzel}, {Maiolino},
  {Sternberg}, {Davies}, and {Verma}]{HerreraCamus2018b}
R.~{Herrera-Camus}, E.~{Sturm}, J.~{Graci{\'a}-Carpio}, D.~{Lutz},
  A.~{Contursi}, S.~{Veilleux}, J.~{Fischer}, E.~{Gonz{\'a}lez-Alfonso},
  A.~{Poglitsch}, L.~{Tacconi}, R.~{Genzel}, R.~{Maiolino}, A.~{Sternberg},
  R.~{Davies}, and A.~{Verma}.
\newblock {SHINING, A Survey of Far-infrared Lines in Nearby Galaxies. II.
  Line-deficit Models, AGN Impact, [C II]-SFR Scaling Relations, and
  Mass-Metallicity Relation in (U)LIRGs}.
\newblock \emph{\apj}, 861:\penalty0 95, July 2018{\natexlab{b}}.
\newblock \doi{10.3847/1538-4357/aac0f9}.

\bibitem[{Zhao} et~al.(2016){Zhao}, {Yan}, and {Tsai}]{Zhao2016}
Y.~{Zhao}, L.~{Yan}, and C.-W. {Tsai}.
\newblock {Properties of Interstellar Medium In Infrared-bright QSOs Probed by
  [O I] 63 um and [C II] 158 um Emission Lines}.
\newblock \emph{\apj}, 824:\penalty0 146, June 2016.
\newblock \doi{10.3847/0004-637X/824/2/146}.

\bibitem[{Kaufman} et~al.(1999){Kaufman}, {Wolfire}, {Hollenbach}, and
  {Luhman}]{Kaufman1999}
M.~J. {Kaufman}, M.~G. {Wolfire}, D.~J. {Hollenbach}, and M.~L. {Luhman}.
\newblock {Far-Infrared and Submillimeter Emission from Galactic and
  Extragalactic Photodissociation Regions}.
\newblock \emph{\apj}, 527:\penalty0 795--813, December 1999.
\newblock \doi{10.1086/308102}.

\bibitem[{Kaufman} et~al.(2006){Kaufman}, {Wolfire}, and
  {Hollenbach}]{Kaufman2006}
M.~J. {Kaufman}, M.~G. {Wolfire}, and D.~J. {Hollenbach}.
\newblock {[Si II], [Fe II], [C II], and H$_{2}$ Emission from Massive
  Star-forming Regions}.
\newblock \emph{\apj}, 644:\penalty0 283--299, June 2006.
\newblock \doi{10.1086/503596}.

\bibitem[{Pound} and {Wolfire}(2008)]{PDRToolbox}
M.~W. {Pound} and M.~G. {Wolfire}.
\newblock {The Photo Dissociation Region Toolbox}.
\newblock In R.~W. {Argyle}, P.~S. {Bunclark}, and J.~R. {Lewis}, editors,
  \emph{Astronomical Data Analysis Software and Systems XVII}, volume 394 of
  \emph{Astronomical Society of the Pacific Conference Series}, page 654,
  August 2008.

\bibitem[{Wall}(1996)]{Wall1996}
J.~V. {Wall}.
\newblock {Practical Statistics for Astronomers - II. Correlation,
  Data-modelling and Sample Comparison}.
\newblock \emph{\qjras}, 37:\penalty0 519, December 1996.

\bibitem[{Helou} et~al.(1988){Helou}, {Khan}, {Malek}, and
  {Boehmer}]{Helou1988}
G.~{Helou}, I.~R. {Khan}, L.~{Malek}, and L.~{Boehmer}.
\newblock {IRAS observations of galaxies in the Virgo cluster area}.
\newblock \emph{\apjs}, 68:\penalty0 151--172, October 1988.
\newblock \doi{10.1086/191285}.

\bibitem[{Graci{\'a}-Carpio} et~al.(2011){Graci{\'a}-Carpio}, {Sturm},
  {Hailey-Dunsheath}, {Fischer}, {Contursi}, {Poglitsch}, {Genzel},
  {Gonz{\'a}lez-Alfonso}, {Sternberg}, {Verma}, {Christopher}, {Davies},
  {Feuchtgruber}, {de Jong}, {Lutz}, and {Tacconi}]{GraciaCarpio2011}
J.~{Graci{\'a}-Carpio}, E.~{Sturm}, S.~{Hailey-Dunsheath}, J.~{Fischer},
  A.~{Contursi}, A.~{Poglitsch}, R.~{Genzel}, E.~{Gonz{\'a}lez-Alfonso},
  A.~{Sternberg}, A.~{Verma}, N.~{Christopher}, R.~{Davies}, H.~{Feuchtgruber},
  J.~A. {de Jong}, D.~{Lutz}, and L.~J. {Tacconi}.
\newblock {Far-infrared Line Deficits in Galaxies with Extreme
  L\_$\{$FIR$\}$/M\_$\{$H\_$\{$2$\}$$\}$ Ratios}.
\newblock \emph{\apjl}, 728:\penalty0 L7, February 2011.
\newblock \doi{10.1088/2041-8205/728/1/L7}.

\bibitem[{Contursi} et~al.(2017){Contursi}, {Baker}, {Berta}, {Magnelli},
  {Lutz}, {Fischer}, {Verma}, {Nielbock}, {Gr{\'a}cia Carpio}, {Veilleux},
  {Sturm}, {Davies}, {Genzel}, {Hailey-Dunsheath}, {Herrera-Camus}, {Janssen},
  {Poglitsch}, {Sternberg}, and {Tacconi}]{Contursi2017}
A.~{Contursi}, A.~J. {Baker}, S.~{Berta}, B.~{Magnelli}, D.~{Lutz},
  J.~{Fischer}, A.~{Verma}, M.~{Nielbock}, J.~{Gr{\'a}cia Carpio},
  S.~{Veilleux}, E.~{Sturm}, R.~{Davies}, R.~{Genzel}, S.~{Hailey-Dunsheath},
  R.~{Herrera-Camus}, A.~{Janssen}, A.~{Poglitsch}, A.~{Sternberg}, and L.~J.
  {Tacconi}.
\newblock {Interstellar medium conditions in z 0.2 Lyman-break analogs}.
\newblock \emph{\aap}, 606:\penalty0 A86, October 2017.
\newblock \doi{10.1051/0004-6361/201730609}.

\bibitem[{Cormier} et~al.(2014){Cormier}, {Madden}, {Lebouteiller}, {Hony},
  {Aalto}, {Costagliola}, {Hughes}, {R{\'e}my-Ruyer}, {Abel}, {Bayet},
  {Bigiel}, {Cannon}, {Cumming}, {Galametz}, {Galliano}, {Viti}, and
  {Wu}]{Cormier2014}
D.~{Cormier}, S.~C. {Madden}, V.~{Lebouteiller}, S.~{Hony}, S.~{Aalto},
  F.~{Costagliola}, A.~{Hughes}, A.~{R{\'e}my-Ruyer}, N.~{Abel}, E.~{Bayet},
  F.~{Bigiel}, J.~M. {Cannon}, R.~J. {Cumming}, M.~{Galametz}, F.~{Galliano},
  S.~{Viti}, and R.~{Wu}.
\newblock {The molecular gas reservoir of 6 low-metallicity galaxies from the
  Herschel Dwarf Galaxy Survey. A ground-based follow-up survey of CO(1-0),
  CO(2-1), and CO(3-2)}.
\newblock \emph{\aap}, 564:\penalty0 A121, April 2014.
\newblock \doi{10.1051/0004-6361/201322096}.

\bibitem[{Cormier} et~al.(2015){Cormier}, {Madden}, {Lebouteiller}, {Abel},
  {Hony}, {Galliano}, {R{\'e}my-Ruyer}, {Bigiel}, {Baes}, {Boselli},
  {Chevance}, {Cooray}, {De Looze}, {Doublier}, {Galametz}, {Hughes},
  {Karczewski}, {Lee}, {Lu}, and {Spinoglio}]{Cormier2015}
D.~{Cormier}, S.~C. {Madden}, V.~{Lebouteiller}, N.~{Abel}, S.~{Hony},
  F.~{Galliano}, A.~{R{\'e}my-Ruyer}, F.~{Bigiel}, M.~{Baes}, A.~{Boselli},
  M.~{Chevance}, A.~{Cooray}, I.~{De Looze}, V.~{Doublier}, M.~{Galametz},
  T.~{Hughes}, O.~{\L}. {Karczewski}, M.-Y. {Lee}, N.~{Lu}, and L.~{Spinoglio}.
\newblock {The Herschel Dwarf Galaxy Survey. I. Properties of the
  low-metallicity ISM from PACS spectroscopy}.
\newblock \emph{\aap}, 578:\penalty0 A53, June 2015.
\newblock \doi{10.1051/0004-6361/201425207}.

\bibitem[{Magdis} et~al.(2014){Magdis}, {Rigopoulou}, {Hopwood}, {Huang},
  {Farrah}, {Pearson}, {Alonso-Herrero}, {Bock}, {Clements}, {Cooray},
  {Griffin}, {Oliver}, {Perez Fournon}, {Riechers}, {Swinyard}, {Scott},
  {Thatte}, {Valtchanov}, and {Vaccari}]{Magdis2014}
G.~E. {Magdis}, D.~{Rigopoulou}, R.~{Hopwood}, J.-S. {Huang}, D.~{Farrah},
  C.~{Pearson}, A.~{Alonso-Herrero}, J.~J. {Bock}, D.~{Clements}, A.~{Cooray},
  M.~J. {Griffin}, S.~{Oliver}, I.~{Perez Fournon}, D.~{Riechers}, B.~M.
  {Swinyard}, D.~{Scott}, N.~{Thatte}, I.~{Valtchanov}, and M.~{Vaccari}.
\newblock {A Far-infrared Spectroscopic Survey of Intermediate Redshift (Ultra)
  Luminous Infrared Galaxies}.
\newblock \emph{\apj}, 796:\penalty0 63, November 2014.
\newblock \doi{10.1088/0004-637X/796/1/63}.

\bibitem[{Carilli} and {Walter}(2013)]{Carilli2013}
C.~L. {Carilli} and F.~{Walter}.
\newblock {Cool Gas in High-Redshift Galaxies}.
\newblock \emph{\araa}, 51:\penalty0 105--161, August 2013.
\newblock \doi{10.1146/annurev-astro-082812-140953}.

\bibitem[{Zanella} et~al.(2018){Zanella}, {Daddi}, {Magdis}, {Diaz Santos},
  {Cormier}, {Liu}, {Cibinel}, {Gobat}, {Dickinson}, {Sargent}, {Popping},
  {Madden}, {Bethermin}, {Hughes}, {Valentino}, {Rujopakarn}, {Pannella},
  {Bournaud}, {Walter}, {Wang}, {Elbaz}, and {Coogan}]{Zanella2018}
A.~{Zanella}, E.~{Daddi}, G.~{Magdis}, T.~{Diaz Santos}, D.~{Cormier},
  D.~{Liu}, A.~{Cibinel}, R.~{Gobat}, M.~{Dickinson}, M.~{Sargent},
  G.~{Popping}, S.~C. {Madden}, M.~{Bethermin}, T.~M. {Hughes}, F.~{Valentino},
  W.~{Rujopakarn}, M.~{Pannella}, F.~{Bournaud}, F.~{Walter}, T.~{Wang},
  D.~{Elbaz}, and R.~T. {Coogan}.
\newblock {The [C II] emission as a molecular gas mass tracer in galaxies at
  low and high redshifts}.
\newblock \emph{\mnras}, 481:\penalty0 1976--1999, December 2018.
\newblock \doi{10.1093/mnras/sty2394}.

\bibitem[{Umehata} et~al.(2017){Umehata}, {Matsuda}, {Tamura}, {Kohno},
  {Smail}, {Ivison}, {Steidel}, {Chapman}, {Geach}, {Hayes}, {Nagao}, {Ao},
  {Kawabe}, {Yun}, {Hatsukade}, {Kubo}, {Kato}, {Saito}, {Ikarashi},
  {Nakanishi}, {Lee}, {Izumi}, {Mori}, and {Ouchi}]{Umehata2017}
H.~{Umehata}, Y.~{Matsuda}, Y.~{Tamura}, K.~{Kohno}, I.~{Smail}, R.~J.
  {Ivison}, C.~C. {Steidel}, S.~C. {Chapman}, J.~E. {Geach}, M.~{Hayes},
  T.~{Nagao}, Y.~{Ao}, R.~{Kawabe}, M.~S. {Yun}, B.~{Hatsukade}, M.~{Kubo},
  Y.~{Kato}, T.~{Saito}, S.~{Ikarashi}, K.~{Nakanishi}, M.~{Lee}, T.~{Izumi},
  M.~{Mori}, and M.~{Ouchi}.
\newblock {ALMA Reveals Strong [C II] Emission in a Galaxy Embedded in a Giant
  Ly{$\alpha$} Blob at z = 3.1}.
\newblock \emph{\apjl}, 834:\penalty0 L16, January 2017.
\newblock \doi{10.3847/2041-8213/834/2/L16}.

\bibitem[{Bothwell} et~al.(2017){Bothwell}, {Aguirre}, {Aravena}, {Bethermin},
  {Bisbas}, {Chapman}, {De Breuck}, {Gonzalez}, {Greve}, {Hezaveh}, {Ma},
  {Malkan}, {Marrone}, {Murphy}, {Spilker}, {Strandet}, {Vieira}, and
  {Wei{\ss}}]{Bothwell2017}
M.~S. {Bothwell}, J.~E. {Aguirre}, M.~{Aravena}, M.~{Bethermin}, T.~G.
  {Bisbas}, S.~C. {Chapman}, C.~{De Breuck}, A.~H. {Gonzalez}, T.~R. {Greve},
  Y.~{Hezaveh}, J.~{Ma}, M.~{Malkan}, D.~P. {Marrone}, E.~J. {Murphy}, J.~S.
  {Spilker}, M.~{Strandet}, J.~D. {Vieira}, and A.~{Wei{\ss}}.
\newblock {ALMA observations of atomic carbon in z $\sim$ 4 dusty star-forming
  galaxies}.
\newblock \emph{\mnras}, 466:\penalty0 2825--2841, April 2017.
\newblock \doi{10.1093/mnras/stw3270}.

\bibitem[{De Breuck} et~al.(2011){De Breuck}, {Maiolino}, {Caselli}, {Coppin},
  {Hailey-Dunsheath}, and {Nagao}]{deBreuck2011}
C.~{De Breuck}, R.~{Maiolino}, P.~{Caselli}, K.~{Coppin},
  S.~{Hailey-Dunsheath}, and T.~{Nagao}.
\newblock {Enhanced [CII] emission in a z = 4.76 submillimetre galaxy}.
\newblock \emph{\aap}, 530:\penalty0 L8, June 2011.
\newblock \doi{10.1051/0004-6361/201116868}.

\bibitem[{Stacey} et~al.(2010){Stacey}, {Hailey-Dunsheath}, {Ferkinhoff},
  {Nikola}, {Parshley}, {Benford}, {Staguhn}, and {Fiolet}]{Stacey2010}
G.~J. {Stacey}, S.~{Hailey-Dunsheath}, C.~{Ferkinhoff}, T.~{Nikola}, S.~C.
  {Parshley}, D.~J. {Benford}, J.~G. {Staguhn}, and N.~{Fiolet}.
\newblock {A 158 {$\mu$}m [C II] Line Survey of Galaxies at z \~{} 1-2: An
  Indicator of Star Formation in the Early Universe}.
\newblock \emph{\apj}, 724:\penalty0 957--974, December 2010.
\newblock \doi{10.1088/0004-637X/724/2/957}.

\bibitem[{Cormier} et~al.(2010){Cormier}, {Madden}, {Hony}, {Contursi},
  {Poglitsch}, {Galliano}, {Sturm}, {Doublier}, {Feuchtgruber}, {Galametz},
  {Geis}, {de Jong}, {Okumura}, {Panuzzo}, and {Sauvage}]{Cormier2010}
D.~{Cormier}, S.~C. {Madden}, S.~{Hony}, A.~{Contursi}, A.~{Poglitsch},
  F.~{Galliano}, E.~{Sturm}, V.~{Doublier}, H.~{Feuchtgruber}, M.~{Galametz},
  N.~{Geis}, J.~{de Jong}, K.~{Okumura}, P.~{Panuzzo}, and M.~{Sauvage}.
\newblock {The effects of star formation on the low-metallicity ISM: NGC 4214
  mapped with Herschel/PACS spectroscopy}.
\newblock \emph{\aap}, 518:\penalty0 L57, July 2010.
\newblock \doi{10.1051/0004-6361/201014699}.

\bibitem[Madden and Cormier(2018)]{Madden2018}
Suzanne~C. Madden and Diane Cormier.
\newblock Dwarf galaxies: Their low metallicity interstellar medium.
\newblock \emph{Proceedings of the International Astronomical Union},
  14\penalty0 (S344):\penalty0 240–254, 2018.
\newblock \doi{10.1017/S1743921318007147}.

\bibitem[{Gao} and {Solomon}(2004{\natexlab{c}})]{Gao2004}
Y.~{Gao} and P.~M. {Solomon}.
\newblock {The Star Formation Rate and Dense Molecular Gas in Galaxies}.
\newblock \emph{\apj}, 606:\penalty0 271--290, May 2004{\natexlab{c}}.
\newblock \doi{10.1086/382999}.

\bibitem[{Jim{\'e}nez-Donaire} et~al.(2017){Jim{\'e}nez-Donaire}, {Bigiel},
  {Leroy}, {Cormier}, {Gallagher}, {Usero}, {Bolatto}, {Colombo},
  {Garc{\'{\i}}a-Burillo}, {Hughes}, {Kramer}, {Krumholz}, {Meier}, {Murphy},
  {Pety}, {Rosolowsky}, {Schinnerer}, {Schruba}, {Tomi{\v c}i{\'c}}, and
  {Zschaechner}]{JimenezDonaire2017b}
M.~J. {Jim{\'e}nez-Donaire}, F.~{Bigiel}, A.~K. {Leroy}, D.~{Cormier},
  M.~{Gallagher}, A.~{Usero}, A.~{Bolatto}, D.~{Colombo},
  S.~{Garc{\'{\i}}a-Burillo}, A.~{Hughes}, C.~{Kramer}, M.~R. {Krumholz}, D.~S.
  {Meier}, E.~{Murphy}, J.~{Pety}, E.~{Rosolowsky}, E.~{Schinnerer},
  A.~{Schruba}, N.~{Tomi{\v c}i{\'c}}, and L.~{Zschaechner}.
\newblock {Optical depth estimates and effective critical densities of dense
  gas tracers in the inner parts of nearby galaxy discs}.
\newblock \emph{\mnras}, 466:\penalty0 49--62, April 2017.
\newblock \doi{10.1093/mnras/stw2996}.

\bibitem[{Juneau} et~al.(2009){Juneau}, {Narayanan}, {Moustakas}, {Shirley},
  {Bussmann}, {Kennicutt}, and {Vanden Bout}]{Juneau2009}
S.~{Juneau}, D.~T. {Narayanan}, J.~{Moustakas}, Y.~L. {Shirley}, R.~S.
  {Bussmann}, R.~C. {Kennicutt}, Jr., and P.~A. {Vanden Bout}.
\newblock {Enhanced Dense Gas Fraction in Ultraluminous Infrared Galaxies}.
\newblock \emph{\apj}, 707:\penalty0 1217--1232, December 2009.
\newblock \doi{10.1088/0004-637X/707/2/1217}.

\bibitem[Puschnig(2020)]{DGT}
Johannes Puschnig.
\newblock {Dense Gas Toolbox}, February 2020.
\newblock URL \url{https://doi.org/10.5281/zenodo.3686329}.

\bibitem[{van der Tak} et~al.(2007){van der Tak}, {Black}, {Sch{\"o}ier},
  {Jansen}, and {van Dishoeck}]{vanderTak2007}
F.~F.~S. {van der Tak}, J.~H. {Black}, F.~L. {Sch{\"o}ier}, D.~J. {Jansen}, and
  E.~F. {van Dishoeck}.
\newblock {A computer program for fast non-LTE analysis of interstellar line
  spectra. With diagnostic plots to interpret observed line intensity ratios}.
\newblock \emph{\aap}, 468:\penalty0 627--635, June 2007.
\newblock \doi{10.1051/0004-6361:20066820}.

\bibitem[{Leroy} et~al.(2017){Leroy}, {Usero}, {Schruba}, {Bigiel},
  {Kruijssen}, {Kepley}, {Blanc}, {Bolatto}, {Cormier}, {Gallagher}, {Hughes},
  {Jim{\'e}nez-Donaire}, {Rosolowsky}, and {Schinnerer}]{Leroy2017}
A.~K. {Leroy}, A.~{Usero}, A.~{Schruba}, F.~{Bigiel}, J.~M.~D. {Kruijssen},
  A.~{Kepley}, G.~A. {Blanc}, A.~D. {Bolatto}, D.~{Cormier}, M.~{Gallagher},
  A.~{Hughes}, M.~J. {Jim{\'e}nez-Donaire}, E.~{Rosolowsky}, and
  E.~{Schinnerer}.
\newblock {Millimeter-wave Line Ratios and Sub-beam Volume Density
  Distributions}.
\newblock \emph{\apj}, 835:\penalty0 217, February 2017.
\newblock \doi{10.3847/1538-4357/835/2/217}.

\bibitem[{Messa} et~al.(2019){Messa}, {Adamo}, {{\~A}-stlin}, {Melinder},
  {Hayes}, {Bridge}, and {Cannon}]{Messa2019}
Matteo {Messa}, Angela {Adamo}, G{\"o}ran {{\~A}-stlin}, Jens {Melinder},
  Matthew {Hayes}, Johanna~S. {Bridge}, and John {Cannon}.
\newblock {Star-forming clumps in the Lyman Alpha Reference Sample of galaxies
  - I. Photometric analysis and clumpiness}.
\newblock \emph{\mnras}, 487\penalty0 (3):\penalty0 4238--4260, Aug 2019.
\newblock \doi{10.1093/mnras/stz1337}.

\bibitem[{Dekel} et~al.(2009){Dekel}, {Sari}, and {Ceverino}]{Dekel2009}
A.~{Dekel}, R.~{Sari}, and D.~{Ceverino}.
\newblock {Formation of Massive Galaxies at High Redshift: Cold Streams, Clumpy
  Disks, and Compact Spheroids}.
\newblock \emph{\apj}, 703:\penalty0 785--801, September 2009.
\newblock \doi{10.1088/0004-637X/703/1/785}.

\bibitem[{Kimm} et~al.(2019){Kimm}, {Blaizot}, {Garel}, {Michel-Dansac},
  {Katz}, {Rosdahl}, {Verhamme}, and {Haehnelt}]{Kimm2019}
Taysun {Kimm}, J{\'e}r{\'e}my {Blaizot}, Thibault {Garel}, L{\'e}o
  {Michel-Dansac}, Harley {Katz}, Joakim {Rosdahl}, Anne {Verhamme}, and Martin
  {Haehnelt}.
\newblock {Understanding the escape of LyC and Ly{\ensuremath{\alpha}} photons
  from turbulent clouds}.
\newblock \emph{\mnras}, 486\penalty0 (2):\penalty0 2215--2237, June 2019.
\newblock \doi{10.1093/mnras/stz989}.

\bibitem[{Guaita} et~al.(2015){Guaita}, {Melinder}, {Hayes}, {{\"O}stlin},
  {Gonzalez}, {Micheva}, {Adamo}, {Mas-Hesse}, {Sandberg}, {Ot{\'\i}-Floranes},
  {Schaerer}, {Verhamme}, {Freeland}, {Orlitov{\'a}}, {Laursen}, {Cannon},
  {Duval}, {Rivera-Thorsen}, {Herenz}, {Kunth}, {Atek}, {Puschnig}, {Gruyters},
  and {Pardy}]{Guaita2015}
L.~{Guaita}, J.~{Melinder}, M.~{Hayes}, G.~{{\"O}stlin}, J.~E. {Gonzalez},
  G.~{Micheva}, A.~{Adamo}, J.~M. {Mas-Hesse}, A.~{Sandberg},
  H.~{Ot{\'\i}-Floranes}, D.~{Schaerer}, A.~{Verhamme}, E.~{Freeland},
  I.~{Orlitov{\'a}}, P.~{Laursen}, J.~M. {Cannon}, F.~{Duval},
  T.~{Rivera-Thorsen}, E.~C. {Herenz}, D.~{Kunth}, H.~{Atek}, J.~{Puschnig},
  P.~{Gruyters}, and S.~A. {Pardy}.
\newblock {The Lyman alpha reference sample. IV. Morphology at low and high
  redshift}.
\newblock \emph{\aap}, 576:\penalty0 A51, Apr 2015.
\newblock \doi{10.1051/0004-6361/201425053}.

\bibitem[{Leroy} et~al.(2011){Leroy}, {Bolatto}, {Gordon}, {Sandstrom},
  {Gratier}, {Rosolowsky}, {Engelbracht}, {Mizuno}, {Corbelli}, {Fukui}, and
  {Kawamura}]{Leroy2011}
A.~K. {Leroy}, A.~{Bolatto}, K.~{Gordon}, K.~{Sandstrom}, P.~{Gratier},
  E.~{Rosolowsky}, C.~W. {Engelbracht}, N.~{Mizuno}, E.~{Corbelli}, Y.~{Fukui},
  and A.~{Kawamura}.
\newblock {The CO-to-H$_{2}$ Conversion Factor from Infrared Dust Emission
  across the Local Group}.
\newblock \emph{\apj}, 737:\penalty0 12, August 2011.
\newblock \doi{10.1088/0004-637X/737/1/12}.

\end{thebibliography}


\appendix

\FloatBarrier
\section{Bayesian Joint Posterior Probability Distribution for Draine \& Li model parameters}\label{sec:appendix_corner}

\rotatebox{90}{\begin{minipage}{0.95\textheight}
   \vspace*{3cm}
   \includegraphics[width=11.9cm]{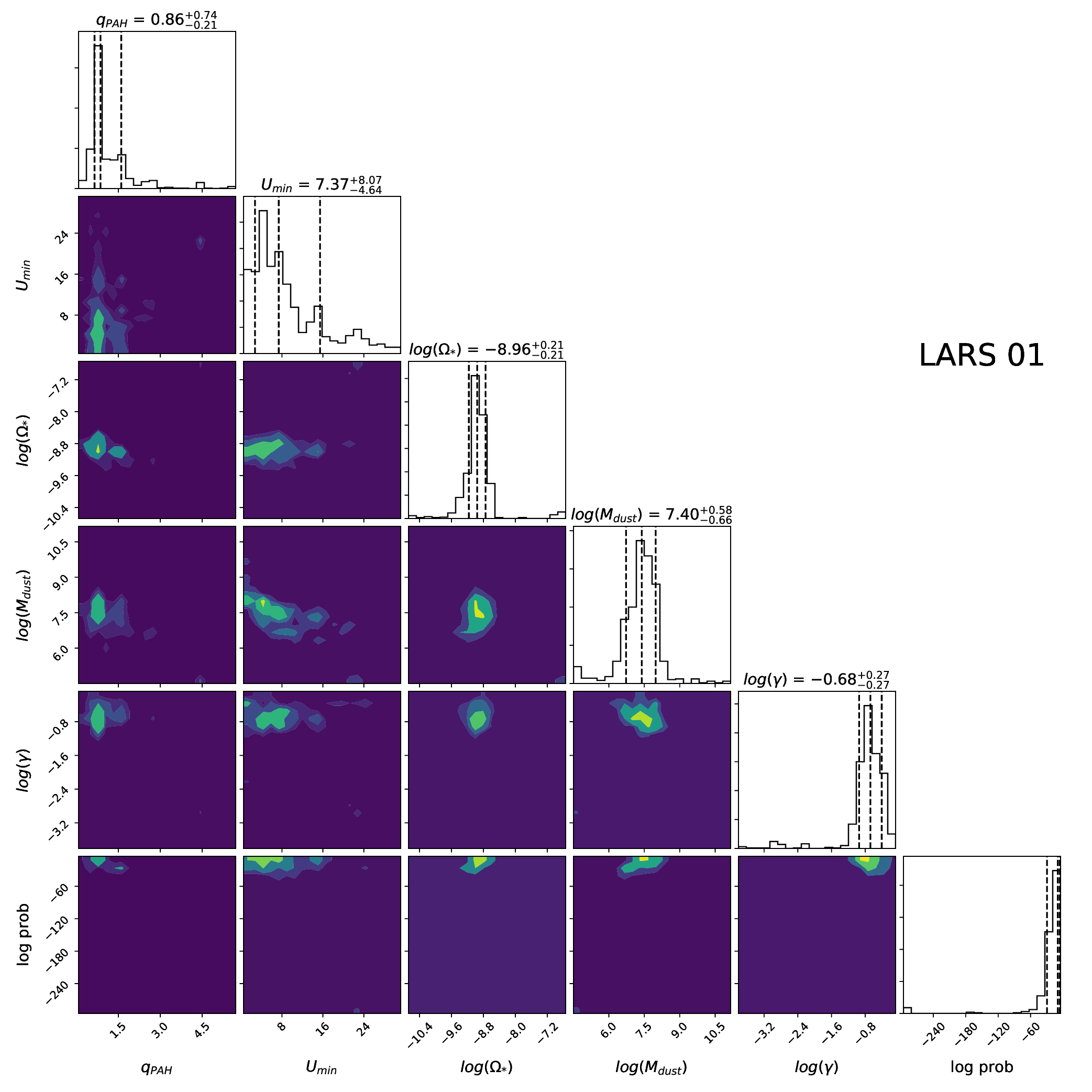}
   \includegraphics[width=11.9cm]{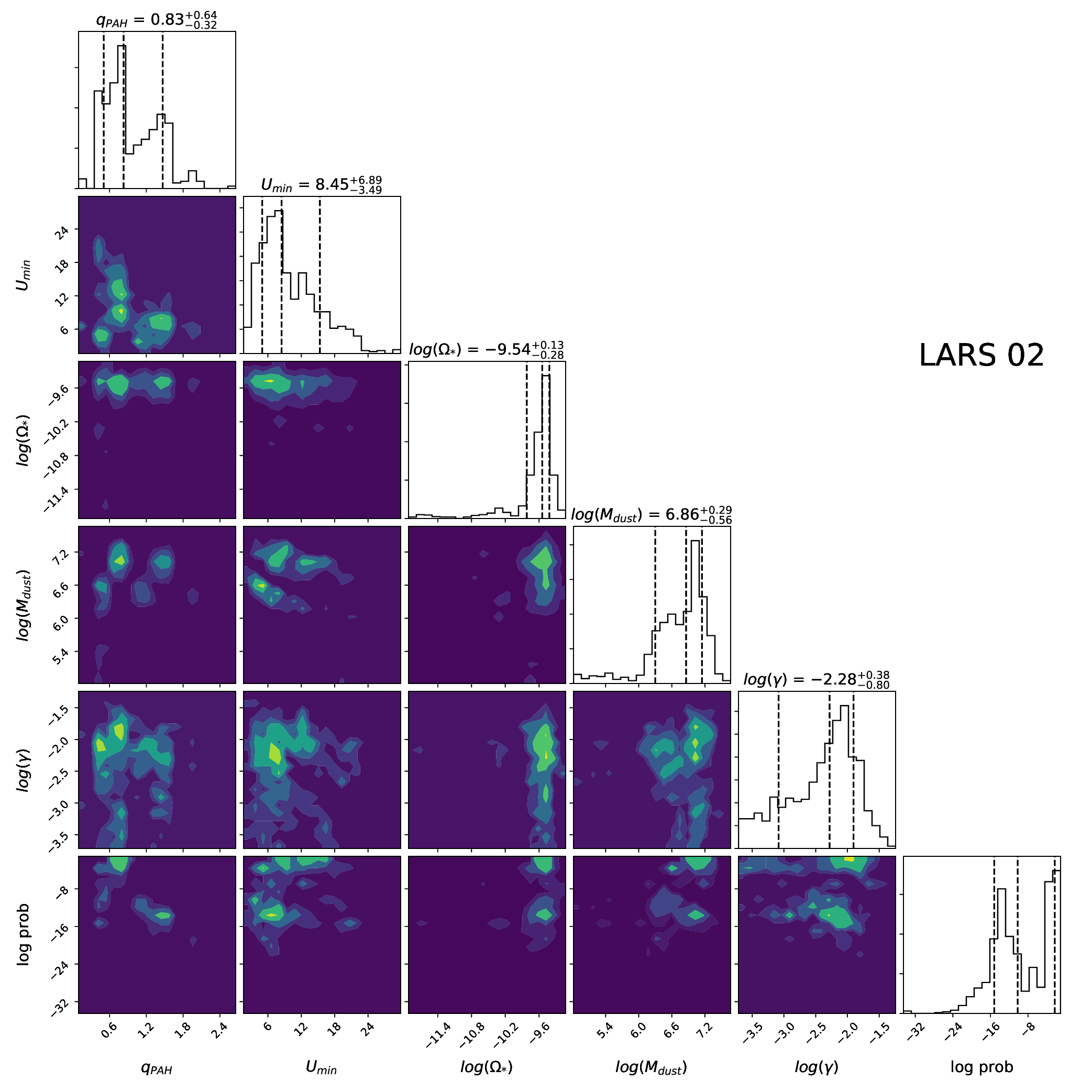}
   \captionof{figure}{Corner plots showing the posterior probability distribution for the derived Draine \& Li model parameters.}
   \label{fig:bayes_corner1}
\end{minipage}}

\rotatebox{90}{\begin{minipage}{0.95\textheight}
   \vspace*{3cm}
   \includegraphics[width=11.9cm]{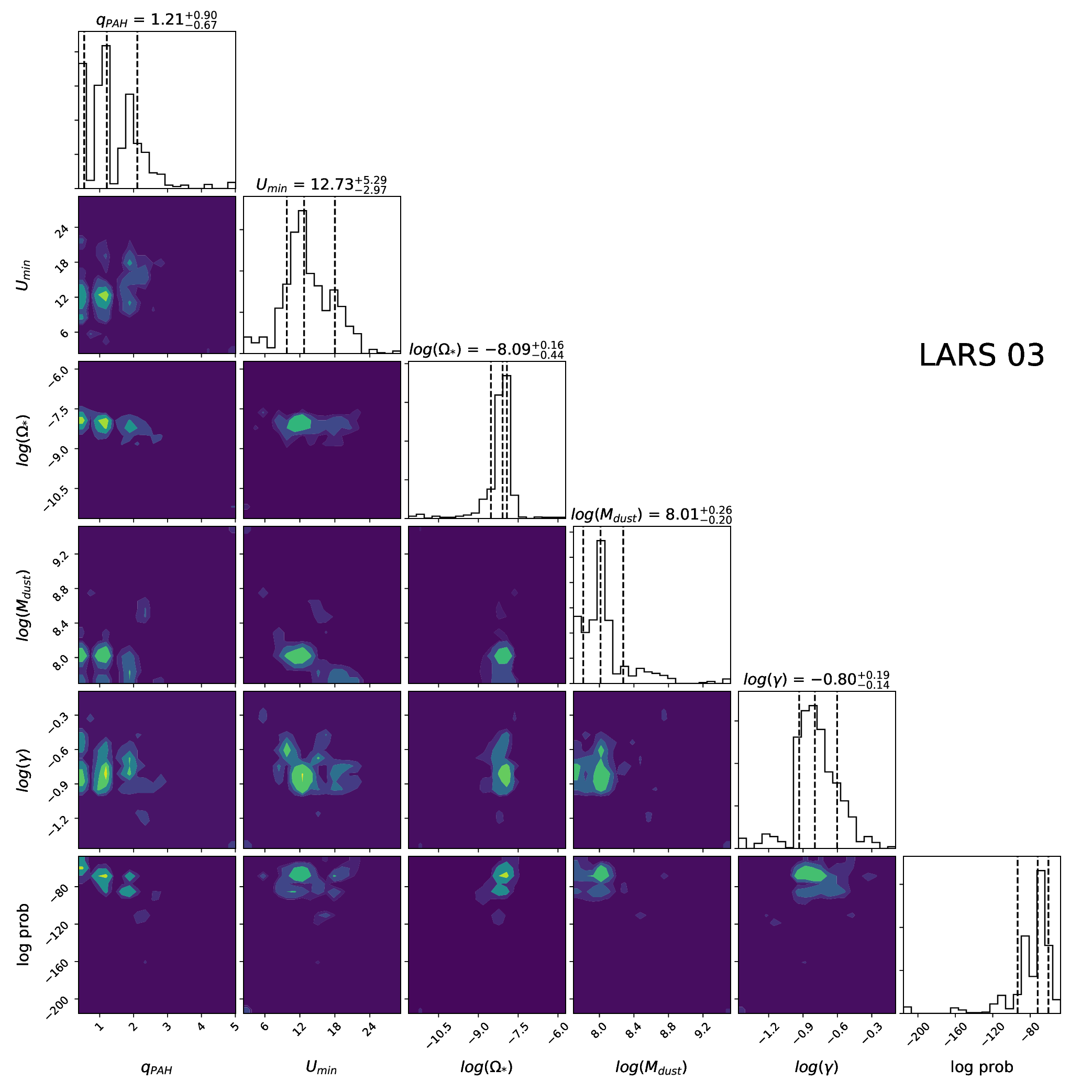}
   \includegraphics[width=11.9cm]{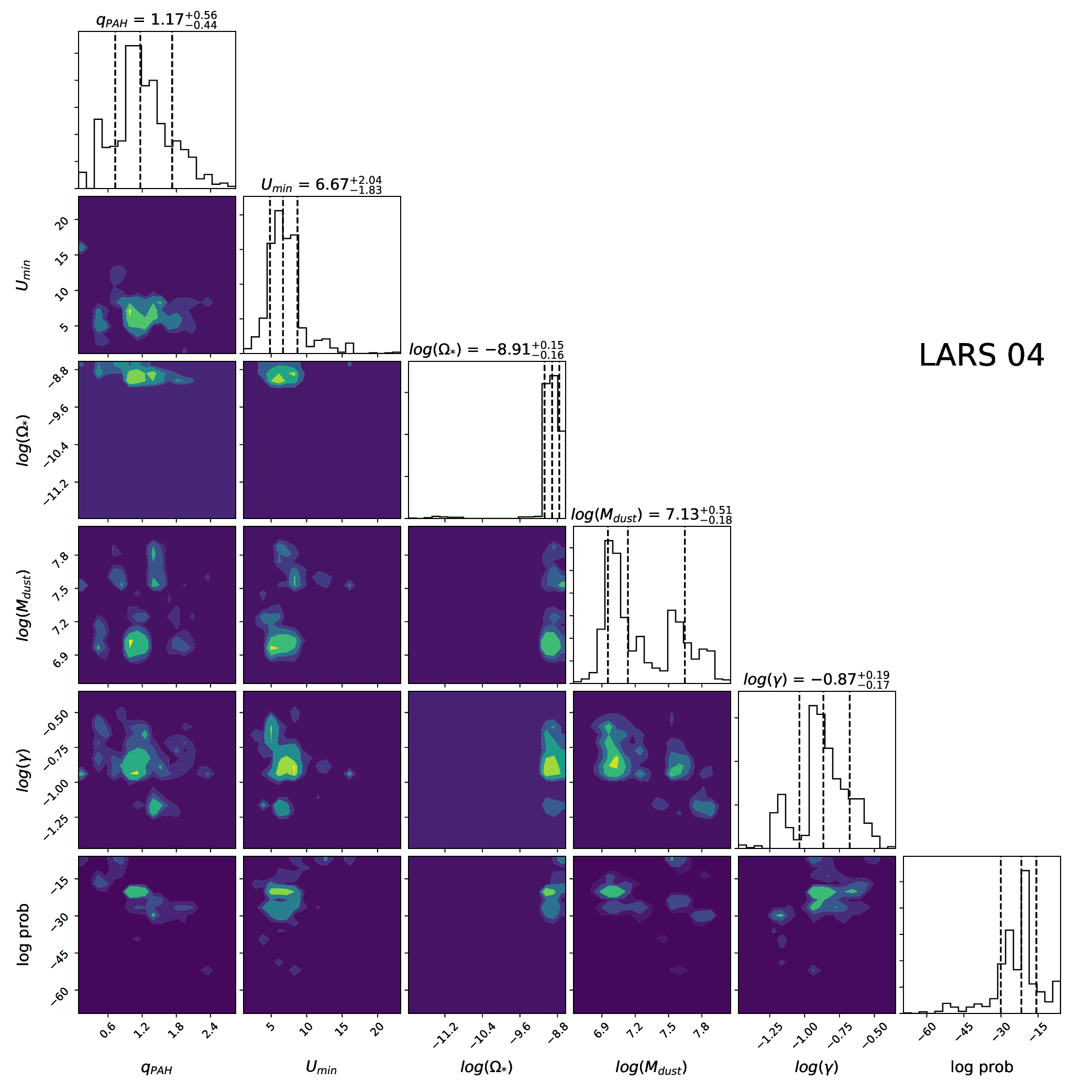}
   \captionof{figure}{Corner plots showing the posterior probability distribution for the derived Draine \& Li model parameters.}
   \label{fig:bayes_corner1}
\end{minipage}}

\rotatebox{90}{\begin{minipage}{0.95\textheight}
   \vspace*{3cm}
   \includegraphics[width=11.9cm]{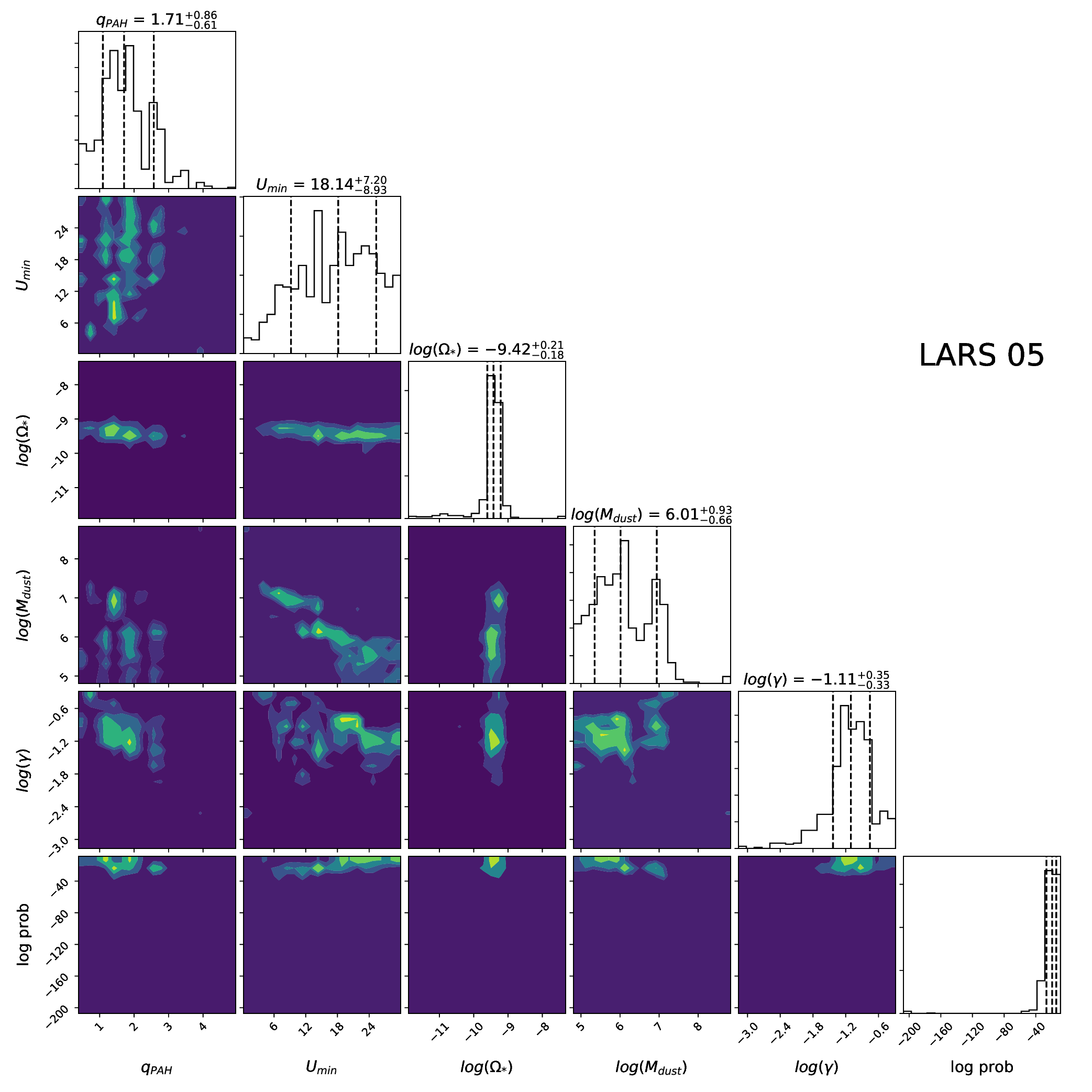}
   \includegraphics[width=11.9cm]{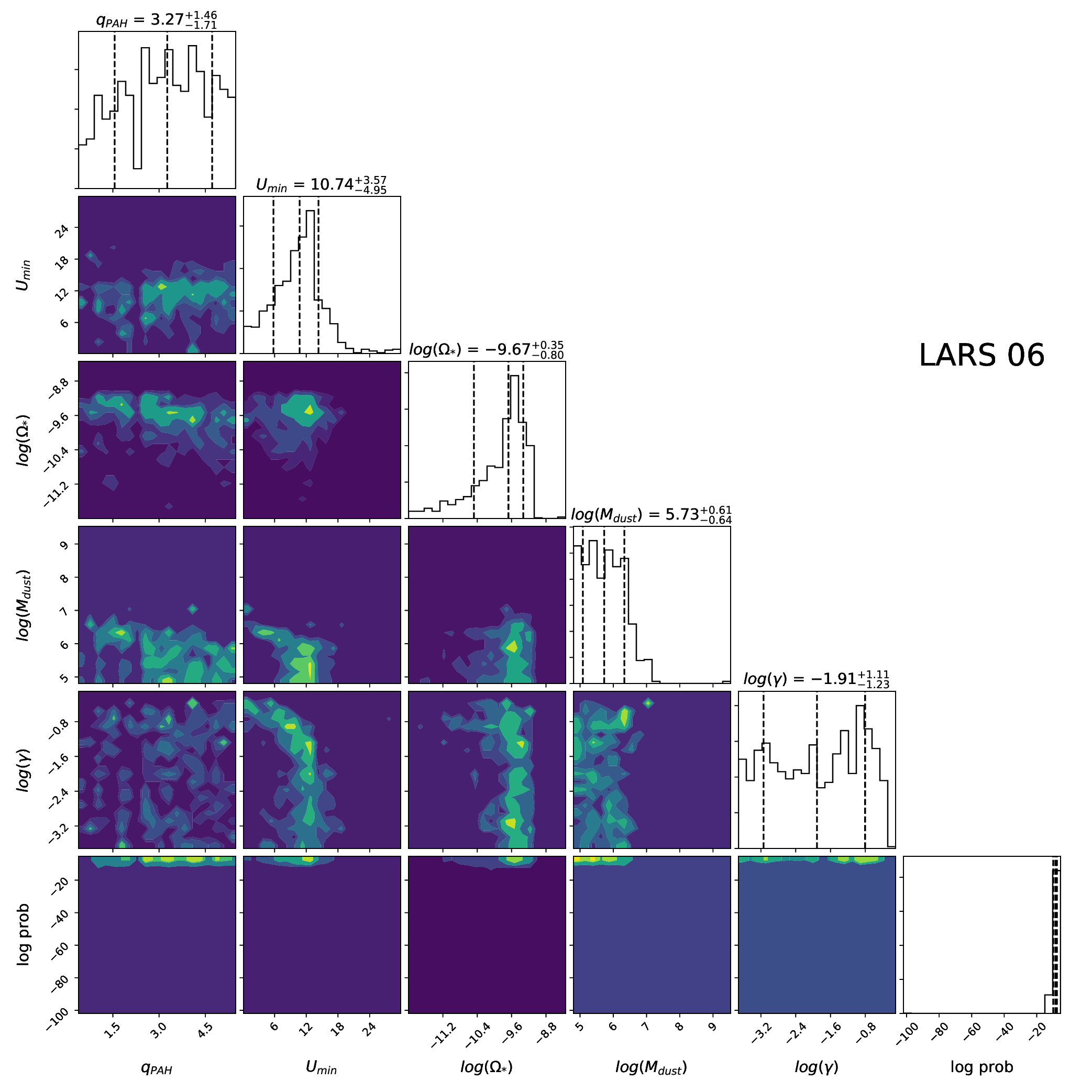}
   \captionof{figure}{Corner plots showing the posterior probability distribution for the derived Draine \& Li model parameters.}
   \label{fig:bayes_corner1}
\end{minipage}}

\rotatebox{90}{\begin{minipage}{0.95\textheight}
   \vspace*{3cm}
   \includegraphics[width=11.9cm]{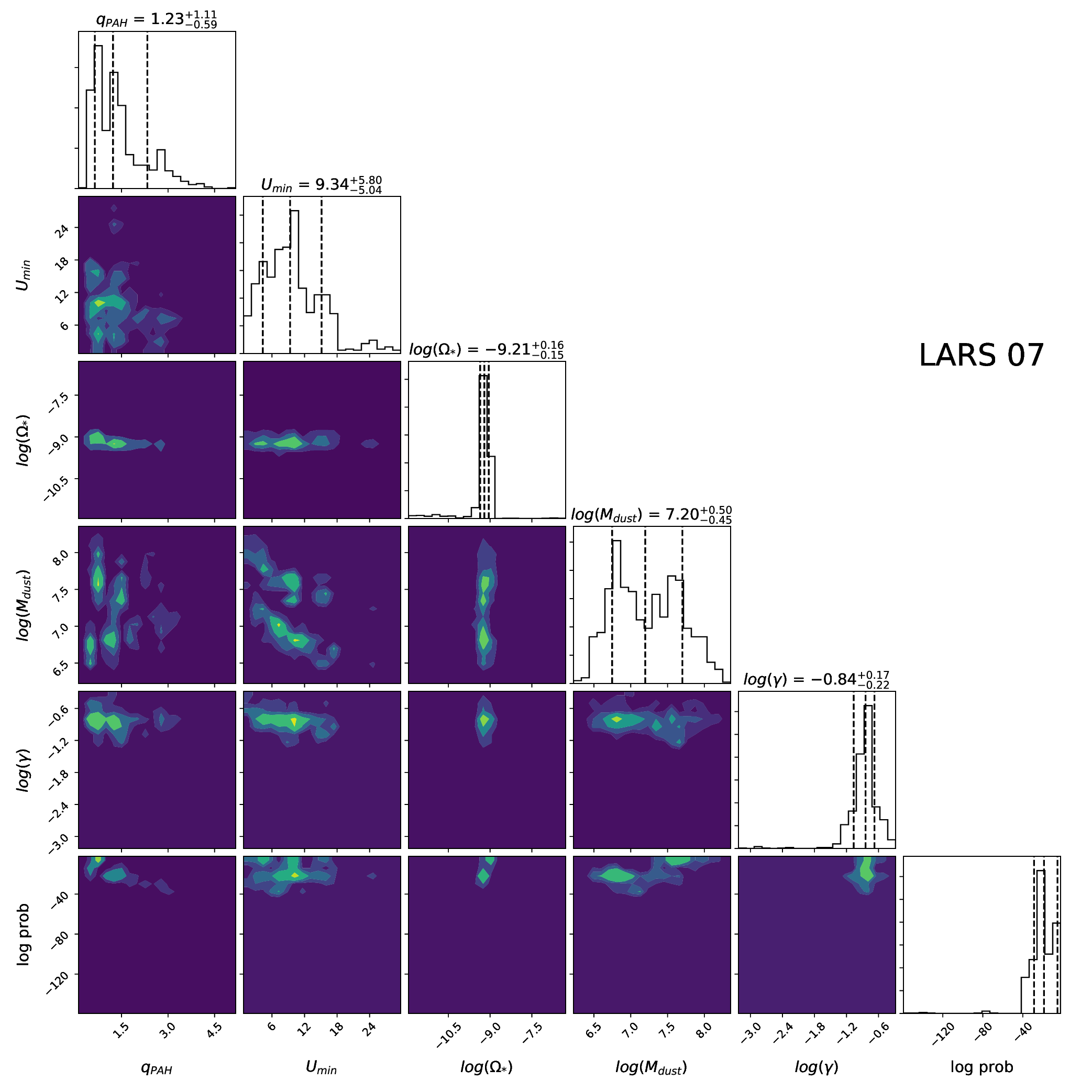}
   \includegraphics[width=11.9cm]{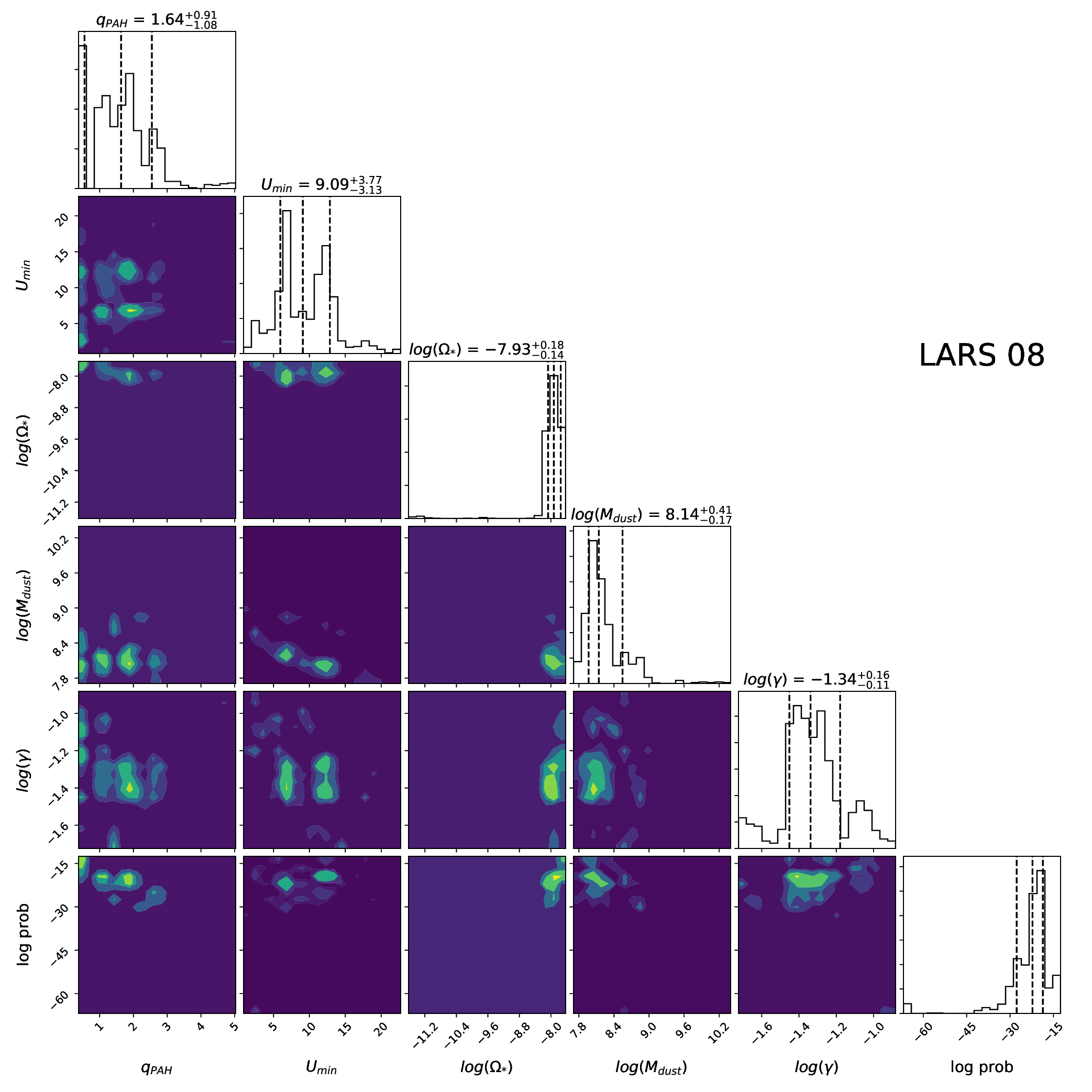}
   \captionof{figure}{Corner plots showing the posterior probability distribution for the derived Draine \& Li model parameters.}
   \label{fig:bayes_corner1}
\end{minipage}}

\rotatebox{90}{\begin{minipage}{0.95\textheight}
   \vspace*{3cm}
   \includegraphics[width=11.9cm]{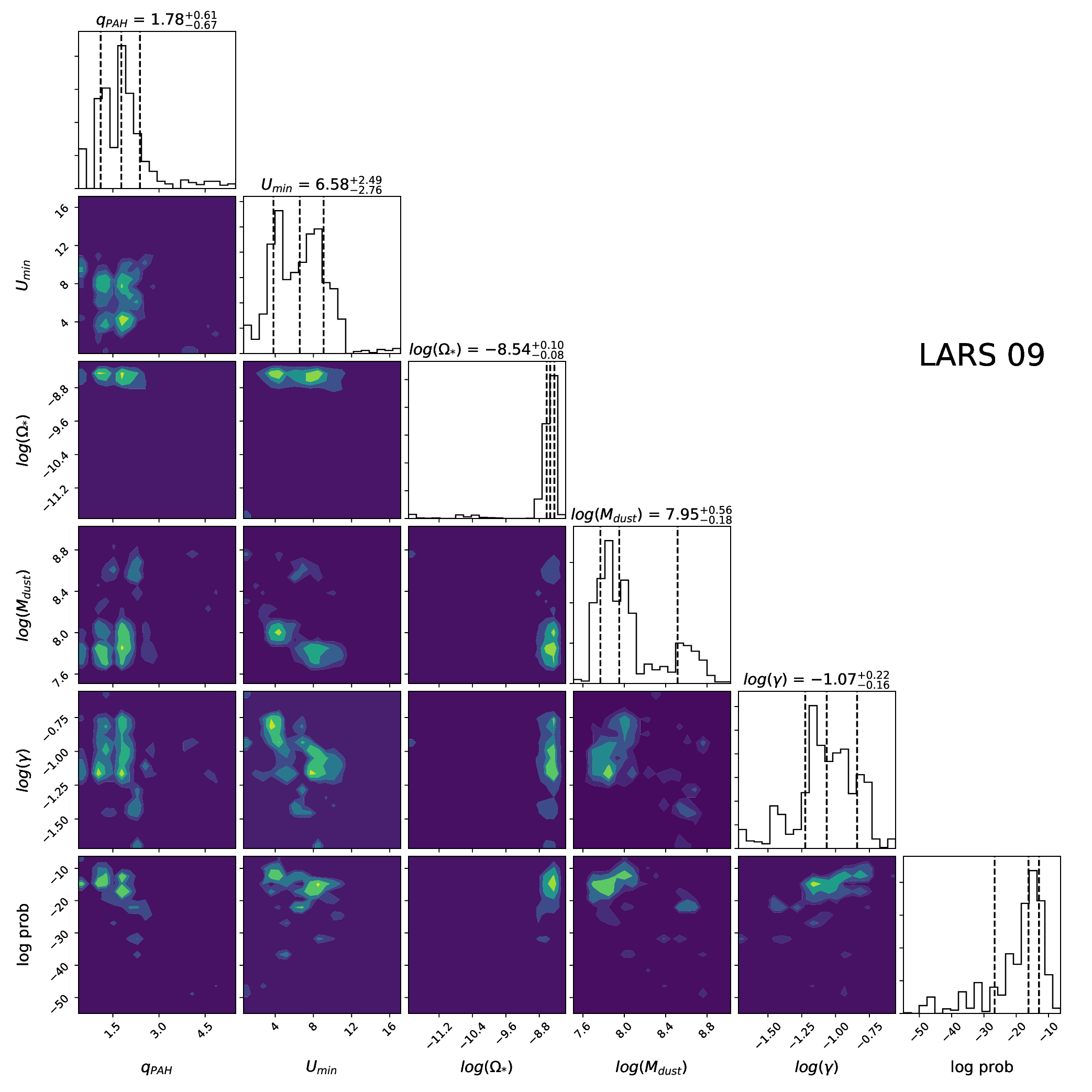}
   \includegraphics[width=11.9cm]{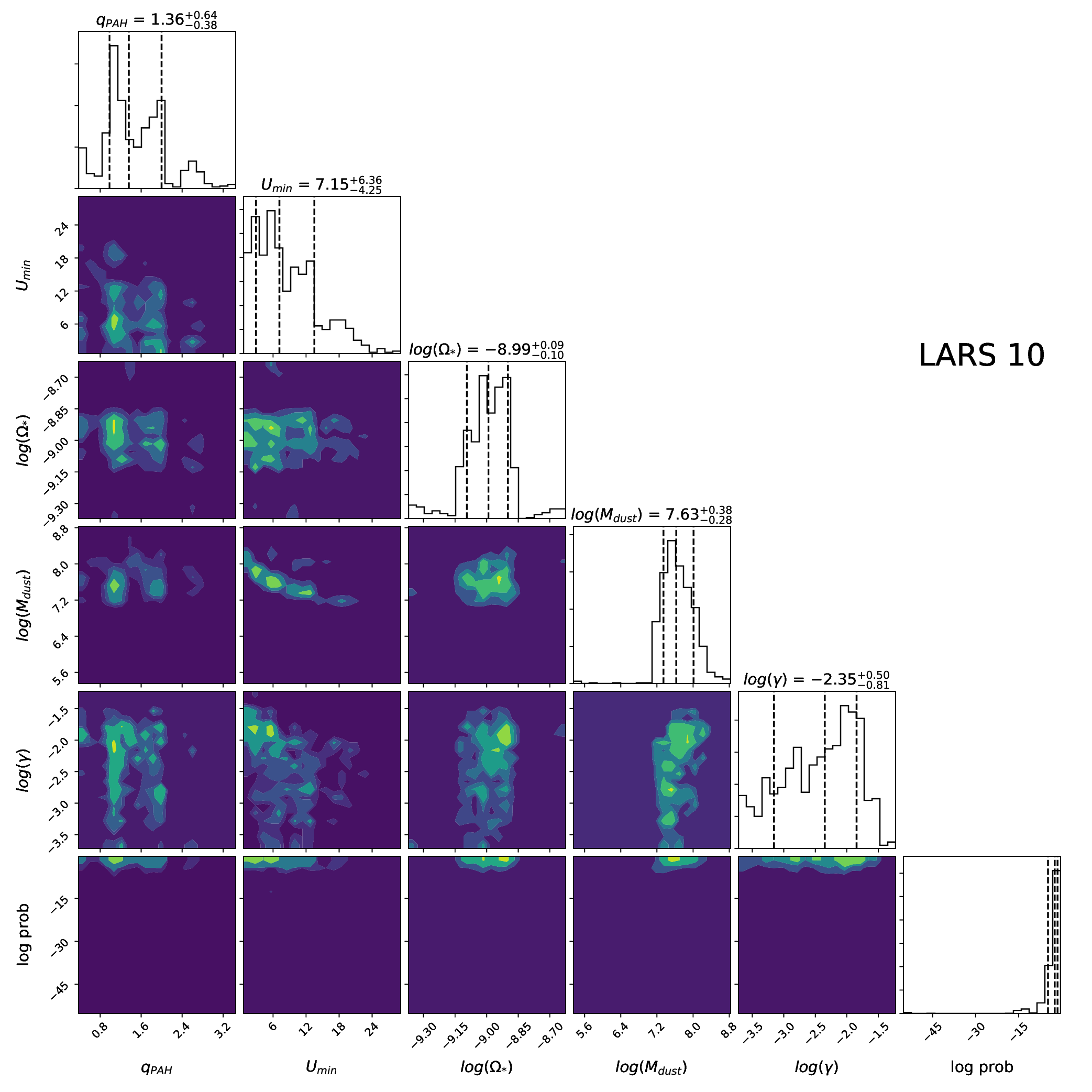}
   \captionof{figure}{Corner plots showing the posterior probability distribution for the derived Draine \& Li model parameters.}
   \label{fig:bayes_corner1}
\end{minipage}}

\rotatebox{90}{\begin{minipage}{0.95\textheight}
   \vspace*{3cm}
   \includegraphics[width=11.9cm]{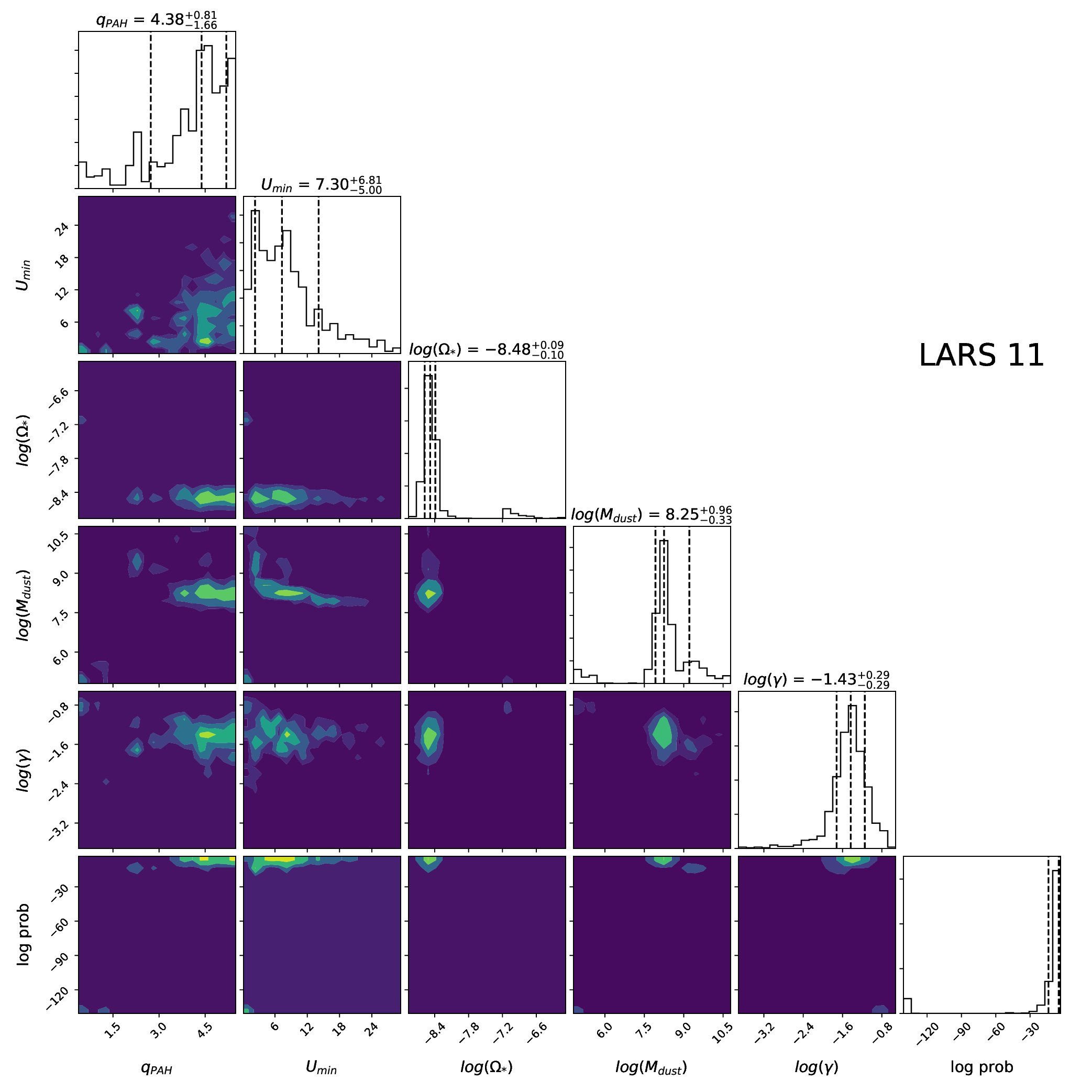}
   \includegraphics[width=11.9cm]{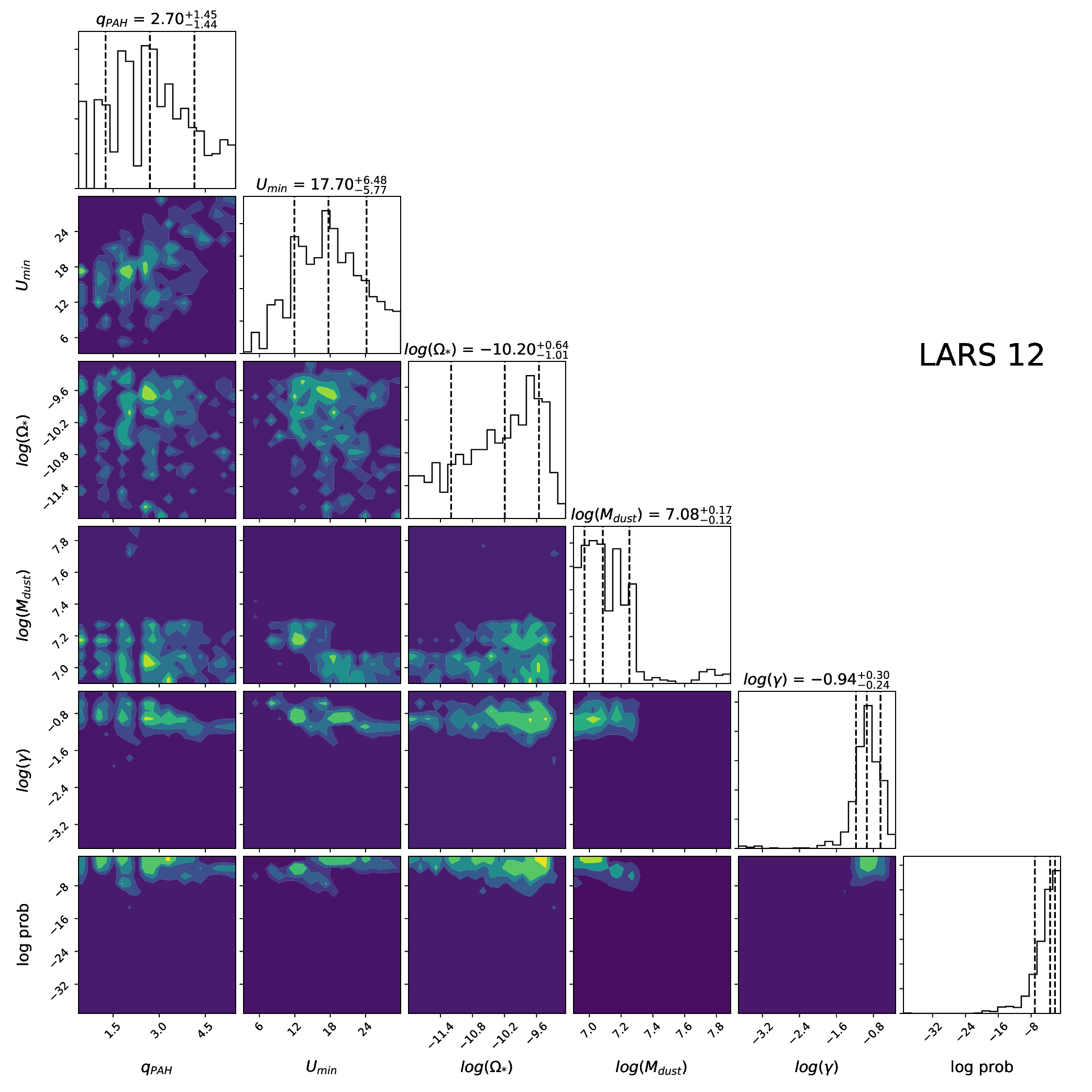}
   \captionof{figure}{Corner plots showing the posterior probability distribution for the derived Draine \& Li model parameters.}
   \label{fig:bayes_corner1}
\end{minipage}}

\rotatebox{90}{\begin{minipage}{0.95\textheight}
   \vspace*{3cm}
   \includegraphics[width=11.9cm]{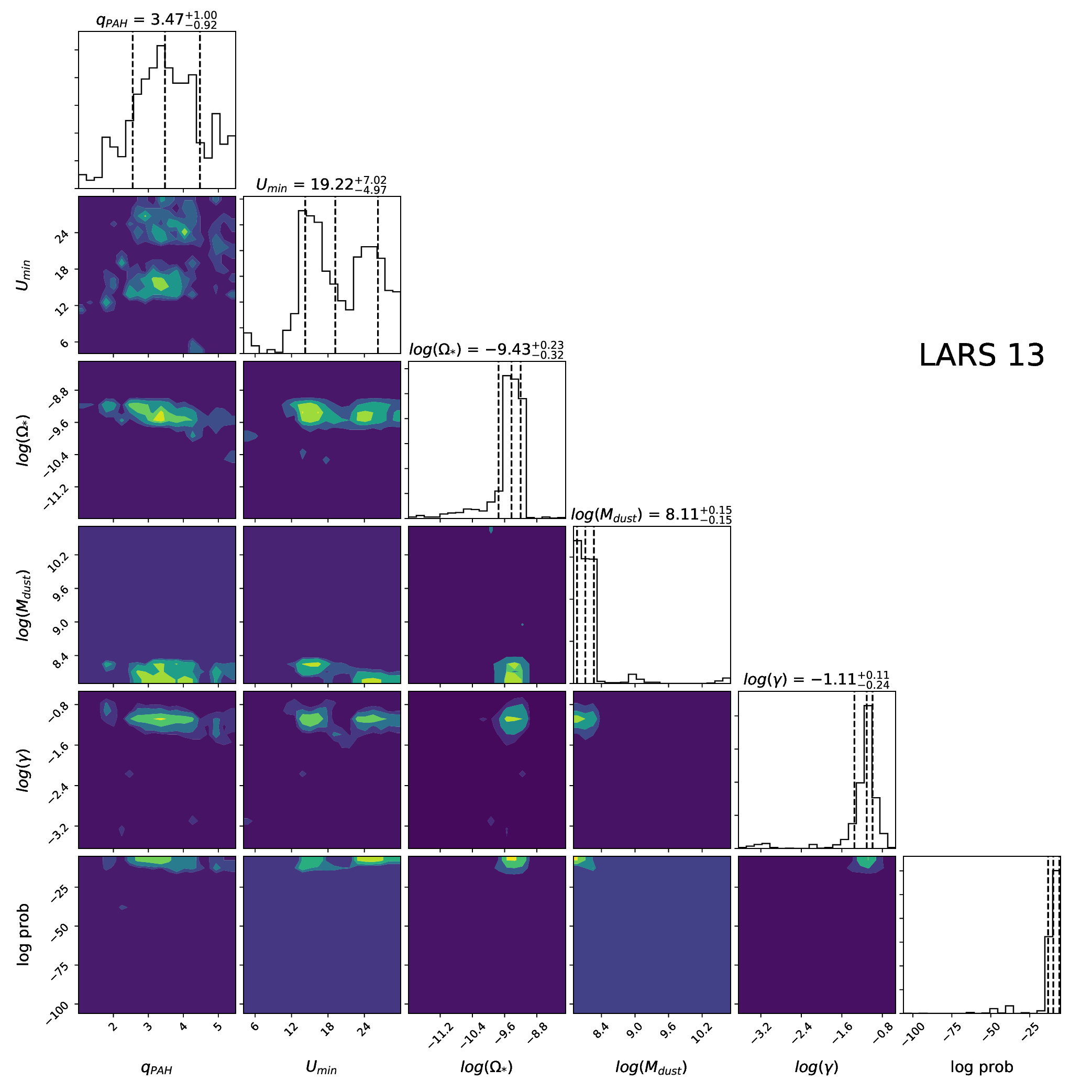}
   \includegraphics[width=11.9cm]{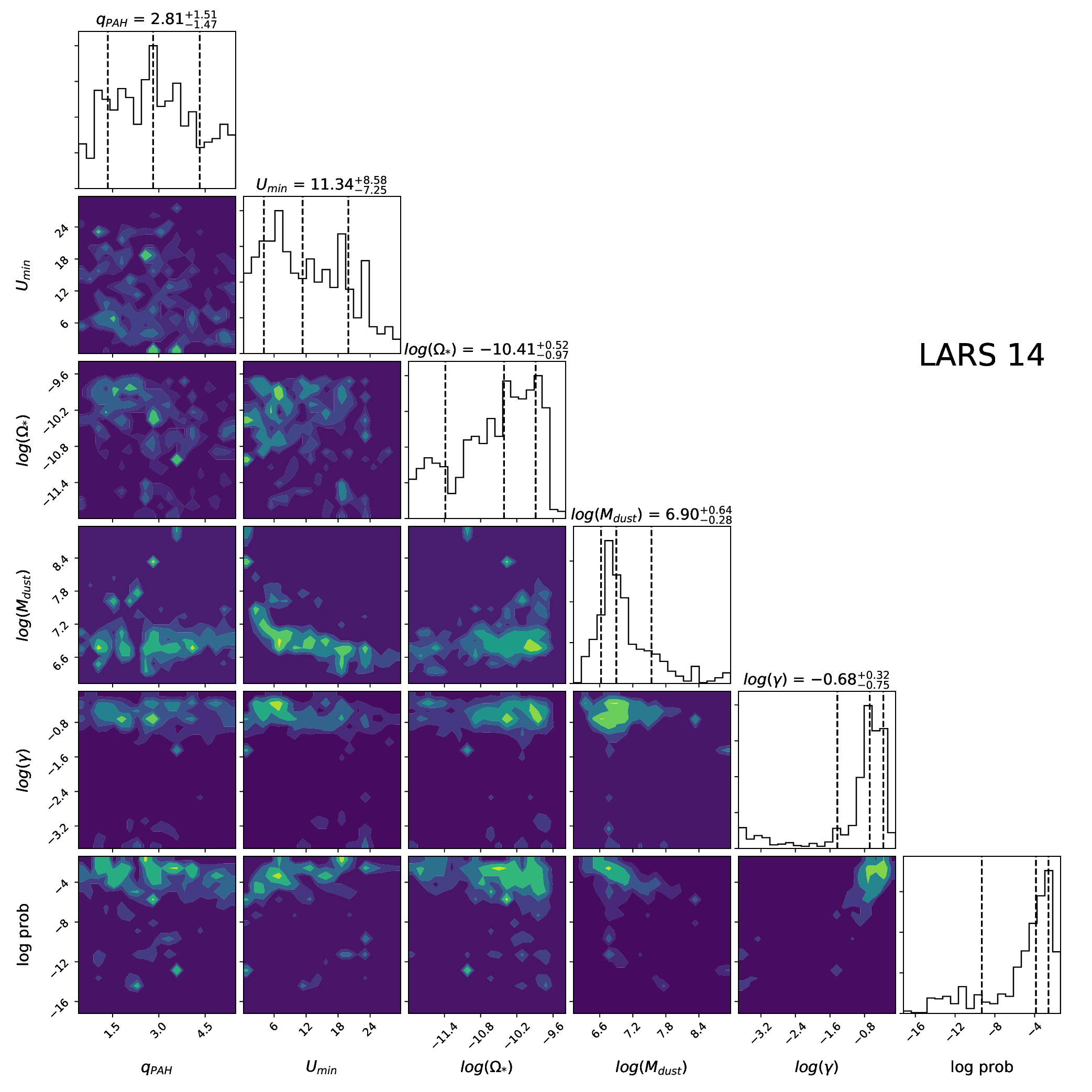}
   \captionof{figure}{Corner plots showing the posterior probability distribution for the derived Draine \& Li model parameters.}
   \label{fig:bayes_corner1}
\end{minipage}}

\newpage
\FloatBarrier
\section{Draine \& Li SED fits using the MCMC method}\label{sec:appendix_sed}
\begin{figure*}[!ht]
   \subfloat{\includegraphics[width=9cm]{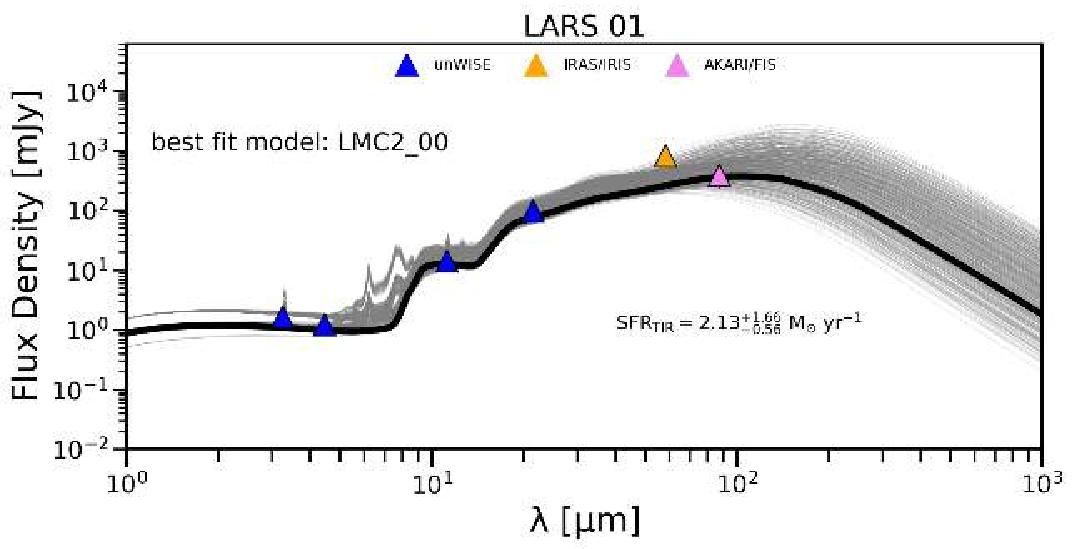}}\hspace*{\fill}
   \subfloat{\includegraphics[width=9cm]{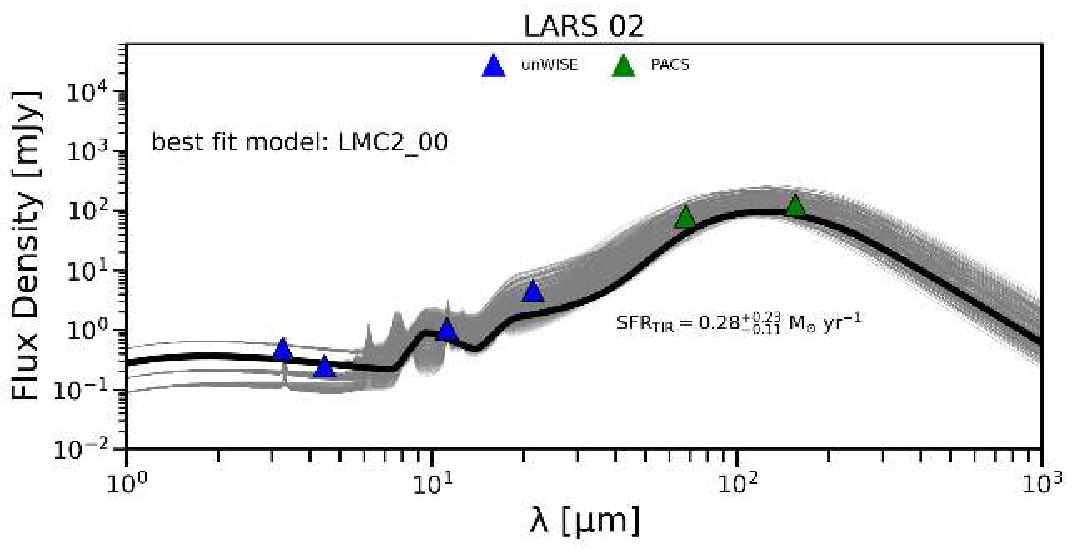}}\hspace*{\fill}

   \subfloat{\includegraphics[width=9cm]{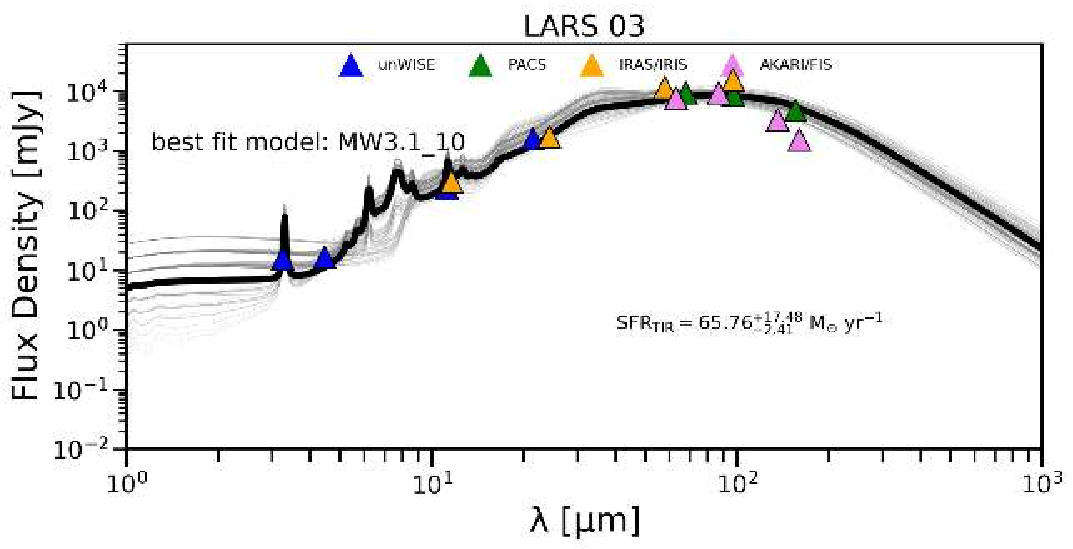}}\hspace*{\fill}
   \subfloat{\includegraphics[width=9cm]{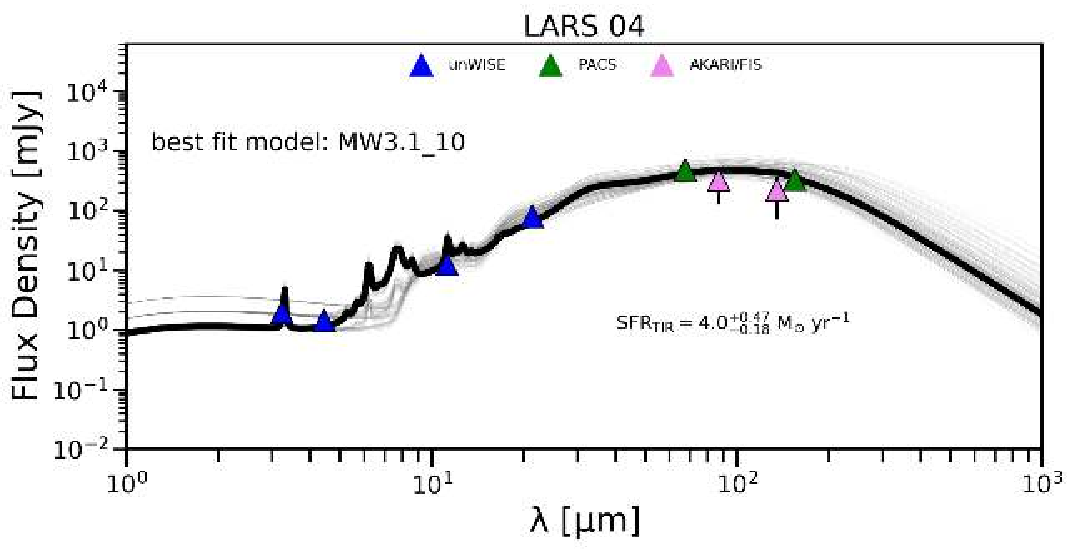}}\hspace*{\fill}

   \subfloat{\includegraphics[width=9cm]{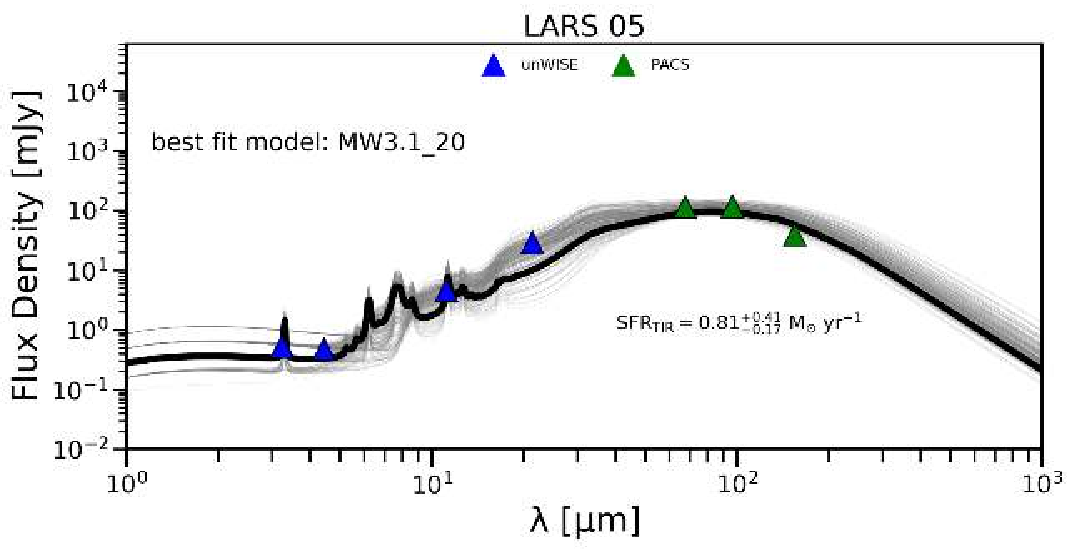}}\hspace*{\fill}
   \subfloat{\includegraphics[width=9cm]{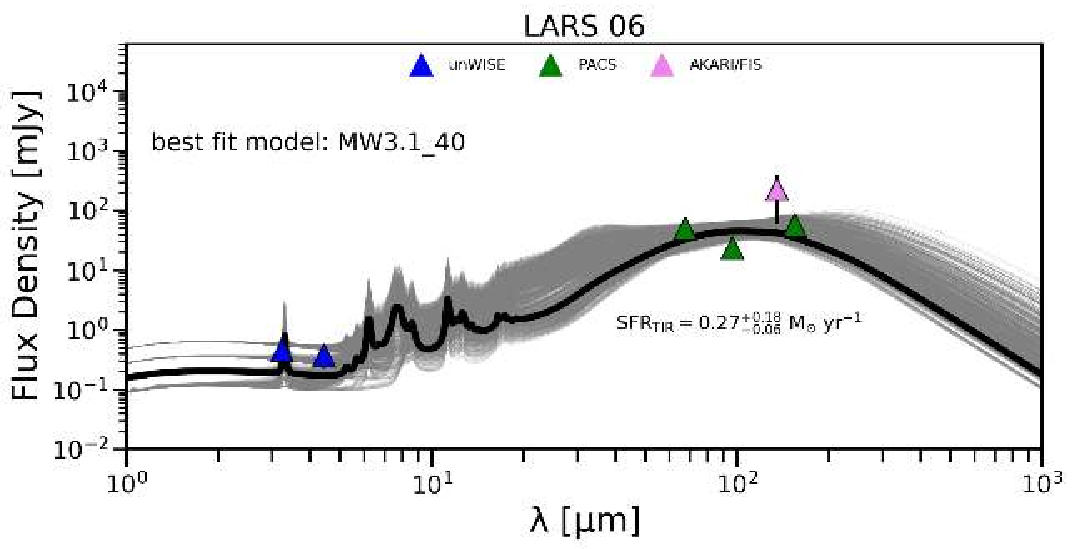}}\hspace*{\fill}

   \subfloat{\includegraphics[width=9cm]{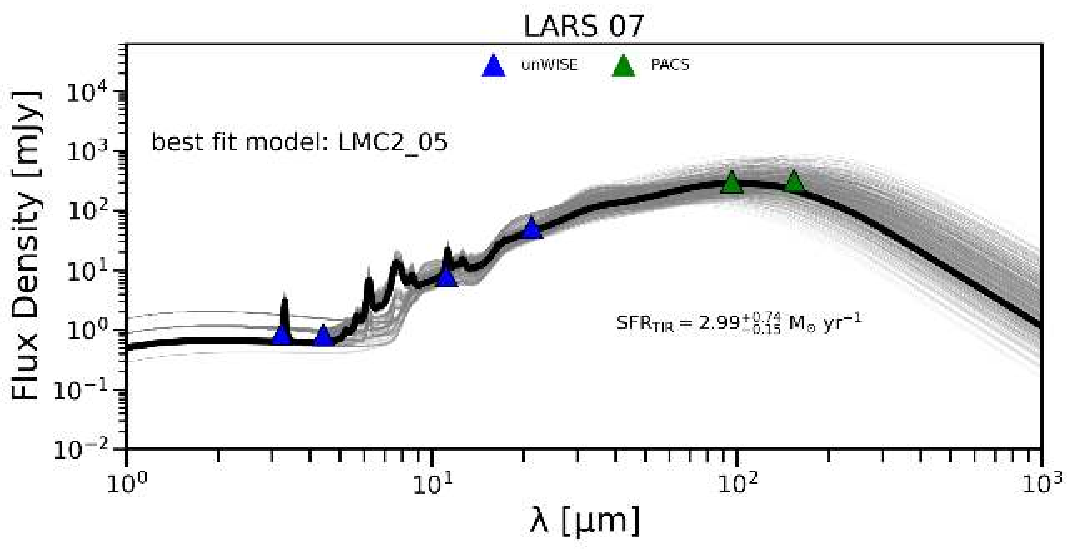}}\hspace*{\fill}
   \subfloat{\includegraphics[width=9cm]{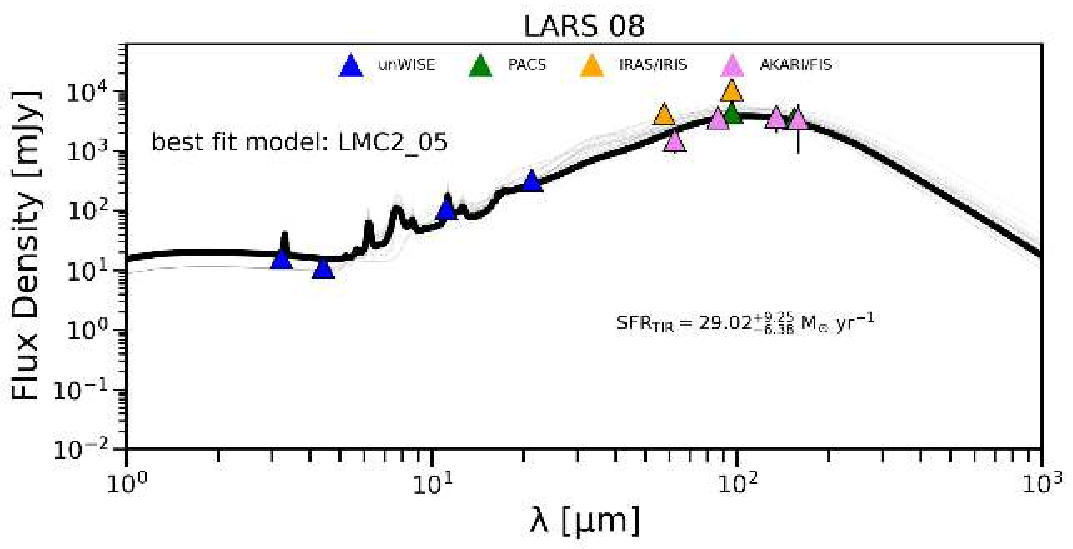}}\hspace*{\fill}
   \caption{
	Infrared spectral energy distribution of LARS galaxies inferred from exploration of the parameter space using the Markov-Chain Monte Carlo method.
	The best fit SED is indicated by a \textit{thick black curve} and was found from the 50\% quantile of the logarithmic posterior distribution.
	The other fits that are shown (\textit{thin grey lines}) are those lying within the quantile-based credible interval corresponding to 16\% and 84\%.
	They are thus representing our uncertainties.
	}
   \label{fig:irsed1}
\end{figure*}

\begin{figure*}[ht]
   \subfloat{\includegraphics[width=9cm]{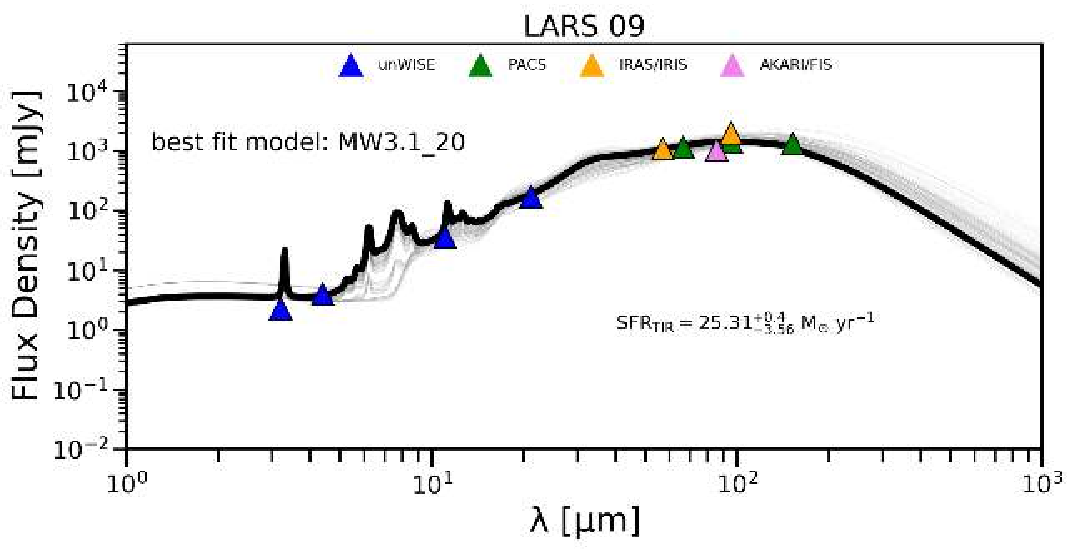}}\hspace*{\fill}
   \subfloat{\includegraphics[width=9cm]{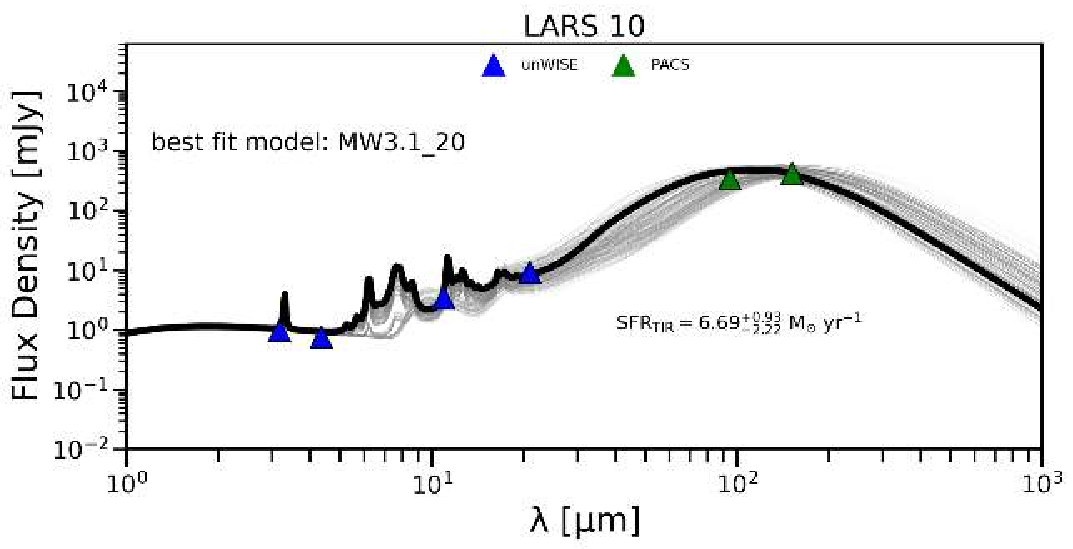}}\hspace*{\fill}

   \subfloat{\includegraphics[width=9cm]{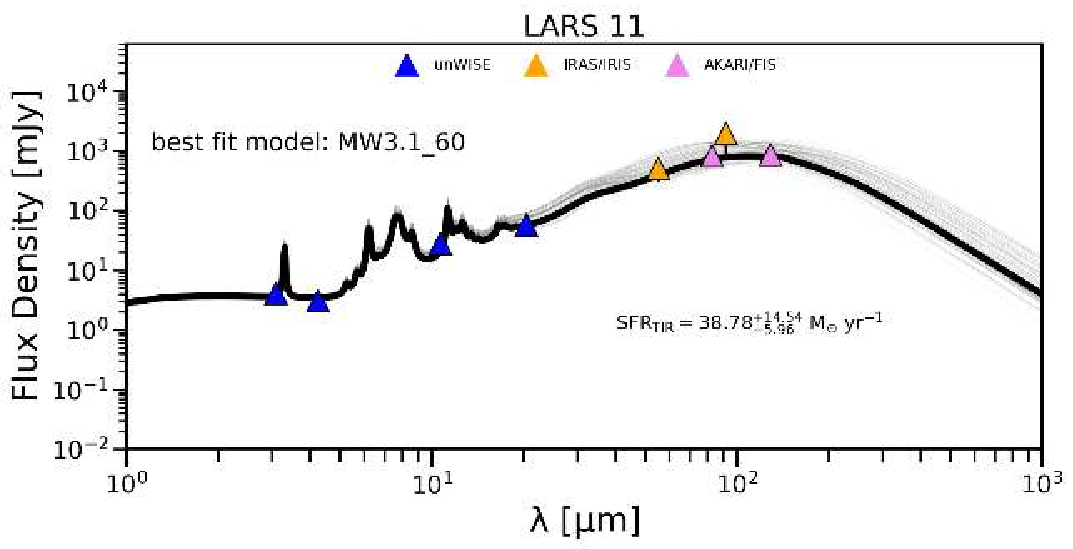}}\hspace*{\fill}
   \subfloat{\includegraphics[width=9cm]{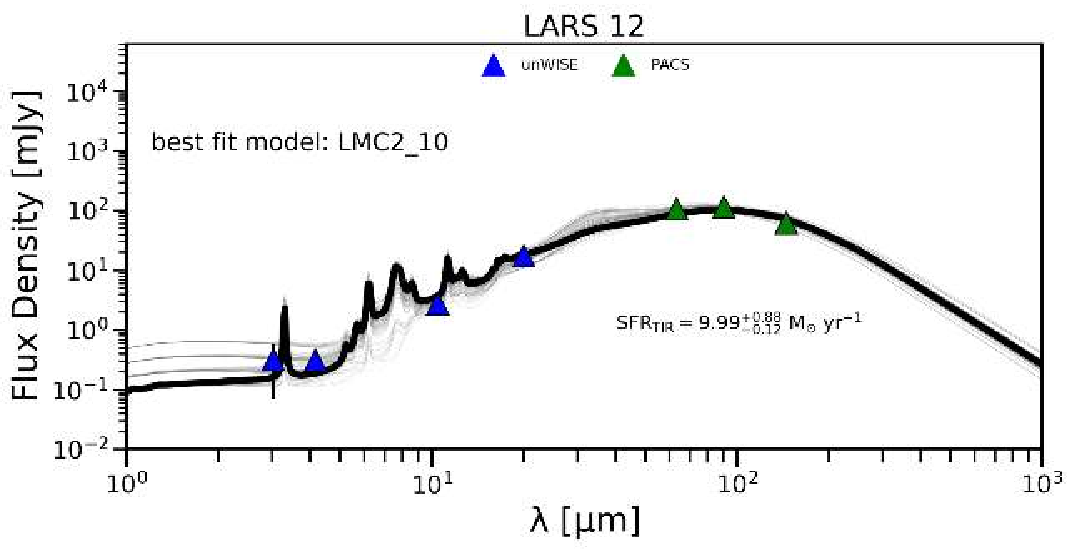}}\hspace*{\fill}

   \subfloat{\includegraphics[width=9cm]{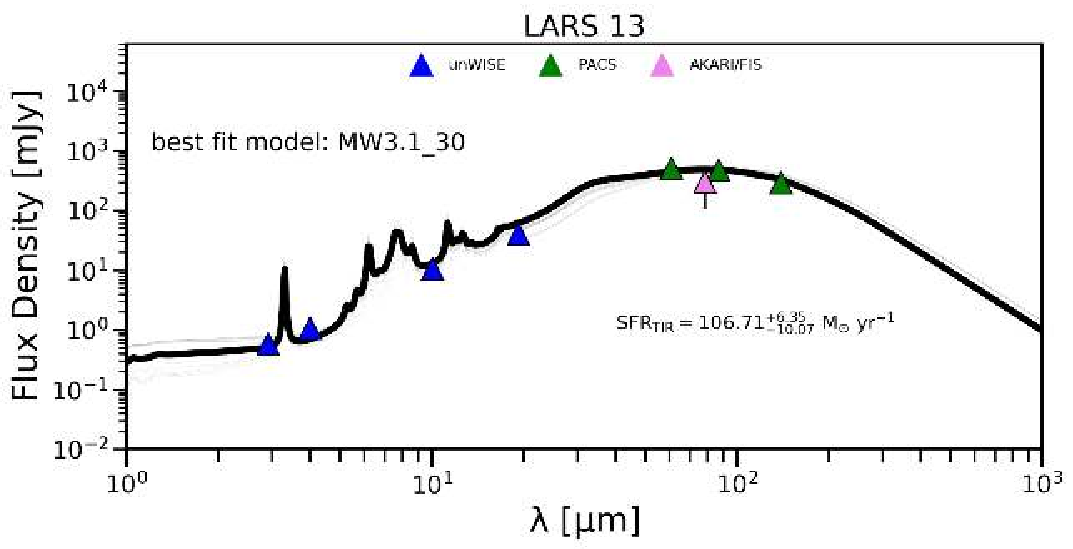}}\hspace*{\fill}
   \subfloat{\includegraphics[width=9cm]{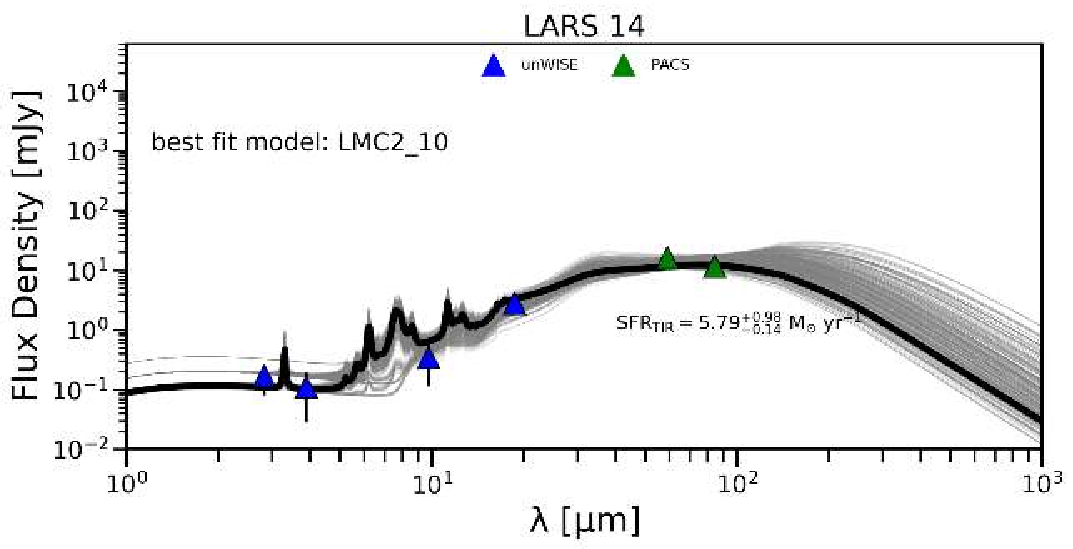}}\hspace*{\fill}
   \caption{Same as in Figure \ref{fig:irsed1}.}
   \label{fig:irsed2}
\end{figure*}

\FloatBarrier
\section{PDR models Fit Quality}\label{sec:appendix_pdr}

\begin{figure*}[!ht]
\centering
        \includegraphics[width=1\textwidth]{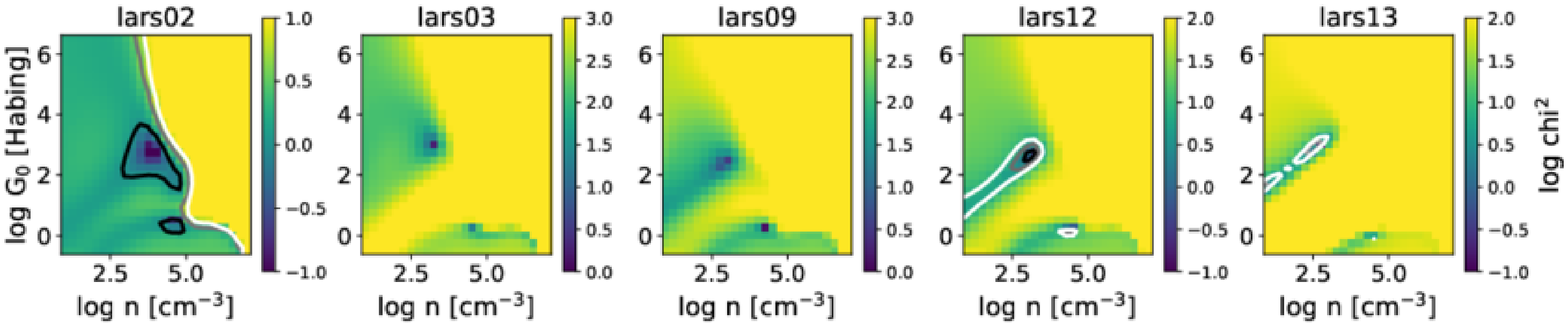}
\caption{
	PDR model logarithmic $\chi^2$ distribution across the (n,G$_0$) solution plane.
	The contours show 1, 2 and 3$\upsigma$ significance regions around the best model fits.
	Note that for LARS 3 and 9 (both merging galaxies) the models are not able to reproduce
	the observations.
	}
\label{fig:pdr_chi2}
\end{figure*}

\newpage
\FloatBarrier
\section{Molecular Gas Density and Temperature from Line Radiative Transfer Modeling}\label{sec:appendix_nT}

\begin{table*}[!ht]
\begin{threeparttable}
\caption{Calculated molecular line ratios used as input for radiative transfer modeling of LARS 3 and 8}
\label{tab:lineratios}
\begin{tabular}{llrrrr}
\hline \hline
ID & Line   & I$_{mb}$     & C$_{AP}$ & $\theta$     & Line/CO(1–0) \\
   &        & [K km/s] &          & [arcsec] &              \\
3  & CO~(1--0)   & 7.13         & 1.94     & 22.1         & 1.00         \\
3  & CO~(2--1)   & 15.93        & 3.58     & 11.0         & 1.02         \\
3  & HCN~(1--0)  & 1.06         & 1.00     & 28.7         & 0.13         \\
3  & HCO+~(1--0) & 1.08         & 1.00     & 28.6         & 0.13         \\
   &        &              &          &              &              \\
8  & CO~(1--0)   & 6.85         & 1.36     & 22.2         & 1.00         \\
8  & CO~(2--1)   & 11.70        & 1.10     & 11.1         & 0.35         \\
8  & CO~(3--2)   & 6.50         & 1.00     & 18.5         & 0.48         \\
8  & HCN~(1--0)  & 0.29         & 1.00     & 28.9         & 0.05         \\
8  & HCO+~(1--0) & 0.19         & 1.00     & 28.7         & 0.03        
\end{tabular}
\begin{tablenotes}
\item The galaxy id is given in \textit{column 1}, followed by the observed integrated main beam brightness temperature
in \textit{column 2} and the aperture correction factor as well as the beam size in \textit{columns 3 and 4}.
Normalized line ratios (\textit{column 5}) are then calculated as: ($I_{mb}\ *\ C_{AP}\ *\ \theta^2)/(I_{mb,CO10}\ *\ C_{AP,CO10}\ *\ \theta^2_{CO10})$.
Note that the squared beam-size needs to be taken into account, as explained in \cite{Solomon2005}.
\end{tablenotes}
\end{threeparttable}
\end{table*}

\begin{figure}[!ht]
\centering
   \includegraphics[width=10cm]{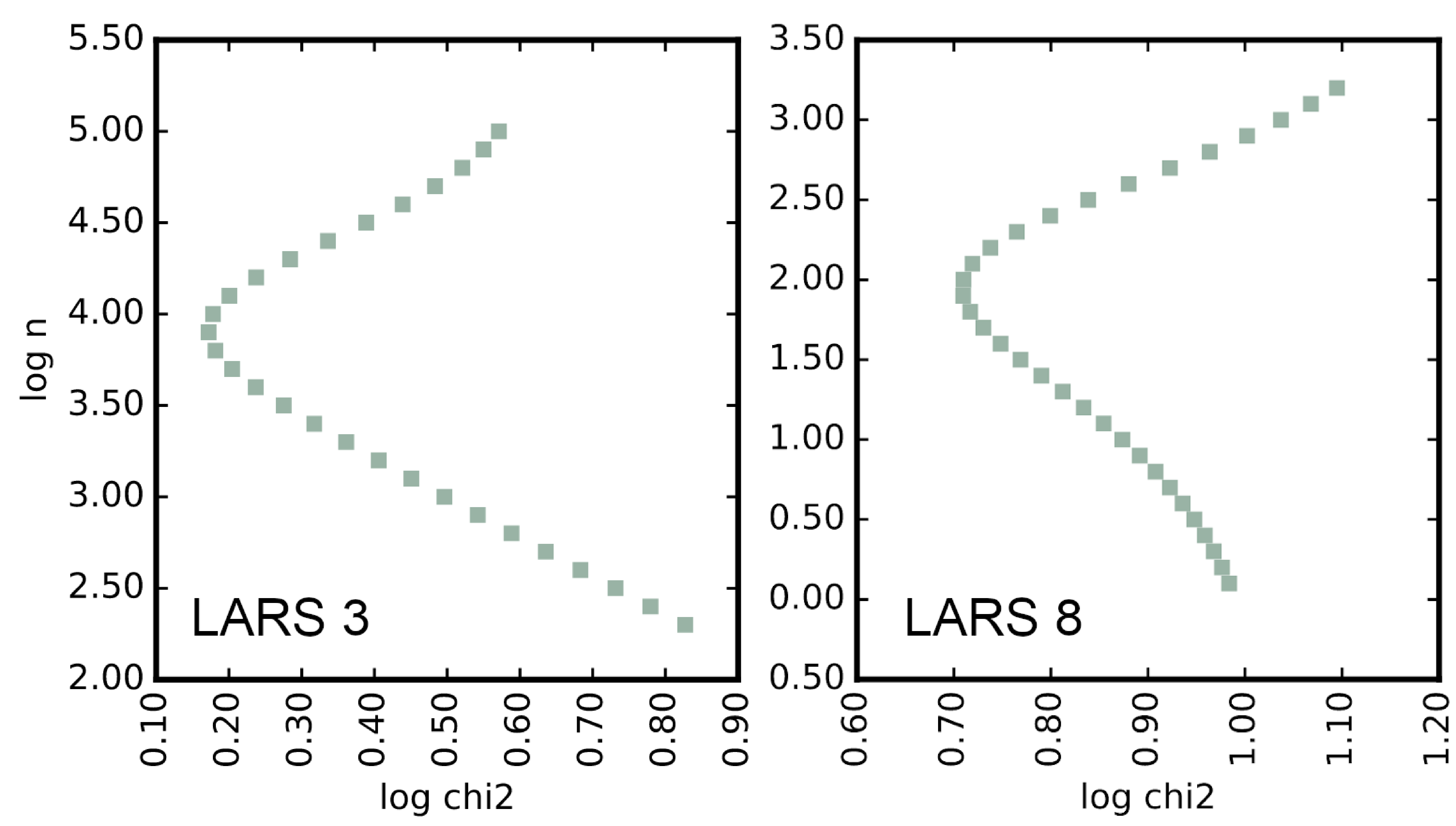}
     \caption{
     Mass-weighted mean molecular gas density vs. $\chi^2$ for LARS 3 and 8.
     The (mass-weighted) mean density (log n) of the emitting gas could be derived using a novel radiative transfer approach.
     The models (Puschnig et al. in prep) were minimized against observed line luminosities of CO (1-0), CO (2-1), HCN (1-0) and HCO+ (1-0) for LARS 3
     and additional CO (3-2) for LARS 8.
     }
     \label{fig:lars_molecular_radtrans}
\end{figure}

\section{Choice of GDR and $\upalpha_{CO}$}\label{sec:discuss_gdr_alpha}
Given the fact that our galaxy sample is very heterogeneous in terms of metallicity,
stellar mass and star formation activity,
we adopt a metallicity-dependent gas-to-dust ratio and CO-to-H$_2$ conversion factor
when calculating total and molecular gas masses respectively.
For the galaxies with atomic gas mass estimates from H~I 21 cm observations,
we compare available recipies for GDR and $\upalpha_{CO}$. The results are
shown in Table \ref{tab:gdr_recipies}. 
We find that the combination of the prescriptions in \cite{Narayanan2012} (N12) for $\upalpha_{CO}$ and the linear GDR
scalings in \cite{RemyRuyer2014} give the best fit for our sample, with a deviation of only
$\sim0.1$dex
for the total gas mass (see Figure \ref{fig:gas_obs_vs_est}). 
The sublinear GDR scaling of \cite{Leroy2011} (L11) for example underestimates our observed molecular masses. The $\upalpha_{CO}$ prescriptions
of \cite{Schruba2012} (S12) and \cite{Accurso2017} (A17) lead to deviations of $\sim0.3$dex and $\sim0.2$dex compared to our
dust-based estimates. It is important to note that the prescriptions for $\upalpha_{CO}$ of N12, S12 and A17 are fundamentally
different and rely on different assumptions. S12 assume a constant star formation efficiency or gas depletion time
to derive their relation for the metallicity-dependence of $\upalpha_{CO}$. A17 use combined CO (1-0) and [C~II]158$\upmu$m observations
and relate those to $\upalpha_{CO}$. In their prescription, $\upalpha_{CO}$ depends not only on metallicity, but also on the offset of the
galaxy from the main sequence. The metallicity basically describes the total dust content available to shield CO from UV
radiation and the offset from the main sequence describes the strength of this radiation field. On the other hand, N12
define $\upalpha_{CO}$ to depend on metallicity and CO surface brightness I$_{CO}$ (i.e. velocity integrated brightness temperature).
While the metallicity dependence in N12 has
basically the same meaning as in S12 or A17, i.e. to account for CO-dark gas at low dust contents, the dependence on
I$_{CO}$ makes $\upalpha_{CO}$ sensitive to environmental variations, i.e. variations in density and temperature of
the molecular gas. For example, it is known that ULIRGs (mergers) have lower $\upalpha_{CO}$ values,
because in nuclear starbursts the molecular gas is denser and hotter with molecular gas being mostly non-virialized
compared to normal star forming regions.
This leads to brighter CO emission (and thus lower $\upalpha_{CO}$) due to higher excitation temperatures.
The fact that 1) the peak brightness temperature of the emission is sensitive to excitation temperature, and
2) the line width increases in the presence of non-virialized, dense, hot gas, makes I$_{CO}$ sensitive
to changing environmental conditions.
Hence, the reason why the $\upalpha_{CO}$ prescription of N12 best fits our observations is an indication for
changing environmental conditions among LARS galaxies that have strong implications for $\upalpha_{CO}$.

\begin{table*}[htb]
\caption{Mean deviation in \textit{dex} of dust-based estimates of the molecular gas mass $\Delta M_{H_2}$,
i.e. dust-based total gas minus M$_{HI}$ from \cite{Pardy2014}, see Table \ref{tab:lars_general},
and mean deviation in \textit{dex} of dust-based estimates of the total gas mass $\Delta M_{gas}$,
i.e. dust-based total gas minus M$_{HI}$+M$_{H_2}$. Note that deviations could only be calculated for
galaxies with both, $M_{HI}$ and $M_{H_2}$ measured (LARS 3, 8, 9 and 11), and are biased towards higher metallicities.
We evaluate the deviations for a combination of prescriptions for the gas-to-dust ratio (R14: \cite{RemyRuyer2014}, L11: \cite{Leroy2011},
and for the CO-to-H$_2$ conversion factor (N12: \cite{Narayanan2012}, S12: \cite{Schruba2012}, A17: \cite{Accurso2017}).
}
\label{tab:gdr_recipies}
\begin{tabular}{lllllll}
\hline \hline
          & N12              & N12              & S12              & S12              & A17              & A17              \\
          & $\Delta M_{H_2}$ & $\Delta M_{g}$ & $\Delta M_{H_2}$ & $\Delta M_{g}$ & $\Delta M_{H_2}$ & $\Delta M_{g}$ \\
R14,ref   & 0.23             & 0.11             & 0.48             & 0.32             & 0.34             & 0.19             \\
R14,PL    & 0.61             & 0.36             & 0.31             & 0.22             & 0.31             & 0.28             \\
R14,BPL   & \textbf{0.23}    & \textbf{0.11}    & 0.48             & 0.32             & 0.34             & 0.19             \\
L11       & neg              & 0.14             & neg              & 0.52             & neg              & 0.14            
\end{tabular}
\end{table*}


\end{document}